\documentclass[final,3p,times,pdflatex]{elsarticle}
\usepackage{amsmath}
\usepackage{amssymb}
\usepackage{graphicx}
\usepackage{color} 
\usepackage{slashed}
\usepackage{pifont}
\usepackage{caption}
\usepackage{morefloats}
\usepackage{longtable}
\usepackage{float}
\usepackage{array}
\usepackage{subfig}
\definecolor{Red}{cmyk}{0,1,1,0}
\definecolor{Blue}{cmyk}{1,0.75,0,0}

\newcommand\beq{\begin{eqnarray}}
\newcommand\eeq{\end{eqnarray}}

\def\SOFTSUSY{{\tt SOFTSUSY}}
\def\code#1{{\tt #1}}

\journal{Computer Physics Communications}

\begin{document}

\begin{frontmatter}

\begin{flushright}
DAMTP-2017-13\\
\end{flushright}

\title{The Calculation of Sparticle and Higgs Decays in the Minimal and Next-to-Minimal Supersymmetric Standard Models: {\tt SOFTSUSY4.0}}

\author[damtp]{B.C.~Allanach}
\cortext[cor1]{Corresponding author}
\author[damtp]{T.~Cridge}
\ead{t.cridge@damtp.cam.ac.uk}
\address[damtp]{DAMTP, CMS, University of Cambridge, Wilberforce road,
  Cambridge, CB3  0WA, United Kingdom}
\begin{abstract}
We describe a major extension of the {\tt SOFTSUSY} spectrum calculator to
include 
the calculation of the decays, branching ratios and lifetimes of sparticles
into lighter 
sparticles, covering  the next-to-minimal supersymmetric standard model
(NMSSM) as 
well as the minimal supersymmetric standard model (MSSM).
This document
acts as a manual for the
new version of {\tt SOFTSUSY}, which includes the calculation of sparticle
decays. We present a comprehensive collection of explicit expressions used by
the program for the various partial widths of the different decay modes in the
appendix.  
\end{abstract}

\end{frontmatter}

\section{Program Summary}
\noindent{\em Program title:} \SOFTSUSY{} \\
{\em Program obtainable   from:} {\tt http://softsusy.hepforge.org/} \\
{\em Distribution format:}\/ tar.gz \\
{\em Programming language:} {\tt C++}, {\tt fortran} \\
{\em Computer:}\/ Personal computer. \\
{\em Operating system:}\/ Tested on Linux 4.4.0-36-generic, Linux 3.13.0-93-generic
\\
{\em Word size:}\/ 64 bits. \\
{\em External routines:}\/ None \\
{\em Typical running time:}\/ 0.1-1 seconds per parameter point. \\
{\em Nature of problem:}\/ Calculating supersymmetric particle partial decay
widths in the 
MSSM or the NMSSM\@, given the parameters and spectrum which has already been
calculated by \SOFTSUSY{}. \\
{\em Solution method:}\/ Analytic expressions for tree-level 2 body decays and loop-level decays and
one-dimensional numerical integration for 3 body decays.\\
{\em Restrictions:}\/ Decays are calculated in the real $R-$parity conserving
MSSM  or the real $R-$parity conserving
NMSSM only. No additional charge-parity violation (CPV) relative to the
Standard Model (SM). Sfermion mixing has only been accounted for in the third
generation of sfermions in the decay calculation. Decays in the MSSM are
2-body and 3-body, whereas decays in the NMSSM are 2-body only. \\
{\em CPC Classification:}\/ 11.1 and 11.6. \\
{\em Does the new version supersede the previous version?:}\/ Yes. \\
{\em Reasons for the new version:}\/ Significantly extended functionality. The
decay rates and branching ratios of sparticles are particularly useful for
collider searches. Decays calculated in the NMSSM will be a particularly
useful check of the other programs in the literature, of which there are few.\\
{\em Summary of revisions:}\/
Addition of the calculation of sparticle and Higgs decays. 
All 2-body and important 3-body tree-level
decays, including phenomenologically important loop-level decays (notably,
Higgs decays to $gg$, $\gamma \gamma$ and $Z \gamma$). Next-to-leading order corrections
are added to neutral Higgs  decays to $q \bar q$ for quarks $q$ of any flavour and to the neutral Higgs
decays to $gg$.

\section{Introduction}
The phenomenology of simple supersymmetric extensions of the Standard Model
has become something of an industry in particle physics of late. The potential
of such models to explain the technical hierarchy problem (i.e.\ the relative
smallness of the Higgs mass as compared to the Planck or gauge unification
scale) has motivated its study. The models have potential implications for
cosmology as well as for collider physics. Several different computational
tools are necessary for the study of the models'
phenomenology~\cite{Allanach:2008zn}. The initial step is a calculation of the
supersymmetric spectrum, as well as particle couplings. This is the step 
that the computer program \SOFTSUSY~has previously performed, in the
$R-$parity conserving MSSM~\cite{Allanach:2001kg}, the $R-$parity conserving
NMSSM~\cite{Allanach:2013kza} and the $R-$parity violating
MSSM~\cite{Allanach:2009bv}. By adding one-loop corrections to neutrino
masses, the program 
has been extended~\cite{Allanach:2011de} to calculate neutrino masses and
mixings in the presence of lepton number violating $R-$parity
violation. Further non-trivial 
additions of higher order corrections result in increased precision for the
gauge 
and Yukawa couplings~\cite{Allanach:2014nba} and the squark and gluino pole
masses~\cite{Allanach:2016rxd}. Other publicly available computer programs
exist which perform the task of computing the supersymmetric spectrum: for the
MSSM, there is \code{FLEXIBLESUSY}~\cite{Athron:2014yba}, 
\code{ISASUSY}~\cite{Paige:2003mg}, \code{SUSEFLAV},
\code{SUSPECT}~\cite{Djouadi:2002ze}
and \code{sPHENO}~\cite{Porod:2003um}, whilst \code{FEYNHIGGS} can calculate the
higgs masses \cite{Heinemeyer:1998yj}. In the NMSSM, there is only one
stand-alone dedicated tool for spectrum calculation:
\code{NMSSMTools}~\cite{Ellwanger:2004xm,Ellwanger:2006rn}, whereas the
\code{SARAH}~\cite{Staub:2013tta} framework can be combined with
\code{FLEXIBLESUSY} or \code{sPHENO} in order to calculate the spectrum. Meanwhile,
\code{NMSSMCALC} \cite{Baglio:2013iia} can be used for the computation of the Higgs masses
and decays in the NMSSM\@.
The plethora of computer programs is useful: some of them use different
approximations and extend to different models. Even the programs having the
same apparent approximations differ in their numerical output because the
higher order corrections not included are implicitly different: both for the
sparticle spectrum~\cite{Allanach:2003jw} and the Higgs
masses~\cite{Allanach:2004rh,Staub:2015aea,deFlorian:2016spz}. Thus, the size of
differences between the programs for observables calculated at the same order
or approximation serve as an rough estimate of the size of higher order
corrections. In some cases, different approximations are used and these can
help investigate different r\'{e}gimes of parameter space. For example, one
can deal with sparticle threshold effects differently: either at fixed order
at say $M_Z$ using the MSSM or NMSSM as an effective field theory above $M_Z$
(this is the approach taken by \code{SOFTSUSY}, \code{sPHENO} and
\code{SUSPECT} for instance), or one could integrate the sparticles out in the
renormalisation group equations at some higher scale (the effective field theory approach taken by 
\code{ISAJET} and \code{NMSSMTools}). The former approach includes finite
terms of order $M_Z^2/(16 \pi^2 M_p^2)$, where $p$ is the mass of some
sparticle, but takes the sparticle mass splittings $\Delta M_p$ into account
at some level approximation, typically $\sim {\mathcal O}\left(\log [\Delta M_p^2] / (16 \pi^2)\right)$,
whereas the latter 
approach often misses the finite terms but re-sums the mass splitting terms. 
One generically expects the former approach to be more accurate when
sparticles are not too heavy, and the latter when the sparticles are very
heavy and when the splittings between them are large. Which one is more
accurate given current lower bounds upon sparticle masses is a quantitative
question that is observable dependent. 

In order to provide predictions for future sparticle and Higgs searches at
colliders, or indeed in order to interpret searches at them in terms 
of the MSSM and NMSSM, cross-section estimates as well as 
simulations of collisions are required, in order to
estimate acceptances and efficiencies. For the collision simulations,
estimates of the 
various decay partial widths for the sparticles and Higgs
particles are required. Some Monte-Carlo event generators perform this task in
the MSSM, for example \code{PYTHIA}~\cite{Sjostrand:2014zea} or
\code{HERWIG7}~\cite{Bahr:2008pv}, but 
there also exist dedicated tools like
\code{SUSYHIT}~\cite{Djouadi:2006bz} (a combination of \code{HDECAY}~\cite{Djouadi:1997yw} and
\code{SDECAY}~\cite{Djouadi:2005}) and
\code{FeynHiggs}~\cite{Heinemeyer:1998yj} (the latter specialising in the 
Higgs boson decays). \code{sPHENO}~also contains a decay
calculation for the MSSM\@. In the NMSSM however, the options for calculating
sparticle and 
Higgs decays are rather slim: \code{NMSSMTools}~is the only stand-alone
option, whereas \code{SARAH}\footnote{After submission of the present paper to
the electronic {\tt arXiv}, a new calculation in the \code{SARAH}~framework
was presented for the generic calculation of two-body partial decay widths at
the full one-loop level~\protect\cite{Goodsell:2017pdq}.}~can be combined with
\code{sPHENO}. For calculating Higgs decays in the NMSSM (including CP
violation) though, one can use \code{NMSSMCALC}~\cite{Baglio:2013iia}, which
calculates 
Higgs decay widths and branching ratios including some dominant QCD loop
corrections. 
Previous 
versions of \code{SOFTSUSY}~contained an interface to 
\code{NMSSMTools}~so that the NMSSM spectrum and couplings could then be fed
into the program in order to predict sparticle and Higgs boson partial decay
widths. 

The present paper describes a significant extension in functionality in
\code{SOFTSUSY}: to calculate and output the various partial widths for the
decays of sparticles and Higgs bosons in the MSSM and in the NMSSM\@. 
Emphasis has been placed on speed of execution, preferring to perform as much
of the calculation analytically as is practicable. 
We hope
that this addition of functionality to \code{SOFTSUSY}~will facilitate
collider studies of sparticle and supersymmetric 
Higgs searches: both through the study of differences with the other programs
as an estimate of the size of theoretical uncertainty in the prediction, and 
through a fast and unified computation. 

The rest of the paper is organised as follows: in Section~\ref{sec:conv}, we 
specify the conventions used in this paper and the assumptions made in the
decay calculator and we provide some details about the method of computation. Following this, in section~\ref{sec:list}, we list the decay modes included in the program. Next, in section~\ref{sec:comp}, 
we provide a few examples of comparison tests 
with a couple of other publicly available tools for the MSSM and the NMSSM,
before summarising in Section~\ref{sec:sum}. We show how to run the program and
provide explicit flags for controlling its behaviour in~\ref{sec:run}, providing some sample output in~\ref{sec:out}. 
We provide explicit formulae for
the partial widths in~\ref{appendix:exp} to~\ref{appendix:QCDcorrdec}; these have not been collected
together in one reference before. 

\section{Conventions, Assumptions and Method \label{sec:conv}}

Throughout $\tilde{Z}_i$ and $\tilde{W}_j$ are used for neutralinos
($i=1,2,3,4$ in the MSSM or $i=1,2,3,4,5$ in the NMSSM) and charginos
($j=1,2$), respectively. This  is different to the commonly used
$\tilde{\chi}_{i}^{0}$ and $\tilde{\chi}_{j}^+$ notation for ease of reading, particularly
when they appear in subscripts. The notation for the mass-ordered CP even and
CP odd neutral Higgs bosons is that $h_i \in \{h, H, H3\}$ for $i=1,2,3$ are the CP even neutral Higgs
bosons in order of increasing mass, whilst $A_i \in \{A, A2\}$ for $i=1,2$ are the CP odd neutral
Higgs bosons again in order of increasing mass, remembering that $H3$ and $A2$
occur only in the NMSSM. 

The partial width formulae for all of the decay modes included in the {\tt
  SOFTSUSY} decay calculation\footnote{The source code for the calculations
  is in the file
  \code{src/decays.cpp}, which is in the \code{C++} programming language.} are
listed in~\ref{appendix:exp} 
onwards, many of these were rederived and have been written in one consistent
set of conventions. 

\subsection{MSSM} \label{MSSMconventions}
While the conventions used in the decays code are largely those used in {\tt
  SOFTSUSY}~\cite{Allanach:2001kg}, there are differences in a few places in
order to allow easier comparison with partial width (PW) formulae provided
elsewhere. The few differences with respect to Ref.~\cite{Allanach:2001kg} are
as listed below\footnote{In the decay code itself, the neutralino mixing
  matrix used ($N$ in {\tt SOFTSUSY notation}) is transposed.}: 
\begin{itemize}
	\item[$\bullet$] In our calculations, it is convenient to work in a
          basis where the third generation sfermions are mass
          ordered with $m_{\tilde{f}_1} < m_{\tilde{f}_2}$. In order to ensure
          this, the mixing angle $\theta_f$ is transformed accordingly
          ($\theta_f \rightarrow \theta_f + \pi/2$) in the case where the {\tt
            SOFTSUSY} spectrum generator has $m_{\tilde{f}_1} >
          m_{\tilde{f}_2}$. 
	\item[$\bullet$] The mixing angles for the charginos are transformed
          with respect to the {\tt SOFTSUSY} spectrum generator in order to
          match conventions used elsewhere
          (e.g. \cite{Djouadi:2002ze}). Therefore $\theta_{L/R}$ as indicated
          below is given by $\theta_{L/R}^{decays} = -\theta_{L/R}^{spectrum}
          + \pi/2$. 
\end{itemize}

\subsection{NMSSM} \label{NMSSMconventions}
The conventions used in the decay code are predominantly those described previously in the {\tt SOFTSUSY} NMSSM manual \cite{Allanach:2013kza}, but there are differences in a few places. As well as those listed above, there are a few changes specific to the NMSSM, to allow straightforward comparison with {\tt NMSSMTools}~\cite{Ellwanger:2004xm,Ellwanger:2006rn,Ellwanger:2012dd,Ellwanger:2006ch}:
\begin{itemize}
	\item[$\bullet$] The Charge Parity (CP) even neutral Higgs mixing matrix is altered relative to the matrix $R$ provided by {\tt SOFTSUSY}~\cite {Allanach:2013kza}. The matrix $S$ used in the decay formulae is obtained via an orthogonal transformation exchanging eigenstates:  
\begin{equation}
S = R \left( \begin{array}{ccc}
	0 & 1 & 0 \\
	1 & 0 & 0 \\
	0 & 0 & 1 \end{array} \right)
  =	\left( \begin{array}{ccc}
	R(1,2) & R(1,1) & R(1,3) \\
	R(2,2) & R(2,1) & R(2,3) \\
	R(3,2) & R(3,1) & R(3,3) \end{array} \right)
\end{equation}
i.e.\ the first two columns are interchanged 
	\item[$\bullet$] The CP odd neutral Higgs mixing matrix is altered
          relative to the matrix provided by {\tt SOFTSUSY}
          \cite{Allanach:2013kza}, the matrix $P$ detailed in the decay
          formulae (different to the $P$ in Ref.~\cite{Allanach:2013kza} which
          we write here as $P^{prov}$) is given below. The differences are
          that the           first 
          row of $P^{prov}$ is 
          dropped (as this refers to the goldstone boson) and the first and
          second columns are interchanged. The ratio of the Higgs vacuum
          expectation values (vevs) $\beta$ and the mixing angle $\theta_A$
          are as used elsewhere in {\tt
            SOFTSUSY}~\cite{Allanach:2001kg,Allanach:2013kza}: 
\begin{equation}
P = \left( \begin{array}{ccc}
	P^{prov}(2,2) & P^{prov}(2,1) & P^{prov}(2,3) \\
	P^{prov}(3,2) & P^{prov}(3,1) & P^{prov}(3,3) \\
	0 & 0 & 0 \end{array} \right)
	= \left( \begin{array}{ccc}
	\cos\beta\cos\theta_A & \sin\beta\cos\theta_A & \sin\theta_A \\
	\cos\beta\sin\theta_A & \sin\beta\sin\theta_A & -\cos\theta_A \\
	0 & 0 & 0 \end{array} \right)
\end{equation}

\end{itemize}

\subsection{Mass Choices and Scales Used} \label{masschoicesandscales}
The input parameters from the {\tt SOFTSUSY} spectrum
generator~\cite{Allanach:2001kg,Allanach:2013kza} can be evaluated at a
variety of different scales. The choice of the renormalisation scale used is
an important consideration and can result in differences between decay
calculator predictions. Different scales effectively correspond to including
different higher order terms in the calculation. For example, consider the
decay of a gluino  into a top and a stop. One must choose a renormalisation
scale for the coupling. The masses of the particles involved could be running
masses evaluated at different scales, or pole masses. Each choice affects the
numerical value of the PW, but are all equivalent at tree-level. In {\tt SOFTSUSY} the following choices are made: 
\begin{itemize}
\item In general, unless explicitly stated otherwise, the masses of the
  supersymmetric (SUSY) and Higgs particles and other parameters, such as
  mixing angles and gauge couplings, are evaluated at the scale $M_{SUSY} =
  x\sqrt{m_{\tilde{t}_1}(M_{SUSY})m_{\tilde{t}_2}(M_{SUSY})}$, where x can be
  set by the user but by default is taken to be $1$. Here,
  $m_{\tilde{t}_i}(M_{SUSY})$ is the running $i^{th}$ stop mass evaluated at
  a modified dimensional reduction~\cite{Jones:1994} ($\overline{DR}$) renormalisation scale
  $M_{SUSY}$.  
\item For Higgs loop decays the gauge coupling strengths $\alpha_s$ and $\alpha$ are evaluated at the mass of the decaying Higgs.
\item For Higgs loop decays to $\gamma\gamma$ or $Z\gamma$ the masses of the
 important  quarks (i.e. $m_{t}$, $m_{b}$,  $m_{c}$) are evaluated at
 the mass of the decaying Higgs. Below $M_Z$, these are run in 3 loop
 QCD and 1 loop in QED\@. 
In the calculation of decays of $H$, $H3$, $A$ and $A2$ quark masses  are run
to $m_H$, $m_{H3}$, $m_{A}$, $m_{A2}$ 
in the (N)MSSM as appropriate, whilst for the lightest CP even Higgs $h$ the
$m_{t}$, $m_{b}$ and $m_{c}$ are run to $m_h$ in 1 loop QED and 3 loop QCD. 
\item Throughout the program, unless otherwise stated here, we generally use two
  different quark masses; ``kinematic masses'' for the kinematics (i.e.\ for
  masses of particles in the initial or final states) and ``running masses''
  for the evaluation of couplings.  This hopefully allows a large part of some
  higher order  corrections 
  to be incorporated into the quark legs via the mass
  running. The way in which these masses are evaluated is listed in Table~\ref{massestable} below. 
\item In addition to the above quark masses, there are extra masses {\tt
    mcpole} and {\tt mspole} defined in {\tt decays.h} which are set
  to particular values used in the QCD corrections to neutral
  Higgs boson decays to $q \bar{q}$ or $gg$. 
\item If the QCD corrections to these decays are turned off in the program then
  the running masses for the quarks are used in order to attempt to hopefully
  incorporate some of the Next-to-Leading-Order (NLO) corrections to the quark
  legs. 
\end{itemize}

\begin{center}
\begin{table}
\centering
\begin{tabular}{|c|c|c|c|} \hline
\multicolumn{2}{|c|}{kinematic masses} & \multicolumn{2}{c|}{running (coupling) masses} \\ \hline
 {\tt mtPole} & pole mass from propagator & {\tt runmt} & $\overline{DR}$ mass at $M_Z$ \\ \hline
 {\tt mbPole} & pole mass from propagator & {\tt runmb} & $\overline{DR}$ mass at $M_Z$ \\ \hline 
 {\tt mtauPole} & pole mass from propagator & {\tt runmtau} & $\overline{DR}$ mass at $M_Z$\\ \hline
 {\tt mc} & $\overline{MS}$ mass at $M_Z$ & {\tt runmc} & Yukawa-extracted mass at $M_Z$ \\ \hline
 {\tt ms} & $\overline{MS}$ mass at $M_Z$ & {\tt runms} & Yukawa-extracted mass at $M_Z$ \\ \hline
 {\tt mup} & $\overline{MS}$ mass at $M_Z$ & {\tt runmu} & Yukawa-extracted mass at $M_Z$ \\ \hline
 {\tt mdo} & $\overline{MS}$ mass at $M_Z$ & {\tt runmd} & Yukawa-extracted mass at $M_Z$ \\ \hline
 {\tt mel} & $\overline{MS}$ mass at $M_Z$ & {\tt runmel} & Yukawa-extracted mass at $M_Z$ \\ \hline
 {\tt mmu} & $\overline{MS}$ mass at $M_Z$ & {\tt runmmu} & Yukawa-extracted mass at $M_Z$ \\ \hline
 {\tt polemw} & pole mass from propagator & {\tt runmw} & running $W$ mass at $M_{SUSY}$ \\ \hline
 {\tt polemz} & pole mass from propagator & {\tt runmz} & running $Z$ mass at $M_{SUSY}$ \\ \hline
\end{tabular}
\caption{The two different types of masses used for the fermions and gauge
  bosons. The {\tt names} given are those used in the code ({\tt
    src/decays.cpp}). ``kinematic'' masses are used for the masses of initial
  and final state particles in the decay formulae whilst ``running
  (coupling)'' masses are used where masses appear in expressions involving
  couplings in the partial 
  width formulae. Note that within {\tt SOFTSUSY}, the $\overline{MS}$ masses
  include 
  only SM corrections whilst the Yukawa-extracted masses ($\overline{DR}$
  masses) include SM and SUSY corrections.} 
\label{massestable}
\end{table}
\end{center}
As detailed in Table~\ref{massestable}, for the third generation sfermions the ``kinematic'' masses are pole
masses obtained from the propagators whilst
the ``running (coupling)'' masses are in the $\overline{DR}$ scheme. For the
``kinematic'' masses of the first two generation fermions, the $\overline{MS}$
mass at $M_Z$ is used, whilst the ``running'' masses are extracted from the
running Yukawa couplings. For the electron and muon, the running is again
small (only QED). The ``kinematic'' masses for the vector bosons are pole
masses, whilst for the ``running(coupling)'' masses they are running
$\overline{DR}$ masses evaluated at $M_{SUSY}$. 

\begin{itemize}
\item $\overline{MS}$ masses include only SM corrections within {\tt SOFTSUSY}; 3-loop QCD and 1-loop QED corrections are included.
\item Yukawa-extracted masses are in the $\overline{DR}$ scheme and include SM and SUSY corrections.
\item Quark input masses can be reset by the user within the {\tt SMINPUTS} block of the SLHA/SLHA2 input file. The kinematic and running masses used will then change accordingly.
\end{itemize}

The different choices of scales for the input parameters is one of the key
sources of differences between different decay calculator programmes. It is
worth noting that an experimental value of Fermi's constant, $G_{F}$, is also
used, this is inconsistent with the tree-level expression ${G_{F} \over
  \sqrt{2}} = {g^2 \over 8m_{W}^2}$ as it is an empirical quantity and so
incorporates higher order terms. 

\subsection{Assumptions Made}

The following assumptions are made in the decay calculator:
\begin{itemize}
\item $R$-parity conservation in the MSSM and in the NMSSM.
\item No additional CP violation relative to the SM.
\item No additional flavour violation relative to the SM.
\item Sfermion mixing has only been accounted for in the third generation of
  sfermions as it is proportional to the Yukawa couplings, which are negligible for the
  first two generations.  
\end{itemize}

\subsection{Method}

In our description in this section and the rest of the paper, we classify decay modes 
according to both the number of daughter particles (with $N$ body meaning $N$ decay products) and 
the order of the corrections included; i.e.\ tree-level, 1-loop or 2-loop.

For 2 body tree-level decay modes, the analytical expressions for the partial widths are explicitly used in order to
provide fast evaluation. Similarly, for the 2 body 1-loop decays the loop integrals were
performed analytically and the resulting formulae used.
For 3 body decay modes (all tree-level), the phase space integral
has been  analytically reduced to a one-dimensional integral, which is then
performed using adaptive Gaussian numerical integration~\cite{numRec}. The 
exception to this is the neutral CP even lightest Higgs decays to a vector
boson 
and an off-shell vector boson, which then decays into a fermion anti-fermion
pair, this calculation was performed explicitly and all integrals were
evaluated analytically. 

The tree-level 3 body decay modes were therefore where most complications arose. 
In general for an $N$-body tree-level decay there are $N$ integrals to perform,
one over the three-momenta of each of the final state particles. One of
these integrals is always trivial to perform using the momentum-conserving
delta function. For the 2 body tree-level decay widths this leaves one
remaining  integral with the energy delta function, this can then be performed
using 
the standard result for the integral of a function of a variable multiplied by
a delta function containing another function of the same variable. For
tree-level 3 body decay
widths however, one has two remaining integrals to perform and in general they are 
non-trivial to determine analytically. In certain cases the symmetry of
the integrands, along with certain assumptions, may allow them to be performed.
For example, in the cases of the SM-like neutral CP even higgs decays to a vector
boson and fermion anti-fermion pair via an off-shell vector boson,
$h \rightarrow WW^* \rightarrow W f' \bar{f}$ and 
$h \rightarrow ZZ^* \rightarrow Z f \bar{f}$, the
mass of the Higgs boson ensures that the outgoing fermions may not be top 
quarks. Therefore one can neglect the masses of the outgoing fermions and
greatly simplify the calculation. Passarino-Veltman reduction \cite{Passarino:1979} can then 
allow reduction of the integrals to a one-dimensional integral, which in 
this case may be determined explicitly analytically; the result given
in both~\ref{Higgs_Sector} and in the ``Higgs Hunter's Guide'' \cite{HHG}.
For a general 3 body decay mode the calculations are however considerably
more involved. There are two approaches that can be taken once the first
trivial integral using the momentum-conserving delta function is performed;
at this stage the partial width can be written as a double differential decay
rate in two (reduced) Mandelstam variables as is the case in {\tt SUSYHIT-1.4},
following the work performed in reference \cite{Mambrini:2001}, these
two dimensional integrals can then be performed numerically. Alternatively, often
one of the two integrals (remaining after the first trivial integral is 
performed) may be evaluated analytically, leaving a single one-dimensional
integral to be performed numerically. This is the approach used in the work
in reference \cite{Baer:1998} and is the method adopted in {\tt sPHENO} 
\cite{Porod:2003um, Porod:2011}, from which the expressions we use
for the 3 body decays originate. The Feynman diagrams involved, effects
included and any assumptions made for each of the 3 body decays are 
given in detail in~\ref{appendix:MSSM3body} with the corresponding partial
width formulae.

The situation is of course more complicated for loop decays than at tree-level,
each loop provides an additional loop integral to be performed, in the case
of the 1-loop decays included in {\tt SOFTSUSY},  they were performed
explicitly with the help of Passarino-Veltman reduction \cite{Passarino:1979}.

\section{Decay Modes Included \label{sec:list}}
The following section provides a list of all the decay modes included in the
decay part of the {\tt SOFTSUSY} package; they are split into MSSM SUSY
tree-level 2 body decays, MSSM Next-to-Lightest Supersymmetric Particle (NLSP) decays
to the gravitino LSP (Lightest SUSY Particle), MSSM Higgs tree-level 2 body decays, MSSM Higgs
1-loop 2 body decays, MSSM tree-level 3 body decays,  NMSSM SUSY and Higgs tree-level 2 body decays, NMSSM
1-loop 2 body decays and decays for which QCD corrections have been
included. A comprehensive list of the formulae for all of the decays included
is given explicitly in~\ref{appendix:exp} to~\ref{appendix:QCDcorrdec} for
ease of reference. To summarise, we include:
\begin{itemize}
\item All MSSM 2 body decays at (at least) tree-level, both sparticle and
  Higgs boson decays. 
\item Next-to-Lightest SUSY Particle (NLSP) 2 body decays to gravitinos in the
  MSSM at tree level.
\item The phenomenologically most relevant three-body decays of gluinos, charginos and neutralinos.
\item Higgs decays to $\gamma\gamma$ and $Z\gamma$ at leading order (i.e.\ one-loop) in the MSSM and NMSSM.
\item QCD corrections to neutral Higgs decays to quarks (1-loop) and to gluons (2-loop) in the MSSM and NMSSM.
\item All NMSSM 2 body decays at tree-level, including both the extended neutralino and extended Higgs sectors.
\end{itemize}

Whilst the majority of the decay modes are therefore calculated at tree-level,
some effects of higher order corrections are approximated via the use of
running masses and couplings, as calculated using the {\tt SOFTSUSY} spectrum
generator \cite{Allanach:2001kg}, the details of the mass choices were given
in Section~\ref{masschoicesandscales}. In~\ref{appendix:exp} there are a
series of tables indicating all the modes included, along with appendix references for
their partial width formulae as used in {\tt SOFTSUSY}.  

The branching ratios for each mode are grouped into decay tables for each
parent SUSY or Higgs particle and are all printed to standard
output in the SLHA/SLHA2 convention~\cite{Skands:2003cj,Allanach:2008qq} in
order to allow it to be passed straightforwardly to other programs
(such as {\tt PYTHIA}~\cite{Sjostrand:2014zea},
\code{HERWIG7}~\cite{Bahr:2008pv}, {\tt MadGraph}~\cite{Alwall:2014}, for
instance).

\subsection{MSSM SUSY Tree-Level Two Body Decays}
The detailed formulae for these modes are in~\ref{appendix:MSSM2body}.
The gluino $\tilde{g}$ decays included are:
\begin{align*}
\tilde{g} &\rightarrow q \tilde{q}_{L/R}^* , \bar{q} \tilde{q}_{L/R}, t \tilde{t}_{1/2}^*, \bar{t} \tilde{t}_{1/2},
b \tilde{b}_{1/2}^*, \bar{b} \tilde{b}_{1/2},
\end{align*}
where $1/2$ means 1 {\em or}\/ 2, and $L/R$ means $L$ {\em or} $R$.
The sfermion $\tilde{f}$ decays included are, for the first two generations
where there is no sfermion mixing: \\ ($f^{'}$ indicates a fermion in the same
generation as the $f$ fermion but with opposite third component of weak
isospin, i.e. $f$ and $f^{'}$ could be $u$ and $d$ or $\nu_{e}$ and $e^{-}$): 
\begin{align*}
\tilde{q}_{L/R} &\rightarrow \tilde{g} q, \\
\tilde{f}_{L} &\rightarrow \tilde{W}_{j} f', \\
\tilde{f}_{L/R} &\rightarrow \tilde{Z}_{i} f.
\end{align*}
For the third generation sfermions:
\begin{align*}
\tilde{b}_{1/2} &\rightarrow \tilde{g} b, \tilde{W}_{j} t, \tilde{Z}_{i} b, \tilde{t}_{1/2} W^-, \tilde{t}_{1/2} H^-, \\
\tilde{t}_{1/2} &\rightarrow \tilde{g} t, \tilde{W}_{j} b, \tilde{Z}_{i} t, \tilde{b}_{1/2} W^+, \tilde{b}_{1/2} H^+,\\
\tilde{b}_{2} &\rightarrow \tilde{b}_{1} Z, \tilde{b}_{1} h/H/A, \\
\tilde{t}_{2} &\rightarrow \tilde{t}_{1} Z, \tilde{t}_{1} h/H/A, \\
\tilde{\tau}_{1/2} &\rightarrow \tilde{W}_{j} \nu_{\tau}, \tilde{Z}_{i} \tau, \tilde{\nu}_{\tau} W^-, \tilde{\nu}_{\tau} H^-, \\
\tilde{\nu}_{\tau} &\rightarrow \tilde{W}_{j} \tau, \tilde{Z}_{i} \nu_{\tau},
                     \tilde{\tau}_{1/2} W^+, \tilde{\tau}_{1/2} H^+, \\
\tilde{\tau}_{2} &\rightarrow \tilde{\tau}_{1} Z, \tilde{\tau}_{1} h/H/A. 
\end{align*}
For charginos, the two-body decay modes included are:
\begin{align*}
\tilde{W}_{j} &\rightarrow \tilde{q}_L \bar{q'}, \tilde{q}_{1/2} \bar{q'}, \tilde{l}_L \bar{\nu}_{l}, \tilde{\nu}_{{l}_{L}} \bar{l}, \tilde{\tau}_{1/2} \bar{\nu}_{\tau}, \tilde{\nu}_{{\tau}_{L}} \bar{\tau},  \tilde{Z}_{i} W^+, \tilde{Z}_{i} H^+,
\\
\tilde{W}_{2} &\rightarrow \tilde{W}_{1} Z, \tilde{W}_{1} h/H/A.
\end{align*}
For neutralinos the two-body decay modes are ($k>i$ as the neutralinos
are mass ordered): 
\begin{align*}
\tilde{Z}_{i} &\rightarrow \tilde{f}_{L/R} \bar{f}, \tilde{f}_{1/2} \bar{f}, \tilde{W}_{j} W^+, \tilde{W}_{j} H^+,\\
\tilde{Z}_{k} &\rightarrow \tilde{Z}_{i} Z, \tilde{Z}_i h/H/A.
\end{align*}

\subsection{MSSM Decays to Gravitinos}
The following NLSP $\rightarrow$ $\tilde{G} + SM$ decays are included,
for when the gravitino $\tilde{G}$ is the lightest supersymmetric particle
(LSP):
\begin{align*}
\tilde{g} \rightarrow g \tilde{G}, \qquad
\tilde{q}_{i} \rightarrow q \tilde{G}, \qquad
\tilde{l} \rightarrow l \tilde{G}, \qquad
\tilde{Z}_{i} \rightarrow \gamma \tilde{G}, \qquad
\tilde{Z}_{i} \rightarrow Z \tilde{G}, \qquad
\tilde{Z}_{i} \rightarrow \phi \tilde{G},
\end{align*}
where $\phi$ denotes one of the neutral Higgs bosons $h$, $H$ or $A$.
The formulae for the partial widths are in~\ref{appendix:gravitinos}.

\subsection{MSSM Higgs Tree-Level Two Body Decays}
The tree-level 2 body decay modes included for the Higgs particles in the
MSSM are as follows (see later for the 1-loop Higgs decays included), the
formulae for the partial widths are explicitly given in~\ref{Higgs_Sector}. Note that for the decays to sfermions any combination of handedness
is allowed, $LL$, $LR$, $RL$, $RR$. Similarly, for the third generation where
there is squark mixing, and for decays to charginos, all combinations $11$,
$12$, $21$, $22$ are possible: 
\begin{align*}
h/H/A &\rightarrow \tilde{f}_{L/R} \tilde{f}_{L/R}^*, \tilde{f}_{1/2} \tilde{f}_{1/2}^*, \tilde{W}_{1/2} \tilde{W}_{1/2}, \tilde{Z}_{i} \tilde{Z}_{l}, l^+ l^-, \\
H^+ &\rightarrow \tilde{Z}_{i} \tilde{W}_{j}, q \bar{q'}, \nu_{l} \bar{l}, \tilde{f}_{L/R} \tilde{f}_{L/R}^*, \tilde{t}_{1/2} \tilde{b}_{1/2}^*, \nu_{\tau} \tilde{\tau}_{1/2}^*, h W^+,\\
h/H &\rightarrow A A, A Z, \\
H &\rightarrow H^+ H^-, hh, \\
A &\rightarrow h/H Z.
\end{align*}

The neutral Higgs decays to quarks are not included in this list as QCD corrections have been incorporated for these, see subsection~\ref{ssec:QCD_Corrected_Decays}.
For $H^+$, decays to CKM suppressed combinations of $q$ and $q'$ are also considered, for example $H^+ \rightarrow u \bar{s}$. Note also that the decays $H^+ \rightarrow H/A W^+$ are not included as they are kinematically forbidden in the MSSM (assuming tree-level mass formulae), these modes are included in the NMSSM.

\subsection{MSSM Higgs 1-loop Two body decays}
The key Higgs 1-loop decays are also included as these are very important 
channels for LHC Higgs discovery and measurement. The explicit expressions for
their partial widths are in~\ref{Higgs_Sector}:
\begin{align*}
h/H/A &\rightarrow \gamma\gamma, Z\gamma.
\end{align*}

The important loop decay to two gluons $gg$ incorporate QCD corrections and so are listed in  subsection~\ref{ssec:QCD_Corrected_Decays}.

\subsection{MSSM Tree-Level Three Body Decays}
The phenomenologically most important three-body decays in the MSSM are included, for the neutralino decays to another neutralino and a fermion anti-fermion pair then $i>j$ as the neutralinos are mass-ordered.:
\begin{align*}
h &\rightarrow V f \bar{f}. \\
\tilde{g} &\rightarrow \tilde{Z}_i q \bar{q}, \tilde{W}_i q \bar{q'}. \\
\tilde{Z}_{i} &\rightarrow \tilde{Z}_j f \bar{f}, \tilde{W}_j f \bar{f'}. \\
\tilde{W}_{j} &\rightarrow \tilde{Z}_i f \bar{f'}.
\end{align*}

As of yet, there are no three-body decays of sfermions included; this may be
resolved in future versions. The explicit formulae used, for which {\tt sPHENO}
\cite{Porod:2003um} provided a useful reference, are given in~\ref{appendix:MSSM3body}. In our decay calculator the 3 body decay modes
are only calculated where no tree level 2 body modes are available for the particular
decaying SUSY or Higgs particle to the specific daughter(s) considered. Often
3 body modes will also not be output, even if they have been calculated,
as in scenarios where other 2 body decay modes are kinematically available
to the decaying parent particle these 2 body modes will dominate the total
decay width. In such cases the 3 body modes often have branching ratios
smaller than {\tt BRTol}, which is the minimum BR output in the decay tables and may be
set by the user in the input file. 

\subsection{NMSSM SUSY and Higgs Tree-Level Two Body decays}
In the NMSSM, decays not involving the extended Higgs or neutralino sectors
are the same as in the MSSM\@. For the extended neutralino and Higgs sectors the
allowed decays are largely as before with the exception that now the
neutralino index $i$ runs from 1 to 5, whilst there is an additional CP even neutral
Higgs and an additional CP odd neutral Higgs. These additional states mix with the 
MSSM states with the same quantum numbers thereby resulting in 5 neutralinos, 3 CP
even neutral Higgs bosons and 2 CP odd neutral Higgs Bosons, all of which we label in order
of increasing mass. Therefore there are now additional mass eigenstates $H3$, $A2$ and $\tilde{Z}_5$.
We use the notation that we label the CP even neutral Higgs states as $h_i \in \{h,H,H3\}$ for $i=1,2,3$ and the CP odd neutral Higgs states as $A_i \in \{A,A2\}$ for $i=1,2$.
These states are of course mixtures of the original MSSM states and the new NMSSM states, therefore the most "NMSSM-type" state need not necessarily be the heaviest.
The $\tilde{Z}_{1,2,3,4}$, $h$ and $H$ (which we now use to label the lightest two CP even neutral Higgs
bosons) and $A$ have the same available modes as listed before; therefore we now list the decay
modes of the additional states. As a guide, the same decays which can occur for the heaviest
of the two CP even Higgs bosons of the MSSM, the $H$, may now also occur for the $H3$; similarly
we can extend the decays of the $A$ to the $A2$, and of the $\tilde{Z}_{1,2,3,4}$ of the MSSM
to the $\tilde{Z}_5$. Additional decay modes in the NMSSM (other than simply extending decays into
neutralinos to include $\tilde Z_5$ or decays into $H3, A2$) are the decays of
the $\tilde{Z}_5$, $H3$ and $A2$. Here, $n=1,2,3,4$ since the $\tilde{Z}_5$
decays into lighter neutralinos: 
\begin{align*}
\tilde{Z}_5 &\rightarrow W \tilde{W}_{1/2}, Z \tilde{Z}_n, H^{\pm}
              \tilde{W}_{1/2}, \tilde{Z}_n h/H/H3/A/A2,  \tilde{f}_{L/R}
              \bar{f}, \tilde{f}_{1/2} \bar{f} , \\
H3 &\rightarrow \tilde{f}_{L/R} \tilde{f}_{L/R}^*, \tilde{f}_{1/2}
     \tilde{f}_{1/2}^*, \tilde{W}_{1/2} \tilde{W}_{1/2},  \tilde{Z}_i
     \tilde{Z}_l, l^+ l^-, AA, A A2, \\
H3 &\rightarrow A2 A2, Z A/A2, H^+ H^-, hh, hH, HH, W^- H^+, VV , \\
A2 &\rightarrow \tilde{f}_{L} \tilde{f}_{R}^*, \tilde{f}_{1} \tilde{f}_{2}^*, \tilde{W}_{1/2} \tilde{W}_{1/2},  \tilde{Z}_i \tilde{Z}_l, l^+ l^-, Z h/H/H3, A h/H/H3, W^- H^+ ,
\end{align*}
where $VV \in \{W^+W^-,\ ZZ \}$.
As before, for the decays to two sfermions, any combination of handedness is
permitted $LL$, $LR$, $RL$, $RR$; similarly for the decays to mixed sfermions
or to charginos $11$, $12$, $21$, $22$ are all allowed. For the decays to
quarks only $c$, $s$, $t$, $b$ are considered. For the $A2$ there are fewer
decays than the $H3$ as many decays are ruled out by CP conservation. Decays 
of the $H3$ or $A2$ to $q \bar{q}$ or $gg$ are listed in subsection~\ref{ssec:QCD_Corrected_Decays}
as QCD corrections are included in these channels.

The explicit partial width expressions used within the decay calculator {\tt
  SOFTSUSY} are given in~\ref{appendix:NMSSMdec}, the expressions in {\tt
  NMSSMTools} \cite{Ellwanger:2004xm,Ellwanger:2012dd,Ellwanger:2006ch} were
used, with appropriate changes. 

\subsection{NMSSM 1-loop Two Body Decays}
Just as in the MSSM, in the NMSSM the phenomenologically important 1-loop decays of Higgs bosons are included:
\begin{align*}
h/H/H3/A/A2 &\rightarrow \gamma\gamma, Z\gamma.
\end{align*}
Again the decay to two gluons is listed later as it includes QCD corrections, see subsection~\ref{ssec:QCD_Corrected_Decays}.
See~\ref{appendix:NMSSMdec} for the detailed formulae used within the code for
the partial widths of these modes. 

\subsection{QCD Corrected Decays} \label{ssec:QCD_Corrected_Decays}
NLO QCD corrections have been incorporated for the decays in which such effects
are most important in both the MSSM and NMSSM, these are the neutral Higgs decays to quarks and decays to
gluons:
\begin{align*}
h/H/H3/A/A2 \rightarrow q \bar{q},  gg
\end{align*}
The expressions used are
given in~\ref{appendix:QCDcorrdec} and are based on those provided in the
calculations in \cite{Spira:2016,Djouadi:1996}. Note that the quarks which are considered for neutral Higgs
decays are only $c$, $s$, $b$ for the lightest CP even neutral higgs $h$, whilst $t$ is also included for the heavier CP
even neutral Higgs Boson(s) and for the CP odd neutral Higgs Boson(s) of the (N)MSSM\@. 
Decays to $u$ and $d$ are negligible.

\section{Validation and Comparison with other programs \label{sec:comp}}

We implemented a number of different approaches to validate the program and the formulae used.
\begin{enumerate}
\item Most of the formulae used in the \SOFTSUSY~program were rederived in
  order to determine any issues, the exception being the 3 body decays for
  which there are very complicated expressions for the integrals to be
  performed, for these the formulae are as in {\tt sPHENO-3.3.8}
  \cite{Porod:2003um}. Meanwhile, NMSSM decay expressions were generalised from
  MSSM expressions and checked against {\tt NMSSMTools}. 
\item MSSM 2 body partial widths were also checked against both Baer and
  Tata ``Weak Scale Supersymmetry'' \cite{TataBaer} and The Higgs Hunter's
  Guide \cite{HHG}. Some small differences were found, in particular relative
  to Baer and Tata, these differences were largely simple typographical errors
  although there were several other differences in the 3 body decay
  formulae, where additional contributions are included. The full formulae for
  the partial widths used in {\tt SOFTSUSY} are given in~\ref{appendix:exp}
  to~\ref{appendix:QCDcorrdec}.  
\item The partial width of every channel was compared against corresponding
  channels in other programs where possible; for example the 2 body MSSM
  decays were compared against results from {\tt SUSYHIT}, the 3 body MSSM
  decays against {\tt sPHENO} and the 2 body NMSSM decays against {\tt
    NMSSMTools}. This ensured that the formulae were correct and any
  differences were down to different input parameters resulting from different
  schemes of running, different mass choices and other similar effects. These checks were performed at several parameter space points for each partial width.  
\item Once all the decay modes of a given initial state were implemented the
  branching ratios were compared against those of a relevant program, i.e.\
  {\tt SUSYHIT} for the MSSM 2 body decays, {\tt sPHENO} for MSSM 3 body
  decays, and {\tt NMSSMTools} (as linked with the {\tt SOFTSUSY} spectrum
  generator) for NMSSM decays. 
\item Many of the modes have been analysed in detail by scanning over the mass
  of the decaying particle and comparing the branching ratio structure with
  known results and with corresponding results in other decay calculator
  programs. 
\end{enumerate}

As described above we therefore performed specific and fairly extensive tests for particular benchmark
points with \code{sPHENO} and \code{SUSYHIT}. Comparisons for some of these benchmark points are provided here for a selection of decaying SUSY and Higgs particles, in addition scans over the mass of the decaying particle are given for the decays of the lightest SM-like Higgs and for the decays of a gluino $\tilde{g}$. This allowed both a qualitative check of the behaviour of the decays in the program and a quantitative comparison of the level of agreement with other programs. In particular the level of agreement with the same input parameters and with our set of input parameters is detailed in some specific cases. This method uses the results the \SOFTSUSY~program produces to validate it.

\subsection{Supersymmetric Decays - $\tilde{t}_1$}

First of all consider the decays of the lightest stop, $\tilde{t}_1$, the
verbatim output of the {\tt SOFTSUSY} decay calculator is given in
Table~\ref{st1verbatim} and was generated with the {\tt lesHouchesInput}
file. The comparison of the results for this benchmark point between the new
{\tt SOFTSUSY} decay calculator and those of {\tt SUSYHIT-1.4} is given in
Table~\ref{stop1table}.

\begin{table}
\begin{verbatim}
#     PDG         Width             
DECAY 1000006     5.72622764e+00   # Stop1 decays
#     BR                  NDA   PDG1   PDG2       Comments                PW                
      3.20176449e-01      2     5      1000024    # ~t_1 -> b ~chi_1+     1.83340323e+00    
      2.21769170e-01      2     5      1000037    # ~t_1 -> b ~chi_2+     1.26990075e+00    
      2.27726880e-01      2     6      1000022    # ~t_1 -> t ~chi_10     1.30401596e+00    
      1.25396681e-01      2     6      1000023    # ~t_1 -> t ~chi_20     7.18049943e-01    
      1.04930819e-01      2     6      1000025    # ~t_1 -> t ~chi_30     6.00857758e-01    
\end{verbatim}
\caption{The decays of a $\tilde{t}_1$ at the parameter point given by the {\tt
    lesHouchesInput} file provided with {\tt SOFTSUSY}. For reference this has
  $m_{\tilde{t}_1} = 808.7$ GeV, $m_{\tilde{W}_1} = 385.0$ GeV,
  $m_{\tilde{W}_2} = 637.5$ GeV, $m_{\tilde{Z}_1} = 204.0$ GeV,
  $m_{\tilde{Z}_2} = 385.0$ GeV, $m_{\tilde{Z}_3} = -622.7$ GeV,
  $m_{\tilde{Z}_4} = 637.2$ GeV. The partial widths are given in GeV units.} \label{st1verbatim}
\end{table}

In this the input values used for the various masses are: top pole mass \code{mtPole}=
174.3 GeV, bottom pole mass \code{mbPole}=4.985 GeV, running top mass 
\code{runmt}=145.555 GeV and running bottom mass \code{runmb}=2.576 GeV. These
differ from the default values used for these quantities in {\tt SUSYHIT},
Table~\ref{stop1table} 
illustrates the differences observed between {\tt SOFTSUSY} and {\tt
  SUSYHIT-1.4}, as well as the differences when {\tt SOFTSUSY} has the {\tt
  SUSYHIT} mass inputs inserted by hand. This demonstrates that the level of
agreement between the programs is around $10\%$, dropping down to $1\%$ when
the same input masses and coupling constants are used in both programs. These
differences result from the different mass and scheme choices,
as outlined in Section~\ref{masschoicesandscales}. 

\begin{center}
\begin{table}
\centering
\begin{tabular}{|c|c|c|c|c|c|c|} \hline
\multicolumn{2}{|c|}{{\tt SOFTSUSY} standard inputs} & \multicolumn{2}{c|}{{\tt SOFTSUSY} {\tt SUSYHIT}'s inputs} & \multicolumn{2}{c|}{{\tt SUSYHIT}} & mode \\ \hline
PW/GeV & BR & PW/GeV & BR & PW/GeV & BR &  \\ \hline
1.8334e+00 & 3.2018e-01 & 1.7080e+00 & 3.2111e-01 & 1.7080e+00 & 3.2181e-01 & $\tilde{t}_1 \rightarrow b \tilde{W}_1$ \\ \hline
1.2699e+00 & 2.2177e-01 & 1.1027e+00 & 2.0730e-01 & 1.1027e+00 & 2.0775e-01 & $\tilde{t}_1 \rightarrow b \tilde{W}_2$ \\ \hline
1.3040e+00 & 2.2773e-01 & 1.2986e+00 & 2.4414e-01 & 1.2992e+00 & 2.4478e-01 & $\tilde{t}_1 \rightarrow t \tilde{Z}_1$ \\ \hline
7.1805e-01 & 1.2540e-01 & 6.8479e-01 & 1.2874e-01 & 6.7286e-01 & 1.2677e-01 & $\tilde{t}_1 \rightarrow t \tilde{Z}_2$ \\ \hline
6.0086e-01 & 1.0493e-01 & 5.2503e-01 & 9.8706e-02 & 5.2485e-01 & 9.8887e-02 & $\tilde{t}_1 \rightarrow t \tilde{Z}_3$ \\ \hline
\end{tabular}
\caption{The $\tilde{t}_1$ decay partial widths and branching ratios as output
  by {\tt SOFTSUSY} with our mass choices (and corresponding Yukawa couplings) and 
  with the masses and Yukawa couplings in {\tt SUSYHIT}, compared with the results of
  {\tt SUSYHIT-1.4}. This illustrates the differences of order $10\%$ or more
  that may arise depending upon mass (``kinematic" and ``running") choices,
  the differences here reduce to order $1\%$ once the same masses are taken. 
  The {\tt lesHouchesInput} file provided with {\tt SOFTSUSY} gives the parameter
  point here, it has a common scalar mass $m_0 = 125$ GeV, a common gaugino
  mass $m_{1/2} = 500$ GeV, ratio of Higgs vacuum expectation values 
  $\tan\beta = 10$, sign of the superpotential $\mu$ parameter $sign(\mu) = +1$
  and common soft SUSY breaking trilinear parameter $A_0 = 0$ in the
  constrained MSSM (CMSSM).  
Here, as elsewhere in this paper,
  we present numerical results in mantissa-exponent notation (i.e.\ e-0a=$\ldots \times 10^{-a}$
  and e+0b=$\ldots \times 10^b$ for $a,b \in \{0,1,2,3, \ldots\}$).} 
\label{stop1table}
\end{table}
\end{center}

\subsection{Higgs Decays - $h$}

Now, in Table~\ref{h0SMliketable}, we can perform similar comparisons between
{\tt SOFTSUSY} and {\tt HDECAY-3.4} of {\tt SUSYHIT-1.4} for Higgs
decays. Here we have taken a SM-like Higgs, in the decoupling limit so all the
SUSY decays are kinematically forbidden, given by a point in the pMSSM parameter space
which has Higgs mass $125$ GeV, this point we call {\tt pmssm1} and the SLHA \cite{Skands:2003cj} form of the input file is given verbatim in Table~\ref{PMSSM1Table}. The results of our decay calculator without QCD
corrections included, with QCD corrections included, and with the same input
quark and gauge boson masses and same input gauge couplings as {\tt SUSYHIT},
again with QCD corrections, are compared with {\tt HDECAY-3.4}. Note that the 
comparisons are done against the non-current version {\tt HDECAY-3.4} as this is the 
version included in the {\tt SUSYHIT-1.4} package. This allowed straightforward comparisons to be done between the new decay
calculator and {\tt SUSYHIT'S} version of \code{HDECAY} as one can input the spectrum as calculated by {\tt SOFTSUSY} straight into {\tt SUSYHIT}. This allowed the effects of the spectrum generator to be isolated as much as possible from the decay calculator which is being tested.

\begin{table}
\begin{small}
\begin{verbatim}
# pMSSM1 input
Block MODSEL                 # Select model
    1    0                   # non universal
    1    1                   # sugra input
Block SMINPUTS               # Standard Model inputs
    1   1.279340000e+02	     # alpha^(-1) SM MSbar(MZ)
    2   1.166370000e-05	     # G_Fermi
    3   1.172000000e-01	     # alpha_s(MZ) SM MSbar
    4   9.118760000e+01	     # MZ(pole)
    5   4.250000000e+00	     # mb(mb) SM MSbar
    6   1.733000000e+02	     # mtop(pole)
    7   1.777000000e+00	     # mtau(pole)
Block MINPAR		     # Input parameters
    1   1.000000000e+03      # m0
    2   3.000000000e+02      # m12
    3   3.000000000e+01	     # tanb
Block SOFTSUSY               # Optional SOFTSUSY-specific parameters
    0   1.000000000e+00      # Calculate decays in output (only for RPC (N)MSSM)
    1   1.000000000e-03      # Numerical precision: suggested range 10^(-3...-6)
    2   0.000000000e+00	     # Quark mixing parameter: see manual
    5   1.000000000e+00      # Include 2-loop scalar mass squared/trilinear RGEs
   24   1.000000000e-09      # If decay BR is below this number, don't output
   25   1.000000000e+00	     # If set to 0, don't calculate 3-body decays (1=default)
   26   1.000000000e+00      # Output PWs
Block EXTPAR          # non-universal SUSY breaking parameters
      0   -1.000000000000000e+00         # Set MX=MSUSY 
      3    1.000000000000000e+03         # M_3(MX)
     11   -7.700000000000000e+03         # At(MX)
     12    1.000000000000000e+03         # Ab(MX)
     13   -3.000000000000000e+03         # Atau(MX)
     23    3.000000000000000e+02         # mu(MX)
     26    3.000000000000000e+03         # mA(pole)
     33    3.000000000000000e+03         # mtauL(MX)
     36    3.000000000000000e+03         # mtauR(MX)
     43    3.500000000000000e+03         # mqL3(MX)
     46    3.800000000000000e+03         # mtR(MX)
\end{verbatim}
\end{small}
\caption{The pMSSM parameter space point used for the $h$ decay comparisons in Table~\ref{h0SMliketable} and Figure~\ref{h0SMBRsbarplot}, given in SLHA form \cite{Skands:2003cj}.} \label{PMSSM1Table}
\end{table} 

\newcolumntype{C}[1]{>{\centering\let\newline\\\arraybackslash\hspace{0pt}}m{#1}}

\begin{center}
\begin{table} 
\centering
\begin{tabular}{|c|c|c|c|c|c|c|c|c|} \hline
\multicolumn{2}{|C{3.0cm}|}{{\tt SOFTSUSY} no QCD corrections} & \multicolumn{2}{C{3.0cm}|}{{\tt SOFTSUSY} with QCD corrections} & \multicolumn{2}{C{3.0cm}|}{SOFTSUSY with {\tt SUSYHIT}'s masses and QCD corrections} & \multicolumn{2}{C{3.0cm}|}{{\tt HDECAY-3.4} with same QCD corrections} & mode \\ \hline
PW/GeV & BR & PW/GeV & BR & PW/GeV & BR & PW/GeV & BR & \\ \hline
1.04e-04 & 3.30e-02 & 2.25e-04 & 4.03e-02 & 2.25e-04 & 4.31e-02 & 2.25e-04 & 4.24e-02 & $h \rightarrow c c$ \\ \hline
8.00e-07 & 2.55e-04 & 1.62e-06 & 2.91e-04 & 1.62e-06 & 3.11e-04 & 1.63e-06 & 3.06e-04 & $h \rightarrow s s$ \\ \hline
1.75e-03 & 5.56e-01 & 3.96e-03 & 7.10e-01 & 3.60e-03 & 6.90e-01 & 3.61e-03 & 6.80e-01 & $h \rightarrow b b$ \\ \hline
8.52e-07 & 2.71e-04 & 8.52e-07 & 1.53e-04 & 9.17e-07 & 1.76e-04 & 9.19e-07 & 1.73e-04 & $h \rightarrow \mu \mu$ \\ \hline
2.61e-04 & 8.30e-02 & 2.61e-04 & 4.67e-02 & 2.59e-04 & 4.97e-02 & 2.60e-04 & 4.90e-02 & $h \rightarrow \tau \tau$ \\ \hline
1.06e-05 & 3.36e-03 & 1.06e-05 & 1.89e-03 & 9.24e-06 & 1.77e-03 & 9.24e-06 & 1.74e-03 & $h \rightarrow \gamma \gamma$ \\ \hline
1.65e-04 & 5.27e-02 & 2.71e-04 & 4.86e-02 & 2.72e-04 & 5.22e-02 & 2.72e-04 & 5.13e-02 & $h \rightarrow g g$ \\ \hline
6.74e-06 & 2.15e-03 & 6.74e-06 & 1.21e-03 & 5.88e-06 & 1.13e-03 & 6.11e-06 & 1.15e-03 & $h \rightarrow Z \gamma$ \\ \hline
7.61e-04 & 2.42e-01 & 7.61e-04 & 1.36e-01 & 7.61e-04 & 1.46e-01 & 8.22e-04 & 1.55e-01 & $h \rightarrow W W^*$ \\ \hline
8.44e-05 & 2.69e-02 & 8.44e-05 & 1.51e-02 & 8.44e-05 & 1.62e-02 & 1.02e-04 & 1.92e-02 & $h \rightarrow Z Z^*$ \\ \hline
\end{tabular}
\caption{The $h$ decay partial widths and branching ratios as output by {\tt
    SOFTSUSY} without QCD corrections, with QCD corrections, with {\tt
    SUSYHIT}'s quark and gauge boson masses and gauge couplings and with QCD
  corrections, and the results of {\tt HDECAY-3.4} from {\tt
    SUSYHIT-1.4}. This illustrates the necessity of including QCD corrections
  for decays to quarks and decays to gluons, as well as the fact that
  differences in mass choices are the primary source of differences between
  {\tt SOFTSUSY} and {\tt HDECAY-3.4}. This is for a pMSSM point given by {\tt pmssm1}
  listed in SLHA~\cite{Skands:2003cj} format in Table~\ref{PMSSM1Table}; it
  has $m_h = 125$GeV. Note that the masses
  and gauge couplings are taken from {\tt SUSYHIT} and inserted into the {\tt
    SOFTSUSY}\/ decay calculator in columns 5 and 6 are $\alpha_s = 0.11$ and
  $m_c = 
  1.40$GeV,   $m_s = 0.19$GeV, $m_b = 4.77$GeV, $m_t = 173.30$GeV for
  the $h \rightarrow qq$ 
  and $h \rightarrow gg$ decays; $m_{\mu} = 0.11$GeV, $m_{\tau} = 1.78$GeV for
  $h \rightarrow l^+ l^-$ decays; $\alpha(M_Z) = 7.29\times10^{-3}$ and $m_W =
  80.35$GeV, $m_t = 188.72$GeV, $m_b = 3.47$GeV, $m_c = 0.74$GeV and $m_{\tau}
  = 1.78$GeV for $h \rightarrow \gamma \gamma$; $m_Z = 91.19$GeV, $m_t =
  173.30$GeV, $m_b = 4.77$GeV, $\alpha = 7.29\times10^{-3}$ and $m_W =
  80.36$GeV for $h \rightarrow Z \gamma$ and 
 for $h  \rightarrow VV^*$.}  
\label{h0SMliketable}
\end{table}
\end{center}


\begin{figure}
\centerline{\includegraphics[scale=1]{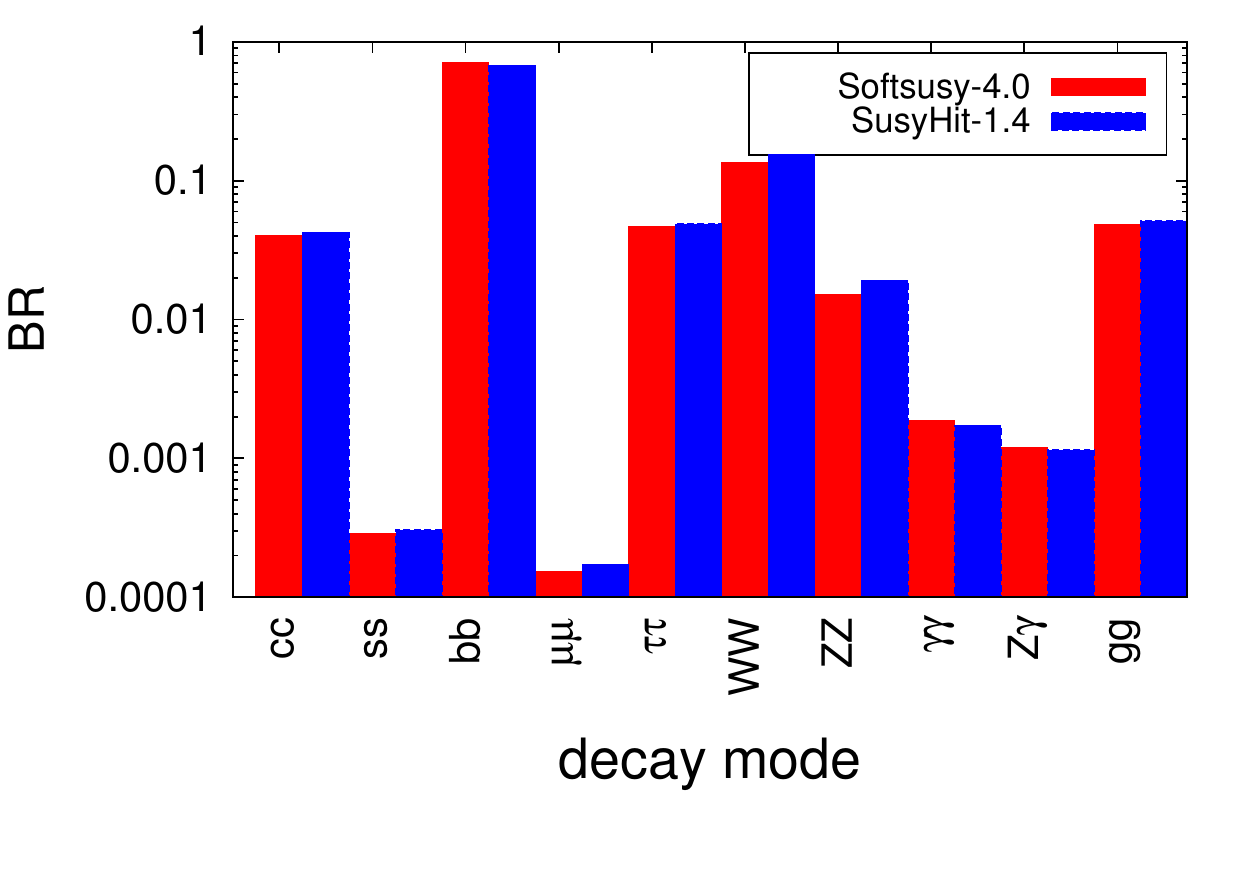}} 
\caption{Branching ratios for a SM-like Higgs predicted by {\tt SOFTSUSY} and
  by {\tt HDECAY-3.4} in {\tt SUSYHIT-1.4} for $m_h=125$
  GeV. This is for a pMSSM point given by {\tt pmssm1}, see Table~\ref{PMSSM1Table}.} \label{h0SMBRsbarplot} 
\end{figure}

In Table~\ref{h0SMliketable} the comparison of the partial widths with QCD corrections
switched on and switched off clearly demonstrates the significant difference
these corrections make to neutral Higgs decays to quarks and to gluons, as is
widely known in the literature
\cite{Djouadi:1996,Djouadi:1995,Spira:1998,Djouadi:2008}. Furthermore, it is
clear that the main source of differences in partial widths between the decay
calculator of {\tt SOFTSUSY-4.0} and {\tt HDECAY} is in the choice of masses
used. Remaining differences tend to be small and are due largely to
differences in other inputs, the exception being the decays to two vector
bosons, where order $10\%$ differences are observed. This is due to {\tt
  HDECAY} incorporating additional effects such as the width of the resonance
and NLO corrections which are not included in {\tt SOFTSUSY}. It
should also be noted here that {\tt HDECAY} performs a numerical integration
whilst {\tt SOFTSUSY} has an explicit expression with no integration required
so the calculation methods are different. A comparison of the branching ratios
output for this SM-like Higgs are given in Figure~\ref{h0SMBRsbarplot}.

\begin{figure}
  \centering
  \subfloat[{\tt SOFTSUSY}]{\includegraphics[height = 10cm]{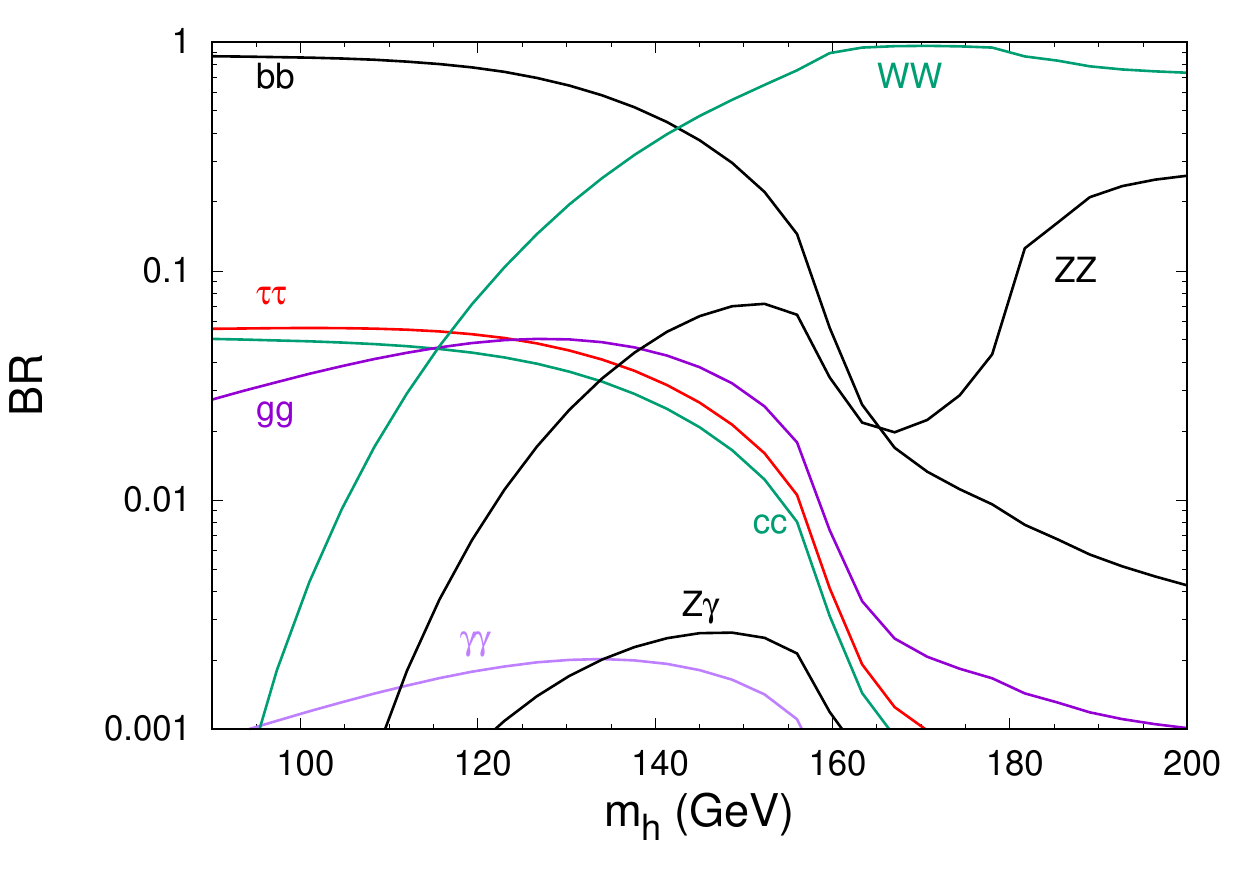}\label{fig:f1}}
  \hfill
  \subfloat[LHC Higgs Cross-Section Working Group]{\includegraphics[height = 10cm]{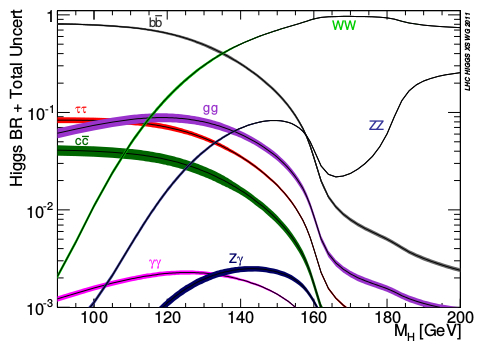}\label{fig:f2}}
  \caption{Branching ratios for a SM(-like) Higgs as calculated in (a) by {\tt SOFTSUSY} and in (b) by the LHC Higgs Cross-Section Working Group in \cite{Denner:2011}. This demonstrates a verification of the partial widths output by {\tt SOFTSUSY}.} \label{h0scancomp}
\end{figure}

In order to provide a qualitative demonstration that the decay calculator is
functioning correctly one may also scan the mass of the decaying particle and
investigate how the partial widths and branching ratios change. 
Figure~\ref{h0scancomp} shows how the branching ratios of a SM-like Higgs
change as its 
mass is scanned from the $Z^0$ boson mass up to $200$ GeV as calculated in
$(a)$ {\tt 
  SOFTSUSY} and in $(b)$ a well-known plot produced by the LHC Higgs Cross
Section Working Group \cite{Denner:2011} in 2011. 
This shows good level of agreement, with small differences due to effects
detailed previously in the quantitative comparison at $m_h = 125$ GeV. 

So far we have demonstrated the validation of SUSY 2 body and Higgs MSSM
decays, including the loop-decays, QCD corrections and Higgs 3 body
decays. Similar validation and comparison was also performed for the MSSM
3 body decays and the NMSSM decays.

\subsection{Three-Body Decays - $\tilde{g}$} \label{glu3bodysec}

\begin{figure}
\centerline{\includegraphics[scale=1.2]{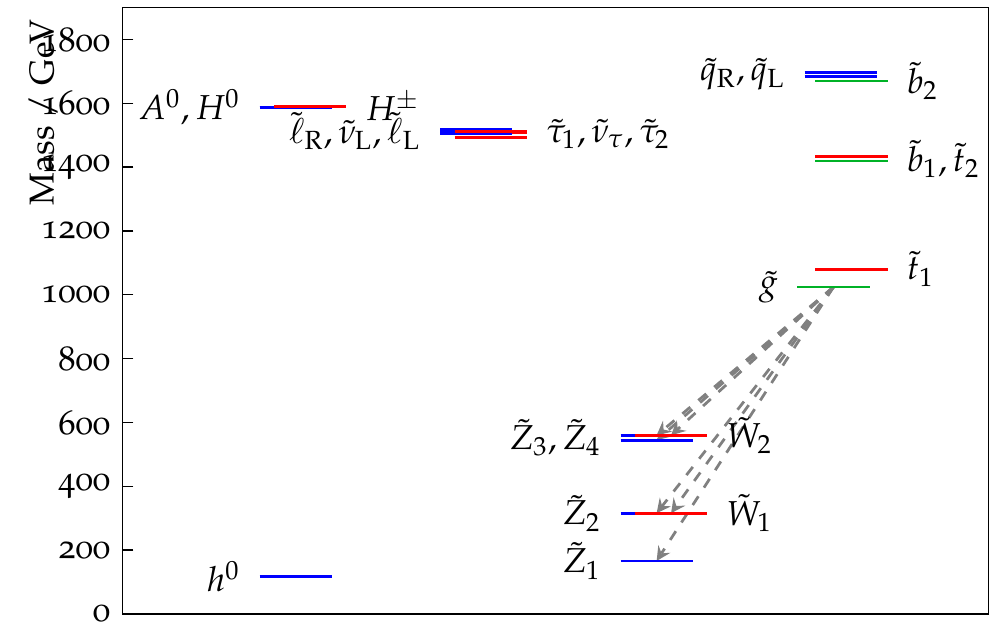}} 
\caption{Spectrum used for quantitative comparison of gluino $\tilde{g}$ three
  body decays. Here the arrows indicate only the 3 body decay modes of the
  gluino: these are those investigated. This CMSSM spectrum has $m_0 = 1500$
  GeV, $m_{1/2} 
  = 400$ GeV, $\tan\beta = 10.37$, sign$(\mu) = +1$, $A_0 = -80$ GeV and was
  generated 
  in {\tt sPHENO}. 
The figure was produced using a modified version {\tt slhaplot-3.0.4} from
  {\tt PySLHA} \cite{Buckley:2013}.} \label{gluino3bodyspc} 
\end{figure}

First, consider the MSSM 3 body decays: an explicit comparison can be
performed for the gluino 3 body decays with the spectrum given in
Figure~\ref{gluino3bodyspc}, the gluino three-body decays to neutralinos and
quark anti-quark pairs are indicated. A comparison of the partial widths and
branching ratios given by {\tt SOFTSUSY}, {\tt sPHENO-3.3.8} and {\tt
  SUSYHIT-1.4} for 
this spectrum is presented in Table~\ref{gluino3bodycomp}. This was performed
taking the mass, coupling and other input decay parameters from {\tt sPHENO}
and inputting these directly by hand 
into the {\tt SOFTSUSY} decay calculator in order to evaluate only differences
due to 
the decay calculation, not any differences which might arise as a result of
differing parameters from the spectrum generators. The agreement
between the three programs is generally very good, in particular the agreement
between {\tt SOFTSUSY} and {\tt sPHENO-3.3.8}, upon which the calculations of
the 3 body decays is based, is usually between $1$ and $5\%$ with the
larger differences often occurring where there are larger differences between
{\tt SUSYHIT-1.4} and {\tt sPHENO-3.3.8}. The exceptions to this are the
decays to third generation quark-anti-quark pairs and the third and fourth
heaviest neutralinos; i.e. $\tilde{g} \rightarrow t \bar{t} \tilde{Z}_{3}$,
$\tilde{g} \rightarrow t \bar{t} \tilde{Z}_{4}$, $\tilde{g} \rightarrow b
\bar{b} \tilde{Z}_{3}$ and $\tilde{g} \rightarrow b \bar{b}
\tilde{Z}_{4}$. Here the differences observed are $10-20\%$ and they arise
because of differences in the Yukawa couplings taken, for example for the $b$
quark here the Yukawa coupling used in {\tt SOFTSUSY} is determined by a
running bottom mass of \code{runmb}=2.63 GeV, whereas {\tt sPHENO} has a
Yukawa coupling corresponding to a mass of \code{runmb}=2.37 GeV. In order to
show this results in the differences observed, the running $b$ mass in {\tt
  SOFTSUSY} was temporarily set to that of {\tt sPHENO} and the comparison for
$\tilde{g} \rightarrow b \bar{b} \tilde{Z}_i$ is provided in
Table~\ref{glutoneutbbnewyuk}. This demonstrates that the decays to
$\tilde{Z}_1$ and $\tilde{Z}_2$ are not significantly altered by the new
Yukawa coupling whereas the decays to $\tilde{Z}_3$ and $\tilde{Z}_4$
(i.e.\ those which showed differences with respect to {\tt sPHENO}) now have
significantly altered partial widths which are in much closer agreement with
{\tt sPHENO}, back down to the few percent level agreement seen in the other
3 body decays. 
\begin{center}
\begin{table}
\centering
\begin{tabular}{|c|c|c|c|c|c|c|} \hline
\multicolumn{2}{|C{3.0cm}|}{{\tt SOFTSUSY}} & \multicolumn{2}{C{3.0cm}|}{{\tt sPHENO-3.38}} & \multicolumn{2}{C{3.0cm}|}{{\tt SUSYHIT-1.4}} & mode \\ \hline
PW/GeV & BR & PW/GeV & BR & PW/GeV & BR & \\ \hline
2.90e-04 & 2.26e-02 & 2.89e-04 & 2.32e-02 & 2.89e-04 & 2.32e-02 & $\tilde{g} \rightarrow \tilde{Z}_1 u \bar{u}$ \\ \hline
3.21e-04 & 2.51e-02 & 3.19e-04 & 2.56e-02 & 3.19e-04 & 2.56e-02 & $\tilde{g} \rightarrow \tilde{Z}_2 u \bar{u}$ \\ \hline
1.35e-07 & 1.06e-05 & 1.35e-07 & 1.08e-05 & 1.35e-07 & 1.08e-05 & $\tilde{g} \rightarrow \tilde{Z}_3 u \bar{u}$ \\ \hline
5.52e-06 & 4.31e-04 & 5.49e-06 & 4.40e-04 & 5.49e-06 & 4.40e-04 & $\tilde{g} \rightarrow \tilde{Z}_4 u \bar{u}$ \\ \hline
9.06e-05 & 7.07e-03 & 9.02e-05 & 7.22e-03 & 9.02e-05 & 7.24e-03 & $\tilde{g} \rightarrow \tilde{Z}_1 d \bar{d}$ \\ \hline
3.07e-04 & 2.40e-02 & 3.06e-04 & 2.45e-02 & 3.06e-04 & 2.45e-02 &  $\tilde{g} \rightarrow \tilde{Z}_2 d \bar{d}$ \\ \hline
1.75e-07 & 1.36e-05 & 1.74e-07 & 1.39e-05 & 1.74e-07 & 1.40e-05 & $\tilde{g} \rightarrow \tilde{Z}_3 d \bar{d}$ \\ \hline
6.67e-06 & 5.21e-04 & 6.64e-06 & 5.31e-04 & 6.64e-06 & 5.33e-04 & $\tilde{g} \rightarrow \tilde{Z}_4 d \bar{d}$ \\ \hline
2.90e-04 & 2.26e-02 & 2.89e-04 & 2.32e-02 & 2.89e-04 & 2.32e-02 & $\tilde{g} \rightarrow \tilde{Z}_1 c \bar{c}$ \\ \hline
3.21e-04 & 2.51e-02 & 3.19e-04 & 2.56e-02 & 3.19e-04 & 2.56e-02 & $\tilde{g} \rightarrow \tilde{Z}_2 c \bar{c}$ \\ \hline
1.35e-07 & 1.05e-05 & 1.41e-07 & 1.13e-05 & 1.35e-07 & 1.08e-05 & $\tilde{g} \rightarrow \tilde{Z}_3 c \bar{c}$ \\ \hline
5.52e-06 & 4.31e-04 & 5.50e-06 & 4.40e-04 & 5.49e-06 & 4.40e-04  & $\tilde{g} \rightarrow \tilde{Z}_4 c \bar{c}$ \\ \hline
9.06e-05 & 7.07e-03 & 9.02e-05 & 7.22e-03 & 9.02e-05 & 7.24e-03  & $\tilde{g} \rightarrow \tilde{Z}_1 s \bar{s}$ \\ \hline
3.07e-04 & 2.40e-02 & 3.06e-04 & 2.45e-02 & 3.06e-04 & 2.45e-02 & $\tilde{g} \rightarrow \tilde{Z}_2 s \bar{s}$ \\ \hline
1.75e-07 & 1.36e-05 & 1.77e-07 & 1.42e-05 & 1.74e-07 & 1.40e-05 & $\tilde{g} \rightarrow \tilde{Z}_3 s \bar{s}$ \\ \hline
6.67e-06 & 5.21e-04 & 6.64e-06 & 5.32e-04 & 6.64e-06 & 5.33e-04  & $\tilde{g} \rightarrow \tilde{Z}_4 s \bar{s}$ \\ \hline
1.47e-03 & 1.15e-01 & 1.47e-03 & 1.17e-01 & 1.44e-03 & 1.15e-01 & $\tilde{g} \rightarrow \tilde{Z}_1 t \bar{t}$ \\ \hline
2.56e-04 & 1.99e-02 & 2.46e-04 & 1.97e-02 & 2.67e-04 & 2.15e-02 & $\tilde{g} \rightarrow \tilde{Z}_2 t \bar{t}$ \\ \hline
3.48e-04 & 2.71e-02 & 3.10e-04 & 2.48e-02 & 3.34e-04 & 2.68e-02 & $\tilde{g} \rightarrow \tilde{Z}_3 t \bar{t}$ \\ \hline
6.13e-04 & 4.79e-02 & 5.66e-04 & 4.53e-02 & 5.21e-04 & 4.18e-02 & $\tilde{g} \rightarrow \tilde{Z}_4 t \bar{t}$ \\ \hline
1.27e-04 & 9.93e-03 & 1.25e-04 & 1.00e-02 & 1.25e-04 & 1.00e-02 & $\tilde{g} \rightarrow \tilde{Z}_1 b \bar{b}$ \\ \hline
7.80e-04 & 6.09e-02 & 7.74e-04 & 6.20e-02 & 7.74e-04 & 6.21e-02 & $\tilde{g} \rightarrow \tilde{Z}_2 b \bar{b}$ \\ \hline
2.20e-05 & 1.72e-03 & 1.77e-05 & 1.42e-03 & 1.78e-05 & 1.43e-03 & $\tilde{g} \rightarrow \tilde{Z}_3 b \bar{b}$ \\ \hline
3.48e-05 & 2.72e-03 & 3.24e-05 & 2.60e-03 & 3.23e-05 & 2.60e-03 & $\tilde{g} \rightarrow \tilde{Z}_4 b \bar{b}$ \\ \hline
6.28e-04 & 4.90e-02 & 6.24e-04 & 5.00e-02 & 6.24e-04 & 5.01e-02 & $\tilde{g} \rightarrow \tilde{W}_1^- u \bar{d}$ \\ \hline
6.28e-04 & 4.90e-02 & 6.24e-04 & 5.00e-02 & 6.24e-04 & 5.01e-02 & $\tilde{g} \rightarrow \tilde{W}_1^+ d \bar{u}$ \\ \hline
6.28e-04 & 4.90e-02 & 6.24e-04 & 5.00e-02 & 6.24e-04 & 5.01e-02 & $\tilde{g} \rightarrow \tilde{W}_1^- c \bar{s}$ \\ \hline
6.28e-04 & 4.90e-02 & 6.24e-04 & 5.00e-02 & 6.24e-04 & 5.01e-02 & $\tilde{g} \rightarrow \tilde{W}_1^+ s \bar{c}$ \\ \hline
1.20e-05 & 9.36e-04 & 1.19e-05 & 9.56e-04 & 1.19e-05 & 9.58e-04 & $\tilde{g} \rightarrow \tilde{W}_2^- u \bar{d}$ \\ \hline
1.20e-05 & 9.36e-04 & 1.19e-05 & 9.56e-04 & 1.19e-05 & 9.58e-04 & $\tilde{g} \rightarrow \tilde{W}_2^+ d \bar{u}$ \\ \hline
1.20e-05 & 9.36e-04 & 1.19e-05 & 9.56e-04 & 1.19e-05 & 9.58e-04 & $\tilde{g} \rightarrow \tilde{W}_2^- c \bar{s}$ \\ \hline
1.20e-05 & 9.36e-04 & 1.19e-05 & 9.56e-04 & 1.19e-05 & 9.58e-04 & $\tilde{g} \rightarrow \tilde{W}_2^+ s \bar{c}$ \\ \hline
9.29e-04 & 7.25e-02 & 9.21e-04 & 7.38e-02 & 9.21e-04 & 7.39e-02 & $\tilde{g} \rightarrow \tilde{W}_1^- t \bar{b}$ \\ \hline
9.29e-04 & 7.25e-02 & 9.21e-04 & 7.38e-02 & 9.21e-04 & 7.39e-02 & $\tilde{g} \rightarrow \tilde{W}_1^+ b \bar{t}$ \\ \hline
1.35e-03 & 1.05e-01 & 1.27e-03 & 1.01e-01 & 1.27e-03 & 1.02e-01 & $\tilde{g} \rightarrow \tilde{W}_2^- t \bar{b}$ \\ \hline
1.35e-03 & 1.05e-01 & 1.27e-03 & 1.01e-01 & 1.27e-03 & 1.02e-01 & $\tilde{g} \rightarrow \tilde{W}_2^+ b \bar{t}$ \\ \hline
\end{tabular}
\caption{The $\tilde{g}$ decay partial widths and branching ratios as output
  by {\tt SOFTSUSY}, {\tt sPHENO-3.3.8}~\cite{Porod:2003um} and {\tt
    SUSYHIT-1.4}~\cite{Djouadi:2006bz} for the spectrum given in
  Figure~\ref{gluino3bodyspc}, for which the gluino only has 3 body decay
  modes kinematically available. 
This CMSSM spectrum has $m_0 = 1500$
  GeV, $m_{1/2} 
  = 400$ GeV, $\tan\beta = 10.37$, sign$(\mu) = +1$, $A_0 = -80$ GeV and was
  generated 
  in {\tt sPHENO}.
} 
\label{gluino3bodycomp}
\end{table}
\end{center}
\begin{center}
\begin{table}
\centering
\begin{tabular}{|c|c|c|c|c|} \hline
\multicolumn{2}{|C{3.0cm}|}{{\tt SOFTSUSY} with altered {\tt runmb}} & \multicolumn{2}{C{3.0cm}|}{{\tt sPHENO-3.38}} & mode \\ \hline
PW/GeV & BR & PW/GeV & BR & \\ \hline
1.27e-04 & 9.93e-03 & 1.25e-04 & 1.00e-02 & $\tilde{g} \rightarrow \tilde{Z}_1 b \bar{b}$ \\ \hline
7.78e-04 & 6.10e-02 & 7.74e-04 & 6.20e-02 & $\tilde{g} \rightarrow \tilde{Z}_2 b \bar{b}$ \\ \hline
1.81e-05 & 1.42e-03 & 1.77e-05 & 1.42e-03 & $\tilde{g} \rightarrow \tilde{Z}_3 b \bar{b}$ \\ \hline
3.18e-05 & 2.50e-03 & 3.24e-05 & 2.60e-03 & $\tilde{g} \rightarrow \tilde{Z}_4 b \bar{b}$ \\  \hline
\end{tabular}
\caption{The $\tilde{g}$ decay partial widths and branching ratios to
  $\tilde{Z}_i b \bar{b}$ as output by {\tt SOFTSUSY} with \code{runmb} taken
  so that the $b$ Yukawa coupling in {\tt SOFTSUSY} matches that in {\tt
    sPHENO}. These decays showed significant differences between the two
  programs, particularly for $\tilde{Z}_3$ and $\tilde{Z}_4$, see
  Table~\ref{gluino3bodycomp}. The agreement is now much improved,
  demonstrating that the differences result from a choice of the running $b$
  mass \code{runmb} and hence in the Yukawa coupling. This again used the
  parameters from {\tt sPHENO}'s spectrum file directly in the {\tt SOFTSUSY}
  decay calculator, with the addition relative to Table~\ref{gluino3bodycomp}
  of {\tt sPHENO}'s running $b$ mass.} 
\label{glutoneutbbnewyuk}
\end{table}
\end{center}

\begin{figure} 
\centerline{\includegraphics[width = 15.5cm]{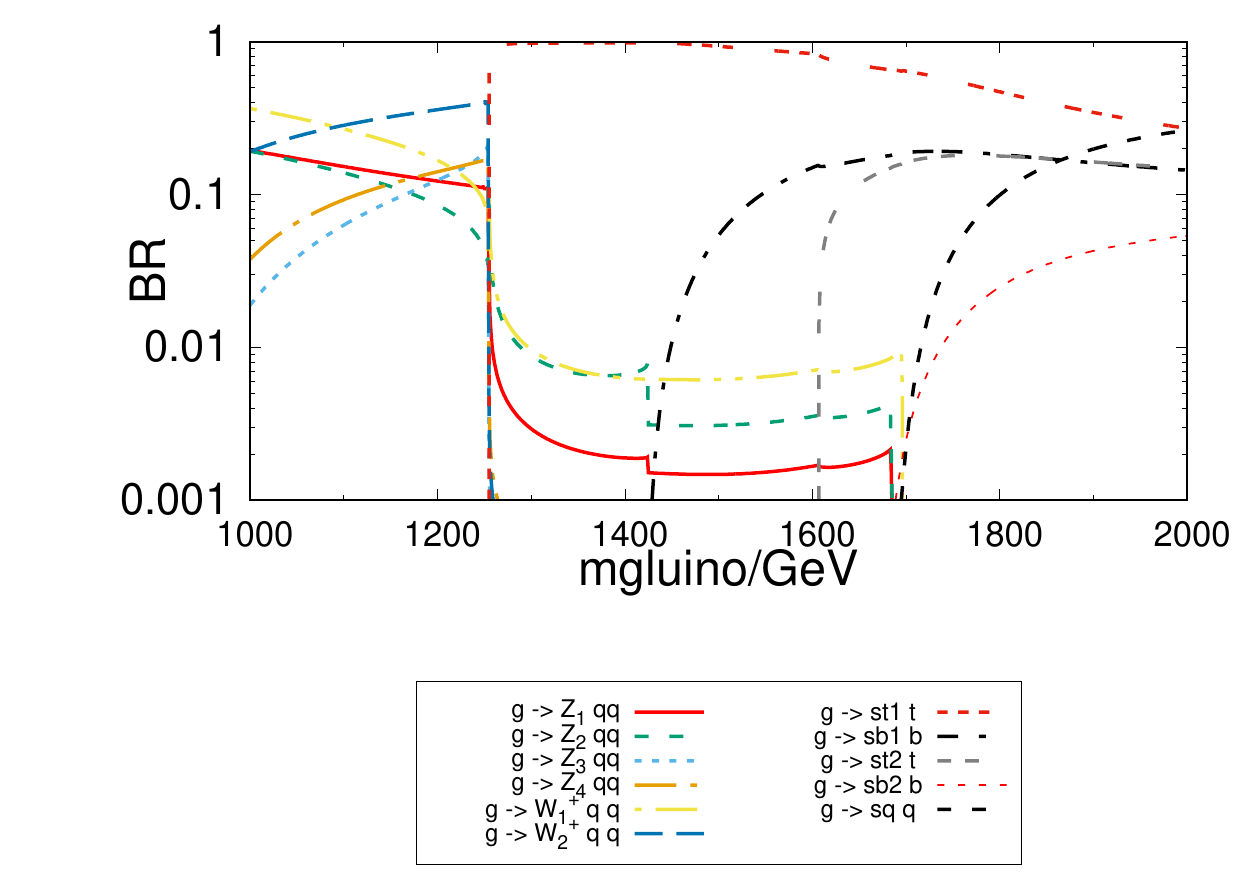}} 
\caption{Branching ratios for the gluino $\tilde{g}$ as its mass is increased
  from $1$ to $2$ TeV. The suppression of 3 body modes relative to 2 body tree
  level modes is clearly evident in the sudden drop in the 3 body branching ratios once the first 2 body mode $\tilde{g} \rightarrow \tilde{t}_1 t$ is kinematically available. Note the ``$g$" indicated in the plot are $\tilde{g}$ (i.e.\ gluinos), whilst ``$Z_i$" are $\tilde{Z}_i$ (i.e.\ neutralinos) and ``$W_j$" are $\tilde{W}_j$ (i.e.\ charginos). ``st" indicates stops $\tilde{t}_i$, ``sb" indicate sbottoms $\tilde{b}_i$, ``sq" are $\tilde{q}$ squarks of the first two generations and ``$q$" here are quarks of the first two generations. } \label{gluino3body}
\end{figure}

A scan over the mass of the gluino to demonstrate the expected suppression of
3 body decays relative to 2 body decays was also performed, see
Figure~\ref{gluino3body}. The result of this is that phenomenologically, 3
body modes are only important when 2 body tree level modes are
unavailable. For this 
reason, {\tt SOFTSUSY} only calculates 3 body modes when there are no
two-body modes kinematically available and does not output the 3 body
modes if 
the total 3 body decay width (sum of all the 3 body decay widths
available) is less than the BR tolerance. 
More details on the other 3 body modes, the contributions included,
approximations made and the level of agreement seen between {\tt SOFTSUSY} and
other decay calculators for each mode are given in~\ref{appendix:MSSM3body}. There, the relevant expressions used by our decay
calculator to determine their partial widths are also provided. 

\subsection{NMSSM Decays - $H$}

\begin{figure} 
\centerline{\includegraphics[scale=1.2]{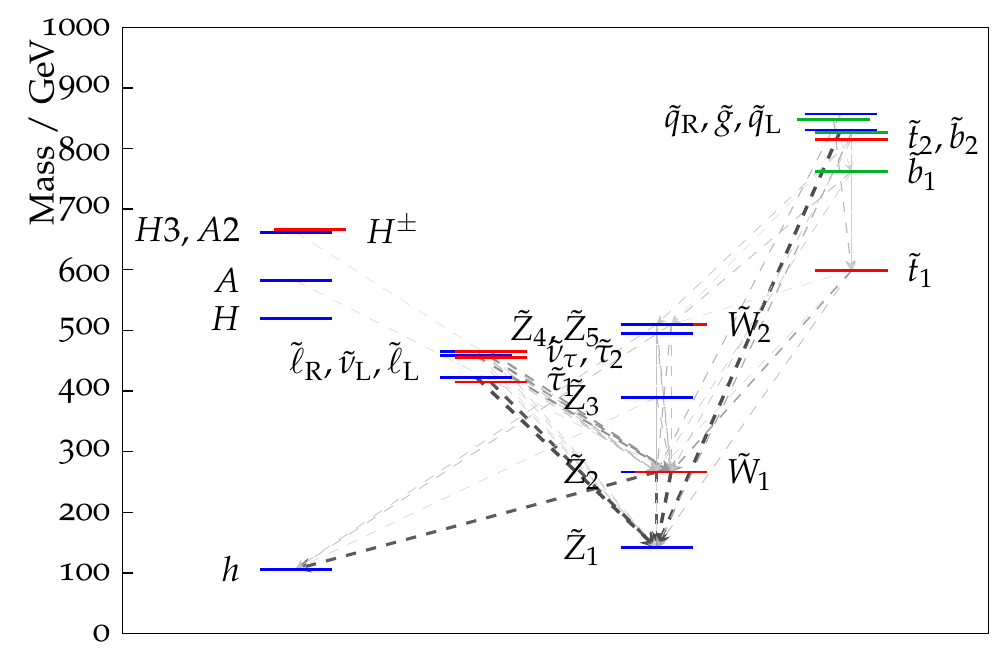}} 
\caption{Mass spectrum and branching ratios for the constrained NMSSM
  parameter point in Table~\protect\ref{nmssmSLHAnoZ3Inputspctable} with $m_0 =
  400$ GeV, $m_{1/2} = 350$ GeV, 
  $\tan\beta = 10$, sign$(\mu) = +1$, $A_0 = -300$ GeV, $\lambda = 0.1$,
  $\kappa = 
  0.1$, $\lambda \langle S \rangle = 200$ GeV and $\xi_F = 100$ GeV
used for the comparison with {\tt NMSSMTools}. The
  arrows represent decay modes of branching ratios (BRs) greater than
  $0.1$, with thicker, bolder arrows representing larger BRs. 
The figure was produced with the aid of {\tt slhaplot-3.0.4} of {\tt PySLHA}
\cite{Buckley:2013}. } \label{nmssmSLHAnoZ3Inputspectrum} 
\end{figure}

Similar detailed checks to those above were performed in the NMSSM fields'
sector and 
we provide some 
details here. In Table~\ref{nmssmSLHAnoZ3Inputspctable} we
present a quantitative comparison of the decays of the second heaviest neutral
CP 
even Higgs of the NMSSM, the $H$. The spectrum is given in
Figure~\ref{nmssmSLHAnoZ3Inputspectrum} with decay modes of branching ratios
(BRs) greater than $0.1$ indicated by arrows, with thicker, bolder arrows
representing larger BRs. For this parameter point, $H$ is the CP even Higgs
which has the largest singlet component. Although validation was carried out
for all the 
particle decays of 
the extended Higgs and neutralinos, for brevity we just provide that of the
$H$ here. 
The comparison in Table~\ref{nmssmSLHAnoZ3Inputspctable} demonstrates that the
level of 
agreement is usually better than $10\%$ with the
exception of a few of the decay modes. The decay modes which show larger
differences are the decays to "down-type" fermions, i.e. fermions with third
component of weak isospin $T_3 = -{1 \over 2}$ and the 1-loop decay to two
photons $H \rightarrow \gamma \gamma$. 
Note that the decays to quarks and to two
gluons here show good agreement with {\tt NMSSMTools}: the scale of the
decaying 
Higgs $m_H = 519.3$ GeV is relatively close to $M_{SUSY} =
\sqrt{m_{\tilde{t}_1}m_{\tilde{t}_2}} = 675.5$ GeV so any differences in the
running between the two programs have little effect. 
{\tt SOFTSUSY} and {\tt
  NMSSMTools} both run the gauge couplings to $m_H$: however there are
potential differences in the running order and 
scheme. {\tt SOFTSUSY} matches at $m_Z$ and then runs $\alpha_s$ in the full
NMSSM at 2-loops with 1-loop threshold corrections at $m_Z$. 
For the case of
the decays $H \rightarrow s \bar{s}$, $b \bar{b}$, $\mu^+ \mu^-$ and $\tau^+
\tau^-$, differences are seen between the default {\tt SOFTSUSY} 
partial widths and those of {\tt NMSSMTools}. Some of these differences can be
explained by the use of different values for the masses from which Higgs
couplings are extracted, particularly in the case
of the decays to $b$, $\mu$ 
and $\tau$ pairs. {\tt SOFTSUSY} uses $m_b(pole) = 4.97$ GeV, $m_{\mu}(M_{SUSY}) =
0.103$ GeV and $m_{\tau}(M_{SUSY}) = 1.80$ GeV; meanwhile {\tt NMSSMTools} uses
$m_b = 4.54$ GeV, $m_{\mu} = 0.106$ GeV and $m_{\tau} = 1.78$ GeV. However,
most of the differences are due to the definition of the CP 
even 
mixing matrix $S$: the coupling of the singlet-like $H$ to ``down-type"
fermions is given by $[S(2,2)/\cos(\beta)]^2$. {\tt SOFTSUSY}
obtains $S(2,2)
= 2.71 \times 10^{-2}$, whilst {\tt NMSSMTools} has $S(2,2) = 2.87 \times
10^{-2}$. Given that the partial widths are proportional to the square of the
mixing matrix element, this results in an approximate $12\%$ difference. 
The {\tt SOFTSUSY} decay calculation uses the tree-level value $S(M_{SUSY})$,
whereas {\tt NMSSMTools} uses $S$ as extracted from the loop-corrected pole
mass matrix. The two choices are equivalent to leading order, and so the
numerical difference between the programs is simply a higher order effect.
For
this reason in columns 5 and 6 of Table~\ref{nmssmSLHAnoZ3Inputspctable}, the
CP even mixing matrix elements have also been set to those of {\tt NMSSMTools}
to demonstrate that this makes up much of the remaining differences. The other
significant difference observed in the partial widths between the default {\tt
  SOFTSUSY} results and those of {\tt NMSSMTools} is in the $\gamma \gamma$
channel. By default {\tt SOFTSUSY} runs $\alpha$ {\em and}
quark masses, whereas {\tt NMSSMTools} runs $\alpha$
but not the quark masses to calculate the Higgs couplings. The quark masses
used by {\tt  
  SOFTSUSY} for this point are $m_t(m_{H}) = 144.5$ GeV, $m_{b}(m_{H}) =
2.40$ GeV, $m_c(m_{H}) 
= 0.57$ GeV whereas {\tt NMSSMTools} uses $m_t = 170.9$ GeV, $m_b = 4.54$ GeV,
$m_c = 1.40$ GeV; meanwhile {\tt SOFTSUSY} uses $\alpha(m_{H}) =
7.88\times10^{-3}$ whereas {\tt NMSSMTools} obtains
$\alpha(m_{H}) = 7.30\times10^{-3}$. The difference in the values of
$\alpha(m_H)$ 
is presumably due to a difference in the scheme\footnote{{\tt
  SOFTSUSY} matches at $m_Z$ and then runs $\alpha$ in the full NMSSM at
2-loops.}. With the quark masses and $\alpha$ used by {\tt NMSSMTools} inserted
into the {\tt SOFTSUSY} decay code the difference between the two programs is
dramatically reduced, with them now showing excellent agreement. 
This clearly
demonstrates that the difference observed is due to different quark masses and
coupling constants taken, in particular it is the quark masses which have the
largest effect here. The reason for such sensitivity to the masses taken is
that for this parameter point there is a large cancellation between the $t$, $W$
and other loop contributions. The degree of the cancellation is consequently
heavily dependent upon the top mass used. With {\tt SOFTSUSY}'s choices
then the real part of the top loop contribution is $\mathcal{R}[I_t] = 8.99 \times
10^{-2}$ and the real part of the $W$ loop contribution $\mathcal{R}[I_W] = -0.114$
whilst the other significant contribution is that of the heaviest chargino
$\tilde{W}_2$: $\mathcal{R}[I_{\tilde{W}_2}] = 5.53 \times 10^{-2}$, resulting
in significant cancellation such that the total of all the particle loop
contributions is $(2.65-6.62i) \times 10^{-2}$. With the quark mass choices of
{\tt NMSSMTools} instead one obtains $\mathcal{R}[I_t] = 0.135$ and so the
total cancellation is much smaller and the total of all the loop contributions
is $(7.16-7.30i) \times 10^{-2}$, which has a modulus much larger than that
obtained using the usual {\tt SOFTSUSY} choices. Once these are
squared this explains the significant discrepancy. Differences
seen between the two programs for this channel should be interpreted as an
indication of a large theoretical error in the calculation at this order for
this parameter point. 
Note that the comparison is carried out
against an old version of {\tt NMSSMTools} ({\tt NMSSMTools-4.2.1}) since 
there only exists an interface
between the  
{\tt SOFTSUSY} spectrum generator and this old version of {\tt
  NMSSMTools}. Therefore comparing with {\tt 
  NMSSMTools-4.2.1} allowed the effects of the spectrum 
generator to be isolated as much as possible from other differences in the
decay calculations for validation.  

\begin{center}
\begin{table} 
\centering
\begin{tabular}{|c|c|c|c|c|c|c|c|c|} \hline
\multicolumn{2}{|C{3.0cm}|}{{\tt SOFTSUSY} default with its quark masses and coupling constants run to $m_{H}$} & \multicolumn{2}{C{3.0cm}|}{{\tt SOFTSUSY} with {\tt NMSSMTools-4.2.1} quark masses and running coupling constants} &
\multicolumn{2}{C{3.0cm}|}{{\tt SOFTSUSY} with {\tt NMSSMTools-4.2.1} quark masses, running coupling constants and $S$} & \multicolumn{2}{C{3.0cm}|}{{\tt NMSSMTools-4.2.1} with same QCD corrections} & mode \\ \hline
PW/GeV & BR & PW/GeV & BR & PW/GeV & BR & PW/GeV & BR & \\ \hline
2.98e-06 & 9.22e-06 & 2.99e-06 & 9.24e-06 & 3.04e-06 & 9.22e-06 & 3.04e-06 & 9.21e-06 & $H \rightarrow c \bar{c}$ \\ \hline
3.65e-07 & 1.13e-06 & 3.67e-07 & 1.14e-06 & 4.10e-07 & 1.24e-06 & 4.33e-07 & 1.31e-06 & $H \rightarrow s \bar{s}$ \\ \hline
9.44e-04 & 2.92e-03 & 7.73e-04 & 2.39e-03 & 8.64e-04 & 2.62e-03 & 8.93e-04 & 2.71e-03 & $H \rightarrow b \bar{b}$ \\ \hline
5.58e-02 & 1.73e-01 & 5.58e-02 & 1.73e-01 & 5.68e-02 & 1.72e-01 & 5.68e-02 & 1.72e-01 & $H \rightarrow t \bar{t}$ \\ \hline
2.52e-07 & 7.79e-07 & 2.68e-07 & 8.27e-07 & 2.99e-07 & 9.06e-07 & 3.16e-07 & 9.56e-07 & $H \rightarrow \mu^- \mu^+$ \\ \hline
7.75e-05 & 2.40e-04 & 7.57e-05 & 2.34e-04 & 8.45e-05 & 2.56e-04 & 8.92e-05 & 2.70e-04 & $H \rightarrow \tau^- \tau^+$ \\ \hline
1.21e-05 & 3.73e-05 & 1.21e-05 & 3.74e-05 & 1.33e-05 & 4.04e-05 & 1.22e-05 & 3.70e-05 & $H \rightarrow \tilde{Z}_1 \tilde{Z}_1$  \\ \hline
3.25e-05 & 1.00e-04 & 3.25e-05 & 1.01e-04 & 3.60e-05 & 1.09e-04 & 3.44e-05 & 1.04e-04 & $H \rightarrow \tilde{Z}_1 \tilde{Z}_2$ \\ \hline
2.00e-02 & 6.18e-02 & 2.00e-02 & 6.18e-02 & 2.00e-02 & 6.06e-02 & 2.07e-02 & 6.27e-02 & $H \rightarrow h h$ \\ \hline
9.03e-08 & 2.97e-07 & 1.61e-07 & 4.97e-07 & 1.62e-07 & 4.91e-07 & 1.68e-07 & 5.09e-07 & $H \rightarrow \gamma \gamma$ \\ \hline
1.47e-04 & 4.54e-04 & 1.47e-04 & 4.54e-04 & 1.49e-04 & 4.53e-04 & 1.53e-04 & 4.63e-04 & $H \rightarrow g g$ \\ \hline
2.07e-06 & 6.39e-06 & 1.93e-06 & 5.98e-06 & 2.14e-06 & 6.47e-06 & 2.21e-06 & 6.69e-06 & $H \rightarrow Z \gamma$ \\ \hline
1.67e-01 & 5.15e-01 & 1.67e-01 & 5.15e-01 & 1.70e-01 & 5.16e-01 & 1.70e-01 & 5.15e-01 & $H \rightarrow W^+ W^-$ \\ \hline
8.00e-02 & 2.47e-01 & 8.00e-02 & 2.47e-01 & 8.17e-02 & 2.48e-01 & 8.16e-02 & 2.47e-01 & $H \rightarrow Z Z$ \\ \hline
3.24e-01 & 1.00e+00 & 3.23e-01 & 1.00e+00 & 3.30e-01 & 1.00e+00 & 3.30e-01 & 1.00e+00 & Column totals \\ \hline
\end{tabular}
\caption{$H$ decay partial widths and branching ratios as output by {\tt
    SOFTSUSY} with the quark masses and coupling constants run to $m_{H}$
  as is the default (except in cases where QCD corrections are applied - here pole masses
  must be used), {\tt SOFTSUSY} with the quark masses and coupling
  constants set temporarily to those of {\tt NMSSMTools} for comparison of
  results, {\tt SOFTSUSY} with the quark masses and coupling
  constants and CP even Higgs mixing matrix ($S$) set temporarily to that
  of {\tt NMSSMTools} for comparison of
  results and {\tt NMSSMTools-4.2.1}
  \cite{Ellwanger:2004xm,Ellwanger:2012dd,Ellwanger:2006ch}. For columns 3 and
  4 this  
  meant setting $\alpha_s(m_{H})= 9.41\times10^{-2}$, $m_b = 4.54$ GeV for
  $H  
  \rightarrow 
  bb$, $m_{\mu}= 0.106$ GeV for $H \rightarrow \mu \mu$, $m_{\tau}=
  1.78$ GeV for $H 
  \rightarrow  
  \tau \tau$, and the same values were used appropriately for $H 
  \rightarrow gg$. These should be compared with the values {\tt SOFTSUSY}
  obtains from 
  running to the $H$ mass of $\alpha_s(m_{H}) = 9.42\times 10^{-2}$,
  $m_{\mu} = 0.103$ GeV, $m_{\tau} = 1.80$~GeV and its $m_b(pole) = 4.97$~GeV. Meanwhile for $H \rightarrow \gamma \gamma$, {\tt SOFTSUSY}
  uses $m_t(m_{H}) = 144.5$~GeV, $m_{b}(m_{H}) = 2.40$~GeV, $m_c(m_{H}) =
  0.57$~GeV whereas {\tt NMSSMTools}
  has $m_t = 170.9$~GeV, $m_b = 4.54$~GeV, $m_c = 1.40$~GeV. These were
  therefore input into {\tt SOFTSUSY}  
  for columns 3 and 4. In {\tt SOFTSUSY} 
  $\alpha(m_{H}) = 7.88\times10^{-3}$ whereas {\tt NMSSMTools} obtains
  $\alpha(m_{H}) = 7.30\times10^{-3}$. This 
  is therefore also set in the decay calculator code for columns 3 and 4. Meanwhile it was noticed there were small
  differences in the smaller elements of the CP even mixing matrix ($S$) between the two codes, therefore $S$ 
  was additionally set to that of {\tt NMSSMTools}, on top of the changes described for columns 3 and 4, in columns
  5 and 6. The 
  parameter point considered is the
constrained NMSSM
  parameter point  with $m_0 =
  400$ GeV, $m_{1/2} = 350$ GeV, 
  $\tan\beta = 10$, sign$(\mu) = +1$, $A_0 = -300$ GeV, $\lambda = 0.1$,
  $\kappa = 
  0.1$, $\lambda \langle S \rangle = 200$ GeV and $\xi_F = 100$ GeV
used for the comparison with {\tt NMSSMTools}.
There is good agreement between the two
  programs with differences around $10\%$. Differences can be larger for the
  decays to quarks and 1-loop decays to $\gamma \gamma$ and $gg$ due to
  differences in the quark masses and coupling constants taken, or indeed due to small
  differences in elements of the CP even mixing matrix $S$. Using the same
  values for these as in {\tt NMSSMTools} significantly reduces these
  differences to around $5\%$ or less. This is illustrated in the improved
  agreement of the 3rd 
  and 4th columns, and 5th and 6th columns, with the 7th and 8th columns. A
  more detailed description of 
  the level of agreement and source of differences is given in the text.} 
\label{nmssmSLHAnoZ3Inputspctable}
\end{table}
\end{center}

\section{Summary} \label{sec:sum}
The fast automated computation of the spectrum and decays of particles in the
MSSM and the NMSSM is now possible all within \code{SOFTSUSY}, and they are
necessary steps in the simulation of collider signatures, required for both
the prediction and interpretation of the collider signatures. The inclusion
of sparticle and Higgs decay partial widths and branching ratios should aid in
estimating 
theoretical uncertainties, particularly in regard to decays in the NMSSM,
where there are few other publicly available tools. 
\code{SOFTSUSY}~has been routinely used by the ATLAS and CMS experiments to
interpret their searches for supersymmetric particles, and so having the
decays calculated within the same package as the spectrum should make their
calculation easier. 
The usual SLHA and SLHA2
conventions for input and output have been followed in order to facilitate
`joining up' \code{SOFTSUSY}~with other observable calculating programs in a
bug-free manner, for example with programs that perform Monte-Carlo event
simulation. 

\section*{Acknowledgements}
This work has been partially supported by STFC consolidated grant 
ST/L000385/1. We thank the Cambridge SUSY working group for helpful
discussions and to F Staub for helpful communications regarding
\code{SARAH}~and \code{sPHENO}.

\appendix

\section{Running \SOFTSUSY~to Calculate Sparticle Decays} \label{sec:run}

\SOFTSUSY~produces an executable called \code{softpoint.x}. It can be run by
the command
\begin{verbatim}
./softpoint.x leshouches < inOutFiles/lesHouchesInput
\end{verbatim}
where the file \code{inOutFiles/lesHouchesInput}~contains an ASCII file
for input
prepared in SUSY Les Houches Accord (SLHA)~\cite{Skands:2003cj} or
SLHA2~\cite{Allanach:2008qq} format. 
A \code{SOFTSUSY}-specific \code{Block}~of the SLHA input file is provided in
the case that decays are required:
\begin{verbatim}
Block SOFTSUSY               # Optional SOFTSUSY-specific parameters
    0   1.000000000e+00      # Calculate decays in output (only for RPC (N)MSSM)
\end{verbatim}
If this block is absent, or if the numerical value is instead
\code{0.000000000e+00}, then decays will not be calculated. 
Other input options are available for changing the behaviour of the program as
regards sparticle decays:
\begin{verbatim}
Block SOFTSUSY               # Optional SOFTSUSY-specific parameters
   24   1.000000000e-06      # If decay BR is below this number, don't output
   25   1.000000000e+00      # If set to 0, don't calculate 3-body decays (1=default)
   26   0.000000000e+00      # If set to 1, output partial widths (0=default)
   
\end{verbatim}
Thus, parameter \code{24}~under this block sets the smallest decay branching
ratio that will be output (the default is, as listed, $10^{-6}$) whereas 
parameter \code{25}~can be used to instruct \code{SOFTSUSY}~not to calculate
the 3-body modes, which are more computationally intensive, requiring
numerical integration. It should be noted that 3-body modes are not currently
included 
in decays of the NMSSM so for such calculations, regardless of the value
of \code{25},~there will be no 3-body modes calculated. Parameter \code{26}~is used to
instruct 
\code{SOFTSUSY}~to output the partial widths in addition to the branching
ratios, this 
may be useful in making theoretical calculations or performing comparisons
with other 
codes. These partial widths are output as a column beyond the comments $\#$
column so 
as to not interfere with the SLHA conventions. 

If command-line input is required (see
Refs.~\cite{Allanach:2001kg,Allanach:2013kza} 
for a full description), the user can use the argument 
\code{--decays}~to tell \SOFTSUSY~to perform the decay calculation. 
The command line argument \code{--minBR=<value>}~tells {\tt SOFTSUSY} which minimum
branching ratio to print out in each decay table (where \code{<value>} is
replaced by a numerical value between \code{0}~and \code{1}), whereas the
command line argument \code{--dontCalculateThreeBody}~tells {\tt SOFTSUSY} {\em
  not}\/ to calculate the 3-body decays of sparticles in order to save time. 
In addition the command line argument \code{--outputPartialWidths}~is used 
to instruct {\tt SOFTSUSY} to output the partial widths in the final column of
the decay tables, beyond the comments so as to allow the output to be read
directly into SLHA programs.

In the decay program {\tt SOFTSUSY} itself there are also additional flags and switches
which may be helpful to the user. There are flags at the start of the code named 
\code{flag<name>}~where one must replace \code{<name>}~with the particle name, when
these flags have value $1$ the particle decays are calculated, therefore by default
all such flags are set to $1$. These flags allow the user to turn off irrelevant
decays for their analyses; for example in producing the scanning plots, such as
Figure~\ref{gluino3body} in Section~\ref{glu3bodysec}, all decays apart from those of the
gluino $\tilde{g}$ were turned off by setting all such flags to $0$ apart from
{\tt flaggluino}, this ensured all decays output were gluino decays, allowing the
plots to be produced more straightforwardly. Similarly there is a Boolean variable
{\tt QCDcorr}, which by default is {\tt true}, which may be used to turn off QCD
corrections. In case the user should want to run the parameters used to different
scales, for example in performing comparisons with other decay calculators, it
should be noted that running in {\tt SOFTSUSY} is implemented using {\tt MssmSoftsusy}
and {\tt NmssmSoftsusy} objects (detailed in references \cite{Allanach:2001kg} and
\cite{Allanach:2013kza} respectively) and the {\tt runto} command. If one alters the
running scales within {\tt SOFTSUSY} one must remember to instruct {\tt SOFTSUSY}
to recalculate the $\overline{DR}$ parameters at this scale using {\tt calcDRbarPars()}.
Nonetheless, any changes made to the code are at the user's risk. Finally, given the
dependence of many of the partial widths on the input parameters, and in particular
on the quark masses used in the cases highlighted previously, users may wish to 
alter the quark masses $m_q$, this may be done in the {\tt BLOCK SMINPUTS} of the
SLHA/SLHA2 input file, the masses used within the decay calculator (``kinematic''
and ``running'') will change accordingly.

\section{Sample Output} \label{sec:out}
The output comes in the standard SLHA/SLHA2
format~\cite{Skands:2003cj,Allanach:2008qq}. A sample of the part relevant for
decays is shown in Table~\ref{sampleoutput}. In concordance with SLHA conventions, the widths/partial widths
(\code{PW}) are output 
in units 
of GeV. The partial widths are output after the comments column so as not to interfere
with reading in the decay tables into other programs; this outputting of partial widths can be switched off,
as described in~\ref{sec:run}, so the final column of partial widths in Table~\ref{sampleoutput} would then not appear.
\code{NDA}~lists the number of daughter particles and
\code{PDGi}~lists the
Particle Data Group~(PDG) code of daughter \code{i} (Section 43 of
Ref.~\cite{Olive:2016xmw}). 
The comment at the end of each line after \code{\#}~lists the decay
mode for easy perusal by the eye.
\begin{center}
\begin{table} 
\centering
{\small
\begin{verbatim}
#     PDG         Width             
DECAY 1000021     1.41325020e+01   # Gluino decays
#     BR                  NDA   PDG1        PDG2       Comments                          PW                
      2.13728751e-02      2     1          -1000001    # ~g -> d ~d_L*              3.02052200e-01    
      2.13728751e-02      2    -1           1000001    # ~g -> db ~d_L              3.02052200e-01    
      4.53465485e-02      2     1          -2000001    # ~g -> d ~d_R*              6.40860189e-01    
      4.53465485e-02      2    -1           2000001    # ~g -> db ~d_R              6.40860189e-01    
      2.26614800e-02      2     2          -1000002    # ~g -> u ~u_L*              3.20263412e-01    
      2.26614800e-02      2    -2           1000002    # ~g -> ub ~u_L              3.20263412e-01    
      4.32957172e-02      2     2          -2000002    # ~g -> u ~u_R*              6.11876811e-01    
      4.32957172e-02      2    -2           2000002    # ~g -> ub ~u_R              6.11876811e-01    
      2.13739410e-02      2     3          -1000003    # ~g -> s ~s_L*              3.02067264e-01    
      2.13739410e-02      2    -3           1000003    # ~g -> sb ~s_L              3.02067264e-01    
      4.53479787e-02      2     3          -2000003    # ~g -> s ~s_R*              6.40880401e-01    
      4.53479787e-02      2    -3           2000003    # ~g -> sb ~s_R              6.40880401e-01    
      2.26621068e-02      2     4          -1000004    # ~g -> c ~c_L*              3.20272270e-01    
      2.26621068e-02      2    -4           1000004    # ~g -> cb ~c_L              3.20272270e-01    
      4.32969654e-02      2     4          -2000004    # ~g -> c ~c_R*              6.11894452e-01    
      4.32969654e-02      2    -4           2000004    # ~g -> cb ~c_R              6.11894452e-01    
      7.37354606e-02      2     5          -1000005    # ~g -> b ~b_1*              1.04206655e+00    
      7.37354606e-02      2    -5           1000005    # ~g -> bb ~b_1              1.04206655e+00    
      4.83477101e-02      2     5          -2000005    # ~g -> b ~b_2*              6.83274111e-01    
      4.83477101e-02      2    -5           2000005    # ~g -> bb ~b_2              6.83274111e-01    
      1.12559217e-01      2     6          -1000006    # ~g -> t ~t_1*              1.59074336e+00    
      1.12559217e-01      2    -6           1000006    # ~g -> tb ~t_1              1.59074336e+00    

[ ... ]
\end{verbatim}   
\caption{Sample Output produced by the {\tt SOFTSUSY} decay calculator, here only the 
gluino, $\tilde{g}$, decay table is shown. The input file used was {\tt lesHouchesInput}
which is provided with the {\tt SOFTSUSY} package and has $m_0 = 125$ GeV, $m_{1/2} = 500$ GeV,
$\tan\beta = 10$, sign$(\mu) = +1$ and $A_0 = 0$ in the CMSSM.} \label{sampleoutput}
}
\end{table}
\end{center}

\section{Particle Class} \label{sec:particle}
As part of the extension of \code{SOFTSUSY} to include decay calculation a
new class has been written, {\tt Particle}. This is a container for all the
relevant decay information of a particle and is used to output the decay
tables. We display the class in Table~\ref{particleclass}.

\newcolumntype{C}{>{\centering\arraybackslash} m{4cm} }
\newcolumntype{D}{>{\centering\arraybackslash} m{11cm} }

\begin{center}
\begin{table} 
\centering
\begin{tabular}{|C|D|} \hline
data variable & description \\ \hline
{\tt string name} & particle name \\ \hline
{\tt double mass} & particle mass \\ \hline
{\tt double PDG} & particle PDG code \\ \hline
{\tt double No{\_}of{\_}Decays} & Total Number of possible decays of particle \\ \hline
{\tt double No{\_}1to2{\_}Decays} & Total Number of possible 2 body decays of particle \\ \hline
{\tt double No{\_}1to3{\_}Decays} & Total Number of possible 3 body decays of particle \\ \hline
{\tt double No{\_}grav{\_}Decays} & Total Number of possible decays of particle to LSP gravitinos \\ \hline
{\tt double No{\_}NMSSM{\_}Decays} & Total Number of possible decays of particle in the NMSSM \\ \hline
{\tt double total{\_}width} & Total Decay Width of the particle \\ \hline
{\tt double two{\_}width} & Two-body decay partial width of the particle \\ \hline
{\tt double three{\_}width} & Three-body decay partial width of the particle \\ \hline
{\tt vector $<$vector$<$double$>>$ Array{\_}Decays} & A Nx6 array, where N = {\tt No{\_}of{\_}Decays}, containing the PDGs of the daughter particles in columns $0$ and $1$ (and $4$ for 3-body decays), the partial widths of the decay modes in column $2$, the number of daughters (NDA) in column $3$ and the branching ratio for each decay mode in column $5$. \\ \hline
{\tt vector $<$string$>$ Array{\_}Comments} & A Nx1 array (vector), where N = {\tt No{\_}of{\_}Decays}, containing a comment for each decay mode which is output with the decay table, detailing the decay mode involved, e.g. $\tilde{g} \rightarrow \bar{u} \tilde{u}_L$ \\ \hline
\end{tabular}
\caption{The information contained in the {\tt Particle} object for each of
  the decaying particles. PDG codes are given in the reference
  \cite{Allanach:2008qq}. Note the numbers of decays contained in {\tt double
    No{\_}\ldots{\_}Decays} are the total number of such decays assuming non are
  kinematically forbidden. All these decays are checked by the program to see
  if they are allowed kinematically and calculated if so. All the numbers of
  decays are in the MSSM unless stated otherwise.} 
\label{particleclass}
\end{table}
\end{center}

\section{Glossary: Reference Tables for Decays\label{appendix:exp}} 

\begin{center}
\begin{table}
\centering
\begin{tabular}{|c|c|c|c|} \hline
\multicolumn{2}{|c|}{\bf{Gluino Decays}} &  \multicolumn{2}{c|}{\bf{Slepton of first 2 generations decays}} \\ \hline
$\tilde{g} \rightarrow q \tilde{q}_{L/R}$ &~\ref{gluqqLR} &  $\tilde{l}_L \rightarrow l \tilde{Z}_i$ &~\ref{lLlneut} \\ \hline
$\tilde{g} \rightarrow q \tilde{q}_{1/2}$ &~\ref{gluqq12} & $\tilde{l}_R \rightarrow l \tilde{Z}_i$ &~\ref{lRlneut} \\ \hline
\multicolumn{2}{|c|}{\bf{Squark of first 2 generations decays}} & $\tilde{\nu}_l \rightarrow \nu_l \tilde{Z}_i$ &~\ref{snulnulneut} \\ \hline
$\tilde{q}_L \rightarrow \tilde{Z}_i q$ &~\ref{qLqneut} &  $\tilde{l}_L \rightarrow \nu_l \tilde{W}_{j}^{-}$ &~\ref{lLnulch1} \\ \hline
$\tilde{q}_{L/R} \rightarrow q \tilde{g}$ &~\ref{qLRqglu} & $\tilde{\nu}_l \rightarrow l \tilde{W}_{j}^{+}$ &~\ref{nullch1} \\ \hline
$\tilde{q}_{L} \rightarrow q^{'} \tilde{W}_{1/2}^-$ &~\ref{qLqpch1} & & \\ \hline
$\tilde{q}_R \rightarrow \tilde{Z}_i q$ &~\ref{qRqneut} &  & \\ \hline
\multicolumn{2}{|c|}{\bf{Squark of 3rd generation decays}} & \multicolumn{2}{c|}{\bf{Slepton of 3rd generation decays}} \\ \hline
$\tilde{q}_{1/2} \rightarrow q \tilde{g}$ &~\ref{q12qglu} & $\tilde{\tau}_{1/2} \rightarrow \tau \tilde{Z}_i$ &~\ref{stau1tauneut} \\ \hline
$\tilde{b}_{1/2} \rightarrow \tilde{W}_{j}^- t$ &~\ref{b1tch1} & $\tilde{\tau}_{1/2} \rightarrow \nu_\tau \tilde{W}_{j}^{-}$ &~\ref{stau1nutauch1} \\ \hline
$\tilde{t}_{1/2} \rightarrow \tilde{W}_{j}^+ b$ &~\ref{t1bch1} & $\tilde{\nu}_\tau \rightarrow \tau \tilde{W}_{j}^{+}$ &~\ref{snutautauch1} \\ \hline
$\tilde{t}_{1/2} \rightarrow \tilde{Z}_i t$ &~\ref{t1tneut} & $\tilde{\tau}_{1/2} \rightarrow \tilde{\nu}_\tau H^-$ &~\ref{stau1nutauHpm} \\ \hline
$\tilde{b}_{1/2} \rightarrow \tilde{Z}_i b$ &~\ref{b1bneut} & $\tilde{\tau}_{1/2} \rightarrow \tilde{\nu}_\tau W^-$ &~\ref{stau1nutauW} \\ \hline
$\tilde{t}_{1/2} \rightarrow \tilde{b}_{1/2} W^+$ &~\ref{t1b1W} & $\tilde{\nu}_{\tau} \rightarrow \tilde{Z}_i \nu_{\tau}$ &~\ref{snulnulneut} \\ \hline
$\tilde{t}_{1/2} \rightarrow \tilde{b}_{1/2} H^+$ &~\ref{t1b1Hpm} & $\tilde{\nu}_{\tau} \rightarrow \tilde{\tau}_{1/2} W^+$ &~\ref{stau1nutauW} \\ \hline
$\tilde{t}_2 \rightarrow \phi \tilde{t}_1$ &~\ref{t2t1phi} & $\tilde{\nu}_{\tau} \rightarrow \tilde{\tau}_{1/2} H^+$ &~\ref{stau1nutauHpm} \\ \hline
$\tilde{b}_2 \rightarrow \phi \tilde{b}_1$ &~\ref{b2b1phi} & $\tilde{\tau}_2 \rightarrow \tilde{\tau}_1 Z$ &~\ref{stau2stau1Z} \\ \hline
$\tilde{q}_2 \rightarrow Z \tilde{q}_1$ &~\ref{q2q1Z} & $\tilde{\tau}_2 \rightarrow \tilde{\tau}_1 \phi$ &~\ref{stau2stau1phi} \\ \hline
\multicolumn{2}{|c|}{\bf{Chargino decays}} & \multicolumn{2}{c|}{\bf{Neutralino decays}} \\ \hline
$\tilde{W}_{i}^+ \rightarrow \bar{q} \tilde{q}_L'$ &~\ref{ch1qqL} & $\tilde{Z}_i \rightarrow \bar{q} \tilde{q}_L/R$ &~\ref{neutuuLR} \\ \hline
$\tilde{W}_{i}^+ \rightarrow \bar{b} \tilde{t}_{1/2}$ &~\ref{ch1bst1} & $\tilde{Z}_i \rightarrow \bar{l} \tilde{l}_L/R$ &~\ref{neutllLR} \\ \hline
$\tilde{W}_{i}^+ \rightarrow \bar{t} \tilde{b}_{1/2}$ &~\ref{ch1tsb1} & $\tilde{Z}_i \rightarrow \bar{t} \tilde{t}_{1/2}$ &~\ref{neuttst1} \\ \hline
$\tilde{W}_i^+ \rightarrow \bar{l} \tilde{l}_{L}$ &~\ref{chllL} & $\tilde{Z}_i \rightarrow \bar{b} \tilde{b}_{1/2}$ &~\ref{neutbsb1} \\ \hline
$\tilde{W}_i^+ \rightarrow \bar{\tau} \tilde{\nu}_{\tau}$ &~\ref{chtaunutau} & $\tilde{Z}_i \rightarrow \bar{\tau} \tilde{\tau}_{1/2}$ &~\ref{neuttaustau1} \\ \hline
$\tilde{W}_i^+ \rightarrow \bar{\tilde{\tau}}_{1/2} \nu_{\tau}$ &~\ref{chstau1nutau} & $\tilde{Z}_i \rightarrow W \tilde{W}_{1/2}$ &~\ref{neutWch1} \\ \hline
$\tilde{W}_1^+ \rightarrow W \tilde{Z}_j$ &~\ref{ch1Wneut} & $\tilde{Z}_j \rightarrow H^+ \tilde{W}_{1/2}$ &~\ref{neutHpmch1} \\ \hline
$\tilde{W}_1^+ \rightarrow H^- \tilde{Z}_j$ &~\ref{ch1Hpmneut} & $\tilde{Z}_i \rightarrow Z \tilde{Z}_{j}$ &~\ref{neutZneut} \\ \hline
$\tilde{W}_2 \rightarrow Z \tilde{W}_1$ &~\ref{ch2Zch1} & $\tilde{Z}_i \rightarrow h \tilde{Z}_{j}$ &~\ref{neuthneut} \\ \hline
$\tilde{W}_2 \rightarrow \phi \tilde{W}_1$ &~\ref{ch2phich1} & $\tilde{Z}_i \rightarrow A \tilde{Z}_{j}$ &~\ref{neutAneut} \\ \hline
\end{tabular}
\caption{MSSM 2 body decays included in the {\tt SOFTSUSY} decay program,
  the references for the formulae in the appendices are given. $\phi$ here is
  $h/H/A$ i.e.\ any of the neutral Higgs bosons. The same references may be
  given for different decays in cases where the underlying formulae are the
  same and the necessary replacements for different outgoing particles are
  given with the formulae.} 
\label{MSSMdecaysreftable}
\end{table}
\end{center}
We begin by listing the various decay modes included in {\tt SOFTSUSY} in some
tables, along with 
equation numbers for quick reference. We group the listings into several
tables, each comprising a set of decays: in Table~\ref{MSSMdecaysreftable}, we
list 2 body MSSM tree-level decays, in Table~\ref{Higgsdecaysreftable}, we
list the MSSM Higgs boson decays, in Table~\ref{1to3decaysreftable}, the MSSM
3 body decays included, in Table~\ref{Gravitinodecaysreftable} we list the
decays into gravitinos, in Table~\ref{NMSSMNeutdecaysreftable} we list the
NMSSM decays of neutralinos not already listed above and in
Table~\ref{NMSSMHiggsdecaysreftable} we list NMSSM Higgs decay modes. 
\begin{center}
\begin{table}
\centering
\begin{tabular}{|c|c|c|c|} \hline
\multicolumn{2}{|c|}{\bf{CP Even Higgs decays} } & \multicolumn{2}{c|}{\bf{CP Odd Higgs decays} } \\ \hline
$h/H \rightarrow f \bar{f}$ &~\ref{hqq},~\ref{hqqQCDcorr} & $A \rightarrow f \bar{f}$ &~\ref{Aqq},~\ref{AqqQCDcorr} \\ \hline
$h/H \rightarrow \tilde{Z}_i \tilde{Z}_j$ &~\ref{hneutneut} & $A \rightarrow \tilde{Z}_i \tilde{Z}_j$ &~\ref{Aneutneut} \\ \hline
$h/H \rightarrow A Z$ &~\ref{htoAZ} & $A \rightarrow h Z$ &~\ref{AhZ} \\ \hline
$h \rightarrow A A$ &~\ref{htoAA} & $A \rightarrow \tilde{f}_i \tilde{f}_j^*$ &~\ref{Afifj} \\ \hline
$H \rightarrow h h$ &~\ref{Htohh} & \multicolumn{2}{c|}{\bf{CP Even/Odd Higgs decays} } \\ \hline
$H \rightarrow A A$ &~\ref{HtoAA} & $\phi \rightarrow \tilde{W}_i^+ \tilde{W}_i^-$ &~\ref{phichch} \\ \hline
$H \rightarrow H^+ H^-$ &~\ref{HHpmHpm} & $\phi \rightarrow \tilde{W}_i^+ \tilde{W}_j^-$ &~\ref{phichchdif} \\ \hline
$h \rightarrow \tilde{q}_{L/R} \tilde{q}_{L/R}^*$ &~\ref{hqLRqLR} & $\phi \rightarrow \gamma \gamma$ &~\ref{phigamgam} \\ \hline
$H \rightarrow \tilde{q}_{L/R} \tilde{q}_{L/R}^*$ &~\ref{HqLRqLR} & $\phi \rightarrow gg$ &~\ref{phigg},~\ref{phiggQCDcorr},~\ref{AggQCDcorr} \\ \hline
$h \rightarrow \tilde{l}_{L/R} {\tilde{l}}_{L/R}^*$ &~\ref{hlLRlLR} & $\phi \rightarrow Z \gamma$ &~\ref{phiZgam} \\ \hline
$h \rightarrow \tilde{t}_{i} \tilde{t}_{j}^*$ &~\ref{hstst} & \multicolumn{2}{c|}{\bf{Charged Higgs decays} } \\ \hline
$h \rightarrow \tilde{b}_i {\tilde{b}}_j^*$ &~\ref{hsbsb} & $H^+ \rightarrow f \bar{f'}$ &~\ref{Hpmqq} \\ \hline
$h \rightarrow \tilde{\tau}_i {\tilde{\tau}}_j^*$ &~\ref{hstau1stau1} & $H^+ \rightarrow \tilde{Z}_i \tilde{W}_j$ &~\ref{Hpmneutch} \\ \hline
$h/H \rightarrow ZZ^*$ &~\ref{hZZstar} & $H^+ \rightarrow W^+ h$ &~\ref{HpmWh} \\ \hline
$h/H \rightarrow WW^*$ &~\ref{hWWstar} & $H^+ \rightarrow \tilde{f}_{L/R} \tilde{f'}_{L/R}$ &~\ref{Hpmeqsq1} \\ \hline
$h/H \rightarrow WW$ &~\ref{hWW} & $H^+ \rightarrow \tilde{f}_i \tilde{f'}_j$ &~\ref{Hpmeqsq2} \\ \hline
$h/H \rightarrow ZZ$ &~\ref{hZZ} & & \\ \hline
\end{tabular}
\caption{Higgs decays included in the {\tt SOFTSUSY} decay program, the references for the formulae in the appendices are given. The same references may be given for different decays in cases where the underlying formulae are the same and the necessary replacements for different outgoing particles are described with the formulae. Multiple references are given for decays where QCD corrections are included, the first reference is the non-QCD corrected decay and the remainder are once QCD corrections are included.}
\label{Higgsdecaysreftable}
\end{table}
\end{center}
\begin{center}
\begin{table}
\centering
\begin{tabular}{|c|c|c|c|} \hline
\multicolumn{2}{|c|}{\bf{Gluino $1 \rightarrow 3$ decays} } & \multicolumn{2}{c|}{\bf{Neutralino $1 \rightarrow 3 $ decays} } \\ \hline
$\tilde{g} \rightarrow q \bar{q} \tilde{Z}_i$ &~\ref{gqqneut} & $\tilde{Z}_i \rightarrow \tilde{Z}_j f \bar{f}$ &~\ref{neutneutff} \\ \hline
$\tilde{g} \rightarrow t \bar{t} \tilde{Z}_i$ &~\ref{gttneut} & $\tilde{Z}_i \rightarrow \tilde{W}_j f' \bar{f}$ &~\ref{neutchff} \\ \hline
$\tilde{g} \rightarrow b \bar{b} \tilde{Z}_i$ &~\ref{gttneut} & \multicolumn{2}{c|}{\bf{Chargino $1 \rightarrow 3 $ decays} } \\ \hline
$\tilde{g} \rightarrow q \bar{q'} \tilde{W}_i^-$ &~\ref{gchqqp} & $\tilde{W}_j \rightarrow \tilde{Z}_i \bar{f'} f$ &~\ref{neutchff} \\ \hline
$\tilde{g} \rightarrow t \bar{b} \tilde{W}_i^-$ &~\ref{gchqqp} & & \\ \hline
\end{tabular}
\caption{Three body decays included in the {\tt SOFTSUSY} decay program, the
  references for the formulae in the appendices are given. Not all 3 body
  decays are included as they are naturally suppressed with respect to the 2
  body tree level decays. For this reason we have aimed only to incorporate
  the most phenomenologically relevant 3 body decays, however more may be
  added in future versions. The same reference is given for neutralino decays
  to a chargino, fermion and anti-fermion as for the ``reverse'' decays of a
  chargino to a neutralino, fermion and anti-fermion as this just results in
  minus signs in several places in the partial width formulae, which are given
  in the appendix.} 
\label{1to3decaysreftable}
\end{table}
\end{center}
\begin{center}
\begin{table}
\centering
\begin{tabular}{|c|c|c|c|} \hline
$\tilde{g} \rightarrow g \tilde{G}$ &~\ref{gluggrav} & $\tilde{q} \rightarrow q \tilde{G}$ &~\ref{sqqgrav} \\ \hline
$\tilde{Z}_i \rightarrow \gamma \tilde{G}$ &~\ref{neutgamgrav} & $\tilde{l} \rightarrow l \tilde{G}$ &~\ref{sqqgrav} \\ \hline
$\tilde{Z}_i \rightarrow \phi \tilde{G}$ &~\ref{neutphigrav} & $\tilde{Z}_i \rightarrow Z \tilde{G}$ &~\ref{neutZgrav} \\ \hline 
\end{tabular}
\caption{The Next-to-Lightest Susy Particle (NLSP) decays to gravitinos included in the program along with the appendix references for their formulae.}
\label{Gravitinodecaysreftable}
\end{table}
\end{center}
\begin{center}
\begin{table} 
\centering
\begin{tabular}{|c|c|c|c|} \hline
\multicolumn{2}{|c|}{\bf{Neutralino Decays} } & \multicolumn{2}{c|}{\bf{Decays into Neutralinos} } \\ \hline
$\tilde{Z}_i \rightarrow \tilde{q}_{L/R} \bar{q}$ &~\ref{neutqLRqNMSSM} & $\tilde{q}_{L/R} \rightarrow q \tilde{Z}_i$ &~\ref{qLRqneutNMSSM} \\ \hline
$\tilde{Z}_i \rightarrow \tilde{t}_{1/2} \bar{t}$ &~\ref{neutsttNMSSM} & $\tilde{t}_{1/2} \rightarrow t \tilde{Z}_i$ &~\ref{sttneutNMSSM} \\ \hline
$\tilde{Z}_i \rightarrow \tilde{b}_{1/2} \bar{b}$ &~\ref{neutsbbNMSSM} & $\tilde{b}_{1/2} \rightarrow b \tilde{Z}_i$ &~\ref{sbbneutNMSSM} \\ \hline
$\tilde{Z}_i \rightarrow \tilde{\tau}_{1/2} \bar{\tau}$ &~\ref{neutstautauNMSSM} & $\tilde{\tau}_{1/2} \rightarrow \tau \tilde{Z}_i$ &~\ref{stautauneutNMSSM} \\ \hline
$\tilde{Z}_i \rightarrow W \tilde{W}_{1/2}$ &~\ref{neutWchNMSSM} & $\tilde{\nu}_{\tau_{1/2}} \rightarrow \nu_{\tau} \tilde{Z}_i$ &~\ref{snutuanutauneutNMSSM} \\ \hline
$\tilde{Z}_i \rightarrow H^{\pm} \tilde{W}_{1/2}$ &~\ref{neutHpmchNMSSM} & $\tilde{W}_1 \rightarrow H^{\pm} \tilde{Z}_j$ &~\ref{chHpmneutNMSSM} \\ \hline
$\tilde{Z}_i \rightarrow Z \tilde{Z}_j$ &~\ref{neutZneutNMSSM} & $\tilde{W}_1 \rightarrow W \tilde{Z}_j$ &~\ref{chWneutNMSSM} \\ \hline
$\tilde{Z}_i \rightarrow h_{k} \tilde{Z}_j$ &~\ref{neuthneutNMSSM} & & \\ \hline
$\tilde{Z}_i \rightarrow A_{k} \tilde{Z}_j$ &~\ref{neutAneutNMSSM} & & \\ \hline
\end{tabular}
\caption{The NMSSM decays involving neutralinos that are included in the {\tt SOFTSUSY} decay program. Note any decays not involving neutralinos or neutral Higgs Bosons are the same as in the MSSM.}
\label{NMSSMNeutdecaysreftable}
\end{table}
\end{center}
\begin{center}
\begin{table} 
\centering
\begin{tabular}{|c|c|c|c|} \hline
\multicolumn{2}{|c|}{\bf{CP Even Higgs Decays} } & \multicolumn{2}{c|}{\bf{CP Odd Higgs Decays} } \\ \hline
$h_i \rightarrow f \bar{f}$ &~\ref{hiqqNMSSM},~\ref{hqqQCDcorr} & $A_{i} \rightarrow f \bar{f}$ &~\ref{AqqNMSSM},~\ref{AqqQCDcorr} \\ \hline
$h_{i} \rightarrow \tilde{f}_{L/R} \tilde{f}_{L/R}$ &~\ref{hqLRqLRNMSSM} & $A_{i} \rightarrow \tilde{f}_L \tilde{f}_R$ &~\ref{AqLqRNMSSM} \\ \hline
$h_{i} \rightarrow \tilde{f}_{L/R} \tilde{f}_{R/L}$ &~\ref{hqLRqRLNMSSM} & $A_i \rightarrow \tilde{Z}_j \tilde{Z}_k$ &~\ref{AneutneutNMSSM} \\ \hline
$h_{i} \rightarrow \tilde{t}_{j} \tilde{t}_{j}$ &~\ref{hststNMSSM} & $A_i \rightarrow \tilde{W_{j}} \tilde{W_{j}}$ &~\ref{AchchNMSSM} \\ \hline
$h_{i} \rightarrow \tilde{t}_{1} \tilde{t}_{2}$ &~\ref{hst1st2NMSSM} & $A_i \rightarrow \tilde{W_{1}} \tilde{W_{2}}$ &~\ref{Ach1ch2NMSSM} \\ \hline
$h_{i} \rightarrow \tilde{b}_{j} \tilde{b}_{j}$ &~\ref{hststNMSSM} & $A_i \rightarrow h_j Z$ &~\ref{AhZNMSSM} \\ \hline
$h_{i} \rightarrow \tilde{b}_{1} \tilde{b}_{2}$ &~\ref{hst1st2NMSSM} & $A_i \rightarrow H^{\pm} W$ &~\ref{AHpmWNMSSM} \\ \hline
$h_{i} \rightarrow \tilde{\tau}_{j} \tilde{\tau}_{k}$ &~\ref{hstaustauNMSSM} & $A_i \rightarrow \gamma \gamma$ &~\ref{AgamgamNMSSM} \\ \hline
$h_{i} \rightarrow \tilde{W}_j \tilde{W}_j$ &~\ref{hchchNMSSM} & $A_i \rightarrow Z \gamma$ &~\ref{AZgamNMSSM} \\ \hline
$h_{i} \rightarrow \tilde{W}_1 \tilde{W}_2$ &~\ref{hch1ch2NMSSM} & $A_i \rightarrow gg $ &~\ref{AggNMSSM},~\ref{AggQCDcorrNMSSM} \\ \hline
$h_{i} \rightarrow \tilde{Z}_j \tilde{Z}_k$ &~\ref{hneutneutNMSSM} & \multicolumn{2}{c|}{\bf{Decays into Higgs Bosons} } \\ \hline
$h_{i} \rightarrow A_{j} A_{k}$ &~\ref{hAANMSSM} & $\tilde{b}_2 \rightarrow \tilde{b}_1 h_{i}$ &~\ref{b2b1hNMSSM} \\ \hline
$h_i \rightarrow A_j Z$ &~\ref{hAZNMSSM} & $\tilde{t}_2 \rightarrow \tilde{t}_1 h_{i}$ &~\ref{t2t1hNMSSM} \\ \hline
$h_i \rightarrow H^+ H^-$ &~\ref{hHpmHpmNMSSM} & $\tilde{\tau}_2 \rightarrow \tilde{\tau}_1 h_{i}$ &~\ref{stau2stau1hNMSSM} \\ \hline
$h_i \rightarrow W^+ H^-$ &~\ref{hWHpmNMSSM} & $\tilde{b}_2 \rightarrow \tilde{b}_1 A_{i}$ &~\ref{sb2sb1ANMSSM} \\ \hline
$h_i\rightarrow ZZ^*$ &~\ref{hZZstarNMSSM} & $\tilde{t}_2 \rightarrow \tilde{t}_1 A_{i}$ &~\ref{sb2sb1ANMSSM} \\ \hline
$h_i \rightarrow WW^*$ &~\ref{hWWstarNMSSM} & $\tilde{\tau}_2 \rightarrow \tilde{\tau}_1 A_{i}$ &~\ref{sb2sb1ANMSSM} \\ \hline
$h_i \rightarrow ZZ$ &~\ref{hZZstarNMSSM} & $\tilde{W}_2 \rightarrow \tilde{W}_1 h_{i}$ &~\ref{ch2ch1hNMSSM} \\ \hline
$h_i \rightarrow WW$ &~\ref{hWWstarNMSSM} & $\tilde{W}_2 \rightarrow \tilde{W}_1 A_{i}$ &~\ref{ch2ch1ANMSSM} \\ \hline
$h_i \rightarrow \gamma \gamma$ &~\ref{hgamgamNMSSM} & $h_i \rightarrow h_j h_k$ &~\ref{hHH3NMSSM} \\ \hline
$h_i \rightarrow gg$ &~\ref{hggNMSSM}~\ref{hggQCDcorrNMSSM} & $h_i \rightarrow h_j h_k$ &~\ref{hHH3NMSSM} \\ \hline
$h_i \rightarrow Z\gamma$ &~\ref{hZgamNMSSM} & $A_2 \rightarrow A h_i$ &~\ref{A2AhiNMSSM} \\ \hline
\end{tabular}
\caption{The NMSSM decays involving neutral Higgs Bosons that are included in the {\tt SOFTSUSY} decay program, the references for the formulae in the appendices are given, where two references are given the first is for the leading order case and the second for the QCD-corrected case. Note any decays not involving neutralinos or neutral Higgs Bosons are the same as in the MSSM.}
\label{NMSSMHiggsdecaysreftable}
\end{table}
\end{center}

\section{Kinematic Functions} \label{kin}

Here we begin the list of partial width expressions used in calculating the decay branching ratios in {\tt SOFTSUSY}, we hope this provides a useful reference. With the exception of the 3 body decays, the majority of these widths were rederived as a form of validation. 

The following are a list of commonly occurring functions that arise from the kinematics of the decays:
$\tilde{\lambda}^{1/2}$ appears as a result of the phase space integration:
\begin{equation}
\tilde{\lambda}^{1/2} (m_{1}, m_2, m_{3}) = \sqrt{\left(1- ({m_2 + m_3 \over m_1})^2\right)\left(1-({m_2 - m_3 \over m_1})^2\right)}.
\end{equation}
For loop integrals the real and imaginary parts of the loop give the following kinetic factor, where $\tau_{a} = 4({m_{a} \over m_{h_i}})^2$:
\begin{equation} \label{ftau}
f(\tau) = \begin{cases}
			[\sin^{-1} ({1 \over \sqrt{\tau}})]^2, \: $for$ & \: \: \tau \geq 1, \\
			-{1 \over 4} [\ln({1+\sqrt{1-\tau} \over 1-\sqrt{1-\tau}}) - i \pi]^2, \: $for$ & \: \: \tau < 1, \\
			\end{cases}
\end{equation}
this is where the real and imaginary parts come from.

For the $Z\gamma$ decay loops, the kinetic factor $g(\tau)$ also occurs:
\begin{equation}
g(\tau) = \begin{cases} \label{gtau}
			\sqrt{\tau - 1} \sin^{-1} ({1 \over \sqrt{\tau}}), \: $for$ & \: \: \tau \geq 1, \\
			{1 \over 2}\sqrt{1-\tau} [\ln\left({1+\sqrt{1-\tau} \over 1-\sqrt{1-\tau}}\right) - i \pi], \: $for$ & \: \: \tau < 1 \\
			\end{cases}
\end{equation}

\section{MSSM Two Body Decay Formulae} \label{appendix:MSSM2body}

Here we list for ease of reference the formulae for the partial widths of each of the $1 \rightarrow 2$ decay modes
incorporated into the decay calculator “{\tt {\tt SOFTSUSY}}”. The $1 \rightarrow 2$ decay widths were all rederived, the book by Baer and Tata \cite{TataBaer} was used as a guide, however differences exist relative to their formulae. The formulae provided in {\tt SUSYHIT} \cite{Djouadi:2006bz,Djouadi:1997yw,Djouadi:2005} also provided a useful check.

\subsection{Gluinos} \label{Gluinos}
The partial widths for the decays of the gluinos to squarks and quarks are:
\begin{equation}\label{gluqqLR}
\Gamma(\tilde{g} \rightarrow q \tilde{q}_{L/R}) = {\alpha_{S} \over 4} {1 \over 2m_{\tilde{g}}} (1 + {m_{q}^2 \over m_{\tilde{g}}^2} - {m_{\tilde{q}_{L/R}}^2 \over m_{\tilde{g}}^2}) \tilde{\lambda}^{1/2} (m_{\tilde{g}}, m_q, m_{\tilde{q}_{L/R}}), 
\end{equation} 
\begin{equation}\label{gluqq12}
\Gamma(\tilde{g} \rightarrow q \tilde{q}_{1/2}) = {\alpha_{S} \over 4} {1 \over 2m_{\tilde{g}}} [1 + {m_{q}^2 \over m_{\tilde{g}}^2} - {m_{\tilde{q}_{L/R}}^2 \over m_{\tilde{g}}^2} \mp 2\sin2\theta_q {m_q \over m_{\tilde{g}}}]  \tilde{\lambda}^{1/2} (m_{\tilde{g}}, m_q, m_{\tilde{q}_{L/R}}),
\end{equation} 
where the minus/plus sign applies for $\tilde{q}_1$/$\tilde{q}_2$ respectively.

\subsection{Squarks} \label{Squarks}

The partial widths for the decays of the squarks to quarks are:
\begin{equation}\label{qLRqglu}
\Gamma(\tilde{q}_{L/R} \rightarrow q \tilde{g} ) = {4\alpha_{S} \over 3} {1 \over 2m_{\tilde{q}_{L/R}} } (1 - {m_{q}^2 \over m_{\tilde{q}_{L/R}^2}} - {m_{\tilde{g}}^2 \over m_{\tilde{q}_{L/R}^2}}) \tilde{\lambda}^{1/2} (m_{\tilde{q}_{L,R}}, m_q, m_{\tilde{g}}),
\end{equation} 
\begin{equation}\label{q12qglu}
\Gamma(\tilde{q}_{1/2} \rightarrow q \tilde{g} ) = {4\alpha_{S} \over 3} {1 \over 2m_{\tilde{q}_{1/2}} } (1 - {m_{q}^2 \over m_{\tilde{q}_{1/2}^2}} - {m_{\tilde{g}}^2 \over m_{\tilde{q}_{1/2}^2}} \pm 2\sin2\theta_q{m_qm_{\tilde{g}} \over m_{\tilde{q}_{1,2}}^2}) \tilde{\lambda}^{1/2} (m_{\tilde{q}_{1,2}}, m_q, m_{\tilde{g}}),
\end{equation} 
where the minus/plus sign applies for $\tilde{q}_1$/$\tilde{q}_2$ respectively.
\begin{equation}\label{qLqpch1}
\Gamma(\tilde{q}_{L} \rightarrow q^{'} \tilde{W}_{1}^- ) = {g^2 \over \sin\theta_{L/R}^2 } {m_{\tilde{q}_{L}} \over 16\pi} ( 1 - {m_{\tilde{W}_{1}^{-}}^2 \over m_{\tilde{q}_{L}}^2} - {m_{q}^2 \over m_{\tilde{q}_{L}}^2 }) \tilde{\lambda}^{1/2} (m_{\tilde{q}_{L}}, m_q, m_{\tilde{W_{1}^-}}).
\end{equation} 
Note here the $'$ indicates that the quark produced is the opposite type to the squark (so $\tilde{d}_L$
produces an up quark for example) and $\theta_L$ is for when up-type quarks (i.e.\ up or charm) are produced
and $\theta_R$ is for when down-type quarks are produced (i.e.\ down or strange). The expression \eqref{qLqpch1} applies for the first two generations of quarks as no mixing has been accounted for. The formula for decay to $\tilde{W}_{2}^{-}$ is similar but $\sin \theta_{L/R} \rightarrow \cos \theta_{L/R}$). The expressions with sfermion mixing, for the third generation of squarks, are given below.
\begin{equation}\label{b1tch1}
\begin{aligned}
\Gamma(\tilde{b}_{1} \rightarrow \tilde{W}_{1}^- t ) = & {m_{\tilde{b}_1} \over 16\pi} \Big[\{ ({-g\sin\theta_L\cos\theta_b + f_b\cos\theta_L\sin\theta_b})^2 + ({-f_t\cos\theta_R\cos\theta_b})^2 \} \\ & \times (1 - {m_{\tilde{W}_{1}}^2 \over m_{\tilde{b}_{1}}^2 } - {m_{t}^2 \over m_{\tilde{b}_{1}}^2 } ) + 4 {m_t m_{\tilde{W}_1} \over m_{\tilde{b}_1}^2 }({-g\sin\theta_L\cos\theta_b + f_b\cos\theta_L\sin\theta_b})\\ &  \times (-f_{t}\cos\theta_R\cos\theta_b)\Big]  \tilde{\lambda}^{1/2}(m_{\tilde{b}_1},m_t,m_{\tilde{W}_1}) ,
\end{aligned}
\end{equation} 
\begin{equation}\label{t1bch1}
\begin{aligned}
\Gamma(\tilde{t}_{1} \rightarrow \tilde{W}_{1}^+ b ) = & {m_{\tilde{t}_1} \over 16\pi} \Big[\{ ({-g\sin\theta_R\cos\theta_t + f_t\cos\theta_R\sin\theta_t})^2 + ({-f_b\cos\theta_L\cos\theta_t})^2\} \\ & \times (1 - {m_{\tilde{W}_{1}}^2 \over m_{\tilde{t}_{1}}^2 } - {m_{b}^2 \over m_{\tilde{t}_{1}}^2 } ) + 4 {m_b m_{\tilde{W}_1} \over m_{\tilde{t}_1}^2 }({-g\sin\theta_R\cos\theta_t + f_t\cos\theta_R\sin\theta_t})\\ & \times (-f_{b}\cos\theta_L\cos\theta_t)\Big]  \tilde{\lambda}^{1/2}(m_{\tilde{t}_1},b,m_{\tilde{W}_1}) ,
\end{aligned}
\end{equation} 
where 
\begin{equation} \label{ftfb}
f_t = {g m_t^{run} \over \sqrt{2}m_W\sin\beta }, \quad \quad f_b = {g m_b^{run} \over \sqrt{2}m_W\cos\beta }.
\end{equation}
(For decays of stops and sbottoms to $\tilde{W}_{2}^+$, $\sin
\theta_{L/R}\rightarrow \cos \theta_{L/R}$ and $\cos \theta_{L/R}\rightarrow
-\sin \theta_{L/R}$, and for decays of $\tilde{b}_2$ 
and $\tilde{t}_2$, $\sin \theta_{t/b} \rightarrow
\cos \theta_{t/b}$ and $\cos \theta_{t/b}\rightarrow -\sin \theta_{t/b}$.)
The squark decays to neutralinos are given by:
\begin{equation} \label{qLqneut}
\begin{aligned}
\Gamma(\tilde{q}_L \rightarrow \tilde{Z}_i q ) = & {1 \over 2} {(\pm g N_{2i} + {g' \over 3} N_{1i})}^2 {m_{\tilde{q}_L}  \over 16\pi} (1 - {m_{\tilde{Z}_i}^2 \over m_{\tilde{q}_L}^2 } - {m_{q}^2 \over m_{\tilde{q}_L}^2 } ) \tilde{\lambda}^{1/2}(m_{\tilde{q}_L}, m_q, m_{\tilde{Z}_i}),
\end{aligned}
\end{equation} 
\begin{equation} \label{qRqneut}
\begin{aligned}
\Gamma(\tilde{q}_R \rightarrow \tilde{Z}_i q ) = & {1 \over 2} {( {a \over 3} g'  N_{1i})}^2 {m_{\tilde{q}_R}  \over 16\pi} (1 - {m_{\tilde{Z}_i}^2 \over m_{\tilde{q}_R}^2 } - {m_{q}^2 \over m_{\tilde{q}_R}^2 } ) \tilde{\lambda}^{1/2}(m_{\tilde{q}_R}, m_q, m_{\tilde{Z}_i}),
\end{aligned}
\end{equation}
where $a = 4$ for up type squarks and $a = -2$ for down type squarks). $N_{ji}$ are neutralino mixing matrix elements.
Decays of $\tilde{t}_1$, $\tilde{b}_1$, $\tilde{t}_2$ and $\tilde{b}_2$ are
similar except for the mixing of the $L$ and $R$ parts so we get a linear
combination of the two pre-factors involving the neutralino mixing matrix
elements $N_{ji}$ with weights which are sines and cosines of the mixing angle
$\theta_{t/b}$. In addition the Higgsino components of the neutralinos become
important: 
\begin{equation}\label{t1tneut}
\begin{aligned}
\Gamma(\tilde{t}_1 \rightarrow \tilde{Z}_i t ) = & {m_{\tilde{t}_1} \over 8\pi}  \tilde{\lambda}^{1/2}(m_{\tilde{t}_1}, m_t, m_{\tilde{Z}_i}) \Big[\frac{1}{2}\{a^2(1-({m_t + m_{\tilde{Z}_i} \over m_{\tilde{t}_1}})^2) + b^2(1-({m_t - m_{\tilde{Z}_i} \over m_{\tilde{t}_1}})^2)\} \Big],
\end{aligned}
\end{equation} 
where
\begin{equation}
a = \frac{1}{2}\left[{1 \over \sqrt{2} }\cos\theta_{t}[-g N_{2i} - {g' \over 3} N_{1i}] - f_u \sin\theta_t N_{4i}  +  {4 \over 3\sqrt{2}} g' N_{1i} sin\theta_t - f_u N_{4i}\cos\theta_t\right],
\end{equation}
\begin{equation}
b = \frac{1}{2}\left[{1 \over \sqrt{2} }\cos\theta_{t}[-g N_{2i} - {g' \over 3} N_{1i}] - f_u \sin\theta_t N_{4i}  -  {4 \over 3\sqrt{2}} g' N_{1i}\sin\theta_t + f_u N_{4i}\cos\theta_t\right].
\end{equation}
Whilst
\begin{equation}\label{b1bneut}
\Gamma (\tilde{b}_1 \rightarrow \tilde{Z}_i b ) = {m_{\tilde{b}_1} \over 8\pi} \tilde{\lambda}^{1/2} (m_{\tilde{b}_1}, m_b, m_{\tilde{Z}_i}) \left[a^2 (1- ({m_{\tilde{Z}_i} + m_b \over m_{\tilde{b}_1}})^2) + b^2 (1- ({m_{\tilde{Z}_i} - m_b \over m_{\tilde{b}_1}})^2)\right],
\end{equation} 
where
\begin{equation}
\begin{aligned}
a = \frac{1}{2}\left[{1 \over \sqrt{2}}(\cos\theta_b[-N_{1i}{g' \over 3} + N_{3i}g)] - \sin\theta_b N_{3i}f_d - {2 \over 3\sqrt{2}}\sin{\theta_b}N_{1i}g' - \cos\theta_bf_d N_{3i}\right],
\end{aligned}
\end{equation}
\begin{equation}
\begin{aligned}
b = \frac{1}{2}\left[{1 \over \sqrt{2}}(\cos\theta_b[-N_{1i}{g' \over 3} + N_{3i}g)] - \sin\theta_b N_{3i}f_d + {2 \over 3\sqrt{2}}\sin{\theta_b}N_{1i}g' + \cos\theta_bf_d N_{3i}\right].
\end{aligned}
\end{equation}
As usual if we instead consider $\tilde{q}_2$, make the changes $m_{q_1} \rightarrow m_{q_2}$, $\cos\theta_q \rightarrow \sin\theta_q$ and $\sin\theta_q \rightarrow -\cos\theta_q$.
\begin{equation}\label{t1b1W}
\begin{aligned}
\Gamma (\tilde{t}_1 \rightarrow \tilde{b}_1 W^+ ) = {g^2 \over 32\pi} {m_{\tilde{t}_1}^3 \over m_{W}^2} {\tilde{\lambda}^{3/2}}(m_{\tilde{t}_1},m_W,m_{\tilde{b}_1}) \cos^2\theta_t \cos^2\theta_b.
\end{aligned}
\end{equation} 
For $\tilde{t}_2$, $\cos\theta_t \rightarrow \sin\theta_t$, whereas for
$\tilde{b}_2$ then $\cos\theta_b \rightarrow \sin\theta_b$. 
If the sbottoms are the initial states and stops are in the final state then
exchange $m_{\tilde{t}_i}$ and $m_{\tilde{b}_i}$. 
For the decays to charged Higgs bosons:
\begin{equation}\label{t1b1Hpm}
\begin{aligned}
\Gamma (\tilde{t}_1 \rightarrow \tilde{b}_1 H^+ ) = {g^2 \over 32\pi m_{\tilde{t}_1} m_{W}^{2}} A^2 \tilde{\lambda}^{1/2} ( m_{\tilde{t}_1}, m_{H^+}, m_{\tilde{b}_1} ),
\end{aligned}
\end{equation} 
where here
\begin{equation}
\begin{aligned}
A = & m_t m_b (\tan\beta + \cot\beta)\sin\theta_t\sin\theta_b +  m_t(\mu + A_t \cot\beta)\sin\theta_t\cos\theta_b + m_b(\mu + A_b \tan\beta)\sin\theta_b\cos\theta_t \\ &  + (m_b^2 \tan\beta + m_t^2 \cot\beta - m_W^2 \sin2 \beta) \cos\theta_t\cos\theta_b.
\end{aligned}
\end{equation}
If instead we have $\tilde{t}_2$ then $\cos\theta_t \rightarrow \sin\theta_t$ and if we have $\tilde{b}_2$ then $\cos\theta_b \rightarrow \sin\theta_b$ and again if the sbottoms are the initial states and the stops are the final state then exchange $m_{\tilde{t}_i}$ and $m_{\tilde{b}_i}$.
\begin{equation}\label{t2t1phi}
\begin{aligned}
\Gamma (\tilde{t}_2 \rightarrow \phi \tilde{t}_1) = {A_{\phi}^2 \over 16\pi m_{\tilde{t}_2} } \tilde{\lambda}^{1/2} (m_{\tilde{t}_2}, m_\phi, m_{\tilde{t}_1}),
\end{aligned}
\end{equation} 
where
\begin{equation}
\begin{aligned}
A_h = {g m_W \over 4} \sin(\beta+\alpha)\Big[1- {5 \over 3}{g'^2 \over g^2}\Big]\sin2\theta_t + {g m_t \over 2m_{W}\sin\beta}\cos2\theta_t (A_t \cos\alpha + \mu \sin\alpha),
\end{aligned}
\end{equation}
$A_H$ is similar but we must transform $\cos\alpha \rightarrow -\sin\alpha$ and $\sin\alpha \rightarrow \cos\alpha$, whilst
\begin{equation}
\begin{aligned}
A_A = {g m_t \over 2 m_W} (A_t \cot\beta + \mu).
\end{aligned}
\end{equation}
For $\tilde{b}_2$ decaying to a Higgs and $\tilde{b}_1$:
\begin{equation}\label{b2b1phi}
\begin{aligned}
\Gamma (\tilde{b}_2 \rightarrow \phi \tilde{b}_1) = {B_{\phi}^2 \over 16\pi m_{\tilde{b}_2} } \tilde{\lambda}^{1/2} (m_{\tilde{b}_2}, m_\phi, m_{\tilde{b}_1}),
\end{aligned}
\end{equation} 
where
\begin{equation}
\begin{aligned}
B_h = g m_W \sin(\alpha + \beta) {1 \over 4} [-1 + {1 \over 3}{g'^2 \over g^2} \sin2 \theta_b] + g m_b \cos2 \theta_b {1 \over 2 m_W \cos{\beta}} [ -A_b \sin\alpha - \mu \cos\alpha],
\end{aligned}
\end{equation}
$B_H$ is similar but again we must transform $\cos\alpha \rightarrow -\sin\alpha$ and $\sin\alpha \rightarrow \cos\alpha$, whilst
\begin{equation}
\begin{aligned}
B_A = {g m_b \over 2 m_W} (A_b \tan\beta +\mu).
\end{aligned}
\end{equation}
For third generation squark decays to Z bosons we have the following (note
that the amplitude is proportional to the sine squared of the mixing so this
does not occur for the first two generations): 
\begin{equation}\label{q2q1Z}
\begin{aligned}
\Gamma (\tilde{q}_2 \rightarrow Z \tilde{q}_1) = {g^2 m_{\tilde{q}_2}^3 \over 64\pi m_{Z}^2 \cos^2 \theta_W} \tilde{\lambda}^{3/2}(m_{\tilde{q}_2}, m_{\tilde{q}_1}, m_Z) \cos^2 \theta_q \sin^2 \theta_q.
\end{aligned}
\end{equation} 

\subsection{Sleptons} \label{Sleptons}
\begin{equation}\label{lLlneut}
\begin{aligned}
\Gamma (\tilde{l}_L \rightarrow l \tilde{Z}_i) = \frac{1}{2}[g N_{2i} + g' N_{1i}]^2 {m_{\tilde{l}_L} \over 16\pi} (1 - {m_{\tilde{Z}_i}^2 \over m_{\tilde{l}_L}^2} - {m_{l}^2 \over m_{\tilde{l}_L}^2}) \tilde{\lambda}^{1/2}(m_{\tilde{l}_L}, m_l, m_{\tilde{Z}_i}).
\end{aligned} 
\end{equation} 
\begin{equation}\label{lRlneut}
\begin{aligned}
\Gamma (\tilde{l}_R \rightarrow l \tilde{Z}_i) = \frac{1}{2}[g' N_{1i}]^2 {m_{\tilde{l}_R} \over 16\pi} (1 - {m_{\tilde{Z}_i}^2 \over m_{\tilde{l}_R}^2} - {m_{l}^2 \over m_{\tilde{l}_R}^2}) \tilde{\lambda}^{1/2}(m_{\tilde{l}_R}, m_l, m_{\tilde{Z}_i}).
\end{aligned}
\end{equation} 
\begin{equation}\label{snulnulneut}
\begin{aligned}
\Gamma (\tilde{\nu}_l \rightarrow \nu_l \tilde{Z}_i) = \frac{1}{2}[g N_{2i} - g' N_{1i}]^2 {m_{\tilde{\nu}_l} \over 16\pi} (1 - {m_{\tilde{Z}_i}^2 \over m_{\tilde{\nu}_l}^2} )^2.
\end{aligned}
\end{equation} 
\begin{equation}\label{lLnulch1}
\begin{aligned}
\Gamma (\tilde{l}_L \rightarrow \nu_l \tilde{W}_{1}^{-}) = {g^2 \sin^2 \theta_L \over 16\pi} m_{\tilde{l}_L} (1 - {m_{\tilde{W}_1}^2 \over m_{\tilde{l}_L}^2} )^2.
\end{aligned}
\end{equation} 
For decays to $\tilde{W}_2$ make the replacement $\sin\theta_L \rightarrow \cos\theta_L$.
\begin{equation}\label{nullch1}
\begin{aligned}
\Gamma (\tilde{\nu}_l \rightarrow l \tilde{W}_{1}^{+}) = {g^2 \sin^2 \theta_R \over 16\pi} m_{\tilde{\nu}_l} (1 - {m_{\tilde{W}_1}^2 \over m_{\tilde{\nu}_l}^2} - {m_{l}^2 \over m_{\tilde{\nu}_l}^2} )^2 \tilde{\lambda}^{1/2} (m_{\tilde{\nu}_l}, m_l, m_{\tilde{W}_1}).
\end{aligned}
\end{equation} 
For decays to $\tilde{W}_2$ make the replacement $\sin\theta_R \rightarrow \cos\theta_R$.
\begin{equation}\label{stau1tauneut}
\begin{aligned}
\Gamma (\tilde{\tau}_1 \rightarrow \tau \tilde{Z}_i) = {m_{\tilde{\tau}_1} \over 8\pi} \left[a^2 (1- ({m_\tau + m_{\tilde{Z}_i} \over m_{\tilde{\tau}_1}})^2) + b^2 (1- ({m_\tau - m_{\tilde{Z}_i} \over m_{\tilde{\tau}_1}})^2)\right] \tilde{\lambda}^{1/2}(m_{\tilde{\tau}_1}, m_\tau, m_{\tilde{Z}_i}),
\end{aligned}
\end{equation} 
where
\begin{equation}
a = \frac{1}{2}\left[{1 \over \sqrt{2}}\sin\theta_\tau (g N_{2i} + g' N_{1i}) + f_\tau N_{3i} \cos\theta\tau - \sqrt{2} g' N_{1i}\cos\theta\tau + f_\tau N_{3i} \sin\theta\tau\right],
\end{equation}
\begin{equation}
b = \frac{1}{2}\left[{1 \over \sqrt{2}}\sin\theta_\tau (g N_{2i} + g' N_{1i}) + f_\tau N_{3i} \cos\theta\tau + \sqrt{2} g' N_{1i}\cos\theta\tau + f_\tau N_{3i} \sin\theta\tau\right],
\end{equation}
and
\begin{equation} \label{funderscoretau}
f_\tau = {g m_\tau \over \sqrt{2}m_W \cos\beta}.
\end{equation}	
For $\tilde{\tau}_2$ decaying replace $m_{\tilde{\tau}_1} \rightarrow m_{\tilde{\tau}_2}$, $\cos\theta_\tau \rightarrow \sin\theta_\tau$ and $\sin\theta\tau \rightarrow -\cos\theta\tau$.
\begin{equation}\label{stau1nutauch1}
\begin{aligned}
\Gamma (\tilde{\tau}_1 \rightarrow \nu_\tau \tilde{W}_{1}^{-}) = [-g\sin\theta_L\sin\theta_\tau - f_\tau \cos\theta_L\cos\theta_\tau]^2 m_{\tilde{\tau}_1} (1 - {m_{\tilde{W}_1}^2 \over m_{\tilde{\tau}_1}^2})^2.
\end{aligned}
\end{equation} 
For decays to $\tilde{W}_2$ make the replacements $m_{\tilde{W}_1} \rightarrow m_{\tilde{W}_2}$, $\sin\theta_L \rightarrow \cos\theta_L$ and $\cos\theta_L \rightarrow -\sin\theta_L$, meanwhile for $\tilde{\tau}_2$ decays change $m_{\tilde{\tau}_1} \rightarrow m_{\tilde{\tau}_2}$, $\cos\theta_\tau \rightarrow \sin\theta_\tau$ and $\sin\theta_\tau \rightarrow -\cos\theta_\tau$.
\begin{equation}\label{snutautauch1}
\begin{aligned}
\Gamma (\tilde{\nu}_\tau \rightarrow \tau \tilde{W}_{1}^{+}) = {m_{\tilde{\nu}_\tau} \over 16\pi} \Big[& (g^2 \sin^2\theta_R + f_\tau^2 \cos^2\theta_L)(1 - {m_{\tilde{W}_1} \over m_{\tilde{\nu}_\tau}^2} -  {m_{\tau}^2 \over m_{\tilde{\nu}_\tau}^2})  \\ & - 4{m_\tau m_{\tilde{W}_1} \over m_{\tilde{\nu}_\tau}^2} g\sin\theta_R f_\tau \cos\theta_L \Big] \tilde{\lambda}^{1/2} (m_{\tilde{\nu}_\tau}, m_\tau, m_{\tilde{W}_1}).
\end{aligned}
\end{equation} 
For decays to $\tilde{W}_2$ then make the replacements $m_{\tilde{W}_1} \rightarrow m_{\tilde{W}_2}$, $\sin\theta_R \rightarrow \cos\theta_R$ and $\cos\theta_L \rightarrow -\sin\theta_L$.
\begin{equation}\label{stau1nutauHpm}
\begin{aligned}
\Gamma (\tilde{\tau}_1 \rightarrow \tilde{\nu}_\tau H^-) = {g^2 \over 32\pi m_W^2 m_{\tilde{\tau}_1}} \Big[m_\tau^2 \tan\beta\sin\theta_{\tau} - m_\tau (\mu + A_\tau \tan\beta) \cos\theta_{\tau}\Big]^2 \tilde{\lambda}^{1/2} (m_{\tilde{\tau}_1}, m_{\tilde{\nu}_\tau}, m_H^-).
\end{aligned}
\end{equation} 
For $\tilde{\tau}_2$ decays then one must make the changes $m_{\tilde{\tau}_1} \rightarrow m_{\tilde{\tau}_2}$, $\cos\theta_\tau \rightarrow \sin\theta_\tau$ and $\sin\theta_\tau \rightarrow -\cos\theta_\tau$ as usual.
\begin{equation}\label{stau1nutauW}
\begin{aligned}
\Gamma (\tilde{\tau}_1 \rightarrow \tilde{\nu}_\tau W^-) = {g^2 \sin^2\theta_\tau m_{\tilde{\tau}_1}^3 \over 32\pi m_W^2 } \tilde{\lambda}^{3/2} (m_{\tilde{\tau}_1}, m_{\tilde{\nu}_\tau}, m_W).
\end{aligned}
\end{equation} 
The equations for $\tilde{\tau}_{1/2} \rightarrow \tilde{\nu}_\tau W^-$ and $\tilde{\tau}_{1/2} \rightarrow \tilde{\nu}_\tau H^-$ can be used for $\tilde{\nu}_{\tau} \rightarrow \tilde{\tau}_{1/2} W^-$ and $\tilde{\nu}_{\tau} \rightarrow \tilde{\tau}_{1/2} H^-$ be interchanging the $m_{\tilde{\tau}_{1/2}} \leftrightarrow m_{\tilde{\nu}_{\tau}}$.
\begin{equation}\label{stau2stau1Z}
\begin{aligned}
\Gamma (\tilde{\tau}_2 \rightarrow \tilde{\tau}_1 Z) = {g^2 \sin^2\theta_{\tau} \cos^2 \theta_{\tau} m_{\tilde{\tau}_2}^3 \over 64\pi m_Z^2 \cos^2 \theta_W } \tilde{\lambda}^{3/2} (m_{\tilde{\tau}_2}. m_{\tilde{\tau}_1}, m_Z) ,
\end{aligned}
\end{equation} 
\begin{equation}\label{stau2stau1phi}
\begin{aligned}
\Gamma (\tilde{\tau}_2 \rightarrow \tilde{\tau}_1 \phi) = {{\tilde{A}_{\phi}}^2 \over 16 \pi m_{\tilde{\tau}_2} } \tilde{\lambda}^{1/2} (m_{\tilde{\tau}_2}, m_{\tilde{\tau}_1}, m_{\phi}) ,
\end{aligned}
\end{equation} 
where $\tilde{A}_h$ is
\begin{equation}
\begin{aligned}
\tilde{A}_h = {-g m_w \over 4} \sin(\alpha + \beta) \sin2\theta_{\tau}\Big[-1 + 3{g'^2 \over  g^2}\Big] + {g m_\tau \over 2 m_W \cos\beta}\cos2\theta_{\tau}(\mu\cos\alpha + A_\tau \sin\alpha),
\end{aligned}
\end{equation}
$\tilde{A}_H$ is the same as $\tilde{A}_h$ but with the changes $\cos\alpha \rightarrow -\sin\alpha$ and $\sin\alpha \rightarrow \cos\alpha$, meanwhile $\tilde{A}_A$ is:
\begin{equation}
\begin{aligned}
\tilde{A}_A = {g m_\tau \over 2 m_W} (\mu + A_\tau \tan\beta).
\end{aligned}
\end{equation}

\subsection{Charginos} \label{Charginos}

\begin{equation}\label{ch1qqL}
\begin{aligned}
\Gamma (\tilde{W}_1 \rightarrow \bar{q} \tilde{q}_L') = {3 m_{\tilde{W}_1} \over 32\pi} (g^2\sin^2\theta_{L/R}) (1- {m_{\tilde{q}_L}^2 \over m_{\tilde{W}_1}^2} + {{m_q}^2 \over m_{\tilde{W}_1}^2 }) \tilde{\lambda}^{1/2}(m_{\tilde{W}_1},m_q, m_{\tilde{q}_L}).
\end{aligned}
\end{equation} 
Note here the ' on the squark indicates it's of the opposite $SU(2)_L$ type to
the quark, e.g.\ if the quark 
is an up then the squark is a $\tilde{d}_L$. Also note that $\theta_L$ occurs when up-type quarks (i.e.\ up or charm)
are produced and $\theta_R$ is when down-type quarks are produced (i.e.\ down or strange).
(The formula for decay of $\tilde{W}_{2}^{-}$ is similar but we must change $\sin \theta_{L/R} \rightarrow \cos\theta_{L/R}$).
\begin{equation}\label{ch1bst1}
\begin{aligned}
\Gamma (\tilde{W}_1^+ \rightarrow \bar{b} \tilde{t}_{1}) = {3 m_{\tilde{W}_1} \over 32\pi} \left[(\mathcal{A}^2 + C^2 \sin^2\theta_t) (1 - {m_{\tilde{t}_1}^2 \over {m_{\tilde{W}_1}^2}} + {m_{b}^2 \over {m_{\tilde{W}_1}^2} }) + 4\mathcal{A}C\sin\theta_t {m_b \over m_{\tilde{W}_1}}\right] \tilde{\lambda}^{1/2}(m_{\tilde{W}_1},m_b, m_{\tilde{t}_1}),
\end{aligned}
\end{equation} 
where
\begin{equation}
\begin{aligned}
\mathcal{A} = g\sin\theta_R\cos\theta_t - f_u \cos\theta_R\sin\theta_t,
\end{aligned}
\end{equation}
\begin{equation}
\begin{aligned}
C = -f_d \cos\theta_L.
\end{aligned}
\end{equation}
For $\tilde{t}_2$ take $\cos\theta_t \rightarrow \sin\theta_t$, $\sin\theta_t \rightarrow -\cos\theta_t$ and $m_{\tilde{t}_1} \rightarrow m_{\tilde{t}_2}$.
For $\tilde{W}_2$ take $\cos\theta_R \rightarrow -\sin\theta_R$ , $\cos\theta_L \rightarrow -\sin\theta_L$ and $\sin\theta_R \rightarrow \cos\theta_R$, and also $m_{\tilde{W}_{1}} \rightarrow m_{\tilde{W}_{2}}$.
\begin{equation}\label{ch1tsb1}
\begin{aligned}
\Gamma (\tilde{W}_1^+ \rightarrow \bar{t} \tilde{b}_{1}) = {3 m_{\tilde{W}_1} \over 32\pi} \left[(\mathcal{A}^2 + C^2 \cos^2\theta_b) (1 - {m_{\tilde{b}_1}^2 \over {m_{\tilde{W}_1}^2}} + {m_{t}^2 \over {m_{\tilde{W}_1}^2} }) + 4\mathcal{A}C\cos\theta_b {m_t \over m_{\tilde{W}_1}}\right] \tilde{\lambda}^{1/2}(m_{\tilde{W}_1},m_t, m_{\tilde{b}_1}),
\end{aligned}
\end{equation} 
where now
\begin{equation}
\begin{aligned}
\mathcal{A} = -g\sin\theta_L\cos\theta_b + f_d \cos\theta_L\sin\theta_b,
\end{aligned}
\end{equation}
\begin{equation}
\begin{aligned}
C = f_u \cos\theta_R,
\end{aligned}
\end{equation}
\begin{equation}
\begin{aligned}
f_t = {g m_{t}^{run} \over \sqrt{2} m_W \sin{\beta} }, \: \: \: \: \: \: \: \: \: \: \: \: \: \: \: \: \: \: \: \: \: \: \: \: \:   f_b = {g m_{b}^{run} \over \sqrt{2} m_W \cos\beta}.
\end{aligned}
\end{equation}
For $\tilde{b}_2$ take $\cos\theta_b \rightarrow \sin\theta_b$, $\sin\theta_b \rightarrow -\cos\theta_b$ and o $m_{\tilde{b}_1} \rightarrow m_{\tilde{b}_2}$.
For $\tilde{W}_2$ take $\cos\theta_R \rightarrow -\sin\theta_R$, $\cos\theta_L \rightarrow -\sin\theta_L$ and $\sin\theta_L \rightarrow \cos\theta_L$, and also $m_{\tilde{W}_1} \rightarrow m_{\tilde{W}_2}$.
\begin{equation}\label{chllL}
\begin{aligned}
\Gamma (\tilde{W}_i^+ \rightarrow \bar{l} \tilde{l}_{L}) = {m_{\tilde{W}_i} \over 32\pi} A^2 (1- {m_{\tilde{l}_L}^2 \over m_{\tilde{W}_i}^2} + {m_{l}^2 \over m_{\tilde{W}_i}^2})  \tilde{\lambda}^{1/2}(m_{\tilde{W}_i},m_l, m_{\tilde{l}_L}),
\end{aligned}
\end{equation} 
where
\begin{equation}
  A=\begin{cases}
    -g \sin\theta_{L/R}, & \text{for $\tilde{W}_1$}.\\
    -g \cos\theta_{L/R}, & \text{for $\tilde{W}_2$}.
  \end{cases}
\end{equation}
$\theta_L$ is used for decays to $\nu_l$ and $\theta_R$ for decays to $\tilde{\nu}_{l_L}$.
\begin{equation}
\begin{aligned}\label{chtaunutau}
\Gamma (\tilde{W}_i^+ \rightarrow \bar{\tau} \tilde{\nu}_{\tau}) = {m_{\tilde{W}_i} \over 32\pi} \left[(A^2 + B^2)( 1 - {m_{\tilde{\nu}_{\tau}}^2 \over m_{\tilde{W}_{i}}^2} + {m_{\tau}^2 \over m_{\tilde{W}_{i}}^2 }) + 4AB {m_{\tau} \over m_{\tilde{W}_i}}\right] \tilde{\lambda}^{1/2}(m_{\tilde{W}_i},m_{\tau}, m_{\tilde{\nu}_{\tau}}),
\end{aligned}
\end{equation} 
where
\begin{equation}
  A=\begin{cases}
    g \sin\theta_{R}, & \text{for $\tilde{W}_1$},\\
    g \cos\theta_{R}, & \text{for $\tilde{W}_2$},
  \end{cases}
\end{equation}
\begin{equation}
  B=\begin{cases}
    -f_{\tau} \cos\theta_{L}, & \text{for $\tilde{W}_1$},\\
    f_{\tau}  \sin\theta_{L}, & \text{for $\tilde{W}_2$},
  \end{cases}
\end{equation}
and $f_\tau$ has been given before in \eqref{funderscoretau}.
\begin{equation}\label{chstau1nutau}
\begin{aligned}
\Gamma (\tilde{W}_i^+ \rightarrow \bar{\tilde{\tau}}_1 \nu_{\tau}) = {m_{\tilde{W}_i} \over 32\pi} \mathcal{A}^2 ( 1 - {m_{\tilde{\tau}_{1}}^2 \over m_{\tilde{W}_{i}}^2})^2 ,
\end{aligned}
\end{equation} 
where
\begin{equation}
\begin{aligned}
\mathcal{A} = -g\sin\theta_L\sin\theta_\tau - f_\tau \cos\theta_L\cos\theta_\tau.
\end{aligned}
\end{equation}
For $\tilde{W}_2$ make the replacements $\cos\theta_L \rightarrow -\sin\theta_L$, $\sin\theta_L \rightarrow \cos\theta_L$ and $m_{\tilde{W}_1} \rightarrow m_{\tilde{W}_2}$.
For $\tilde{\tau}_2$ make the replacements $\cos\theta_\tau \rightarrow \sin\theta_\tau$, $\sin\theta_\tau \rightarrow -\cos\theta_\tau$ and $m_{\tilde{\tau}_1} \rightarrow m_{\tilde{\tau}_2}$.
\begin{equation}\label{ch1Wneut}
\begin{aligned}
\Gamma (\tilde{W}_1^+ \rightarrow W \tilde{Z}_j) = {g^2 \over 16\pi|m_{\tilde{W}_1}| } \tilde{\lambda}^{1/2} (m_{\tilde{W}_1}, m_W, m_{\tilde{Z}_j}) \Big[(X^2 + Y^2)\Big(m_{\tilde{W}_1}^2 + m_{\tilde{Z}_j}^2 - m_{W}^2  \\ +{1 \over m_{W}^2}\{(m_{\tilde{W}_{1}}^2 - m_{\tilde{Z}_{j}}^2)^2 - m_{W}^4\}\Big) - 6(X^2 - Y^2)m_{\tilde{W}_1} m_{\tilde{Z}_j}) \Big] ,
\end{aligned}
\end{equation} 
where
\begin{equation}
\begin{aligned}
X = {1 \over 2} [\cos\theta_R N_{4j} {1 \over \sqrt{2}} - \sin\theta_R N_{2j} - \cos\theta_L N_{3j} {1 \over \sqrt{2}} - \sin\theta_L N_{2j}].
\end{aligned}
\end{equation}
$Y$ is the same as $X$ except the first two terms change sign.
For $\tilde{W}_2$ transform $\cos\theta_L \rightarrow -\sin\theta_L$, $\sin\theta_L \rightarrow \cos\theta_L$, $\cos\theta_R \rightarrow -\sin\theta_R$, $\sin\theta_R \rightarrow \cos\theta_R$ and change $m_{\tilde{W}_{1}} \rightarrow m_{\tilde{W}_2}$.
\begin{equation}\label{ch1Hpmneut}
\begin{aligned}
\Gamma (\tilde{W}_1^+ \rightarrow H^+ \tilde{Z}_j) = {1 \over 16\pi|m_{\tilde{W}_1}| } \tilde{\lambda}^{1/2} (m_{\tilde{W}_1}, m_{H^-}, m_{\tilde{Z}_j}) [(a^2 + b^2)(m_{\tilde{W}_{1}}^2 + m_{\tilde{Z}_{j}}^2 - m_{H^-}^2) \\ + 2(a^2 - b^2)m_{\tilde{W}_{1}}m_{\tilde{Z}_{j}}],
\end{aligned}
\end{equation} 
where
\begin{equation}
\begin{aligned}
a = {1 \over 2} (-\cos\beta A_2 + \sin\beta A_4), & \\
b = {1 \over 2} (-\cos\beta A_2 - \sin\beta A_4),
\end{aligned}
\end{equation}
and
\begin{equation}
\begin{aligned}
A_2 = -{1 \over \sqrt{2}} [g N_{2j} + g' N_{1j}]\cos\theta_{R} - g N_{4j} \sin\theta_{R},  & \\
A_4 = -{1 \over \sqrt{2}} [g N_{2j} + g' N_{1j}]\cos\theta_{L} + g N_{3j} \sin\theta_{L}.
\end{aligned}
\end{equation}
For $\tilde{W}_2$ change $\cos\theta_L \rightarrow -\sin\theta_L$, $\sin\theta_L \rightarrow \cos\theta_L$, $\cos\theta_R \rightarrow -\sin\theta_R$, $\sin\theta_R \rightarrow \cos\theta_R$ and $m_{\tilde{W}_{1}} \rightarrow m_{\tilde{W}_2}$.
\begin{multline}\label{ch2Zch1}
\Gamma (\tilde{W}_2 \rightarrow Z \tilde{W}_1) = {1 \over 64\pi m_{\tilde{W}_2} } \tilde{\lambda}^{1/2} (m_{\tilde{W}_2}, m_Z, m_{\tilde{W}_1}) {(gg')^2 \over g^2 + g'^2} \Big[({g \over g'} + {g' \over g})^2  [(x^2 + y^2) \Big(m_{\tilde{W}_{2}}^2  + m_{\tilde{W}_{1}}^2 - m_{Z}^2  + {1 \over m_{W}^2}(( m_{\tilde{W}_{1}}^2 - m_{\tilde{W}_{2}}^2)^2 -m_{Z}^4)\Big) \\  +6(x^2 - y^2)m_{\tilde{W}_{1}}m_{\tilde{W}_{2}}\Big] , 
\end{multline} 
where
\begin{equation}
x = {1 \over 2}(\sin \theta_L \cos \theta_L - \sin \theta_R \cos \theta_R),
\end{equation}
and $y$ is the same as $x$ except the second term changes sign.
\begin{equation}\label{ch2phich1}
\Gamma (\tilde{W}_2 \rightarrow \phi \tilde{W}_1) = {g^2 \over 32\pi m_{\tilde{W}_2} } \tilde{\lambda}^{1/2} (m_{\tilde{W}_2}, m_{\phi}, m_{\tilde{W}_1}) \Big[(S_{\phi}^2 + P_{\phi}^2) (m_{\tilde{W}_{2}}^2  + m_{\tilde{W}_{1}}^2 - m_{\phi}^2)  + 2(S_{\phi}^2 -P_{\phi}^2)m_{\tilde{W}_{1}}m_{\tilde{W}_{2}}\Big],
\end{equation} 
where
\begin{equation}
S_h = {1 \over 2} (-\sin \theta_R \sin \theta_L \sin \alpha - \cos \theta_L \cos \theta_R \cos \alpha + \sin \theta_L \sin \theta_R \cos \alpha + \cos \theta_L \cos \theta_R \sin \alpha )  ,
\end{equation}
$P_h$ is the same as $S_h$ except the first and second terms gain an extra minus sign (become +).
$S_H$ , $P_H$ are the same as $S_h$ and $P_h$ if you take $\sin\alpha \rightarrow -\cos\alpha$ and $\cos\alpha \rightarrow \sin\alpha$.
\begin{equation}
S_A = {1 \over 2} (\sin \theta_R \sin \theta_L \sin \beta - \cos \theta_L \cos \theta_R \cos \beta - \sin \theta_L \sin \theta_R \cos \beta + \cos \theta_L \cos \theta_R \sin \beta ).
\end{equation}
Again $P_A$ is the same as $S_A$ except the first two terms gain an additional minus sign.
\subsection{Neutralinos} \label{Neutralinos}

\begin{equation}\label{neutuuLR}
\Gamma (\tilde{Z}_i \rightarrow \bar{u} \tilde{u}_L/R) = {3 C^2 |m_{\tilde{Z}_i}| \over 32\pi} \tilde{\lambda}^{1/2} (m_{\tilde{Z}_i}, m_u, m_{\tilde{u}_{L/R}}) (1+ {m_{u}^2 \over m_{\tilde{Z}_i}^2}  - {m_{\tilde{u}_{L/R}}^2 \over m_{\tilde{Z}_i}^2}),
\end{equation} 
where
\begin{equation}
  C=\begin{cases}
    {1 \over \sqrt{2}}(-g N_{2i} - {g' \over 3} N_{1i}), & \text{for $\tilde{u}_L$},\\
    {-4 \over 3\sqrt{2}}g' N_{1i}, & \text{for $\tilde{u}_R$}.
  \end{cases}
\end{equation}
Neutralino decays to charm and $\tilde{c}_{L/R}$ have a similar expression.
For decays to down and $\tilde{d}_{L/R}$ or to strange and $\tilde{s}_{L/R}$ then $C$ for the $L$ component is the
same as above except $g \rightarrow -g$ and $C$ for the $R$ component has a factor of 2 rather than -4 in the
numerator of the pre-factor. The masses must also be changed appropriately. Note the
difference between the left-handed (LH) and right-handed
(RH) squark comes from the LH squark coupling to both the zino and
wino components of the neutralinos whereas the RH squark couples only to the zino components.
\begin{equation}\label{neutllLR}
\Gamma (\tilde{Z}_i \rightarrow \bar{l} \tilde{l}_L/R) = {C^2 |m_{\tilde{Z}_i}| \over 32\pi} \tilde{\lambda}^{1/2} (m_{\tilde{Z}_i}, m_l, m_{\tilde{l}_{L/R}}) (1+ {m_{l}^2 \over m_{\tilde{Z}_i}^2}  - {m_{\tilde{l}_{L/R}}^2 \over m_{\tilde{Z}_i}^2}),
\end{equation} 
where
\begin{equation}
  C=\begin{cases}
    {1 \over \sqrt{2}}(g N_{2i} + g' N_{1i}), & \text{for $\tilde{l}_L$},\\
    \sqrt{2} g' N_{1i}, & \text{for $\tilde{l}_R$}.
  \end{cases}
\end{equation}
Again the difference here between the $L$ and $R$ sleptons is due to the $L$ sleptons coupling to the wino
and zino components of the neutralinos whilst the $R$ sleptons couple only to the zino components.
\begin{equation}\label{neuttst1}
\Gamma (\tilde{Z}_i \rightarrow \bar{t} \tilde{t}_{1}) = {3 |m_{\tilde{Z}_i}| \over 16\pi} \tilde{\lambda}^{1/2} (m_{\tilde{Z}_i}, m_t, m_{\tilde{t}_{1}}) \left[a^2\{(1+{m_t \over m_{\tilde{Z}_i}})^2 - ({m_{\tilde{t}_1} \over m_{\tilde{Z}_i}})^2\} + b^2\{(1-{m_t \over m_{\tilde{Z}_i}})^2 - ({m_{\tilde{t}_1} \over _{\tilde{Z}_i}})^2\}\right],
\end{equation} 
where
\begin{equation}
a = {1 \over 2}(\alpha+\beta), \:  \:   \: \: \:  \: \:\:  \: \: \:  \: \: \:  \: \: \:  \: \: \:  \: \: \:  \: \: \: \:  b = {1 \over 2}(\alpha-\beta),
\end{equation}
\begin{equation}
\alpha = \cos \theta_t {1 \over \sqrt{2}} [-g N_{2i} - {g' \over 3} N_{1i}] - f_t \sin \theta_t N_{4i},
\end{equation}
\begin{equation}
\beta = {4 \over 3\sqrt{2}} g' N_{1i} \sin \theta_t - f_t \cos \theta_t N_{4i}.
\end{equation}
For $\tilde{t}_2$ take $\cos\theta_t \rightarrow \sin\theta_t$, $\sin\theta_t \rightarrow -\cos\theta_t$ and $m_{\tilde{t}_1} \rightarrow m_{\tilde{t}_2}$.
\begin{equation}\label{neutbsb1}
\Gamma (\tilde{Z}_i \rightarrow \bar{b} \tilde{b}_{1}) = {3 |m_{\tilde{Z}_i}| \over 16\pi} \tilde{\lambda}^{1/2} (m_{\tilde{Z}_i}, m_b, m_{\tilde{b}_{1}}) \left[a^2\{(1+{m_b \over m_{\tilde{Z}_i}})^2 - ({m_{\tilde{b}_1} \over m_{\tilde{Z}_i}})^2\} + b^2\{(1-{m_b \over m_{\tilde{Z}_i}})^2 - ({m_{\tilde{b}_1} \over _{\tilde{Z}_i}})^2\}\right],
\end{equation} 
where $a$ and $b$ are as before but the $\alpha$ and $\beta$ are different:
\begin{equation}
\alpha = \cos \theta_b {1 \over \sqrt{2}} [-{g' \over 3} N_{1i} + g N_{2i}] - f_b \sin \theta_b N_{3i},
\end{equation}
\begin{equation}
\beta = -\sin \theta_b {2 \over 3\sqrt{2}}g' N_{1i}  - \cos \theta_b f_b N_{3i}.
\end{equation}
For $\tilde{b}_2$ take $\cos\theta_b \rightarrow \sin\theta_b$, $\sin\theta_b \rightarrow -\cos\theta_b$ and 	$m_{\tilde{b}_1} \rightarrow m_{\tilde{b}_2}$.
\begin{equation}\label{neuttaustau1}
\Gamma (\tilde{Z}_i \rightarrow \bar{\tau} \tilde{\tau}_{1}) = {|m_{\tilde{Z}_i}| \over 16\pi} \tilde{\lambda}^{1/2} (m_{\tilde{Z}_i}, m_\tau, m_{\tilde{\tau}_{1}}) \left[ a^2\{(1-({m_\tau + m_{\tilde{\tau}_1} \over m_{\tilde{Z}_i}})^2\} + b^2\{(1-({m_\tau - m_{\tilde{\tau}_1} \over m_{\tilde{Z}_i}})^2\} \right],
\end{equation} 
where
\begin{equation}
a = {1 \over 2} (\alpha + \beta) ,
\end{equation}
\begin{equation}
b = {1 \over 2} (\beta - \alpha) ,
\end{equation}
\begin{equation}
\alpha = {1 \over \sqrt{2}} \sin \theta_\tau [g N_{2i} + g' N_{1i}] + f_\tau N_{3i} \cos \theta_\tau,
\end{equation}
\begin{equation}
\beta = -\sqrt{2} g' N_{1i} \cos \theta_\tau + f_\tau N_{3i} \sin \theta_\tau.
\end{equation}
For $\tilde{\tau}_2$ make the replacements $\cos\theta_\tau \rightarrow \sin\theta_\tau$, $\sin\theta_\tau \rightarrow -\cos\theta_\tau$ and $m_{\tilde{\tau}_1} \rightarrow m_{\tilde{\tau}_2}$.
\begin{multline}\label{neutWch1}
\Gamma (\tilde{Z}_i \rightarrow W \tilde{W}_{1}) = {g^2 \over 16\pi |m_{\tilde{Z}_i}|} \tilde{\lambda}^{1/2} (m_{\tilde{Z}_i}, m_W, m_{\tilde{W}_{1}}) \Big[(X^2 +Y^2) \Big(m_{\tilde{Z}_{i}}^2 + m_{\tilde{W}_1}^2 -m_W^2 + {1 \over m_W^2}\{(m_{\tilde{Z}_i}^2 - m_{\tilde{W}_1}^2)^2 - m_W^4\}\Big) \\ - 6(X^2 -Y^2)m_{\tilde{Z}_i}m_{\tilde{W}_1}\Big],
\end{multline} 
where
\begin{equation}
X = {1 \over 2}[\cos \theta_R N_{4i} {1 \over \sqrt{2}} - \sin \theta_R N_{2i} - \cos \theta_L N_{3i} {1 \over \sqrt{2}} - \sin \theta_L N_{2i}],
\end{equation}
and $Y$ is the same as $X$ except the first two terms get an extra minus sign. For $\tilde{W}_2$ change $\cos\theta_L \rightarrow -\sin\theta_L$, $\sin\theta_L \rightarrow \cos\theta_L$, $\cos\theta_R \rightarrow -\sin\theta_R$, $\sin\theta_R \rightarrow \cos\theta_R$ and $m_{\tilde{W}_1} \rightarrow m_{\tilde{W}_2}$.
\begin{equation}\label{neutHpmch1}
\Gamma (\tilde{Z}_j \rightarrow H^+ \tilde{W}_{1}) = {1 \over 16\pi |m_{\tilde{Z}_i}|} \tilde{\lambda}^{1/2} (m_{\tilde{Z}_j}, m_{H^+}, m_{\tilde{W}_{1}}) \Big[(a^2 +b^2) (m_{\tilde{Z}_{j}}^2 + m_{\tilde{W}_1}^2 -m_{H^+}^2) + 2(a^2 -b^2)m_{\tilde{Z}_j}m_{\tilde{W}_1}\Big],
\end{equation} 
where $a$ and $b$ and then $A_2$ and $A_4$ are exactly as given for the decay $\tilde{W}_1 \rightarrow H^+ Z_j$ in \eqref{ch1Hpmneut}.
\begin{multline}\label{neutZneut}
\Gamma (\tilde{Z}_i \rightarrow Z \tilde{Z}_{j}) = {g^2 + {g'}^2 \over 64\pi |m_{\tilde{Z}_i}|} \tilde{\lambda}^{1/2} (m_{\tilde{Z}_ij}, m_{Z}, m_{\tilde{Z}_{j}}) \{N_{4i}N_{4j} - N_{3i}N_{3j}\}^2 \\ \times \left[m_{\tilde{Z}_{i}}^2 + m_{\tilde{Z}_j}^2 -m_{Z}^2 +{1 \over m_Z^2}[(m_{\tilde{Z}_i}^2 - m_{\tilde{Z}_j}^2)^2 - m_Z^4] + 6m_{\tilde{Z}_i}m_{\tilde{Z}_j}\right].
\end{multline} 
\begin{equation}\label{neuthneut}
\Gamma (\tilde{Z}_i \rightarrow h \tilde{Z}_{j}) = {(X_{ij}^{h} + X_{ji}^{h})^2 \over 16\pi |m_{\tilde{Z}_i}|} \tilde{\lambda}^{1/2} (m_{\tilde{Z}_ij}, m_{h}, m_{\tilde{Z}_{j}}) [m_{\tilde{Z}_{i}}^2 + m_{\tilde{Z}_j}^2 -m_{h}^2 +2m_{\tilde{Z}_i}m_{\tilde{Z}_j}],
\end{equation} 
where
\begin{equation} \label{Xijh}
X_{ij}^{h} = {1 \over 2} [N_{3i} \sin \alpha + N_{4i}\cos \alpha] (-g N_{2j} + g' N_{1j}),
\end{equation}
$X_{ji}^{h}$ is the same but with $i \leftrightarrow j$.
For $\tilde{Z}_i \rightarrow H \tilde{Z}_j$ the formula is the same except one must change $\sin\alpha \rightarrow -\cos\alpha$, $\cos\alpha \rightarrow \sin\alpha$ and $m_h \rightarrow m_H$.
\begin{equation}\label{neutAneut}
\Gamma (\tilde{Z}_i \rightarrow A \tilde{Z}_{j}) = {(X_{ij}^{A} + X_{ji}^{A})^2 \over 16\pi |m_{\tilde{Z}_i}|} \tilde{\lambda}^{1/2} (m_{\tilde{Z}_ij}, m_{A}, m_{\tilde{Z}_{j}}) [m_{\tilde{Z}_{i}}^2 + m_{\tilde{Z}_j}^2 -m_{A}^2 -2m_{\tilde{Z}_i}m_{\tilde{Z}_j}],
\end{equation} 
where
\begin{equation} \label{XijA}
X_{ij}^{A} = {1 \over 2} [N_{3i} \sin \beta - N_{4i} \cos \beta] (-gN_{2j} + g'N_{1j}),
\end{equation}
and $X_{ji}^{A}$ is the same but with $i \leftrightarrow j$.

\subsection{Higgs Sector} \label{Higgs_Sector}

Once more, the partial widths for all of the Higgs decays incorporated into {\tt SOFTSUSY} were rederived, including for the three-body and 1-loop decays, however the majority of them can also be found in ``The Higgs Hunter's Guide" \cite{HHG}.
\begin{equation}\label{hqq}
\Gamma (h \rightarrow q \bar{q}) = {3g^2 m_h \over 32\pi} \Big({m_q \over m_W}\Big)^2 (1-4{m_q^2 \over m_h^2})^{3 \over 2} \mathcal{J}^2, 
\end{equation} 
where
\begin{equation}
\mathcal{J}=\begin{cases}
    {\cos \alpha \over \sin \beta}, & \text{for up type quarks (u,c,t)},\\
    {\sin \alpha \over \cos \beta}, & \text{for down type quarks (d,s,b)}.
  \end{cases}
\end{equation}
The same formulae apply for the decays to leptons, however without the factor of 3 which arises due to colour.
This is similar for $H \rightarrow qq$ except we must make the replacements $\sin\alpha \rightarrow -\cos\alpha$, $\cos\alpha \rightarrow \sin\alpha$ and $m_h \rightarrow m_H$.

With regards to the {\tt SOFTSUSY} spectrum generator, when the mixing parameter is set to $-1$ it considers
only third family Yukawa couplings to be non zero. This would mean no $h \rightarrow \mu \mu$ decay, which may be important phenomenologically in spite of its small branching ratio. In this case, for
the decay, we use the pole muon mass to calculate the branching ratio. 

\begin{equation}\label{Aqq}
\Gamma (A \rightarrow q \bar{q}) = {3 g^2 \mathcal{J}_A^2 \over 32\pi} \Big({m_q \over m_W}\Big)^2 m_A \sqrt{1 - 4 ({m_q \over m_A})^2)},
\end{equation} 
where
\begin{equation}
\mathcal{J}_A=\begin{cases}
    1 /(\tan \beta), & \text{for up type quarks (u,c,t)},\\
    \tan \beta, & \text{for down type quarks (d,s,b)}.
  \end{cases}
\end{equation}

Again, the same formulae apply for the decays to leptons, however without the factor of 3 which arises due to colour.

\begin{equation}\label{hneutneut}
\Gamma (h \rightarrow \tilde{Z}_i \tilde{Z}_j) = {|m_h| \over 8\pi} (X_{ij}^{h} + X_{ji}^{h})^2 \tilde{\lambda}^{1 \over 2} (m_h,m_{\tilde{Z}_i}, m_{\tilde{Z}_j}) \Big(1 - ({m_{\tilde{Z}_i} + m_{\tilde{Z}_j} \over m_h})^2\Big),
\end{equation} 
with an extra pre-factor of $\frac{1}{2}$ if $i=j$ (as the above formula
includes a pre-factor of 2 from $\tilde{Z}_i \tilde{Z}_j$ being
indistinguishable from $\tilde{Z}_j \tilde{Z}_i$). 
Here $X_{ij}^{h}$ is as in Eq. \eqref{Xijh} and $X_{ji}^{h}$ is the same but with $i \leftrightarrow j$.
Again similar formulae exist for $H \rightarrow \tilde{Z}_i \tilde{Z}_j$ except we transform $\sin\alpha \rightarrow -\cos\alpha$,  $\cos\alpha \rightarrow \sin\alpha$ and $m_h \rightarrow m_H$.
\begin{equation}\label{Aneutneut}
\Gamma (A \rightarrow \tilde{Z}_i \tilde{Z}_j) = {|m_A| \over 8\pi} (X_{ij}^{A} + X_{ji}^{A})^2 \tilde{\lambda}^{1 \over 2} (m_A,m_{\tilde{Z}_i}, m_{\tilde{Z}_j}) \Big(1 - ({m_{\tilde{Z}_i} + m_{\tilde{Z}_j} \over m_A})^2\Big),
\end{equation} 
here $X_{ij}^{A}$ is as given in Eq. \eqref{XijA} and $X_{ji}^{A}$ is the same but with $i \leftrightarrow j$.
\begin{equation}\label{phichch}
\Gamma (\phi \rightarrow \tilde{W}_i^+ \tilde{W}_i^-) = {g^2 |m_\phi| \over 4\pi} S^2  \tilde{\lambda}^{a} (m_\phi,m_{\tilde{W}_i}, m_{\tilde{W}_i}) ,
\end{equation} 
where $a=3/2$ for $\phi= h, H$ or $a=1/2$ for $A$ and $S$ is given by:
\begin{equation}
S = \begin{cases}
  	{1 \over 2} (-\sin \alpha \sin \theta_R \cos \theta_L + \cos \alpha \sin \theta_L \cos \theta_R), & $for$ :\ h \rightarrow \tilde{W}_1 \tilde{W}_1, \\
  	{1 \over 2} (\sin \alpha \cos \theta_R \sin \theta_L - \cos \alpha \cos \theta_L \sin \theta_R), & $for$ :\ h \rightarrow \tilde{W}_2 \tilde{W}_2, \\
  	{1 \over 2} (\cos \alpha \sin \theta_R \cos \theta_L + \sin \alpha \sin \theta_L \cos \theta_R), & $for$ :\ H \rightarrow \tilde{W}_1 \tilde{W}_1, \\
  	-{1 \over 2} (\cos \alpha \cos \theta_R \sin \theta_L + \sin \alpha \cos \theta_L \sin \theta_R), & $for$ :\ H \rightarrow \tilde{W}_2 \tilde{W}_2, \\
  	{1 \over 2} (\sin \beta \sin \theta_R \cos \theta_L + \cos \beta \sin \theta_L \cos \theta_R), & $for$ :\ A \rightarrow \tilde{W}_1 \tilde{W}_1, \\
  	-{1 \over 2} (\sin \beta \cos \theta_R \sin \theta_L + \cos \beta \cos \theta_L \sin \theta_R), & $for$ :\ A \rightarrow \tilde{W}_2 \tilde{W}_2. \\
  \end{cases}
\end{equation}
\begin{equation}\label{phichchdif}
\Gamma (\phi \rightarrow \tilde{W}_i^+ \tilde{W}_j^-) = {g^2 \over 16\pi |m_\phi|} \tilde{\lambda}^{1 \over 2} (m_\phi,m_{\tilde{W}_i}, m_{\tilde{W}_j}) \left[S^2 (1 - ({m_{\tilde{W}_i} + m_{\tilde{W}_j} \over m_\phi})^2) + P^2 (1 - ({m_{\tilde{W}_i} - m_{\tilde{W}_j} \over m_\phi})^2) \right],
\end{equation} 
where
\begin{equation}
S = \begin{cases}
  	{1 \over 2} (\sin \alpha \sin \theta_R \sin \theta_L + \cos \alpha \cos \theta_L \cos \theta_R - \sin \theta_L \sin \theta_R \cos \alpha - \cos \theta_L \cos \theta_R \sin \alpha), & $for$ :\ \phi = h, \\
  	{1 \over 2} (-\cos \alpha \sin \theta_R \sin \theta_L + \sin \alpha \cos \theta_L \cos \theta_R - \sin \theta_L \sin \theta_R \sin \alpha + \cos \theta_L \cos \theta_R \cos \alpha), & $for$ :\ \phi = H, \\
  	{1 \over 2} (-\sin \beta \sin \theta_R \sin \theta_L + \cos \beta \cos \theta_L \cos \theta_R + \sin \theta_L \sin \theta_R \cos \beta - \cos \theta_L \cos \theta_R \sin \beta), & $for$ :\ \phi = A. \\
    \end{cases}
\end{equation}
$P$ is the same as $S$ except the signs of the first two terms are reversed.
\begin{align}
\Gamma (h \rightarrow A A) &= {g^2 m_W^2\over 128\pi |m_h| \cos^4(\theta_W)} \tilde{\lambda}^{1 \over 2} (m_h,m_A, m_A) \sin^2 (\alpha + \beta) \cos^2 2\beta. \label{htoAA} \\  
\Gamma (H \rightarrow h h) &= {g^2 m_W^2 \over 128\pi |m_H| \cos^4(\theta_W)} \tilde{\lambda}^{1 \over 2} (m_H,m_h, m_h) \Big[\cos 2\alpha \cos(\alpha + \beta) - 2 \sin 2\alpha \sin(\alpha+\beta)\Big]^2. \label{Htohh} \\  
\Gamma (H \rightarrow A A) &= {g^2 m_W^2 \over 128\pi |m_H| \cos^4(\theta_W)} \tilde{\lambda}^{1 \over 2} (m_H,m_A, m_A) \cos^2 2\beta \cos^2(\alpha + \beta). \label{HtoAA} \\ 
\Gamma (H \rightarrow H^+ H^-) &= {g^2 m_W^2 \over 16\pi |m_H|} \tilde{\lambda}^{1 \over 2} (m_H,m_{H^+}, m_{H^-}) \left[\cos (\beta -\alpha) - {\cos(\alpha + \beta) \cos 2\beta \over 2\cos^2 \theta_W}\right]^2. \label{HHpmHpm}  \\ 
\Gamma (h \rightarrow A Z) &= {g^2 |m_h^3| \cos^2 (\beta -\alpha) \over 64\pi \cos^2 \theta_w m_{Z}^2} \tilde{\lambda}^{3 \over 2} (m_h,m_Z, m_A). \label{htoAZ}
\end{align} 
The decay $H \rightarrow AZ$ follows the same formula but with the changes $\cos(\beta - \alpha) \rightarrow \sin(\beta - \alpha)$ and $m_h \rightarrow m_H$.
\begin{equation}\label{AhZ}
\Gamma (A \rightarrow h Z) = {g^2 |m_A^3| \cos^2 (\beta -\alpha) \over 64\pi \cos^2 \theta_w m_{Z}^2} \tilde{\lambda}^{3 \over 2} (m_A,m_Z, m_h).
\end{equation} 
The decay $A \rightarrow HZ$ is not included as it's largely ruled out by SUSY mass constraints.
\begin{equation}\label{hqLRqLR}
\Gamma(h \rightarrow \tilde{q}_{L/R} \tilde{q}_{L/R}^*) = {3 \over 16 \pi m_h} \tilde{\lambda}^{1/2}(m_h,m_{\tilde{q}_{L/R}},m_{\tilde{q}_{L/R}}) \mathcal{C}_{h \tilde{q}_{L/R} \tilde{q}_{L/R}}^2,
\end{equation} 
where
\begin{equation}
\mathcal{C}_{h \tilde{q}_{L/R} \tilde{q}_{L/R}} = \begin{cases}
			g\Big[m_W ({1 \over 2} - {1 \over 6}{g'^2 \over g^2} )\sin(\beta+\alpha) - {m_{u}^2 \cos \alpha \over m_W \sin \beta}\Big], & $for: $\: \: \tilde{u}_L\bar{\tilde{u}}_L, \\
			g\Big[m_W (-{1 \over 2} - {1 \over 6}{g'^2 \over g^2} )\sin(\beta+\alpha) + {m_{d}^2 \sin \alpha \over m_W \cos \beta}\Big], & $for: $ \: \: \tilde{d}_L\bar{\tilde{d}}_L, \\
			g\Big[{2 \over 3}m_W{g'^2 \over g^2}\sin(\beta+\alpha) - {m_{u}^2 \cos \alpha \over m_W \sin \beta}\Big], & $for: $ \: \: \tilde{u}_R\bar{\tilde{u}}_R, \\
			g\Big[{-m_W \over 3}{g'^2 \over g^2}\sin(\beta+\alpha) + {m_{d}^2 \sin \alpha \over m_W \cos \beta}\Big], & $for: $ \: \: \tilde{d}_R\bar{\tilde{d}}_R, \\
			{g m_{u} \over 2 m_{W} \sin\beta}(\mu \sin\alpha + A_{u}\cos\alpha), & $for: $ \: \: \tilde{u}_L \bar{\tilde{u}}_R \: $or $\: \tilde{u}_R \bar{\tilde{u}}_L, \\ 
			{g m_{d} \over 2 m_{W} \cos\beta}(-\mu \cos\alpha - A_{d}\sin\alpha), & $for: $ \: \: \tilde{d}_L \bar{\tilde{d}}_R \: $or $\: \tilde{d}_R \bar{\tilde{d}}_L.\\
			\end{cases}
\end{equation}
\begin{equation}\label{HqLRqLR}
\Gamma(H \rightarrow \tilde{q}_{L/R} \tilde{q}_{L/R}^*) = {3 \over 16 \pi m_H} \tilde{\lambda}^{1/2}(m_{H},m_{\tilde{q}_{L/R}}, m_{\tilde{q}_{L/R}}) \mathcal{C}_{H \tilde{q}_{L/R} \tilde{q}_{L/R}}^2,
\end{equation} 
where
\begin{equation}
\mathcal{C}_{H \tilde{q}_{L/R} \tilde{q}_{L/R}} = \begin{cases}
			g\Big[-m_W ({1 \over 2} - {1 \over 6}{g'^2 \over g^2} )\cos(\beta+\alpha) - {m_{u}^2 \sin \alpha \over m_W \sin \beta}\Big], & $for: $\: \: \tilde{u}_L\bar{\tilde{u}}_L, \\
			g\Big[m_W ({1 \over 2} + {1 \over 6}{g'^2 \over g^2} )\cos(\beta+\alpha) - {m_{d}^2 \cos \alpha \over m_W \cos \beta}\Big], & $for: $\: \: \tilde{d}_L\bar{\tilde{d}}_L, \\
			g\Big[{-2m_W \over 3}{g'^2 \over g^2}\cos(\beta+\alpha) - {m_{u}^2 \sin \alpha \over m_W \cos \beta}\Big], & $for:  $\: \: \tilde{u}_R\bar{\tilde{u}}_R, \\
			g\Big[{m_W \over 3}{g'^2 \over g^2}\cos(\beta+\alpha) - {m_{d}^2 \cos \alpha \over m_W \cos \beta}\Big], & $for: $\: \: \tilde{d}_R\bar{\tilde{d}}_R, \\
			{g m_{u} \over 2 m_{W} \sin\beta}(-\mu \cos\alpha + A_{u}\sin\alpha), & $for: $\: \: \tilde{u}_L \bar{\tilde{u}}_R \: $or $\: \tilde{u}_R \bar{\tilde{u}}_L, \\
			{g m_{d} \over 2 m_{W} \cos\beta}(-\mu \sin\alpha + A_{d}\cos\alpha), & $for: $\: \: \tilde{d}_L \bar{\tilde{d}}_R \: $or $\: \tilde{d}_R \bar{\tilde{d}}_L. \\
			\end{cases}
\end{equation}

\begin{equation} \label{hlLRlLR}
\Gamma(h \rightarrow \tilde{l}_{L/R} \bar{\tilde{l}}_{L/R}) = {1 \over 16 \pi m_{h}}\tilde{\lambda}^{1/2}(m_h,m_{\tilde{l}_{L/R}}, m_{\tilde{l}_{L/R}}) \mathcal{C}_{h \tilde{l}_{L/R} \tilde{l}_{L/R}}^2,
\end{equation}
where
\begin{equation}
\mathcal{C}_{h \tilde{l}_{L/R} \tilde{l}_{L/R}} = \begin{cases}
			g\Big[m_{W}({1 \over 2}+{1 \over 2}{g'^2 \over g^2})\Big]\sin(\beta+\alpha), &$for $ \tilde{\nu}_L {\tilde{\nu}}_L^*, \\
			g\Big[m_{W}(-{1 \over 2}+{1 \over 2}{g'^2 \over g^2})\sin(\alpha+\beta) + {m_{\tilde{e}_L}^2 \sin\alpha \over m_{W} \cos\beta}\Big], &$for $ \tilde{e}_L \tilde{{e}}_L^*, \\
			g\Big[-m_{W}{g'^2 \over g^2}\sin(\alpha+\beta) + {m_{\tilde{e}_{L}}^2\sin\alpha \over m_{W} \cos\beta}\Big], &$for $ \tilde{e}_R {\tilde{e}}_R^*, \\
			{g m_{\tilde{e}_L} \over 2 m_{W} \cos \beta} (-\mu \cos \alpha - A_{e}\sin\alpha), &$for $ \tilde{e}_L {\tilde{e}}_R^* \: $or $\: \tilde{e}_R {\tilde{e}}_L^*. \\
			\end{cases}
\end{equation}
For third generation sfermions, the formulae are more complicated as a result of sfermion mixing and Yukawa coupling effects:
\begin{equation} \label{hstst}
\Gamma(h \rightarrow \tilde{t}_{i} \tilde{t}_{j}^*) = {3 \over 16 \pi m_{h}} \tilde{\lambda}^{1/2}(m_h,m_{\tilde{t}_i}, m_{\tilde{t}_j}) \mathcal{C}_{h \tilde{t}_{i} \tilde{t}_{j}}^2,
\end{equation}
where here $i$ and $j$ can each be 1 or 2 independently of each other. The coupling depends on $i$ and $j$, for $\tilde{t}_1 {\tilde{t}}_1^*$ (i.e. $i=j=1$):
\begin{equation}
\mathcal{C}_{h \tilde{t}_{1} \tilde{t}_{1}} = \cos^2 \theta_t \mathcal{C}_{h \tilde{t}_{L} \tilde{t}_{L}} + \sin^2 \theta_t \mathcal{C}_{h \tilde{t}_{R} \tilde{t}_{R}} - 2\sin\theta_t \cos\theta_t \mathcal{C}_{h \tilde{t}_{L} \tilde{t}_{R}},
\end{equation}
where $\mathcal{C}_{h \tilde{t}_{L}\tilde{t}_{L}}$, $\mathcal{C}_{h \tilde{t}_{R} \tilde{t}_{R}}$ and $\mathcal{C}_{h \tilde{t}_{L} \tilde{t}_{R}}$ are the corresponding couplings of $\tilde{u}_L \bar{\tilde{u}}_L$, $\tilde{u}_R \bar{\tilde{u}}_R$ and $\tilde{u}_L \bar{\tilde{u}}_R$, respectively with the changes $m_u \rightarrow m_t$ and $A_u \rightarrow A_t$.
For $\tilde{t}_2 {\tilde{t}}_2^*$ make the replacements $\cos\theta_t \rightarrow \sin\theta_t$, $\sin\theta_t \rightarrow -\cos\theta_t$, $m_{\tilde{t}_1} \rightarrow m_{\tilde{t}_2}$.
For $\tilde{t}_1 {\tilde{t}}_2^*$ or $\tilde{t}_2 {\tilde{t}}_1^*$:
\begin{equation}
\mathcal{C}_{h \tilde{t}_{1} \tilde{t}_{2}} = (\mathcal{C}_{h \tilde{t}_{L} \tilde{t}_{L}} - \mathcal{C}_{h \tilde{t}_{R} \tilde{t}_{R}})\cos\theta_t \sin\theta_t + \mathcal{C}_{h \tilde{t}_{L} \tilde{t}_{R}} \cos 2\theta_t.
\end{equation}
For $H \rightarrow \tilde{t}_i {\tilde{t}}_j^*$ everything is as above but one must transform $\sin\alpha \rightarrow -\cos\alpha$ and $\cos\alpha \rightarrow \sin\alpha$ and $m_h \rightarrow m_H$.
\begin{equation}\label{hsbsb}
\Gamma(h \rightarrow \tilde{b}_i {\tilde{b}}_j^*) = {3 \over 16 \pi m_h} \tilde{\lambda}^{1/2}(m_h, m_{\tilde{b}_i}, m_{\tilde{b}_j}) \mathcal{C}_{h \tilde{b}_{i} \tilde{b}_{j}}^2.
\end{equation} 
For $\tilde{b}_1 {\tilde{b}}_1^*$, i.e. $i=j=1$:
\begin{equation}
\mathcal{C}_{h \tilde{b}_{1} \tilde{b}_{1}} = \mathcal{C}_{h \tilde{b}_{L} \tilde{b}_{L}} \cos^2 \theta_b + \mathcal{C}_{h \tilde{b}_{R} \tilde{b}_{R}} \sin^2 \theta_b - 2 \cos\theta_b \sin\theta_b \mathcal{C}_{h \tilde{b}_{L} \tilde{b}_{R}},
\end{equation}
where $\mathcal{C}_{h \tilde{b}_{L} \tilde{b}_{L}}$, $\mathcal{C}_{h \tilde{b}_{R} \tilde{b}_{R}}$ and $\mathcal{C}_{h \tilde{b}_{L} \tilde{b}_{R}}$ correspond to the couplings given for $\tilde{d}_L \bar{\tilde{d}}_L$, $\tilde{d}_R \bar{\tilde{d}}_R$ and $\tilde{d}_L \bar{\tilde{d}}_R$ with the changes $m_d \rightarrow m_b$ and $A_d \rightarrow A_b$.
For $\tilde{b}_2 {\tilde{b}}_2^*$ it's the same except one must change $\cos\theta_b \rightarrow \sin\theta_b$, $\sin\theta_b \rightarrow -\cos\theta_b$, $m_{\tilde{b}_1} \rightarrow m_{\tilde{b}_2}$.
For $\tilde{b}_1 {\tilde{b}}_2^*$ or $\tilde{b}_2 \bar{\tilde{b}}_1$:
\begin{equation}
\mathcal{C}_{h \tilde{b}_{1} \tilde{b}_{2}} = (\mathcal{C}_{h \tilde{b}_{L} \tilde{b}_{L}} - \mathcal{C}_{h \tilde{b}_{R} \tilde{b}_{R}})\sin\theta_b \cos\theta_b + \mathcal{C}_{h \tilde{b}_{L} \tilde{b}_{R}} \cos 2\theta_b.
\end{equation}
For $H \rightarrow \tilde{b}_i {\tilde{b}}_j^*$ everything is as above with the replacements $\sin\alpha \rightarrow -\cos\alpha$ and $\cos\alpha \rightarrow \sin\alpha$ and $m_h \rightarrow m_H$.
\begin{equation}\label{hstau1stau1}
\Gamma(h \rightarrow \tilde{\tau}_1 {\tilde{\tau}}_1^*) = {1 \over 16 \pi m_h} \tilde{\lambda}^{1 \over 2} \mathcal{C}_{h \tilde{\tau}_{1} \tilde{\tau}_{1}}^2,
\end{equation} 
where
\begin{equation}
\mathcal{C}_{h \tilde{\tau}_{1} \tilde{\tau}_{1}} = \mathcal{C}_{h \tilde{\tau}_{L} \tilde{\tau}_{L}} \sin^2 \theta_\tau + \mathcal{C}_{h \tilde{\tau}_{R} \tilde{\tau}_{R}} \cos^2 \theta_\tau + 2 \cos\theta_\tau \sin\theta_\tau \mathcal{C}_{h \tilde{\tau}_{L} \tilde{\tau}_{R}}.
\end{equation}
$h \rightarrow \tilde{\tau}_2 {\tilde{\tau}}_2^*$ is the same with the replacements $\cos\theta_\tau \rightarrow \sin\theta_\tau$, $\sin\theta_\tau \rightarrow -\cos\theta_\tau$ and $m_{\tilde{\tau}_1} \rightarrow m_{\tilde{\tau}_2}$.
For $h \rightarrow$ $\tilde{\tau}_1 {\tilde{\tau}}_2^*$ or $\tilde{\tau}_2 {\tilde{\tau}}_1^*$ the coupling is instead given by:
\begin{equation}
\mathcal{C}_{h \tilde{\tau}_{1} \tilde{\tau}_{2}} = (\mathcal{C}_{h \tilde{\tau}_{R} \tilde{\tau}_{R}} - \mathcal{C}_{h \tilde{\tau}_{L} \tilde{\tau}_{L}})\cos\theta_\tau \sin\theta_\tau + \mathcal{C}_{h \tilde{\tau}_{L} \tilde{\tau}_{R}} \cos 2\theta_\tau.
\end{equation}
$\mathcal{C}_{h \tilde{\tau}_{L} \tilde{\tau}_{L}}$, $\mathcal{C}_{h \tilde{\tau}_{R} \tilde{\tau}_{R}}$ and $\mathcal{C}_{h \tilde{\tau}_{L} \tilde{\tau}_{R}}$ are identical to the corresponding couplings of $\tilde{e}_L {\tilde{e}}_L^*$, $\tilde{e}_R {\tilde{e}}_R^*$ and $\tilde{e}_L {\tilde{e}}_R^*$ respectively, with the expected replacements.
For $H \rightarrow \tilde{\tau}_i {\tilde{\tau}}_j^*$ everything is as above with the changes $\sin\alpha \rightarrow -\cos\alpha$, $\cos\alpha \rightarrow \sin\alpha$ and $m_h \rightarrow m_H$.
\begin{equation} \label{Afifj}
\Gamma(A \rightarrow \tilde{f}_i {\tilde{f}}_j^*) = {N_c \over 16 \pi m_A} \tilde{\lambda}^{1/2} \mathcal{C}_{A \tilde{f}_{i} \tilde{f}_{j}}^2,
\end{equation}
note $i\neq j$ by CP conservation, and $N_c$ is 3 for squarks and 1 for
sleptons. The coupling is given by:
\begin{equation}
\mathcal{C}_{A \tilde{f}_{i} \tilde{f}_{j}} = \begin{cases}
			{g m_f \over 2 m_W}(\mu + A_f \cot \beta), $  for$ \: \: & u$-type sfermions $\tilde{u}, \tilde{c}, \tilde{t}, \tilde{\nu,} \\
			{g m_f \over 2 m_W}(\mu + A_f \tan \beta), $  for$ \: \: & d$-type sfermions $\tilde{d}, \tilde{s}, \tilde{b}, \tilde{l}. \\
			\end{cases}
\end{equation}
\begin{equation}\label{Hpmqq}
\Gamma(H^+ \rightarrow q \bar{q'}) = {3 g^2 CKM^2 \over 32 \pi m_{W}^2 m_{H^+}} \tilde{\lambda}^{1/2}(m_{H^+},m_{q_1}, m_{q_2}) \Big\{[m_{q_1}^2 \tan^2 \beta + {m_{q_2}^2  \over \tan^2 \beta}] (m_{H^+}^2 - m_{q_1}^2 - m_{q_2}^2) - 4m_{q_1}^2 m_{q_2}^2\Big\},
\end{equation} 
here $m_{q_1}$ is the mass of the $u$-type quark and $m_{q_2}$ is the mass of the $d$-type quark.
\begin{equation}\label{Hpmneutch}
\Gamma(H^+ \rightarrow \tilde{Z}_i \tilde{W}_j) = {1 \over 8 \pi m_{H^+}} \tilde{\lambda}^{1/2}(m_{H^+},m_{\tilde{Z}_i}, m_{\tilde{W}_j})\left[(a^2 + b^2)(m_{H^+}^2 - m_{\tilde{Z}_i}^2 - m_{\tilde{W}_j}^2) -2(a^2 -b^2)m_{\tilde{Z}_i}m_{\tilde{W}_j}\right],
\end{equation} 
where
for $j=1$ i.e. $\tilde{W}_{1}$:
\begin{equation}
a = {1 \over 2} (-\cos\beta A_2 + \sin\beta A_4), \quad \quad \quad
b = {1 \over 2} (-\cos\beta A_2 - \sin\beta A_4),
\end{equation}
for $j=2$ i.e. $\tilde{W}_{2}$:
\begin{equation}
a = {1 \over 2}(-\cos\beta A_1 + \sin\beta A_3), \quad \quad \quad
b = {1 \over 2}(-\cos\beta A_1 - \sin\beta A_3).
\end{equation}
The $A_{i}$ are:
\begin{align}
A_1 &= {1 \over \sqrt{2}}[g N_{2i} + g' N_{1i}]\sin \theta_R - gN_{4i} \cos\theta_R ,\\
A_2 &= {1 \over \sqrt{2}}[-g N_{2i} - g' N_{1i}]\cos \theta_R - gN_{4i} \sin\theta_R, \\
A_3 &= {1 \over \sqrt{2}}[g N_{2i} + g' N_{1i}]\sin \theta_L + gN_{3i} \cos\theta_L, \\
A_4 &= {1 \over \sqrt{2}}[-g N_{2i} - g' N_{1i}]\cos \theta_L + gN_{3i} \sin\theta_L.
\end{align}
\begin{equation}\label{HpmWh}
\Gamma(H^+ \rightarrow W^+ h) = {g^2 \cos^2 (\beta-\alpha) m_{H^+}^3 \over 64 \pi m_{W}^2} \tilde{\lambda}^{3/2}(m_{H^+},m_W,m_h).
\end{equation} 
\begin{equation} \label{Hpmeqsq1}
\Gamma(H^+ \rightarrow \tilde{q}_{L/R} \tilde{q'}_{L/R}) = {3 B \over 16 \pi m_{H^+}} \tilde{\lambda}^{1/2}(m_{H^+},m_{\tilde{q}_{L/R}}, m_{\tilde{q'}_{L/R}}), 
\end{equation} 
where $B$ is the coupling and is given by:
\begin{equation}
B = \begin{cases}
		{g \over \sqrt{2}}\Big[-m_W \sin2\beta + {1 \over m_W}(m_{d}^2 \tan \beta + m_{u}\cot\beta)\Big], &$  for $  \tilde{u}_L \tilde{d}_L, \\
		gm_u m_d (\tan\beta + \cot\beta){1 \over \sqrt{2} m_W}, &$  for $ \tilde{u}_R \tilde{d}_R, \\
		{-g m_d \over \sqrt{2} m_W}(A_d \tan\beta + \mu), &$  for $ \tilde{u}_L \tilde{d}_R, \\
		{-gm_u \over \sqrt{2}m_W}(A_u \cot \beta + \mu) &$  for $ \tilde{u}_R \tilde{d}_L. \\
		\end{cases}
\end{equation}
\begin{equation} \label{Hpmeqsq2}
\Gamma(H^+ \rightarrow \tilde{q}_i \tilde{q'}_j) = {3 \over 16 \pi m_{H^+}} \tilde{\lambda}^{1/2}(m_{H^+},m_{\tilde{q}_i}, m_{\tilde{q}_j}) C^2,
\end{equation} 
note $q$ is the top squark and $q'$ the bottom squark; for $i=j=1$ we have $\tilde{t}_1 \tilde{b}_1$ and:
\begin{equation} \label{HpmstausC}
C = \cos\theta_t \cos\theta_b B_{\tilde{u}_L \tilde{d}_L} + \sin\theta_t \sin\theta_b B_{\tilde{u}_R \tilde{d}_R} - \cos\theta_t \sin\theta_b B_{\tilde{u}_L \tilde{d}_R} - \sin\theta_t \cos\theta_b B_{\tilde{u}_R \tilde{d}_L},
\end{equation}
for a $\tilde{b}_2$ we take $ \cos\theta_b \rightarrow \sin \theta_b$, $\sin\theta_b \rightarrow -\cos\theta_b$;
for a $\tilde{t}_2$ we take $ \cos\theta_t \rightarrow \sin \theta_t$, $\sin\theta_t \rightarrow -\cos\theta_t$.
Note, the same formulae as in Eq.s \eqref{Hpmeqsq1} and \eqref{Hpmeqsq2} can be used for decays of $H^{\pm}$ to sleptons, however for staus, because of the conventions used, one must use $\theta_{\tau} - \pi/2$ which means the replacements  $\cos\theta_{\tau} \rightarrow \sin\theta_{\tau}$ and $\sin\theta_{\tau} \rightarrow -\cos\theta_{\tau}$ are necessary in $C$ in Eq. \eqref{HpmstausC}.

Decays to two vector bosons are somewhat more complicated. Included in {\tt {\tt SOFTSUSY}} are the cases both where the Higgs has mass $m_{h/H} > 2m_V $, and so decays to two on-shell vector bosons, and also the case where the Higgs has mass $m_V < m_{h/H} < 2m_V$, so that it may only undergo a decay to one on-shell vector boson and one off-shell vector boson, which then decays into a fermion anti-fermion pair, i.e. $h/H \rightarrow ZZ^* \rightarrow Zf\bar{f}$ or $h/H \rightarrow WW^* \rightarrow Wf'\bar{f}$. This is technically a $1 \rightarrow 3$ decay but is included here as it is computed exactly without the need for numerical integration, unlike the $1 \rightarrow 3$ decays listed later. To obtain the formulae for $h/H \rightarrow VV^*$, one therefore sums over all possible outgoing $f^{(')}\bar{f}$ into which the $V^*$ may decay.
First consider the case where $m_V < m_{h/H} < 2m_V$ so we have decays $h/H \rightarrow WW^* \rightarrow Wf'\bar{f}$ and $h/H \rightarrow ZZ^* \rightarrow Zf\bar{f}$, this is how the SM-like lightest Higgs, $h$, will decay:
\begin{equation} \label{hZZstar}
\Gamma(h/H \rightarrow ZZ^*) = {G_{F}^2 m_{h/H} m_{W}^4 c_{h/HVV}^2 \over 64 \pi^3 \cos^4 \theta_W} F(\epsilon_{Z}) \left[7 - {40 \over 3} \sin^2 \theta_W + {160 \over 9} \sin^4 \theta_W\right],
\end{equation} 
\begin{equation} \label{hWWstar}
\Gamma(h/H \rightarrow WW^*) = {3 G_{F}^2 m_{W}^4 m_{h/H} c_{h/HVV}^2 \over 16 \pi^3}  F(\epsilon_{W}) ,
\end{equation} 
here
\begin{equation}
\epsilon_{V} = {m_V \over m_{h/H}}, \quad \quad c_{hVV} = \sin(\beta-\alpha), \quad \quad c_{HVV} = \cos(\beta-\alpha),
\end{equation}
and 
\begin{equation}
\begin{aligned}
F(\epsilon_{V}) = {3(1-8\epsilon_{V}^2 + 20\epsilon_{V}^4) \over \sqrt{4\epsilon_{V}^2-1}}\cos^{-1}\Big[{3\epsilon_{V}^2-1 \over 2\epsilon_{V}^3}\Big] - (1-\epsilon_{V}^2)\Big({47 \over 2}\epsilon_{V}^2 -{13 \over 2} + {1 \over \epsilon_{V}^2}\Big) - 3(1-6\epsilon_{V}^2+4\epsilon_{V}^4)\log(\epsilon_{V}).
\end{aligned}
\end{equation}
If however $m_{h/H} > 2m_{V}$ then the decay to two on-shell vector bosons occurs instead and the formulae are:
\begin{equation} \label{hWW}
\Gamma(h/H \rightarrow WW) = {G_{F} m_{h/H}^3 \over 8\pi \sqrt{2}} \tilde{\lambda}^{1 \over 2}(m_{h/H},m_{W},m_{W})(1-r^2 + {3 \over 4}r^4)c_{h/HWW}^2,
\end{equation}
\begin{equation} \label{hZZ}
\Gamma(h/H \rightarrow ZZ) = {G_{F} m_{h/H}^3 \over 16\pi \sqrt{2}} \tilde{\lambda}^{1 \over 2}(m_{h/H},m_{Z},m_{Z})(1-r^2 + {3 \over 4}r^4)c_{h/HZZ}^2,
\end{equation}
where $r = 2{m_{V} \over m_{h/H}}$.

Note throughout many of the decay formulae there is some ambiguity at tree-level about whether one should use $g^2 \over 8 m_W^2$ or $G_{F} \over \sqrt{2}$, whilst these are formally equivalent they are not at a given order of calculation as $G_{F} \over \sqrt{2}$ is a measured value, thereby containing many higher order vertex corrections. In general throughout the program we use $G_{F} \over \sqrt{2}$ as this is found to give better agreement where higher order calculations are available. For example in the di-boson decays comparing with {\tt HDECAY-3.4}, which includes higher orders and finite width effects, we find that our agreement is improved using $G_{F} \over \sqrt{2}$.

\underline{One loop decays to $\gamma \gamma$:}

The function $f(\tau)$ appears, it is given previously in Eq. \eqref{ftau},
\begin{equation} \label{phigamgam}
\Gamma (\phi \rightarrow \gamma \gamma) = {G_{F} \alpha_{em}^2 (m_{\phi}) m_{\phi}^3 \over 128 \pi^3 \sqrt{2}}||\Sigma I_{loop}^{\phi}|^2,
\end{equation}
where the $I_{loop}^{\phi}$ are the amplitudes of the contributions of different particles in the loop to the decay $ \phi \rightarrow \gamma \gamma$.
The top contributions are:
\begin{equation}
I_{t}^{h} = {4 \over 3} {\cos \alpha \over \sin \beta} \Big[-2\tau \{1+(1-\tau)f(\tau)\}\Big],
\end{equation}
\begin{equation}
I_{t}^{H} = {4 \over 3} {\sin \alpha \over \sin \beta} \Big[-2\tau \{1+(1-\tau)f(\tau)\}\Big],
\end{equation}
\begin{equation}
I_{t}^{A} = -{8 \over 3}  \tau f(\tau){\cot \beta}.
\end{equation}
The stop contributions for $h \rightarrow \gamma \gamma$ are:
\begin{equation}
I_{\tilde{t}_1}^h = {4 \over 3} \tau (1- \tau f(\tau))\Big[R_{\tilde{t}_L \tilde{t}_L}^1 \cos^2 \theta_t + R_{\tilde{t}_R \tilde{t}_R}^1 \sin^2 \theta_t - 2R_{\tilde{t}_L \tilde{t}_R}^1 \cos\theta_t \sin\theta_t\Big],
\end{equation}
\begin{equation}
I_{\tilde{t}_2}^h = {4 \over 3} \tau (1- \tau f(\tau))\Big[R_{\tilde{t}_L \tilde{t}_L}^2 \sin^2 \theta_t + R_{\tilde{t}_R \tilde{t}_R}^2 \cos^2 \theta_t + 2R_{\tilde{t}_L \tilde{t}_R}^2 \cos\theta_t \sin\theta_t\Big],
\end{equation}
where
\begin{align}
R_{\tilde{t}_L \tilde{t}_L}^i &= R_{\tilde{t}_L} {m_{W} \over g m_{\tilde{t}_i}} = \left[m_W ({1 \over 2}-{1 \over 6}\tan^2 \theta_W)\sin(\alpha+\beta) - {m_{t}^2 \cos \alpha \over m_{W} \sin \beta}\right]{m_W \over  m_{\tilde{t}_i}}, \\
R_{\tilde{t}_R \tilde{t}_R}^i &= R_{\tilde{t}_R} {m_{W} \over g m_{\tilde{t}_i}} = \left[m_W{2 \over 3} \tan^2 \theta_W  \sin(\alpha+\beta) - {m_{t}^2 \cos \alpha \over m_{W} \sin \beta}\right]{m_W \over m_{\tilde{t}_i}}, \\
R_{\tilde{t}_L \tilde{t}_R}^i &= R_{\tilde{t}_L \tilde{t}_R} {m_W \over g m_{\tilde{t}_i}} = {m_t \over 2 m_W \sin\beta}(\mu \sin\alpha + A_t \cos\alpha){m_W \over m_{\tilde{t}_i}}.
\end{align}
The stop contributions for $H$ are the same but the $R_{\tilde{t}_L \tilde{t}_L}^i$, $R_{\tilde{t}_R \tilde{t}_R}^i$,  $R_{\tilde{t}_L \tilde{t}_R}^i$ (note $i=1,2$ it represents whether the loop is a $\tilde{t}_1$ or a $\tilde{t}_2$ loop) are different:
\begin{align}
R_{\tilde{t}_L \tilde{t}_L}^i &= {m_{W} \over m_{\tilde{t}_i}}\left[-m_W ({1 \over 2} - {1 \over 6}\tan^2 \theta_W)\cos(\alpha+\beta) -{m_{t}^2 \sin\alpha \over m_{W}\sin\beta}\right], \\
R_{\tilde{t}_R \tilde{t}_R}^i &= {m_{W} \over m_{\tilde{t}_i}}\left[-m_W {2 \over 3}\tan^2 \theta_W\cos(\alpha+\beta) -{m_{t}^2 \sin\alpha \over m_{W}\sin\beta}\right], \\
R_{\tilde{t}_L \tilde{t}_R}^i &= {m_W \over m_{\tilde{t}_i}} {m_t \over 2 m_W \sin\beta}(-\mu \cos\alpha + A_t \sin\alpha).
\end{align}
$A \rightarrow \gamma \gamma$ has no stop loop contribution because of CP conservation, i.e. $I_{\tilde{t}_{1/2}}^A = 0$.
The bottom contributions are:
\begin{equation}
I_{b}^h = -{1 \over 3}\Big[-2\tau\{1+(1-\tau)f(\tau)\}\Big]{\sin\alpha \over \cos\beta},
\end{equation}
\begin{equation}
I_{b}^H = {1 \over 3}\Big[-2\tau\{1+(1-\tau)f(\tau)\}\Big]{\cos\alpha \over \cos\beta},
\end{equation}
\begin{equation}
I_{b}^A = -{1 \over 3}\{2\tau f(\tau)\}\tan\beta.
\end{equation}
The sbottom contributions to $h \rightarrow \gamma \gamma$ are:
\begin{equation}
I_{\tilde{b}_1}^h = {1 \over 3}\tau \{1- \tau f(\tau)\}\left[R_{\tilde{b}_L \tilde{b}_L}^1 \cos^2 \theta_b + R_{\tilde{b}_R \tilde{b}_R}^1 \sin^2 \theta_b - 2\sin\theta_b \cos\theta_b R_{\tilde{b}_L \tilde{b}_R}^1\right],
\end{equation}
\begin{equation}
I_{\tilde{b}_2}^h = {1 \over 3}\tau\{1-\tau f(\tau)\}\left[R_{\tilde{b}_L \tilde{b}_L}^2 \sin^2 \theta_b + R_{\tilde{b}_R \tilde{b}_R}^2 \cos^2 \theta_b + 2\sin\theta_b \cos\theta_b R_{\tilde{b}_L \tilde{b}_R}^2\right],
\end{equation}
where here
\begin{align} 
R_{\tilde{b}_L \tilde{b}_L}^i &= {m_{W} \over m_{\tilde{b}_i}} \left[m_W (-{1 \over 2} - {1 \over 6}\tan^2 \theta_W)\sin(\alpha+\beta) + {m_{b}^2 \sin\alpha \over m_{W} \cos \beta}\right], \\
R_{\tilde{b}_R \tilde{b}_R}^i &= {m_{W} \over m_{\tilde{b}_i}}\left[-{1 \over 3}m_W \tan^2 \theta_W \sin(\alpha + \beta) + {m_{b}^2 \sin\alpha \over m_W \cos\beta}\right], \\
R_{\tilde{b}_L \tilde{b}_R}^i &= {m_W \over m_{\tilde{b}_i}}{m_b \over 2m_W \cos\beta}(-\mu \cos\alpha -A_b \sin\alpha).
\end{align}
For $H \rightarrow \gamma \gamma$ the sbottom contributions are the same except the $R_{\tilde{b}_L \tilde{b}_L}^i$, $R_{\tilde{b}_R \tilde{b}_R}^i$, $R_{\tilde{b}_L \tilde{b}_R}^i$ change:
\begin{align}
R_{\tilde{b}_L \tilde{b}_L}^i &= {m_W \over g m_{\tilde{b}_i}}g\left[m_W ({1 \over 2}+{1\over 6}\tan^2 \theta_W)\cos(\alpha+\beta) - {m_{b}^2 \cos\alpha \over m_W \cos\beta}\right], \\
R_{\tilde{b}_R \tilde{b}_R}^i &= {m_W \over g m_{\tilde{b}_i}}g\left[m_W ({1 \over 3}\tan^2 \theta_W)\cos(\alpha+\beta) - {m_{b}^2 \cos\alpha \over m_W \cos\beta}\right], \\
R_{\tilde{b}_L \tilde{b}_R}^i &= {m_W \over g m_{\tilde{b}_i}}g{m_{b} \over 2m_W \cos\beta}[-\mu \sin \alpha + A_b \cos \alpha].
\end{align}
$A \rightarrow \gamma \gamma$ has no sbottom loop contribution because of CP conservation, i.e. $I_{\tilde{b}_{1/2}}^A = 0$.
The charm loop contributions are given by:
\begin{equation}
I_{c}^h = {4 \over 3}\Big[-2\tau\{1+(1-\tau)f(\tau)\}\Big]{\cos\alpha \over \sin\beta},
\end{equation}
\begin{equation}
I_{c}^H = {4 \over 3}\Big[-2\tau\{1+(1-\tau)f(\tau)\}\Big]{\sin\alpha \over \sin\beta},
\end{equation}
\begin{equation}
I_{c}^A = -{4 \over 3}(2\tau f(\tau))\cot \beta.
\end{equation}
$\tau$ loop contributions are given by:
\begin{equation}
I_{\tau}^h = 2\tau[1+(1-\tau)f(\tau)]{\sin\alpha \over \cos\beta},
\end{equation}
\begin{equation}
I_{\tau}^H = -2\tau[1+(1-\tau)f(\tau)]{\cos\alpha \over \cos\beta},
\end{equation}
\begin{equation}
I_{\tau}^A = -2\tau f(\tau) \tan\beta.
\end{equation}
$\tilde{\tau}_i$ contributions to $h \rightarrow \gamma \gamma$ are: 
\begin{equation}
I_{\tilde{\tau}_1} = \tau\{1-\tau f(\tau)\}\left[R_{\tilde{\tau}_L \tilde{\tau}_L}^1 \sin^2 \theta_\tau + R_{\tilde{\tau}_R \tilde{\tau}_R}^1 \cos^2 \theta_\tau + 2 \sin \theta_\tau \cos\theta_\tau R_{\tilde{\tau}_L \tilde{\tau}_R}^1\right],
\end{equation}
\begin{equation}
I_{\tilde{\tau}_2} = \tau\{1-\tau f(\tau)\}\left[R_{\tilde{\tau}_L \tilde{\tau}_L}^2 \cos^2 \theta_\tau + R_{\tilde{\tau}_R \tilde{\tau}_R}^2 \sin^2 \theta_\tau - 2 \sin \theta_\tau \cos\theta_\tau R_{\tilde{\tau}_L \tilde{\tau}_R}^2\right],
\end{equation}
here
\begin{align}
R_{\tilde{\tau}_L \tilde{\tau}_L}^i = {m_W \over m_{\tilde{\tau}_i}} \left[m_W (-{1 \over 2}+{1 \over 2}\tan^2 \theta_W)\sin(\alpha+\beta) + {m_{\tau}^2 \sin\alpha \over m_W \cos \beta}\right], \\
R_{\tilde{\tau}_R \tilde{\tau}_R}^i = {m_W \over m_{\tilde{\tau}_i}} \left[-m_W \tan^2 \theta_W\sin(\alpha+\beta) + {m_{\tau}^2 \sin\alpha \over m_W \cos \beta}\right], \\
R_{\tilde{\tau}_L \tilde{\tau}_R}^i = {m_W \over m_{\tilde{\tau}_i}} {m_\tau \over 2 m_W \cos \beta}(-\mu\cos\alpha -A_\tau \sin\alpha).
\end{align}
For $H \rightarrow \gamma \gamma$ via $\tilde{\tau}_i$ it's the same except the $R_{\tilde{\tau}_L \tilde{\tau}_L}^i$, $R_{\tilde{\tau}_R \tilde{\tau}_R}^i$, $R_{\tilde{\tau}_L \tilde{\tau}_R}^i$ differ:
\begin{align}
R_{\tilde{\tau}_L \tilde{\tau}_L}^i &= {m_W \over m_{\tilde{\tau}_i}}\left[m_W({1 \over 2} - {1 \over 2}\tan^2 \theta_W)\cos(\alpha+\beta) - {m_{\tau}^2 \cos \alpha \over m_W \cos\beta}\right], \\
R_{\tilde{\tau}_R \tilde{\tau}_R}^i &= {m_W \over m_{\tilde{\tau}_i}}\left[m_W \tan^2 \theta_W\cos(\alpha+\beta) - {m_{\tau}^2 \cos \alpha \over m_W \cos\beta}\right], \\
R_{\tilde{\tau}_L \tilde{\tau}_R}^i &= {m_W \over m_{\tilde{\tau}_i}}{m_\tau \over 2m_W \cos\beta}(-\mu \sin \alpha + A_\tau \cos\alpha).
\end{align}
$A \rightarrow \gamma \gamma$ has no stau loop contribution because of CP conservation, i.e. $I_{\tilde{\tau}_{1/2}}^A = 0$.
The $W$ loop contributions are:
\begin{equation}
I_{W}^h = \Big[2+3\tau +3\tau(2-\tau)f(\tau)\Big]\sin(\beta-\alpha),
\end{equation}
\begin{equation}
I_{W}^H = \Big[2+3\tau +3\tau(2-\tau)f(\tau)\Big]\cos(\beta-\alpha).
\end{equation}
$A \rightarrow \gamma \gamma$ has no $W$ loop contribution because of CP conservation, i.e. $I_{W}^A = 0$.
$H^+$ loop contributions are:
\begin{align}
I_{H^+}^h &= \tau\{1-\tau f(\tau)\}{m_{W}^2 \over m_{H^+}^2}\left[\sin(\beta - \alpha) + {\cos 2\beta \sin(\beta+\alpha) \over 2 \cos^2 \theta_W}\right], \\
I_{H^+}^H &= \tau\{1-\tau f(\tau)\}{m_{W}^2 \over m_{H^+}^2}\left[\cos(\beta - \alpha) + {\cos 2\beta \cos(\beta+\alpha) \over 2 \cos^2 \theta_W}\right].
\end{align}
$A \rightarrow \gamma \gamma$ has no $H^+$ loop contribution because of CP conservation, i.e. $I_{H^+}^A = 0$.
$\tilde{W}_{i}^+$ loop contributions are:
\begin{align}
I_{\tilde{W}_{1}^+}^h &= \left[-2\tau\{1+(1-\tau)f(\tau)\}\right]{m_{W} \over m_{\tilde{W}_{1}^+}}\sqrt{2}(-\sin\alpha \sin \theta_R \cos\theta_L + \cos\alpha \sin\theta_L \cos\theta_R), \\
I_{\tilde{W}_{2}^+}^h &= \left[-2\tau\{1+(1-\tau)f(\tau)\}\right]{m_{W} \over m_{\tilde{W}_{2}^+}}\sqrt{2}(\sin\alpha \cos \theta_R \sin\theta_L - \cos\alpha \cos\theta_L \sin\theta_R). 
\end{align}
For $\tilde{W}_{i}^+$ contributions to $H$ everything is the same except the replacements $\cos\alpha \rightarrow \sin\alpha$ and $\sin\alpha \rightarrow -\cos\alpha$ are required.
For $A$ the $\tilde{W}_{i}^+$ contributions have:
\begin{align}
I_{\tilde{W}_{1}^+}^A &= -2\tau f(\tau){m_W \over m_{\tilde{W}_{1}^+}}\sqrt{2}(\sin\theta_R \cos\theta_L \sin\beta + \sin\theta_L \cos\theta_R \cos\beta), \\
I_{\tilde{W}_{2}^+}^A &= 2\tau f(\tau){m_{W} \over m_{\tilde{W}_{2}^+}}\sqrt{2}(\cos\theta_R \sin\theta_L \sin\beta +\cos\theta_L \sin\theta_R \cos\beta).
\end{align}

\underline{$\phi \rightarrow gg$}

The coloured particle loop contribution for $\phi \rightarrow gg$ are exactly the same, except the pre-factor changes and the bottom and sbottom contributions get multiplied by 4 in their amplitudes. There can be no uncoloured particles in the loop so there are no W, charged Higgs, chargino, lepton or slepton contributions; only quark and squark loop contributions:

\begin{equation} \label{phigg}
\Gamma(\phi \rightarrow gg) = {\alpha_{s}^2(m_{\phi}) G_{F} m_{\phi}^3 \over 128 \pi^3 \sqrt{2}} {9 \over 8}\Sigma|I_{loop}^\phi|^2,
\end{equation}
with the $I_{b}^{\phi} \rightarrow 4 I_{b}^{\phi}$, $I_{\tilde{b}_i}^{\phi} \rightarrow 4 I_{\tilde{b}_i}^{\phi}$ and the remaining $I_{loop}^{\phi}$ as in the $\phi \rightarrow \gamma \gamma$ decays.

\underline{Loop decays to $Z \gamma$:}
Throughout the function $g(\tau)$ appears, it is given previously in Eq. \eqref{gtau},
\begin{equation} \label{phiZgam}
\Gamma(\phi \rightarrow Z \gamma) = {m_{\phi}^3 \alpha_{em}^2(m_{\phi}) \over 64 \pi^3} {G_{F} \over \sqrt{2}} \Big(1-{m_Z \over m_\phi}^2\Big)^3  \Sigma|I_{loop}^{\phi}|^2.
\end{equation}
$I_1(\tau_{a},\tau_{aZ})$ and $I_2(\tau_{a},\tau_{aZ})$ also occur frequently, where $\tau_{aZ} = 4({m_{a} \over m_{Z}})^2$ cf $\tau_{a} = 4({m_{a} \over m_{h_i}})^2$, they are as follows:
\begin{equation}
I_{1}(\tau_{a},\tau_{aZ}) = {\tau_{a} \tau_{aZ} \over 2 (\tau_{a}-\tau_{aZ})} + {\tau_{a}^2 \tau_{aZ}^2 \over 2(\tau_{a}-\tau_{aZ})^2}[f(\tau_{a})-f(\tau_{aZ})] + {\tau_{a}^2 \tau_{aZ} \over (\tau_{a}-\tau_{aZ})^2}[g(\tau_{a})-g(\tau_{aZ})],
\end{equation}
\begin{equation}
I_{2}(\tau_{a},\tau_{aZ}) = -{\tau_{a} \tau_{aZ} \over 2(\tau_{a}-\tau_{aZ})}[f(\tau_{a})-f(\tau_{aZ})].
\end{equation}
The fermion loop contributions are:
\begin{equation}
I_{t}^h = 3 {\cos\alpha \over \sin\beta}{-{4\over 3}({1\over2} - {4 \over 3}\sin^2 \theta_W) \over \sin\theta_W \cos\theta_W} \Big(I_{1}(\tau_t,\tau_{tZ}) -I_{2}(\tau_t,\tau_{tZ})\Big),
\end{equation}
The expression for $I_{c}^h$ is analogous.
\begin{equation}
I_{b}^h = -3{\sin\alpha \over \cos\beta}{{2 \over 3}(-{1\over 2}+{2\over 3}\sin^2 \theta_W) \over \sin\theta_w \cos\theta_W}\Big(I_{1}(\tau_b,\tau_{bZ}) -I_{2}(\tau_b,\tau_{bZ})\Big) ,
\end{equation}
A similar expression gives $I_{s}^h$.
For the $I_{f}^H$ transform $\cos\alpha \rightarrow \sin\alpha$, $\sin\alpha \rightarrow -\cos\alpha$.
\begin{equation}
I_{t}^A = 3 \cot\beta{{4\over 3}({1\over2} - {4 \over 3}\sin^2 \theta_W) \over \sin\theta_W \cos\theta_W} I_{2}(\tau_t,\tau_{tZ}),
\end{equation}
Again an analogous expression gives $I_{c}^A$.
\begin{equation}
I_{b}^A = -3\tan\beta{{2 \over 3}(-{1\over 2}+{2\over 3}\sin^2 \theta_W) \over \sin\theta_w \cos\theta_W}I_{2}(\tau_b,\tau_{bZ}).
\end{equation}
$I_{s}^A$ is given similarly.
The $W$ loop contributions are given by:
\begin{equation}
I_{W}^h = -{\sin(\beta - \alpha) \over \tan\theta_W}\left[4(3-\tan^2 \theta_W)I_{2} (\tau_W, \tau_{WZ}) + \{(1+{2 \over \tau_W})\tan^2 \theta_W - (5 + {2 \over \tau_W})\}I_{1}(\tau_W, \tau_{WZ})\right].
\end{equation}
$I_{W}^H$ is the same but with the change $\sin(\beta - \alpha) \rightarrow \cos(\beta - \alpha)$. $I_{W}^A = 0$ by CP conservation.
$H^+$ contributions are:
\begin{equation}
I_{H^+}^h = \left[\sin(\beta - \alpha) + {\cos 2\beta \sin(\beta+\alpha) \over 2 \cos^2 \theta_W}\right]{(1-2\sin^2 \theta_W) \over \cos\theta_W \sin\theta_W} I_{1}(\tau_{H^+},\tau_{H^+ Z}){m_{W}^2 \over m_{H^+}^2}.
\end{equation}
For $H$, the $I_{H^+}^H$ are the same except the replacements $\sin(\beta - \alpha) \rightarrow \cos(\beta - \alpha)$ and $ \sin(\beta+\alpha) \rightarrow -\cos(\beta+\alpha)$ which occur because of the transformations $\cos\alpha \rightarrow \sin\alpha$ and $ \sin\alpha \rightarrow -\cos\alpha$. Meanwhile $I_{H^+}^A = 0$ by CP conservation.

\section{MSSM Three Body Decay Formulae} \label{appendix:MSSM3body}
The following decay modes are included in {\tt {\tt SOFTSUSY}}:
\begin{enumerate}
\item $h \rightarrow WW^* \rightarrow W f' \bar{f}$ 
\item $h \rightarrow ZZ^* \rightarrow Z f \bar{f}$ 
\item $\tilde{g} \rightarrow \tilde{Z}_i q \bar{q}$
\item $\tilde{g} \rightarrow \tilde{W}_i q \bar{q'}$
\item $\tilde{Z}_{i} \rightarrow \tilde{Z}_j f \bar{f}$ where $i>j$
\item $\tilde{Z}_{i} \rightarrow \tilde{W}_j f \bar{f'}$
\item $\tilde{W}_{j} \rightarrow \tilde{Z}_i f \bar{f'}$
\end{enumerate}
The modes included are the most phenomenologically relevant modes, the formulae used were not rederived although are written in our notation and are restructured to match the calculations performed in {\tt SOFTSUSY}. The formulae are as provided in {\tt sPHENO-3.3.8} \cite{Porod:2003um,Porod:2011}, which were based on the calculations in \cite{Baer:1998, Djouadi:2001ma}.

\textbf{\underline{$h \rightarrow VV* \rightarrow V f \bar{f}$}}
Detailed previously, see Eqs. \eqref{hWWstar} \eqref{hZZstar}.

\subsection{Gluino $1 \rightarrow 3$ Decays}

\underline{$\tilde{g} \rightarrow \tilde{Z}_i q \bar{q}$}

First the formulae for the decays to a neutralino and a quark-anti-quark pair of the first two generations; in this formula the Yukawa coupling contributions, squark mixing effects and final state quark masses have been neglected as they are negligible. The formulae after this for the third generation quarks include all such effects.
\begin{equation} \label{gqqneut}
\Gamma(\tilde{g} \rightarrow q \bar{q} \tilde{Z}_i) = {\alpha_s \over 8 \pi^2 }\Big[|A_{{\tilde{Z}_i}}^{q}|^2 (\psi_{L} \pm \phi_{L}) + |B_{{\tilde{Z}_i}}^{q}|^2 (\psi_{R} \pm \phi_{R})\Big]
\end{equation}
Here which of the $\pm$ signs to take depends on the signs of the neutralino and gluino masses; the `$+$' sign applies for the case when both masses have the same sign so $m_{\tilde{Z}_i} > 0$ and $m_{\tilde{g}} > 0$ (or when they are both less than 0) and the `$-$' sign applies when one (but not both) of $m_{\tilde{Z}_i}$ and $m_{\tilde{g}}$ are negative. The signs essentially account for the fact that the couplings should become complex as the masses become negative.
Here the $\psi_{L/R}$ and $\phi_{L/R}$ are integrals related to the $\tilde{\psi}$ and $\tilde{\phi}$ integrals given later in Eq.s \eqref{psiintegral} and \eqref{phitildaintegral} by:
\begin{equation}
\psi_{L/R} = {1 \over \pi^2 m_{\tilde{g}}} \tilde{\psi}(m_{\tilde{g}},m_{\tilde{q}_{L/R}},m_{\tilde{q}_{L/R}},m_{\tilde{Z}_i}).
\end{equation}
\begin{equation}
\phi_{L/R} = {1 \over \pi^2 m_{\tilde{g}}} \tilde{\psi}(m_{\tilde{g}},m_{\tilde{q}_{L/R}},m_{\tilde{q}_{L/R}},m_{\tilde{Z}_i}).
\end{equation}
The $A_{{\tilde{Z}_i}}^{q}$ and $B_{{\tilde{Z}_i}}^{q}$ are couplings which depend upon if the quarks are ``up-type" or ``down-type" in $SU(2)_L$ and are:
\begin{equation}
A_{{\tilde{Z}_i}}^{q} = \begin{cases} 
						{1 \over \sqrt{2}} (-gN_{2i} - {g' \over 3}N_{1i})$, for ``up-type" quarks,$ \\
						{1 \over \sqrt{2}} (gN_{2i} - {g' \over 3}N_{1i})$, for ``down-type" quarks,$
						\end{cases}
\end{equation}
\begin{equation}
B_{{\tilde{Z}_i}}^{q} = \begin{cases} 
						{-4 \over 3 \sqrt{2}} g'N_{1i}$, for ``up-type" quarks,$ \\
						{2 \over 3 \sqrt{2}} g' N_{1i}$, for ``down-type" quarks$.
                                              \end{cases}
\end{equation}
For the more complicated case of decays to third generation quarks; the Yukawa coupling contributions, squark-mixing effects and final state quark masses are all included as they can have significant effects. The decay is mediated by either $\tilde{t}_1$ or $\tilde{t}_2$ in the t or u channel, giving 4 Feynman diagrams (2 shown below as $j=1,2$) and 6 (i.e. $^4 C _2$) interferences. The six interferences can be split into 2 ``diagonal'' contributions ($\tilde{t}_1$ t - $\tilde{t}_1$ u interference and $\tilde{t}_2$ t - $\tilde{t}_2$ u interference) and 4 ``non-diagonal'' contributions ($\tilde{t}_1$ t - $\tilde{t}_2$ t, $\tilde{t}_1$ t - $\tilde{t}_2$ u, $\tilde{t}_1$ u - $\tilde{t}_2$ t and $\tilde{t}_1$ u - $\tilde{t}_2$ u interferences). The possibility of negative neutralino masses (which can be absorbed into imaginary couplings) is also included. The formulae are adopted from {\tt sPHENO}, it should be noted that differences exit between these formulae and those present in Baer and Tata's book \cite{TataBaer}.
\begin{figure}
  \centering {\subfloat[channel a]{\includegraphics[scale =
      0.4]{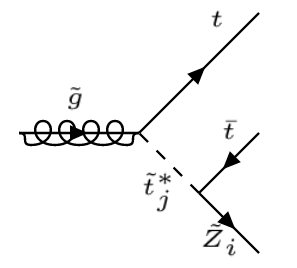}\label{fig:fa}} \: \: \: \: \: \: \: \: \: \: \: \:
    \subfloat[channel b]{\includegraphics[scale = 0.4]{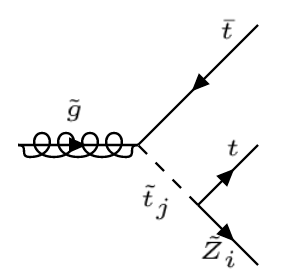}\label{fig:fb}}}
  \caption{Feynman diagrams for the 3 body decay of a gluino into a
    neutralino and a top anti-top pair, as mediated by stops
    $\tilde{t}_{1/2}$. $i=1,2,3,4$ and $j=1,2$.} \label{gneutqqdiag}
\end{figure}

The formulae for the case of $\tilde{g} \rightarrow \tilde{Z}_i t \bar{t}$ are given, from this decays to other quarks can be obtained by making the appropriate replacements.
\begin{equation} \label{gttneut}
\Gamma(\tilde{g} \rightarrow t \bar{t} \tilde{Z}_i) = {\alpha_s \over 8\pi^4 m_{\tilde{g}}}[\Gamma_{\tilde{t}_1} + \Gamma_{\tilde{t}_2} + \Gamma_{\tilde{t}_{1} \tilde{t}_{2}}].
\end{equation}
The $\Gamma_{\tilde{t}_1}$, $\Gamma_{\tilde{t}_2}$, $\Gamma_{\tilde{t}_1 \tilde{t}_2}$ can be split up into different contributions:
\begin{equation}
\Gamma_{\tilde{t}_1} = \Gamma_{LL}(\tilde{t}_1)\cos^2 \theta_t + \Gamma_{RR}(\tilde{t}_1)\sin^2 \theta_t - \sin\theta_t \cos \theta_t \Big[\Gamma_{L_{1} R_{1}} (\tilde{t}_1) + \Gamma_{L_{1} R_{2}} (\tilde{t}_1) + \Gamma_{L_{2} R_{1}} (\tilde{t}_1) + \Gamma_{L_{2} R_{2}} (\tilde{t}_1)\Big].
\end{equation}
\begin{equation}
\Gamma_{\tilde{t}_2} = \Gamma_{LL}(\tilde{t}_2)\sin^2 \theta_t + \Gamma_{RR}(\tilde{t}_2)\cos^2 \theta_t + \sin\theta_t \cos \theta_t \Big[\Gamma_{L_{1} R_{1}} (\tilde{t}_2) + \Gamma_{L_{1} R_{2}} (\tilde{t}_2) + \Gamma_{L_{2} R_{1}} (\tilde{t}_2) + \Gamma_{L_{2} R_{2}} (\tilde{t}_2)\Big].
\end{equation}
\begin{equation}
\Gamma_{\tilde{t}_{1} \tilde{t}_{2}} = \Big[\Gamma_{LL}(\tilde{t}_{1}, \tilde{t}_{2}) + \Gamma_{RR}(\tilde{t}_{1}, \tilde{t}_{2})\Big]\sin\theta_t \cos\theta_t + \Gamma_{LR}(\tilde{t}_1 , \tilde{t}_2) \cos^2 \theta_t + \Gamma_{RL}(\tilde{t}_1 , \tilde{t}_2) \sin^2 \theta_t.
\end{equation}
The moduli of the complex couplings are as follows:
\begin{equation}
|\alpha_{1}^{\tilde{t}_1}| = \tilde{A}^{t}_{\tilde{Z}_{i}} \cos\theta_t - f_{t}N_{4i} \sin\theta_t, \quad \quad |\beta_{1}^{\tilde{t}_1}| = f_{t}N_{4i}\cos\theta_t + \tilde{B}^{t}_{\tilde{Z}_{i}}\sin\theta_t,
\end{equation}
\begin{equation}
|\alpha_{1}^{\tilde{t}_2}| = \tilde{A}^{t}_{\tilde{Z}_i} \sin\theta_t + f_{t}N_{4i}\cos\theta_t, \quad \quad |\beta_{1}^{\tilde{t}_2}| = f_{t}N_{4i}\sin\theta_t - \tilde{B}^{t}_{\tilde{Z}_i}\cos\theta_t,
\end{equation}
where
\begin{equation} 
\tilde{A}^{t}_{\tilde{Z}_i} = -{g \over \sqrt{2}}N_{2i} - {g^{'} \over 3 \sqrt{2}}N_{1i},
\end{equation}
\begin{equation}
\tilde{B}^{t}_{\tilde{Z}_i} = -{4 \over 3}{g^{'} \over \sqrt{2}}N_{1i},
\end{equation}
\begin{equation}
f_{t} = {g m_{t}^{run} \over \sqrt{2} M_{W} \sin\beta}.
\end{equation}
The couplings themselves are then complex and depend upon the sign of the corresponding neutralinos mass. They are of the form $(a,b)$, where this represents the complex number $a+bi$.
For positive masses the couplings are just:
\begin{equation}
a^{\tilde{t}_1} = (|\alpha_{1}^{\tilde{t}_1}|,0) , \quad \quad b^{\tilde{t}_1} = (|\beta_{1}^{\tilde{t}_1}|,0), \quad \quad a^{\tilde{t}_2} = (|\alpha_{1}^{\tilde{t}_2}|,0), \quad \quad b^{\tilde{t}_2} = (|\beta_{1}^{\tilde{t}_2}|,0).
\end{equation}
Meanwhile, for negative neutralino masses the effect of our field redefinition is to multiply the corresponding row of the neutralino mixing matrix by $-i$ therefore the couplings are then:
\begin{equation}
a^{\tilde{t}_1} = (0,-|\alpha_{1}^{\tilde{t}_1}|), \quad \quad b^{\tilde{t}_1} = (0,-|\beta_{1}^{\tilde{t}_1}|), \quad \quad a^{\tilde{t}_2} = (0,-|\alpha_{1}^{\tilde{t}_2}|), \quad \quad b^{\tilde{t}_2} = (0,-|\beta_{1}^{\tilde{t}_2}|).
\end{equation}
In addition to account for differences in interactions for positive and negative neutralino masses as a result of this  coupling difference and the extra associated $\gamma_{5}$ matrices we must also include factors of:
\begin{equation}
(-1)^{\theta_i} = \begin{cases}  +1, $ for positive neutralino masses,$ \\
								 -1, $ for negative neutralino masses$. \\
						\end{cases}
\end{equation}

The formula we use for $\Gamma_{\tilde{t}_1}$ is:
\begin{equation}
\begin{aligned}
\Gamma_{\tilde{t}_1} = & (-1)^{\theta_i}\Big[({a^{\tilde{t}_1}}^2 + {b^{\tilde{t}_1}}^2)\tilde{\psi} (m_{\tilde{g}},m_{\tilde{t}_1}, m_{\tilde{t}_1}, m_{\tilde{Z}_i}) + 4 a^{\tilde{t}_1} b^{\tilde{t}_1}m_{t}m_{\tilde{Z}_i}\tilde{\chi} (m_{\tilde{g}},m_{\tilde{t}_1}, m_{\tilde{t}_1}, m_{\tilde{Z}_i}) \\ & - 4 \sin\theta_t \cos\theta_t({a^{\tilde{t}_1}}^2 + {b^{\tilde{t}_1}}^2) m_{\tilde{g}} m_{t} X(m_{\tilde{g}},m_{\tilde{t}_1}, m_{\tilde{t}_1}, m_{\tilde{Z}_i}) \\ &  -8 \sin\theta_t \cos\theta_t a^{\tilde{t}_1} b^{\tilde{t}_1} m_{\tilde{g}} m_{t}^{2} m_{\tilde{Z}_i}\zeta (m_{\tilde{g}},m_{\tilde{t}_1}, m_{\tilde{t}_1}, m_{\tilde{Z}_i}) \\ & -2\sin\theta_t \cos\theta_t a^{\tilde{t}_1} b^{\tilde{t}_1} Y(m_{\tilde{g}},m_{\tilde{t}_1}, m_{\tilde{t}_1}, m_{\tilde{Z}_i}) + \{{a^{\tilde{t}_1}}^2 \cos^2 \theta_t + {b^{\tilde{t}_1}}^2 \sin^2 \theta_t\}\tilde{\phi} (m_{\tilde{g}},m_{\tilde{t}_1}, m_{\tilde{t}_1}, m_{\tilde{Z}_i}) \\ & - 2 m_{t}^2 \sin\theta_t \cos\theta_t a^{\tilde{t}_1} b^{\tilde{t}_1} \xi (m_{\tilde{g}},m_{\tilde{t}_1}, m_{\tilde{t}_1}, m_{\tilde{Z}_i}) + m_{\tilde{g}} m_{t} a^{\tilde{t}_1} b^{\tilde{t}_1} \xi (m_{\tilde{g}},m_{\tilde{t}_1}, m_{\tilde{t}_1}, m_{\tilde{Z}_i}) \\ & - m_{\tilde{g}} m_{t} a^{\tilde{t}_1} b^{\tilde{t}_1}m_{\tilde{Z}_{i}}^2 \tilde{\rho} (m_{\tilde{g}},m_{\tilde{t}_1}, m_{\tilde{t}_1}, m_{\tilde{Z}_i}) + m_{\tilde{g}}m_{t}^2 m_{\tilde{Z}_i}\{{a^{\tilde{t}_1}}^2 \sin^2 \theta_t  + {b^{\tilde{t}_1}}^2 \cos^2 \theta_t \}\tilde{\rho} (m_{\tilde{g}},m_{\tilde{t}_1}, m_{\tilde{t}_1}, m_{\tilde{Z}_i}) \\ & -m_{\tilde{Z}_i} m_{t}\sin\theta_t \cos\theta_t ({a^{\tilde{t}_1}}^2 + {b^{\tilde{t}_1}}^2)m_{\tilde{g}}^2 \tilde{\rho} (m_{\tilde{g}},m_{\tilde{t}_1}, m_{\tilde{t}_1}, m_{\tilde{Z}_i}) \\ & + m_{\tilde{Z}_i} m_{t} \sin\theta_{t} \cos\theta_{t} ({a^{\tilde{t}_1}}^2 + {b^{\tilde{t}_1}}^2) \xi (m_{\tilde{g}},m_{\tilde{t}_1}, m_{\tilde{t}_1}, m_{\tilde{Z}_i})\Big].
\end{aligned}
\end{equation}

The $\Gamma_{\tilde{t}_2}$ formula we use is:
\begin{equation}
\begin{aligned}
\Gamma_{\tilde{t}_2} = & (-1)^{\theta_i}\Big[({a^{\tilde{t}_2}}^2+{b^{\tilde{t}_2}}^2)\tilde{\psi} (m_{\tilde{g}},m_{\tilde{t}_2}, m_{\tilde{t}_2}, m_{\tilde{Z}_i})  + 4 a^{\tilde{t}_2} b^{\tilde{t}_2} m_{t} m_{\tilde{Z}_i} \tilde{\chi} (m_{\tilde{g}},m_{\tilde{t}_2}, m_{\tilde{t}_2}, m_{\tilde{Z}_i}) \\ & + 4 m_{\tilde{g}}m_{t}\sin\theta_{t}\cos\theta_{t}({a^{\tilde{t}_2}}^2 + {b^{\tilde{t}_2}}^2) X(m_{\tilde{g}},m_{\tilde{t}_2}, m_{\tilde{t}_2}, m_{\tilde{Z}_i}) \\ & + 8\sin\theta_t \cos\theta_t a^{\tilde{t}_2} b^{\tilde{t}_2}m_{\tilde{g}}m_{t}^2 m_{\tilde{Z}_{i}} \zeta (m_{\tilde{g}},m_{\tilde{t}_2}, m_{\tilde{t}_2}, m_{\tilde{Z}_i}) + 2\sin\theta_t \cos\theta_t a^{\tilde{t}_2} b^{\tilde{t}_2} Y(m_{\tilde{g}},m_{\tilde{t}_2}, m_{\tilde{t}_2}, m_{\tilde{Z}_i}) \\ & + \{{a^{\tilde{t}_2}}^2 \sin^2 \theta_t  + \cos^2 \theta_t {b^{\tilde{t}_2}}^2\} \tilde{\phi} (m_{\tilde{g}},m_{\tilde{t}_2}, m_{\tilde{t}_2}, m_{\tilde{Z}_i}) + 2m_{t}^2 \sin\theta_t \cos\theta_t a^{\tilde{t}_2} b^{\tilde{t}_2} \xi (m_{\tilde{g}},m_{\tilde{t}_2}, m_{\tilde{t}_2}, m_{\tilde{Z}_i})  \\ & + m_{\tilde{g}}m_{t} a^{\tilde{t}_2} b^{\tilde{t}_2} \xi (m_{\tilde{g}},m_{\tilde{t}_2}, m_{\tilde{t}_2}, m_{\tilde{Z}_i}) -m_{\tilde{g}}m_{t} m_{\tilde{Z}_i}^2 a^{\tilde{t}_2} b^{\tilde{t}_2} \tilde{\rho} (m_{\tilde{g}},m_{\tilde{t}_2}, m_{\tilde{t}_2}, m_{\tilde{Z}_i}) \\ & + m_{\tilde{g}}m_{\tilde{Z}_i} m_{t}^2 \{{a^{\tilde{t}_2}}^2 \cos^2 \theta_t + {b^{\tilde{t}_2}}^2 \sin^2 \theta_t \} \tilde{\rho} (m_{\tilde{g}},m_{\tilde{t}_2}, m_{\tilde{t}_2}, m_{\tilde{Z}_i}) \\ & - m_{\tilde{Z}_i} m_{t} \sin\theta_t \cos \theta_t ({a^{\tilde{t}_2}}^2 + {b^{\tilde{t}_2}}^2) \xi (m_{\tilde{g}},m_{\tilde{t}_2}, m_{\tilde{t}_2}, m_{\tilde{Z}_i}) \\ & + m_{\tilde{g}}^2 m_{\tilde{Z}_i} m_{t} \sin \theta_t \cos \theta_t ({a^{\tilde{t}_2}}^2 + {b^{\tilde{t}_2}}^2) \tilde{\rho} (m_{\tilde{g}},m_{\tilde{t}_2}, m_{\tilde{t}_2}, m_{\tilde{Z}_i})\Big].
\end{aligned}
\end{equation}

For $\Gamma_{\tilde{t}_1 \tilde{t}_2}$, our formula, again extracted from {\tt sPHENO-3.3.8}, is:
\begin{equation}
\begin{aligned}
\Gamma{\tilde{t}_{1} \tilde{t}_2} = & (-1)^{\theta_i}\Big[4m_{\tilde{g}}m_{t}(\cos^2 \theta_t - \sin^2 \theta_t)(a^{\tilde{t}_1} a^{\tilde{t}_2} + b^{\tilde{t}_1} b^{\tilde{t}_2})X(m_{\tilde{g}},m_{\tilde{t}_1}, m_{\tilde{t}_2}, m_{\tilde{Z}_i}) \\ & + 4m_{\tilde{g}}m_{t}^2 m_{\tilde{Z}_i}(a^{\tilde{t}_1} b^{\tilde{t}_2} + b^{\tilde{t}_1}a^{\tilde{t}_2})(\cos^2 \theta_t - \sin^2 \theta_t)\zeta (m_{\tilde{g}},m_{\tilde{t}_1}, m_{\tilde{t}_2}, m_{\tilde{Z}_i}) \\ & + 2 \{b^{\tilde{t}_1}a^{\tilde{t}_1}\cos^2 \theta_t - \sin^2 \theta_t b^{\tilde{t}_2}a^{\tilde{t}_1} \}Y (m_{\tilde{g}},m_{\tilde{t}_1}, m_{\tilde{t}_2}, m_{\tilde{Z}_i}) \\ & + 2\sin\theta_t \cos\theta_t (a^{\tilde{t}_1} a^{\tilde{t}_2} - b^{\tilde{t}_1}b^{\tilde{t}_2}) \tilde{\phi} (m_{\tilde{g}},m_{\tilde{t}_1}, m_{\tilde{t}_2}, m_{\tilde{Z}_i}) \\ & + 2m_{t}m_{\tilde{Z}_i}(a^{\tilde{t}_1}a^{\tilde{t}_2} - b^{\tilde{t}_1}b^{\tilde{t}_2})\chi^{'} (m_{\tilde{g}},m_{\tilde{t}_1}, m_{\tilde{t}_2}, m_{\tilde{Z}_i}) \\ & + 2 m_{t}^2 \{\cos^2 \theta_t a^{\tilde{t}_1} b^{\tilde{t}_2} - \sin^2 \theta_t b^{\tilde{t}_1}a^{\tilde{t}_2}\}\xi (m_{\tilde{g}},m_{\tilde{t}_1}, m_{\tilde{t}_2}, m_{\tilde{Z}_i}) \\ & + 2m_{\tilde{g}}m_{t}^2 m_{\tilde{Z}_{i}}(b^{\tilde{t}_1} b^{\tilde{t}_2}-a^{\tilde{t}_1} a^{\tilde{t}_2})\sin\theta_t \cos\theta_t \tilde{\rho} (m_{\tilde{g}},m_{\tilde{t}_1}, m_{\tilde{t}_2}, m_{\tilde{Z}_i}) \\ & - 4\sin\theta_t \cos\theta_t m_{\tilde{g}}m_{t}(a^{\tilde{t}_1}b^{\tilde{t}_2}-b^{\tilde{t}_1}a^{\tilde{t}_2})\chi^{'} (m_{\tilde{g}},m_{\tilde{t}_1}, m_{\tilde{t}_2}, m_{\tilde{Z}_i}) \\ & - 2\sin\theta_t \cos\theta_t m_{\tilde{g}}m_{t}(a^{\tilde{t}_1} b^{\tilde{t}_2} - b^{\tilde{t}_1} a^{\tilde{t}_2}) \xi (m_{\tilde{g}},m_{\tilde{t}_1}, m_{\tilde{t}_2}, m_{\tilde{Z}_i}) \\ & + 2m_{t} m_{\tilde{Z}_i}\{\sin^2 \theta_t a^{\tilde{t}_1} a^{\tilde{t}_2} - \cos^2 \theta_t b^{\tilde{t}_1} b^{\tilde{t}_2}\} \xi (m_{\tilde{g}},m_{\tilde{t}_1}, m_{\tilde{t}_2}, m_{\tilde{Z}_i}) \\ & + 2 m_{\tilde{g}}^3 m_{t}(a^{\tilde{t}_1} b_{1}^{\tilde{t}_2} - b^{\tilde{t}_2} a^{\tilde{t}_2})\sin\theta_{t} \cos\theta_t \tilde{\rho} (m_{\tilde{g}},m_{\tilde{t}_1}, m_{\tilde{t}_2}, m_{\tilde{Z}_i}) \\ & - 2 m_{\tilde{g}}^2 m_{t} m_{\tilde{Z}_{i}}\{\sin^2 \theta_t a^{\tilde{t}_1} a^{\tilde{t}_2} - \cos^2 \theta_t b^{\tilde{t}_1}b^{\tilde{t}_2}\} \tilde{\rho} (m_{\tilde{g}},m_{\tilde{t}_1}, m_{\tilde{t}_2}, m_{\tilde{Z}_i})\Big].
\end{aligned}
\end{equation}

For the 3 body decay of $\tilde{g} \rightarrow \tilde{Z}_i b \bar{b}$ then the formulae are exactly as above but with the replacements:
{\centering
$m_{\tilde{t}_{i}} \rightarrow m_{\tilde{b}_{i}}$, \\
$\tilde{A}^{t}_{\tilde{Z}_{i}} \rightarrow \tilde{A}^{b}_{\tilde{Z}_{i}}$, \\
$\tilde{B}^{t}_{\tilde{Z}_{i}} \rightarrow \tilde{B}^{b}_{\tilde{Z}_{i}}$, \\
$f_{t} \rightarrow f_{b}$, \\
$N_{4i} \rightarrow N_{3i}$, \\
$\theta_t \rightarrow \theta_b$, \\
$m_{t} \rightarrow m_{b}$. \\
}
In our program both the $\tilde{g} \rightarrow \tilde{Z}_i t \bar{t}$ and $\tilde{g} \rightarrow \tilde{Z}_i b \bar{b}$ decays are implemented in the same function, just depending on whether its decaying to tops or bottoms different couplings are used as described above.
\begin{equation} 
\tilde{A}^{b}_{\tilde{Z}_i} = {g \over \sqrt{2}}N_{2i} - {g^{'} \over 3 \sqrt{2}}N_{1i},
\end{equation}
\begin{equation}
\tilde{B}^{b}_{\tilde{Z}_i} = {2 \over 3}{g^{'} \over \sqrt{2}}N_{1i},
\end{equation}
\begin{equation}
f_{b} = {g m_{b}^{run} \over \sqrt{2} M_{W} \cos\beta}.
\end{equation}
Note that the integrals used in these equations, $\tilde{\psi}$, $\tilde{\chi}$, $X$, $\zeta$, $\tilde{\phi}$, $\xi$,  $\tilde{\rho}$, $Y$, $\chi^{'}$ are given below in Eq:~\eqref{psiintegral}, \eqref{chitildaintegral}, \eqref{Xintegral}, \eqref{zetaintegral}, \eqref{phitildaintegral}, \eqref{xiintegral}, \eqref{rhotildaintegral} \eqref{Yintegral}, \eqref{chiprimeintegral} respectively, when the formulae for $\tilde{Z}_i \rightarrow \tilde{Z}_j f \bar{f}$ are given.

\textbf{\underline{$\tilde{g} \rightarrow \tilde{W}_i q \bar{q'}$}}

\begin{figure} 
  \centering {\subfloat[]{\includegraphics[scale = 0.25]{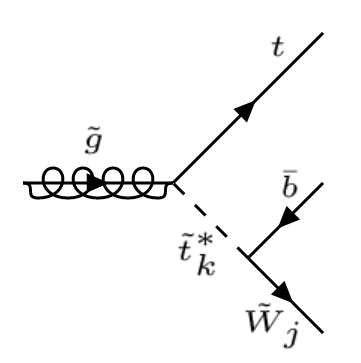}\label{fig:f1a}} \: \: \: \: \: \: \:   \subfloat[]{\includegraphics[scale = 0.25]{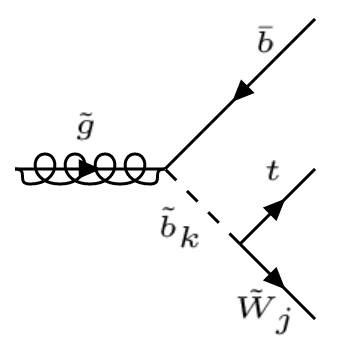}\label{fig:f1b}}} \: \: \: \: \: \: \:
    \centering {\subfloat[]{\includegraphics[scale = 0.25]{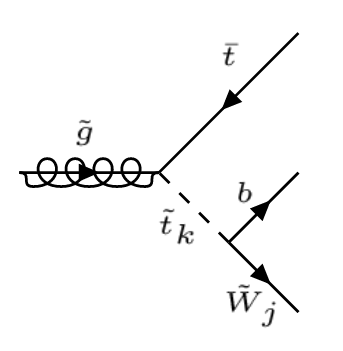}\label{fig:f1c}} \: \: \: \: \: \: \:   \subfloat[]{\includegraphics[scale = 0.25]{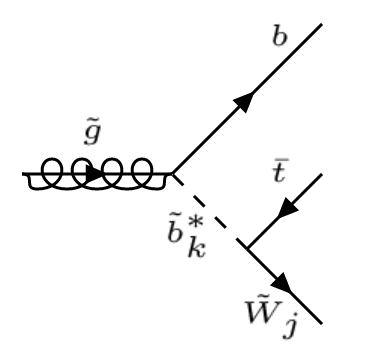}\label{fig:f1d}}}
  \caption{Feynman diagrams for the 3 body decay of a gluino into a chargino and a top-bottom pair, as mediated by stops $\tilde{t}_{1/2}$ and sbottoms $\tilde{t}_{1/2}$. $j,k=1,2$.} \label{gcharqqpdiag}
\end{figure}

The decay of a gluino into a chargino, quark($q$) and anti-quark($\bar{q'}$) can occur via intermediate squarks of either the $q$ or $q'$, therefore there are four possible intermediates in the case where intra-generational squark mixing effects are included. For example, $\tilde{g} \rightarrow \tilde{W}_j t \bar{b}$ may proceed via $\tilde{t}_1$, $\tilde{t}_2$, $\tilde{b}_1$ or $\tilde{b}_2$. Again there are both t and u channel contributions which may contribute to $\tilde{g} \rightarrow \tilde{W}_j t \bar{b}$ or $\tilde{g} \rightarrow \tilde{W}_j b \bar{t}$. The 8 diagrams are therefore shown as a set of 4 ($k = 1,2$ for each intermediate shown) in Fig~\ref{gcharqqpdiag}. There are therefore 4 squared contributions to each of $t \bar{b}$ and $b \bar{t}$ as well as $\tilde{t}_1 \tilde{t}_1$, $\tilde{t}_1 \tilde{t}_2$, $\tilde{t}_2 \tilde{t}_2$, $\tilde{b}_1 \tilde{b}_1$, $\tilde{b}_1 \tilde{b}_2$, $\tilde{b}_2 \tilde{b}_2$, $\tilde{t}_1 \tilde{b}_1$, $\tilde{t}_1 \tilde{b}_2$, $\tilde{t}_2 \tilde{b}_1$ and $\tilde{t}_2 \tilde{b}_2$ interferences. Note ``diagonal'' interferences such as $\tilde{t}_1 \tilde{t}_1$ are included with non-interference squared terms into the $\Gamma_{\tilde{t}_1}$ type contributions.  In the formulae, the Yukawa couplings, intra-generational squark mixing and final state fermion masses are all accounted for; however whilst the bottom quark mass $m_b$ is included in the phase space, it is neglected from the squared matrix element, this drops any $\tilde{b}_k \tilde{b}_l$ interferences as these are proportional to $m_b$. The approach used follows that in Baer and Tata's book \emph{`Weak Scale Supersymmetry'} \cite{TataBaer} with the formulae we use taken from {\tt sPHENO} \cite{Porod:2003um,Porod:2011}, based on the calculations in reference \cite{Baer:1998}. The formulae used, as in {\tt sPHENO}, differ in a few places from those in Baer and Tata.
\begin{equation} \label{gchqqp}
\Gamma(\tilde{g} \rightarrow t \bar{b} \tilde{W}_i^-) = {\alpha_s \over 16 \pi^2 m_{\tilde{g}}} (\Gamma_{\tilde{t}_1} + \Gamma_{\tilde{t}_2} + \Gamma_{\tilde{t}_1 \tilde{t}_2} + \Gamma_{\tilde{b}_1} + \Gamma_{\tilde{b}_2} + \Gamma_{\tilde{t}_1 \tilde{b}_1} + \Gamma_{\tilde{t}_1 \tilde{b}_2} + \Gamma_{\tilde{t}_2 \tilde{b}_1} + \Gamma_{\tilde{t}_2 \tilde{b}_2}).
\end{equation}
The $chargino - quark - squark$ couplings are given below, remember we have the chargino mixing angle $\theta_L$ and $\theta_R$ pre-transformed so that $\theta_{L/R} \rightarrow -\theta_{L/R} + \pi/2$ in order to use the convention where the lightest mass chargino eigenstate $\tilde{W}_1$ appears first in the multiplet:
\begin{equation}
\begin{aligned}
& \alpha_{\tilde{W_1}}^{\tilde{t}_1} = -g \sin\theta_L\cos\theta_t + f_t \cos\theta_R\sin\theta_t, \\ & \beta_{\tilde{W_1}}^{\tilde{t}_1} = - f_b \cos\theta_L\cos\theta_t, \\ & \alpha_{\tilde{W_1}}^{\tilde{b}_1} = -g \sin\theta_L\cos\theta_b + f_b \cos\theta_L\sin\theta_b, \\ & \beta_{\tilde{W_1}}^{\tilde{b}_1} = - f_t \cos\theta_R\cos\theta_b, \\ & \alpha_{\tilde{W_2}}^{\tilde{t}_1} = -g \cos\theta_L\cos\theta_t - f_t \sin\theta_R\sin\theta_t, \\ & \beta_{\tilde{W_2}}^{\tilde{t}_1} =  f_b \sin\theta_L\cos\theta_t, \\ & \alpha_{\tilde{W_2}}^{\tilde{b}_1} = -g \cos\theta_L\cos\theta_b - f_b \sin\theta_L\sin\theta_b, \\ & \beta_{\tilde{W_2}}^{\tilde{b}_1} =  f_t \sin\theta_R\cos\theta_b. \\
\end{aligned}
\end{equation}
We obtain the couplings for $\tilde{t}_2$ and $\tilde{b}_2$ by changing $\cos\theta_{q} \rightarrow \sin\theta_q$ and $\sin\theta_q \rightarrow -\cos\theta_q$. The $\tilde{W}_2$ couplings are obtained from those of $\tilde{W}_1$ by making the replacements $\cos\theta_{L/R} \rightarrow -\sin\theta_{L/R}$ and $\sin\theta_{L/R} \rightarrow \cos\theta_{L/R}$.

The contributions in \eqref{gchqqp} are as follows:
\begin{align}
\Gamma_{\tilde{t}_1} &= [(\alpha_{\tilde{W}_i}^{\tilde{t}_1})^2 + (\beta_{\tilde{W}_i}^{\tilde{t}_1})^2]\left[G_{1}(m_{\tilde{g}},m_{\tilde{t}_1},m_{\tilde{W}_i}) - \sin 2\theta_t G_{8} (m_{\tilde{g}},m_{\tilde{t}_1},m_{\tilde{t}_1}, m_{\tilde{W}_i})\right], \\
\Gamma_{\tilde{t}_2} &= [(\alpha_{\tilde{W}_i}^{\tilde{t}_2})^2 + (\beta_{\tilde{W}_i}^{\tilde{t}_2})^2]\left[G_{1}(m_{\tilde{g}},m_{\tilde{t}_2},m_{\tilde{W}_i}) + \sin 2\theta_t G_{8} (m_{\tilde{g}},m_{\tilde{t}_2},m_{\tilde{t}_2}, m_{\tilde{W}_i})\right], \\
\Gamma_{\tilde{b}_1} &= [(\alpha_{\tilde{W}_i}^{\tilde{b}_1})^2 + (\beta_{\tilde{W}_i}^{\tilde{b}_1})^2]G_{2}(m_{\tilde{g}},m_{\tilde{b}_1},m_{\tilde{W}_i}) + \alpha_{\tilde{W}_i}^{\tilde{b}_1} \beta_{\tilde{W}_i}^{\tilde{b}_1} G_{3} (m_{\tilde{g}},m_{\tilde{b}_1}, m_{\tilde{W}_i}), \\
\Gamma_{\tilde{b}_2} &= [(\alpha_{\tilde{W}_i}^{\tilde{b}_2})^2 + (\beta_{\tilde{W}_i}^{\tilde{b}_2})^2]G_{2}(m_{\tilde{g}},m_{\tilde{b}_2},m_{\tilde{W}_i}) + \alpha_{\tilde{W}_i}^{\tilde{b}_2} \beta_{\tilde{W}_i}^{\tilde{b}_2} G_{3} (m_{\tilde{g}},m_{\tilde{b}_2}, m_{\tilde{W}_i}), \\
\Gamma_{\tilde{t}_1 \tilde{t}_2} &= 2(\alpha_{\tilde{W}_i}^{\tilde{t}_1} \alpha_{\tilde{W}_i}^{\tilde{t}_2} + \beta_{\tilde{W}_i}^{\tilde{t}_1} \beta_{\tilde{W}_i}^{\tilde{t}_2})\cos 2\theta_t G_{8}(m_{\tilde{g}},m_{\tilde{t}_1},m_{\tilde{t}_2},m_{\tilde{W}_i}),
\end{align}
\begin{equation}
\begin{aligned}
\Gamma_{\tilde{t}_1 \tilde{b}_1} = & (\cos\theta_t \sin\theta_b \alpha_{\tilde{W}_i}^{\tilde{b}_1} \beta_{\tilde{W}_i}^{\tilde{t}_1} + \sin\theta_t \cos\theta_b \beta_{\tilde{W}_i}^{\tilde{b}_1} \alpha_{\tilde{W}_i}^{\tilde{t}_1})G_{6}(m_{\tilde{g}},m_{\tilde{t}_1},m_{\tilde{b}_1},m_{\tilde{W}_i}) \\ & - (\cos\theta_t \cos\theta_b \alpha_{\tilde{W}_i}^{\tilde{b}_1}\alpha_{\tilde{W}_i}^{\tilde{t}_1} + \sin\theta_t \sin\theta_b \beta_{\tilde{W}_i}^{\tilde{b}_1}\beta_{\tilde{W}_i}^{\tilde{t}_1}) G_{4}(m_{\tilde{g}},m_{\tilde{t}_1}, m_{\tilde{b}_1},m_{\tilde{W}_i}) \\ & - (\cos\theta_t \cos\theta_b \beta_{\tilde{W}_i}^{\tilde{b}_1} \alpha_{\tilde{W}_i}^{\tilde{t}_1} + \sin\theta_t \sin\theta_b \alpha_{\tilde{W}_i}^{\tilde{b}_1} \beta_{\tilde{W}_i}^{\tilde{t}_1})G_{5}(m_{\tilde{g}},m_{\tilde{t}_1}, m_{\tilde{b}_1}, m_{\tilde{W}_i}) \\ & - (\cos\theta_t\sin\theta_b \beta_{\tilde{W}_i}^{\tilde{b}_1} \beta_{\tilde{W}_i}^{\tilde{t}_1} + \sin\theta_t \cos\theta_b \alpha_{\tilde{W}_i}^{\tilde{b}_1} \alpha_{\tilde{W}_i}^{\tilde{t}_1})G_{7}(m_{\tilde{g}},m_{\tilde{t}_1},m_{\tilde{b}_1},m_{\tilde{W}_i}),
\end{aligned}
\end{equation}
\begin{equation}
\begin{aligned}
\Gamma_{\tilde{t}_1 \tilde{b}_2} = & (-\cos\theta_t \cos\theta_b \alpha_{\tilde{W}_i}^{\tilde{b}_1} \beta_{\tilde{W}_i}^{\tilde{t}_1} + \sin\theta_t \sin\theta_b \beta_{\tilde{W}_i}^{\tilde{b}_1} \alpha_{\tilde{W}_i}^{\tilde{t}_1})G_{6}(m_{\tilde{g}},m_{\tilde{t}_1},m_{\tilde{b}_1},m_{\tilde{W}_i}) \\ & - (\cos\theta_t \sin\theta_b \alpha_{\tilde{W}_i}^{\tilde{b}_1}\alpha_{\tilde{W}_i}^{\tilde{t}_1} - \sin\theta_t \cos\theta_b \beta_{\tilde{W}_i}^{\tilde{b}_1}\beta_{\tilde{W}_i}^{\tilde{t}_1}) G_{4}(m_{\tilde{g}},m_{\tilde{t}_1}, m_{\tilde{b}_1},m_{\tilde{W}_i}) \\ & - (\cos\theta_t \sin\theta_b \beta_{\tilde{W}_i}^{\tilde{b}_1} \alpha_{\tilde{W}_i}^{\tilde{t}_1} - \sin\theta_t \cos\theta_b \alpha_{\tilde{W}_i}^{\tilde{b}_1} \beta_{\tilde{W}_i}^{\tilde{t}_1})G_{5}(m_{\tilde{g}},m_{\tilde{t}_1}, m_{\tilde{b}_1}, m_{\tilde{W}_i}) \\ & - (-\cos\theta_t\cos\theta_b \beta_{\tilde{W}_i}^{\tilde{b}_1} \beta_{\tilde{W}_i}^{\tilde{t}_1} + \sin\theta_t \sin\theta_b \alpha_{\tilde{W}_i}^{\tilde{b}_1} \alpha_{\tilde{W}_i}^{\tilde{t}_1})G_{7}(m_{\tilde{g}},m_{\tilde{t}_1},m_{\tilde{b}_1},m_{\tilde{W}_i}),
\end{aligned}
\end{equation}
\begin{equation}
\begin{aligned}
\Gamma_{\tilde{t}_2 \tilde{b}_1} = & (\sin\theta_t \sin\theta_b \alpha_{\tilde{W}_i}^{\tilde{b}_1} \beta_{\tilde{W}_i}^{\tilde{t}_1} - \cos\theta_t \cos\theta_b \beta_{\tilde{W}_i}^{\tilde{b}_1} \alpha_{\tilde{W}_i}^{\tilde{t}_1})G_{6}(m_{\tilde{g}},m_{\tilde{t}_1},m_{\tilde{b}_1},m_{\tilde{W}_i}) \\ & - (\sin\theta_t \cos\theta_b \alpha_{\tilde{W}_i}^{\tilde{b}_1}\alpha_{\tilde{W}_i}^{\tilde{t}_1} - \cos\theta_t \sin\theta_b \beta_{\tilde{W}_i}^{\tilde{b}_1}\beta_{\tilde{W}_i}^{\tilde{t}_1}) G_{4}(m_{\tilde{g}},m_{\tilde{t}_1}, m_{\tilde{b}_1},m_{\tilde{W}_i}) \\ & - (\sin\theta_t \cos\theta_b \beta_{\tilde{W}_i}^{\tilde{b}_1} \alpha_{\tilde{W}_i}^{\tilde{t}_1} - \cos\theta_t \sin\theta_b \alpha_{\tilde{W}_i}^{\tilde{b}_1} \beta_{\tilde{W}_i}^{\tilde{t}_1})G_{5}(m_{\tilde{g}},m_{\tilde{t}_1}, m_{\tilde{b}_1}, m_{\tilde{W}_i}) \\ & - (\sin\theta_t\sin\theta_b \beta_{\tilde{W}_i}^{\tilde{b}_1} \beta_{\tilde{W}_i}^{\tilde{t}_1} - \cos\theta_t \cos\theta_b \alpha_{\tilde{W}_i}^{\tilde{b}_1} \alpha_{\tilde{W}_i}^{\tilde{t}_1})G_{7}(m_{\tilde{g}},m_{\tilde{t}_1},m_{\tilde{b}_1},m_{\tilde{W}_i}),
\end{aligned}
\end{equation}
\begin{equation}
\begin{aligned}
\Gamma_{\tilde{t}_2 \tilde{b}_2} = & -(\sin\theta_t \cos\theta_b \alpha_{\tilde{W}_i}^{\tilde{b}_1} \beta_{\tilde{W}_i}^{\tilde{t}_1} + \cos\theta_t \sin\theta_b \beta_{\tilde{W}_i}^{\tilde{b}_1} \alpha_{\tilde{W}_i}^{\tilde{t}_1})G_{6}(m_{\tilde{g}},m_{\tilde{t}_1},m_{\tilde{b}_1},m_{\tilde{W}_i}) \\ & - (\sin\theta_t \sin\theta_b \alpha_{\tilde{W}_i}^{\tilde{b}_1}\alpha_{\tilde{W}_i}^{\tilde{t}_1} + \cos\theta_t \cos\theta_b \beta_{\tilde{W}_i}^{\tilde{b}_1}\beta_{\tilde{W}_i}^{\tilde{t}_1}) G_{4}(m_{\tilde{g}},m_{\tilde{t}_1}, m_{\tilde{b}_1},m_{\tilde{W}_i}) \\ & - (\sin\theta_t \sin\theta_b \beta_{\tilde{W}_i}^{\tilde{b}_1} \alpha_{\tilde{W}_i}^{\tilde{t}_1} + \cos\theta_t \cos\theta_b \alpha_{\tilde{W}_i}^{\tilde{b}_1} \beta_{\tilde{W}_i}^{\tilde{t}_1})G_{5}(m_{\tilde{g}},m_{\tilde{t}_1}, m_{\tilde{b}_1}, m_{\tilde{W}_i}) \\ & + (\sin\theta_t\cos\theta_b \beta_{\tilde{W}_i}^{\tilde{b}_1} \beta_{\tilde{W}_i}^{\tilde{t}_1} + \cos\theta_t \sin\theta_b \alpha_{\tilde{W}_i}^{\tilde{b}_1} \alpha_{\tilde{W}_i}^{\tilde{t}_1})G_{7}(m_{\tilde{g}},m_{\tilde{t}_1},m_{\tilde{b}_1},m_{\tilde{W}_i}).
\end{aligned}
\end{equation}
The integrals $G_1$ to $G_8$ are given by the following, where $s_{t} = m_{\tilde{g}}^2 + m_{t}^2 -2 E_{t} m_{\tilde{g}}$ and $s_{b} = m_{\tilde{g}}^2 + m_{b}^2 -2 E_{b} m_{\tilde{g}}$:
\begin{equation}
G_{1}(m_{\tilde{g}},m_{\tilde{t}_k},m_{\tilde{W}_i}) = m_{\tilde{g}} \int {d E_{t} p_{t} E_{t} (s_{t} - m_{\tilde{W}_i}^2)^2 \over (s_{t} - m_{\tilde{t}_k}^2)^2 s_{t}},
\end{equation}
\begin{equation}
G_{2}(m_{\tilde{g}},m_{\tilde{b}_k},m_{\tilde{W}_i}) = m_{\tilde{g}} \int d E_{b} E_{b}^2 \lambda^{1 \over 2} (s_{b},m_{\tilde{W}_i}^2,m_{t}^2) {s_{b} - m_{t}^2 - m_{\tilde{W}_i}^2 \over (s_{b}^2 - m_{\tilde{b}_k}^2)^2 s_b},
\end{equation}
\begin{equation}
G_{3}(m_{\tilde{g}},m_{\tilde{b}_k},m_{\tilde{W}_i}) = \int d E_{b} E_{b}^2 \lambda^{1 \over 2} (s_{b},m_{\tilde{W}_i}^2,m_{t}^2) {4 m_{\tilde{g}} m_{\tilde{W}_i} m_{t} \over (s_{b}^2 - m_{\tilde{b}_k}^2)^2 s_b},
\end{equation}
\begin{equation}
G_{4}(m_{\tilde{g}},m_{\tilde{t}_j},m_{\tilde{b}_k},m_{\tilde{W}_i}) = m_{\tilde{g}} m_{\tilde{W}_i} \int {d E_{t} \over s_{t} - m_{\tilde{t}_j}^2} \left[ E_{b}(max) - E_{b}(min) - {m_{\tilde{b}_j}^2 + m_{t}^2 - 2 E_{t} m_{\tilde{g}} - m_{\tilde{W}_i}^2 \over 2 m_{\tilde{g}}} \log X\right] ,
\end{equation}
\begin{equation}
G_{5}(m_{\tilde{g}},m_{\tilde{t}_j},m_{\tilde{b}_k},m_{\tilde{W}_i}) = {m_{t} \over 2} \int {d E_{t}} {s_t -m_{\tilde{W}_i}^2 \over s_t - m_{\tilde{t}_j}^2} \log X,
\end{equation}
\begin{equation}
G_{6}(m_{\tilde{g}},m_{\tilde{t}_j},m_{\tilde{b}_k},m_{\tilde{W}_i}) = {1 \over 2} \int {d E_t \over s_t - m_{\tilde{t}_j}^2} \left\lbrace [ m_{\tilde{g}} (s_{t} - m_{\tilde{W}_i}^2) + {m_{\tilde{b}_k}^2 - m_{\tilde{g}}^2 \over m_{\tilde{g}}} s_{t}] \log X - 2 s_{t} [E_{b}(max) - E_{b}(min)] \right\rbrace,
\end{equation}
\begin{equation}
G_{7}(m_{\tilde{g}},m_{\tilde{t}_j},m_{\tilde{b}_k},m_{\tilde{W}_i}) = {1 \over 2} m_{\tilde{W}_i} m_{t} \int {d E_t \over s_t - m_{\tilde{t}_j}^2} \left\lbrace 2[E_{b}(max) - E_{b}(min)] - {m_{\tilde{b}_k}^2 - m_{\tilde{g}}^2 \over m_{\tilde{g}}} \log X \right\rbrace,
\end{equation}
\begin{equation}
G_{8}(m_{\tilde{g}},m_{\tilde{t}_1},m_{\tilde{t}_2},m_{\tilde{W}_i}) = m_{\tilde{g}} m_{t} \int {d E_t} {(s_{t} - m_{\tilde{W}_i}^2) [E_{b}(max) - E_{b}(min)] \over (s_{t} - m_{\tilde{t}_1}^2)(s_{t} - m_{\tilde{t}_2}^2)}.
\end{equation}
The limits of integration here are $m_t$ to $(m_{\tilde{g}}^2 + m_{t}^2 - (m_{\tilde{W}_i}+m_{b})^2)/(2m_{\tilde{g}})$ for the $E_t$ integrals, and $m_{b}$ to $(m_{\tilde{g}}^2 - (m_t + m_{\tilde{W}_i})^2)/(2 m_{\tilde{g}})$ for the $E_b$ integrals.
Here $p_t = \sqrt{E_{t}^2 - m_{t}^2}$, $E_{b}(max/min)$ and $X$ are given by:
\begin{equation}
E_{b}(max/min) = {(s_{t} + m_{b}^2 - m_{\tilde{W}_i}^2)(m_{\tilde{g}} - E_{t}) \pm \lambda^{1 \over 2}(s_{t},m_{b}^2,m_{\tilde{W}_i}^2) \over 2 s_{t}},
\end{equation}
\begin{equation}
X = {m_{\tilde{b}_j}^2 + 2 E_{b}(max) m_{\tilde{g}} - m_{\tilde{g}}^2 \over m_{\tilde{b}_j}^2 + 2 E_{b}(min) m_{\tilde{g}} - m_{\tilde{g}}^2}.
\end{equation}
The formulae for the first and second generation quarks can be obtained from those for the third generation straightforwardly, in fact they are simpler as the Yukawa coupling can often be neglected.
\subsection{Neutralino $1 \rightarrow 3$ Decays}

\textbf{\underline{$\tilde{Z}_i \rightarrow \tilde{Z}_j f \bar{f}$}}

\begin{figure}
  \centering {\subfloat[]{\includegraphics[scale=0.3]{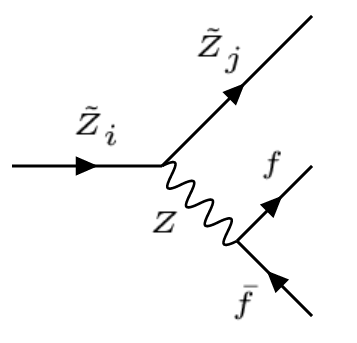}\label{fig:neutneut1}} \: \: \: \: \: \: \:   \subfloat[]{\includegraphics[scale=0.3]{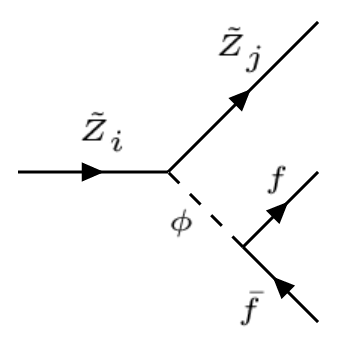}\label{neutneut2}} \: \: \: \: \: \: \:
\subfloat[]{\includegraphics[scale=0.3]{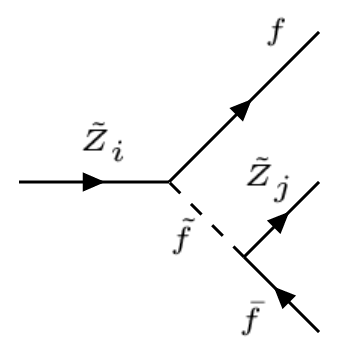}\label{neutneut3}}}
  \caption{Feynman diagrams for the 3 body decay of a neutralino into a lighter neutralino and a fermion anti-fermion pair, as mediated by $Z$ bosons, Higgs bosons $\phi = h,H,A$, or sfermions $\tilde{f}_{1/2}$. $i>j$ and ${i,j}\in{1,2,3,4}$.} \label{neutneutffbardiag}
\end{figure}

For the 3 body decay of a neutralino into a lighter neutralino and a fermion-anti-fermion pair there are three types of contribution, as illustrated in the Feynman diagrams in Fig~\ref{neutneutffbardiag}; $Z$ boson exchange, neutral Higgs boson exchange and sfermion exchange. The effects of Yukawa couplings, sfermion mixing and finite non-zero quark masses in the final state have been included. Similarly, the effects of negative neutralino masses in initial or final states are included. However, we have taken the approach of Baer et al in references \cite{Baer:1998,TataBaer}; whilst we have included the effects of the quark mass in the phase space, the quark mass has been approximated as zero in the squared matrix element. This approximation is justified on the basis that the fermion-anti-fermion pair may not be a $t\bar{t}$ pair as the decay calculator only evaluates the 3 body decays when 2 body decays are absent, given the dominance of 2 body modes over 3 body modes in branching ratios. Whenever the 3 body mode $\tilde{Z}_i \rightarrow \tilde{Z}_j t \bar{t}$ is available then so are the 2 body modes $\tilde{Z}_i \rightarrow \tilde{Z}_j Z$ and $\tilde{Z}_i \rightarrow \tilde{Z}_j h$, which will make the 3 body modes negligible. It is however crucial to include the effects of the non-zero quark masses in the phase space, as has been done, as often the phase space available to these decays is limited (e.g.\ for compressed spectra) and so the reduction in phase space caused by the finite quark masses is important. Similarly, with non-zero quark masses the Higgs intermediate contributions are allowed. Nonetheless the effect of the approximation is just to remove the Higgs boson - $Z$ interferences and CP even - CP odd Higgs boson interferences, which are generally necessarily small compared to other contributions. The formulae for the included contributions themselves are all taken from {\tt sPHENO}, as for the other 3 body modes. It should also be noted that the calculation in {\tt sPHENO} was done in the Feynman gauge so a goldstone contribution, corresponding to the longitudinal component of the $Z$ boson, is required. The included contributions are therefore the squared $Z$ (including goldstone), $\phi$ and $\tilde{f}$ contributions as well as $hH$, $Z\tilde{f}$, $\phi\tilde{f}$ and $ZA$ interferences.\footnote{Calculations of the differential decay rates for this mode are also available, using a different approach to that used here, in reference \cite{Mambrini:2001}.}

\begin{equation} \label{neutneutff}
\begin{aligned}
\Gamma (\tilde{Z}_i \rightarrow \tilde{Z}_j f \bar{f}) = {N_{c} \over 512 \pi^3 |m_{\tilde{Z}_i}|^3}( & \Gamma_Z + \Gamma_h + \Gamma_H + \Gamma_A + \Gamma_{hH} + \Gamma_{\tilde{f}} - 4\Gamma_{h \tilde{f}_1} - 4\Gamma_{h \tilde{f}_2} - 4\Gamma_{H \tilde{f}_1} \\ & - 4\Gamma_{H \tilde{f}_2} - 4\Gamma_{A \tilde{f}_1} - 4\Gamma_{A \tilde{f}_2} + 4\Gamma_{Z \tilde{f}_1} - 4\Gamma_{Z \tilde{f}_2} - 4\Gamma_{ZA} + \Gamma_{G} \\ & + 2\Gamma_{G A} -4\Gamma_{Z G} - 4\Gamma_{G \tilde{f}_1} - 4\Gamma_{G \tilde{f}_2}).
\end{aligned}
\end{equation}
$G$ indicates the goldstone contribution.
In the following contributions we again account for negative neutralino masses via factors of $-1$ corresponding to the effects of absorbing factors of $-i\gamma_5$ into the couplings for negative mass neutralinos:
\begin{equation}
(-1)^{\theta_i} = \begin{cases}
1 $, for $m_{\tilde{Z}_i} > 0, \\
-1 $, for $m_{\tilde{Z}_i} < 0. \\
\end{cases}
\end{equation}
The contributions are as follows:
\begin{equation}
\begin{aligned}
\Gamma_h = 2(X_{ij}^h + X_{ji}^h)^2 f_{q}^2 t_{\alpha_h}^2 \Big[I_{4}^h - 2m_{f}^2 I_{3}^h + 2(-1)^{\theta_j}|m_{\tilde{Z}_i}||m_{\tilde{Z}_j}|I_{2}^h - 4(-1)^{\theta_j}|m_{\tilde{Z}_i}||m_{\tilde{Z}_j}|m_{f}^2 I_{1}^h\Big],
\end{aligned}
\end{equation}

where

\begin{equation}
s = m_{\tilde{Z}_i}^2 + m_{\tilde{Z}_j}^2 - 2|m_{\tilde{Z}_i}|E,
\end{equation}
\begin{equation}
E_{max} = {m_{\tilde{Z}_i}^2 + m_{\tilde{Z}_j}^2 - 4m_{f}^2 \over 2 |m_{\tilde{Z}_i}|},
\end{equation}

\begin{align}
I_{1}^h &= \int_{|m_{\tilde{Z}_j}|}^{E_{max}} dE {2|m_{\tilde{Z}_i}|\sqrt{s-4 m_{f}^2} \sqrt{E^2 - m_{\tilde{Z}_j}^2}2|m_{\tilde{Z}_i}| \over \sqrt{s}(s-m_{h}^2)^2}, \\
I_{2}^h &= \int_{|m_{\tilde{Z}_j}|}^{E_{max}} dE {2|m_{\tilde{Z}_i}|\sqrt{s-4 m_{f}^2} \sqrt{E^2 - m_{\tilde{Z}_j}^2}2| m_{\tilde{Z}_i}| (s - 2 m_{f}^2) \over \sqrt{s}(s-m_{h}^2)^2}, \\
I_{3}^h &= \int_{|m_{\tilde{Z}_j}|}^{E_{max}} dE {2|m_{\tilde{Z}_i}|\sqrt{s-4 m_{f}^2} \sqrt{E^2 - m_{\tilde{Z}_j}^2}2|m_{\tilde{Z}_i}| 2|m_{\tilde{Z}_i}|E \over \sqrt{s}(s-m_{h}^2)^2}, \\
I_{4}^h &= \int_{|m_{\tilde{Z}_j}|}^{E_{max}} dE {2|m_{\tilde{Z}_i}|\sqrt{s-4 m_{f}^2} \sqrt{E^2 - m_{\tilde{Z}_j}^2}2|m_{\tilde{Z}_i}| (s - 2 m_{f}^2) 2|m_{\tilde{Z}_i}|E \over \sqrt{s}(s-m_{h}^2)^2}.
\end{align}

\begin{equation}
\begin{aligned}
\Gamma_H = 2(X_{ij}^H + X_{ji}^H)^2 f_{q}^2 t_{\alpha_H}^2 \Big[I_{4}^H - 2m_{f}^2 I_{3}^H + 2(-1)^{\theta_j}|m_{\tilde{Z}_i}||m_{\tilde{Z}_j}|I_{2}^H - 4(-1)^{\theta_j}|m_{\tilde{Z}_i}||m_{\tilde{Z}_j}|m_{f}^2 I_{1}^H\Big].
\end{aligned}
\end{equation}

where the $I_{1,2,3,4}^H$ are exactly the same as the $I_{1,2,3,4}^h$ with the change $m_{h} \rightarrow m_{H}$.

\begin{equation}
\begin{aligned}
\Gamma_Z = 64 g^2 \sin^2 \theta_W W_{ij}^2 \Big[ & 4(-1)^{\theta_j}|m_{\tilde{Z}_i}||m_{\tilde{Z}_j}|m_{f}^2(\alpha_{f}^2 - \beta_{f}^2)I_{4}^Z + m_{f}^2(\alpha_{f}^2 - \beta_{f}^2)I_{3}^Z \\ & + (-1)^{\theta_i}(-1)^{\theta_j}|m_{\tilde{Z}_i}||m_{\tilde{Z}_j}|(\alpha_{f}^2 + \beta_{f}^2)I_{2}^Z + \frac{1}{2}(\alpha_{f}^2 + \beta_{f}^2)I_{1}^Z\Big],
\end{aligned}
\end{equation}
where the integrals $I_{i}^Z$ are:
\begin{equation}
\begin{aligned}
I_{1}^Z = \int_{4m_{f}^2}^{(|m_{\tilde{Z}_i}| - |m_{\tilde{Z}_j}|)^2} ds & \Big[ {1 \over 3 s^2 (s-m_{Z}^2)^2} \{-2s^4 + (m_{\tilde{Z}_i}^2 + m_{\tilde{Z}_j}^2 + 2m_{f}^2)s^3 + [(m_{\tilde{Z}_i}^2 - m_{\tilde{Z}_j}^2)^2 \\ & - 2(m_{\tilde{Z}_i}^2 + m_{\tilde{Z}_j}^2)2m_{f}^2]s^2 + 2m_{f}^2 (m_{\tilde{Z}_i}^2 - m_{\tilde{Z}_j}^2)^2 s\}{1 \over s} \\ & \times \sqrt{(s-(|m_{\tilde{Z}_i}|-|m_{\tilde{Z}_j}|)^2)((s-(|m_{\tilde{Z}_i}|+|m_{\tilde{Z}_j}|)^2)}\sqrt{s(s-4m_{f}^2)} \Big],
\end{aligned}
\end{equation}

\begin{equation}
\begin{aligned}
I_{2}^Z = \int_{4m_{f}^2}^{(|m_{\tilde{Z}_i}| - |m_{\tilde{Z}_j}|)^2} ds & \Big[{1 \over s (s-m_{Z}^2)^2}(s-2 m_{f}^2) \sqrt{(s-(|m_{\tilde{Z}_i}|-|m_{\tilde{Z}_j}|)^2)((s-(|m_{\tilde{Z}_i}|+|m_{\tilde{Z}_j}|)^2)} \\ & \times \sqrt{s(s-4m_{f}^2)}\Big],
\end{aligned}
\end{equation}

\begin{equation}
\begin{aligned}
I_{3}^Z = \int_{4m_{f}^2}^{(|m_{\tilde{Z}_i}| - |m_{\tilde{Z}_j}|)^2} ds & \Big[{1 \over s (s-m_{Z}^2)^2}(m_{\tilde{Z}_i}^2 + m_{\tilde{Z}_j}^2 - s) \sqrt{(s-(|m_{\tilde{Z}_i}|-|m_{\tilde{Z}_j}|)^2)((s-(|m_{\tilde{Z}_i}|+|m_{\tilde{Z}_j}|)^2)} \\ & \times \sqrt{s(s-4m_{f}^2)}\Big],
\end{aligned}
\end{equation}

\begin{equation}
\begin{aligned}
I_{4}^Z = \int_{4m_{f}^2}^{(|m_{\tilde{Z}_i}| - |m_{\tilde{Z}_j}|)^2} ds & \Big[{1 \over s (s-m_{Z}^2)^2}\sqrt{(s-(|m_{\tilde{Z}_i}|-|m_{\tilde{Z}_j}|)^2)((s-(|m_{\tilde{Z}_i}|+|m_{\tilde{Z}_j}|)^2)}\sqrt{s(s-4m_{f}^2)}\Big].
\end{aligned}
\end{equation}

\begin{equation}
\Gamma_A = (X_{ij}^A + X_{ji}^A)^2 A_{q}^2 \left[I_{4}^A + 2m_{f}^2 I_{3}^A - 2(-1)^{\theta_i}(-1)^{\theta_j}|m_{\tilde{Z}_i}||m_{\tilde{Z}_j}|I_{2}^A - 4(-1)^{\theta_i}(-1)^{\theta_j}m_{f}^2|m_{\tilde{Z}_i}||m_{\tilde{Z}_j}|I_{1}^A \right],
\end{equation}
where the integrals $I_{i}^A$ are:
\begin{align}
I_{1}^A &= \int_{|m_{\tilde{Z}_j}|}^{E_{max}} dE \sqrt{E^2 - m_{\tilde{Z}_j}^2} \sqrt{s - 4m_{f}^2} {4 m_{\tilde{Z}_i}^2 \over \sqrt{s}(s-m_{A}^2)^2}, \\
I_{2}^A &= \int_{|m_{\tilde{Z}_j}|}^{E_{max}} dE \sqrt{E^2 - m_{\tilde{Z}_j}^2} \sqrt{s - 4m_{f}^2} {4 m_{\tilde{Z}_i}^2 (s-2m_{f}^2) \over \sqrt{s}(s-m_{A}^2)^2}, \\
I_{3}^A &= \int_{|m_{\tilde{Z}_j}|}^{E_{max}} dE \sqrt{E^2 - m_{\tilde{Z}_j}^2} \sqrt{s - 4m_{f}^2} {4 m_{\tilde{Z}_i}^2 2|m_{\tilde{Z}_i}|E \over \sqrt{s}(s-m_{A}^2)^2}, \\
I_{4}^A &= \int_{|m_{\tilde{Z}_j}|}^{E_{max}} dE \sqrt{E^2 - m_{\tilde{Z}_j}^2} \sqrt{s - 4m_{f}^2} {4 m_{\tilde{Z}_i}^2 2|m_{\tilde{Z}_i}|E (s-2m_{f}^2) \over \sqrt{s}(s-m_{A}^2)^2},
\end{align}

\begin{equation}
\begin{aligned}
\Gamma_{hH} = 4(X_{ij}^h + X_{ji}^h)(X_{ij}^H + X_{ji}^H)f_{q}^2 t_{\alpha_h} t_{\alpha_H} \Big[ & I_{4}^{hH} - 2m_{f}^2I_{3}^{hH} + 2(-1)^{\theta_i}(-1)^{\theta_j}|m_{\tilde{Z}_i}||m_{\tilde{Z}_j}|I_{2}^{hH} \\ & - 4(-1)^{\theta_i}(-1)^{\theta_j}|m_{\tilde{Z}_i}||m_{\tilde{Z}_j}|m_{f}^2 I_{1}^{hH} \Big],
\end{aligned}
\end{equation}
where:
\begin{equation}
I_{1}^{hH} = \int_{|m_{\tilde{Z}_j}|}^{E_{max}} dE \Big[2|m_{\tilde{Z}_i}|\sqrt{s-4m_{f}^2}\sqrt{E^2 - m_{\tilde{Z}_j}^2} {2|m_{\tilde{Z}_i}| \over \sqrt{s}(s-m_{h}^2)(s-m_{H}^2)}\Big],
\end{equation}

\begin{equation}
I_{2}^{hH} = \int_{|m_{\tilde{Z}_j}|}^{E_{max}} dE \Big[2|m_{\tilde{Z}_i}|\sqrt{s-4m_{f}^2}\sqrt{E^2 - m_{\tilde{Z}_j}^2} {2|m_{\tilde{Z}_i}| (s-2m_{f}^2) \over \sqrt{s}(s-m_{h}^2)(s-m_{H}^2)}\Big],
\end{equation}

\begin{equation}
I_{3}^{hH} = \int_{|m_{\tilde{Z}_j}|}^{E_{max}} dE \Big[2|m_{\tilde{Z}_i}|\sqrt{s-4m_{f}^2}\sqrt{E^2 - m_{\tilde{Z}_j}^2} {4m_{\tilde{Z}_i}^2 E \over \sqrt{s}(s-m_{h}^2)(s-m_{H}^2)}\Big],
\end{equation}

\begin{equation}
I_{4}^{hH} = \int_{|m_{\tilde{Z}_j}|}^{E_{max}} dE \Big[2|m_{\tilde{Z}_i}|\sqrt{s-4m_{f}^2}\sqrt{E^2 - m_{\tilde{Z}_j}^2} {4m_{\tilde{Z}_i}^2 E (s-2m_{f}^2) \over \sqrt{s}(s-m_{h}^2)(s-m_{H}^2)}\Big].	
\end{equation}

\begin{equation}
\Gamma_{\tilde{f}} = 2\Gamma_{\tilde{f}_1 \tilde{f}_1}^{diag} + 2\Gamma_{\tilde{f}_2 \tilde{f}_2}^{diag} + 2\Gamma_{\tilde{f}\tilde{f}}^{nondiag} + \Gamma_{\tilde{f}_1 \tilde{f}_1}^{tu} + \Gamma_{\tilde{f}_2 \tilde{f}_2}^{tu} + 2\Gamma_{\tilde{f}_1 \tilde{f}_2}^{tu},
\end{equation}
There are many sfermion contributions included here, they are as follows:
\begin{equation}
\begin{aligned}
\Gamma_{\tilde{f}_1 \tilde{f}_1}^{diag} = & {8m_{\tilde{Z}_i}^2 \over \pi^2}({\alpha_{\tilde{f}_1}^{\tilde{Z}_i}}^2 + {\beta_{\tilde{f}_1}^{\tilde{Z}_i}}^2) ({\alpha_{\tilde{f}_1}^{\tilde{Z}_j}}^2 + {\beta_{\tilde{f}_1}^{\tilde{Z}_j}}^2)  \tilde{\psi}(\tilde{Z}_i,\tilde{f}_1,\tilde{f}_1,\tilde{Z}_j) \\ & + {32 m_{\tilde{Z}_i}^2 \over \pi^2}(-1)^{\theta_j} ({\alpha_{\tilde{f}_1}^{\tilde{Z}_i}}^2 + {\beta_{\tilde{f}_1}^{\tilde{Z}_i}}^2) {\alpha_{\tilde{f}_1}^{\tilde{Z}_j}} {\beta_{\tilde{f}_1}^{\tilde{Z}_j}} m_{f} |m_{\tilde{Z}_j}|\tilde{\chi}(\tilde{Z}_i,\tilde{f}_1, \tilde{f}_1, \tilde{Z}_j) \\ & - {32 m_{\tilde{Z}_i}^2 \over \pi^2} (-1)^{\theta_i} ({\alpha_{\tilde{f}_1}^{\tilde{Z}_j}}^2 + {\beta_{\tilde{f}_1}^{\tilde{Z}_j}}^2) {\alpha_{\tilde{f}_1}^{\tilde{Z}_i}} {\beta_{\tilde{f}_1}^{\tilde{Z}_i}} m_f |m_{\tilde{Z}_i}|X(\tilde{Z}_i, \tilde{f}_1, \tilde{f}_1, \tilde{Z}_j) \\ & - {64 m_{\tilde{Z}_i}^2 \over \pi^2}(-1)^{\theta_i}(-1)^{\theta_j} {\alpha_{\tilde{f}_1}^{\tilde{Z}_i}} {\beta_{\tilde{f}_1}^{\tilde{Z}_i}} {\alpha_{\tilde{f}_1}^{\tilde{Z}_j}} {\beta_{\tilde{f}_1}^{\tilde{Z}_j}} |m_{\tilde{Z}_i}|m_{f}^2 |m_{\tilde{Z}_j}| \zeta(\tilde{Z}_{i},\tilde{f}_{1}, \tilde{f}_1, \tilde{Z}_j),
\end{aligned}
\end{equation}

\begin{equation}
\begin{aligned}
\Gamma_{\tilde{f}_2 \tilde{f}_2}^{diag} = & {8m_{\tilde{Z}_i}^2 \over \pi^2}({\alpha_{\tilde{f}_2}^{\tilde{Z}_i}}^2 + {\beta_{\tilde{f}_2}^{\tilde{Z}_i}}^2) ({\alpha_{\tilde{f}_2}^{\tilde{Z}_j}}^2 + {\beta_{\tilde{f}_2}^{\tilde{Z}_j}}^2)  \tilde{\psi}(\tilde{Z}_i,\tilde{f}_2,\tilde{f}_2,\tilde{Z}_j) \\ & + {32 m_{\tilde{Z}_i}^2 \over \pi^2}(-1)^{\theta_j} ({\alpha_{\tilde{f}_2}^{\tilde{Z}_i}}^2 + {\beta_{\tilde{f}_2}^{\tilde{Z}_i}}^2) {\alpha_{\tilde{f}_2}^{\tilde{Z}_j}} {\beta_{\tilde{f}_2}^{\tilde{Z}_j}} m_{f} |m_{\tilde{Z}_j}|\tilde{\chi}(\tilde{Z}_i,\tilde{f}_2, \tilde{f}_2, \tilde{Z}_j) \\ & - {32 m_{\tilde{Z}_i}^2 \over \pi^2}(-1)^{\theta_i} ({\alpha_{\tilde{f}_2}^{\tilde{Z}_j}}^2 + {\beta_{\tilde{f}_2}^{\tilde{Z}_j}}^2) {\alpha_{\tilde{f}_2}^{\tilde{Z}_i}} {\beta_{\tilde{f}_2}^{\tilde{Z}_i}} m_f |m_{\tilde{Z}_i}|X(\tilde{Z}_i, \tilde{f}_2, \tilde{f}_2, \tilde{Z}_j) \\ & - {64 m_{\tilde{Z}_i}^2 \over \pi^2}(-1)^{\theta_i}(-1)^{\theta_j} {\alpha_{\tilde{f}_2}^{\tilde{Z}_i}} {\beta_{\tilde{f}_2}^{\tilde{Z}_i}} {\alpha_{\tilde{f}_2}^{\tilde{Z}_j}} {\beta_{\tilde{f}_2}^{\tilde{Z}_j}} |m_{\tilde{Z}_i}|m_{f}^2 |m_{\tilde{Z}_j}| \zeta(\tilde{Z}_i,\tilde{f}_2, \tilde{f}_2, \tilde{Z}_j),
\end{aligned}
\end{equation}

where the $\tilde{\psi}$, $\tilde{\chi}$, $X$, $\zeta$ integrals have been used before for the gluino 3 body decays and are:

\begin{equation}
E^{max}_{f} = {m_{\tilde{Z}_i}^2 - 2m_{f}|m_{\tilde{Z}_j} - m_{\tilde{Z}_j}^2 \over 2|m_{\tilde{Z}_i}|},
\end{equation}

\begin{equation} \label{psiintegral}
\tilde{\psi} (m_{\tilde{Z}_i},m_{\tilde{f}_1},m_{\tilde{f}_2},m_{\tilde{Z}_j}) = \pi^2 m_{\tilde{Z}_i} \int_{m_{f}}^{E^{max}_{f}} dE_f \sqrt{{E_f}^2 - m_{f}^2} E_f {\lambda^{1 \over 2} (s,m_{f}^2, m_{\tilde{Z}_j}^2) \over s} {m_{\tilde{Z}_i}^2 - m_{\tilde{Z}_j}^2 - 2m_{\tilde{Z}_i}E_f \over (s-m_{\tilde{f}_1}^2)(s-m_{\tilde{f}_2}^2)},
\end{equation}

\begin{equation} \label{chitildaintegral}
\tilde{\chi} (m_{\tilde{Z}_i},m_{\tilde{f}_1},m_{\tilde{f}_2},m_{\tilde{Z}_j}) = \pi^2 m_{\tilde{Z}_i} \int_{m_{f}}^{E^{max}_{f}} dE_f \sqrt{{E_f}^2 - m_{f}^2} E_f {\lambda^{1 \over 2} (s,m_{f}^2, m_{\tilde{Z}_j}^2) \over s} {1 \over (s-m_{\tilde{f}_1}^2)(s-m_{\tilde{f}_2}^2)},
\end{equation}

\begin{equation}  \label{Xintegral}
X(m_{\tilde{Z}_i},m_{\tilde{f}_1},m_{\tilde{f}_2},m_{\tilde{Z}_j}) ={\pi^2 \over 2} \int_{m_{f}}^{E^{max}_{f}} dE_f \sqrt{{E_f}^2 - m_{f}^2} {m_{\tilde{Z}_i}^2 - m_{\tilde{Z}_j}^2 - 2m_{\tilde{Z}_i}E_{f} \over m_{\tilde{Z}_i}^2 + m_{f}^2 - 2m_{\tilde{Z}_i}E_{f}} {\lambda^{1 \over 2} (s,m_{f}^2, m_{\tilde{Z}_j}^2) \over (s-m_{\tilde{f}_1}^2)(s-m_{\tilde{f}_2}^2)},
\end{equation}

\begin{equation} \label{zetaintegral}
\zeta(m_{\tilde{Z}_i},m_{\tilde{f}_1},m_{\tilde{f}_2},m_{\tilde{Z}_j}) = \pi^2 \int_{m_{f}}^{E^{max}_{f}} dE_f {[E_{\bar{f}}(max) - E_{\bar{f}}(min)] \over (s-m_{\tilde{f}_1}^2)(s-m_{\tilde{f}_2}^2)}.
\end{equation}

Later the following integrals will also be required:

\begin{equation} \label{phitildaintegral}
\begin{aligned}
\tilde{\phi}(m_{\tilde{Z}_i},m_{\tilde{f}_1},m_{\tilde{f}_2},m_{\tilde{Z}_j}) = {1 \over 2} \pi^2 |m_{\tilde{Z}_i}||m_{\tilde{Z}_j}| \int_{m_{f}}^{E^{max}_{f}} dE_f & {1 \over s-m_{\tilde{f}_1}^2} \Big[-[E_{\bar{f}}(max) - E_{\bar{f}}(min)] - \\ & {m_{\tilde{Z}_j}^2 - m_{f}^2 + 2|m_{\tilde{Z}_i}|E_f - m_{\tilde{f}_2}^2 \over 2|m_{\tilde{Z}_i}|} \log Z(m_{\tilde{f}_2})\Big] ,
\end{aligned}
\end{equation}

where $Z(m_{\tilde{f}_2}) = {m_{\tilde{Z}_i}^2 + m_{f}^2 - 2|m_{\tilde{Z}_i}|E_{\bar{f}}(max) - m_{\tilde{f}_2}^2 \over m_{\tilde{Z}_i}^2 + m_{f}^2 - 2|m_{\tilde{Z}_i}|E_{\bar{f}}(min) - m_{\tilde{f}_2}^2}$.

\begin{equation} \label{xiintegral}
\begin{aligned}
\xi(m_{\tilde{Z}_i},m_{\tilde{f}_1},m_{\tilde{f}_2},m_{\tilde{Z}_j}) = {1 \over 2} \pi^2 \int_{m_{f}}^{E^{max}_{f}} dE_f {1 \over s-m_{\tilde{f}_1}^2} \Big[&[E_{\bar{f}}(max) - E_{\bar{f}}(min)] \\ & - {m_{\tilde{Z}_i}^2 - m_{f}^2 - 2|m_{\tilde{Z}_i}|E_f + m_{\tilde{f}_2}^2 \over 2|m_{\tilde{Z}_i}|} \log Z(m_{\tilde{f}_2})\Big] ,
\end{aligned}
\end{equation}

\begin{equation} \label{rhotildaintegral}
\tilde{\rho}(m_{\tilde{Z}_i},m_{\tilde{f}_1},m_{\tilde{f}_2},m_{\tilde{Z}_j}) = -{\pi^2 \over 2 |m_{\tilde{Z}_i}|} \int_{m_{f}}^{E^{max}_{f}} dE_f {1 \over s-m_{\tilde{f}_1}^2} \log Z(m_{\tilde{f}_2}) ,
\end{equation}

\begin{equation} \label{Yintegral}
\begin{aligned}
Y(m_{\tilde{Z}_i},m_{\tilde{f}_1},m_{\tilde{f}_2},m_{\tilde{Z}_j}) = {\pi^2 \over 2} \int_{m_{f}}^{E^{max}_{f}} & dE_f {1 \over s-m_{\tilde{f}_1}^2}\Big[[E_{\bar{f}}(max) - E_{\bar{f}}(min)]s \\ & + {1 \over 2|m_{\tilde{Z}_i}|}(m_{\tilde{Z}_i}^2 m_{\tilde{Z}_j}^2 - m_{\tilde{Z}_i}^2 m_{\tilde{f}_2}^2 + m_{f}^4 + 2|m_{\tilde{Z}_i}|E_{f}m_{\tilde{f}_2}^2 - m_{\tilde{f}_2}^2 m_{f}^2) \\ & \times \log Z(m_{\tilde{f}_2})\Big] ,
\end{aligned}
\end{equation}

\begin{equation} \label{chiprimeintegral}
\chi^{'}(m_{\tilde{Z}_i},m_{\tilde{f}_1},m_{\tilde{f}_2},m_{\tilde{Z}_j}) = {-\pi^2 \over 2} \int_{m_{f}}^{E^{max}_{f}} {dE_f E_{f} \over s - m_{\tilde{f}_2}^2} \log Z(m_{\tilde{f}_1}),
\end{equation}

where here

\begin{equation}
E_{\bar{f}}(max/min) = {(s + m_{f}^2 - m_{\tilde{Z}_j}^2)(|m_{\tilde{Z}_i}|-E_{f}) \pm \sqrt{(E_{f}^2 - m_{f}^2)(s + m_{f}^2 - m_{\tilde{Z}_j}^2)^2 - 4(E_{f}^2 - m_{f}^2)m_{f}^2 s} \over 2s}.
\end{equation}

\begin{equation}
\begin{aligned}
\Gamma_{\tilde{f} \tilde{f}}^{nondiag} =  {16 m_{\tilde{Z}_i}^2 \over \pi^2}\Big[&({\beta_{\tilde{f}_1}^{\tilde{Z}_i}}{\beta_{\tilde{f}_2}^{\tilde{Z}_i}} + {\alpha_{\tilde{f}_1}^{\tilde{Z}_i}}{\alpha_{\tilde{f}_2}^{\tilde{Z}_i}})({\alpha_{\tilde{f}_1}^{\tilde{Z}_j}}{\alpha_{\tilde{f}_2}^{\tilde{Z}_j}} + {\beta_{\tilde{f}_2}^{\tilde{Z}_j}}{\beta_{\tilde{f}_2}^{\tilde{Z}_j}})\tilde{\psi}(\tilde{Z}_i,\tilde{f}_1, \tilde{f}_2, \tilde{Z}_j) \\ & + 2(-1)^{\theta_j}({\beta_{\tilde{f}_1}^{\tilde{Z}_i}}{\beta_{\tilde{f}_2}^{\tilde{Z}_i}} + {\alpha_{\tilde{f}_1}^{\tilde{Z}_i}} {\alpha_{\tilde{f}_2}^{\tilde{Z}_i}})({\alpha_{\tilde{f}_2}^{\tilde{Z}_j}}{\beta_{\tilde{f}_1}^{\tilde{Z}_j}} + {\alpha_{\tilde{f}_1}^{\tilde{Z}_j}}{\beta_{\tilde{f}_2}^{\tilde{Z}_j}})m_{f} |m_{\tilde{Z}_j}| \tilde{\chi}(\tilde{Z}_i,\tilde{f}_1,\tilde{f}_2,\tilde{Z}_j) \\ & - 2(-1)^{\theta_i}({\alpha_{\tilde{f}_1}^{\tilde{Z}_i}}{\beta_{\tilde{f}_2}^{\tilde{Z}_i}} + {\alpha_{\tilde{f}_2}^{\tilde{Z}_i}}{\beta_{\tilde{f}_1}^{\tilde{Z}_i}})({\alpha_{\tilde{f}_1}^{\tilde{Z}_j}}{\alpha_{\tilde{f}_2}^{\tilde{Z}_j}} + {\beta_{\tilde{f}_2}^{\tilde{Z}_j}}{\beta_{\tilde{f}_1}^{\tilde{Z}_j}})|m_{\tilde{Z}_i}|m_{f}X(\tilde{Z}_i,\tilde{f}_1,\tilde{f}_2,\tilde{Z}_j) \\ & - 2(-1)^{\theta_i}(-1)^{\theta_j}({\alpha_{\tilde{f}_1}^{\tilde{Z}_i}}{\beta_{\tilde{f}_2}^{\tilde{Z}_i}} + {\alpha_{\tilde{f}_2}^{\tilde{Z}_i}} {\beta_{\tilde{f}_1}^{\tilde{Z}_i}})({\alpha_{\tilde{f}_2}^{\tilde{Z}_j}}{\beta_{\tilde{f}_1}^{\tilde{Z}_j}} + {\alpha_{\tilde{f}_1}^{\tilde{Z}_j}} {\beta_{\tilde{f}_2}^{\tilde{Z}_j}})m_{f}^2|m_{\tilde{Z}_i}||m_{\tilde{Z}_j}|\zeta(\tilde{Z}_i,\tilde{f}_1,\tilde{f}_2,\tilde{Z}_j)\Big],
\end{aligned}
\end{equation}

\begin{equation}
\begin{aligned}
\Gamma_{\tilde{f}_1 \tilde{f}_1}^{tu} =  -2\Big\{&8({\alpha_{\tilde{f}_1}^{\tilde{Z}_i}}{\beta_{\tilde{f}_1}^{\tilde{Z}_i}}{\beta_{\tilde{f}_1}^{\tilde{Z}_j}}{\alpha_{\tilde{f}_1}^{\tilde{Z}_j}} + {\beta_{\tilde{f}_1}^{\tilde{Z}_i}}{\alpha_{\tilde{f}_1}^{\tilde{Z}_i}}{\alpha_{\tilde{f}_1}^{\tilde{Z}_j}} {\beta_{\tilde{f}_1}^{\tilde{Z}_j}}){m_{\tilde{Z}_i}^2 \pi^2}(-1)^{\theta_i}Y(\tilde{Z}_i,\tilde{f}_1, \tilde{f}_1, \tilde{Z}_j) \\ & -({\alpha_{\tilde{f}_1}^{\tilde{Z}_i}}^2 {\alpha_{\tilde{f}_1}^{\tilde{Z}_j}}^2 + {\beta_{\tilde{f}_1}^{\tilde{Z}_i}}^2 {\beta_{\tilde{f}_1}^{\tilde{Z}_j}}^2){8 m_{\tilde{Z}_i}^2 \pi^2} (-1)^{\theta_i}(-1)^{\theta_j} \tilde{\phi}(\tilde{Z}_i, \tilde{f}_1, \tilde{f}_1, \tilde{Z}_j) \\ & + ( {\alpha_{\tilde{f}_1}^{\tilde{Z}_i}}{\beta_{\tilde{f}_1}^{\tilde{Z}_i}}{\alpha_{\tilde{f}_1}^{\tilde{Z}_j}}{\beta_{\tilde{f}_1}^{\tilde{Z}_j}}+ {\alpha_{\tilde{f}_1}^{\tilde{Z}_i}}{\beta_{\tilde{f}_1}^{\tilde{Z}_i}} {\alpha_{\tilde{f}_1}^{\tilde{Z}_j}}{\beta_{\tilde{f}_1}^{\tilde{Z}_j}})m_{f}^2 {8 m_{\tilde{Z}_i}^2 \over \pi^2} (-1)^{\theta_i} \xi (\tilde{Z}_i,\tilde{f}_1, \tilde{f}_1, \tilde{Z}_j) \\ & -\{({\alpha_{\tilde{f}_1}^{\tilde{Z}_i}}^2 {\alpha_{\tilde{f}_1}^{\tilde{Z}_j}} {\beta_{\tilde{f}_1}^{\tilde{Z}_j}} + {\beta_{\tilde{f}_1}^{\tilde{Z}_i}}^2 {\alpha_{\tilde{f}_1}^{\tilde{Z}_j}} {\beta_{\tilde{f}_1}^{\tilde{Z}_j}})|m_{\tilde{Z}_i}|m_{f}\}\Big[{8 m_{\tilde{Z}_i}^2 \over \pi^2}\xi(\tilde{Z}_i, \tilde{f}_1, \tilde{f}_1, \tilde{Z}_j) \\ & - {4m_{\tilde{Z}_i}^2 \over \pi^2}(m_{\tilde{Z}_i}^2 + m_{\tilde{Z}_j}^2)(-1)^{\theta_i}\tilde{\rho}(\tilde{Z}_i,\tilde{f}_1,\tilde{f}_1,\tilde{Z}_j) + {8m_{\tilde{Z}_i}^2 \over \pi^2} \chi^{'} (\tilde{Z}_i,\tilde{f}_1,\tilde{f}_1,\tilde{Z}_j)\Big] \\ & + |m_{\tilde{Z}_j}|(-1)^{\theta_j}m_{f}({\alpha_{\tilde{f}_1}^{\tilde{Z}_i}}{\beta_{\tilde{f}_1}^{\tilde{Z}_i}}{\alpha_{\tilde{f}_1}^{\tilde{Z}_j}}^2 + {\alpha_{\tilde{f}_1}^{\tilde{Z}_i}} {\beta_{\tilde{f}_1}^{\tilde{Z}_i}}{\beta_{\tilde{f}_1}^{\tilde{Z}_j}}^2)\Big[{-8m_{\tilde{Z}_i}^2 \over \pi^2}\xi(\tilde{Z}_i,\tilde{f}_1,\tilde{f}_1,\tilde{Z}_j) \\ & + {8 m_{\tilde{Z}_i}^4 \over \pi^2}(-1)^{\theta_i}\tilde{\rho}(\tilde{Z}_i,\tilde{f}_1,\tilde{f}_1,\tilde{Z}_j) - {8 m_{\tilde{Z}_i}^2 \over \pi^2}\chi^{'}(\tilde{Z}_i,\tilde{f}_1,\tilde{f}_1,\tilde{Z}_j)\Big] \\ & - ({\beta_{\tilde{f}_1}^{\tilde{Z}_i}}^2{\beta_{\tilde{f}_1}^{\tilde{Z}_j}}{\alpha_{\tilde{f}_1}^{\tilde{Z}_j}} + {\alpha_{\tilde{f}_1}^{\tilde{Z}_i}}^2{\alpha_{\tilde{f}_1}^{\tilde{Z}_j}} {\beta_{\tilde{f}_1}^{\tilde{Z}_j}})|m_{\tilde{Z}_i}|(-1)^{\theta_i}m_{f}\Big[{4m_{\tilde{Z}_i}^2 \over \pi^2}(m_{\tilde{Z}_i}^2 - m_{\tilde{Z}_j}^2)\tilde{\rho}(\tilde{Z}_i,\tilde{f}_1,\tilde{f}_1,\tilde{Z}_j) \\ & - {8m_{\tilde{Z}_i}^2 \over \pi^2} \chi^{'}(\tilde{Z}_i,\tilde{f}_1,\tilde{f}_1,\tilde{Z}_j)\Big] \\ & + ({\alpha_{\tilde{f}_1}^{\tilde{Z}_i}}{\beta_{\tilde{f}_1}^{\tilde{Z}_i}}{\alpha_{\tilde{f}_1}^{\tilde{Z}_j}}^2 + {\alpha_{\tilde{f}_1}^{\tilde{Z}_i}}{\beta_{\tilde{f}_1}^{\tilde{Z}_i}} {\beta_{\tilde{f}_1}^{\tilde{Z}_j}}^2)m_{f}|m_{\tilde{Z}_j}|{8m_{\tilde{Z}_i}^2 \over \pi^2}(-1)^{\theta_i}(-1)^{\theta_j}\chi^{'}(\tilde{Z}_i,\tilde{f}_1,\tilde{f}_1,\tilde{Z}_j) \\ & -2({\beta_{\tilde{f}_1}^{\tilde{Z}_i}}^2{\alpha_{\tilde{f}_1}^{\tilde{Z}_j}}^2   + {\alpha_{\tilde{f}_1}^{\tilde{Z}_i}}^2 {\beta_{\tilde{f}_1}^{\tilde{Z}_j}}^2)m_{f}^2 (-1)^{\theta_i}(-1)^{\theta_j}|m_{\tilde{Z}_i}||m_{\tilde{Z}_j}|{4 m_{\tilde{Z}_i}^2 \over \pi^2}\tilde{\rho}(\tilde{Z}_i,\tilde{f}_1,\tilde{f}_1,\tilde{Z}_j)\Big\}.
\end{aligned}
\end{equation}

$\Gamma_{\tilde{f}_2 \tilde{f}_2}^{tu}$ is as above but with $\tilde{f}_1 \rightarrow \tilde{f}_2$ everywhere including in the masses and integrals.

\begin{equation}
\begin{aligned}
\Gamma_{\tilde{f}_1 \tilde{f}_2}^{tu} =  -2\Big\{& 8({\alpha_{\tilde{f}_1}^{\tilde{Z}_i}}{\beta_{\tilde{f}_2}^{\tilde{Z}_i}}{\beta_{\tilde{f}_1}^{\tilde{Z}_j}}{\alpha_{\tilde{f}_2}^{\tilde{Z}_j}} + {\beta_{\tilde{f}_1}^{\tilde{Z}_i}}{\alpha_{\tilde{f}_2}^{\tilde{Z}_i}}{\alpha_{\tilde{f}_1}^{\tilde{Z}_j}} {\beta_{\tilde{f}_2}^{\tilde{Z}_j}}){m_{\tilde{Z}_i}^2 \pi^2}Y(\tilde{Z}_i,\tilde{f}_1, \tilde{f}_2, \tilde{Z}_j) \\ & -({\alpha_{\tilde{f}_1}^{\tilde{Z}_i}}{\alpha_{\tilde{f}_2}^{\tilde{Z}_i}} {\alpha_{\tilde{f}_1}^{\tilde{Z}_j}}{\alpha_{\tilde{f}_2}^{\tilde{Z}_j}} + {\beta_{\tilde{f}_1}^{\tilde{Z}_i}}{\beta_{\tilde{f}_2}^{\tilde{Z}_i}} {\beta_{\tilde{f}_1}^{\tilde{Z}_j}}{\beta_{\tilde{f}_2}^{\tilde{Z}_j}}){8 m_{\tilde{Z}_i}^2 \pi^2}(-1)^{\theta_i}(-1)^{\theta_j} \tilde{\phi}(\tilde{Z}_i, \tilde{f}_1, \tilde{f}_2, \tilde{Z}_j) \\ & + ( {\alpha_{\tilde{f}_2}^{\tilde{Z}_i}}{\beta_{\tilde{f}_1}^{\tilde{Z}_i}}{\alpha_{\tilde{f}_2}^{\tilde{Z}_j}}{\beta_{\tilde{f}_1}^{\tilde{Z}_j}}+ {\alpha_{\tilde{f}_1}^{tilde{Z}_i}}{\beta_{\tilde{f}_2}^{\tilde{Z}_i}} {\alpha_{\tilde{f}_1}^{\tilde{Z}_j}}{\beta_{\tilde{f}_2}^{\tilde{Z}_j}})m_{f}^2 {8 m_{\tilde{Z}_i}^2 \over \pi^2}(-1)^{\theta_i} \xi (\tilde{Z}_i,\tilde{f}_1, \tilde{f}_2, \tilde{Z}_j) \\ & -\{({\alpha_{\tilde{f}_1}^{\tilde{Z}_i}}{\alpha_{\tilde{f}_2}^{\tilde{Z}_i}} {\alpha_{\tilde{f}_2}^{\tilde{Z}_j}} {\beta_{\tilde{f}_1}^{\tilde{Z}_j}} + {\beta_{\tilde{f}_1}^{\tilde{Z}_i}}{\beta_{\tilde{f}_2}^{\tilde{Z}_i}} {\alpha_{\tilde{f}_1}^{\tilde{Z}_j}} {\beta_{\tilde{f}_2}^{\tilde{Z}_j}})|m_{\tilde{Z}_i}|m_{f}\}\Big[{8 m_{\tilde{Z}_i}^2 \over \pi^2}\xi(\tilde{Z}_i, \tilde{f}_1, \tilde{f}_2, \tilde{Z}_j) \\ & - {4m_{\tilde{Z}_i}^2 \over \pi^2}(m_{\tilde{Z}_i}^2 + m_{\tilde{Z}_j}^2)(-1)^{\theta_i}\tilde{\rho}(\tilde{Z}_i,\tilde{f}_1,\tilde{f}_2,\tilde{Z}_j) + {8m_{\tilde{Z}_i}^2 \over \pi^2} \chi^{'} (\tilde{Z}_i,\tilde{f}_1,\tilde{f}_2,\tilde{Z}_j)\Big] \\ & + |m_{\tilde{Z}_j}|(-1)^{\theta_j}m_{f}({\alpha_{\tilde{f}_2}^{\tilde{Z}_i}}{\beta_{\tilde{f}_1}^{\tilde{Z}_i}}{\alpha_{\tilde{f}_1}^{\tilde{Z}_j}}{\alpha_{\tilde{f}_2}^{\tilde{Z}_j}} + {\alpha_{\tilde{f}_1}^{\tilde{Z}_i}} {\beta_{\tilde{f}_2}^{\tilde{Z}_i}}{\beta_{\tilde{f}_1}^{\tilde{Z}_j}}{\beta_{\tilde{f}_2}^{\tilde{Z}_j}})\Big[{-8m_{\tilde{Z}_i}^2 \over \pi^2}\xi(\tilde{Z}_i,\tilde{f}_1,\tilde{f}_2,\tilde{Z}_j) \\ & + {8 m_{\tilde{Z}_i}^4 \over \pi^2}(-1)^{\theta_i}\tilde{\rho}(\tilde{Z}_i,\tilde{f}_1,\tilde{f}_2,\tilde{Z}_j) - {8 m_{\tilde{Z}_i}^2 \over \pi^2}\chi^{'}(\tilde{Z}_i,\tilde{f}_1,\tilde{f}_2,\tilde{Z}_j)\Big] \\ & - ({\beta_{\tilde{f}_1}^{\tilde{Z}_i}}{\beta_{\tilde{f}_2}^{\tilde{Z}_i}}{\beta_{\tilde{f}_1}^{\tilde{Z}_j}}{\alpha_{\tilde{f}_2}^{\tilde{Z}_j}} + {\alpha_{\tilde{f}_1}^{\tilde{Z}_i}}{\alpha_{\tilde{f}_2}^{\tilde{Z}_i}}{\alpha_{\tilde{f}_1}^{\tilde{Z}_j}} {\beta_{\tilde{f}_2}^{\tilde{Z}_j}})|m_{\tilde{Z}_i}|(-1)^{\theta_i}m_{f}\Big[{4m_{\tilde{Z}_i}^2 \over \pi^2}(m_{\tilde{Z}_i}^2 - m_{\tilde{Z}_j}^2)\tilde{\rho}(\tilde{Z}_i,\tilde{f}_1,\tilde{f}_2,\tilde{Z}_j) \\ & - {8m_{\tilde{Z}_i}^2 \over \pi^2} \chi^{'}(\tilde{Z}_i,\tilde{f}_1,\tilde{f}_2,\tilde{Z}_j))\Big] \\ & + ({\alpha_{\tilde{f}_1}^{\tilde{Z}_i}}{\beta_{\tilde{f}_2}^{\tilde{Z}_i}}{\alpha_{\tilde{f}_1}^{\tilde{Z}_j}}{\alpha_{\tilde{f}_2}^{\tilde{Z}_j}} + {\alpha_{\tilde{f}_2}^{\tilde{Z}_i}}{\beta_{\tilde{f}_1}^{\tilde{Z}_i}}{\beta_{\tilde{f}_1}^{\tilde{Z}_j}}{\beta_{\tilde{f}_2}^{\tilde{Z}_i}} m_{f}(-1)^{\theta_j}|m_{\tilde{Z}_j})|{8m_{\tilde{Z}_i}^2 \over \pi^2}\chi^{'}(\tilde{Z}_i,\tilde{f}_1,\tilde{f}_2,\tilde{Z}_j) \\ & -2({\beta_{\tilde{f}_1}^{\tilde{Z}_i}}{\beta_{\tilde{f}_2}^{\tilde{Z}_i}}{\alpha_{\tilde{f}_1}^{\tilde{Z}_j}}{\alpha_{\tilde{f}_2}^{\tilde{Z}_j}}   + {\alpha_{\tilde{f}_1}^{\tilde{Z}_i}}{\alpha_{\tilde{f}_2}^{\tilde{Z}_i}} {\beta_{\tilde{f}_1}^{\tilde{Z}_j}}{\beta_{\tilde{f}_2}^{\tilde{Z}_j}})m_{f}^2(-1)^{\theta_i}(-1)^{\theta_j}|m_{\tilde{Z}_i}||m_{\tilde{Z}_j}|{4 m_{\tilde{Z}_i}^2 \over \pi^2}\tilde{\rho}(\tilde{Z}_i,\tilde{f}_1,\tilde{f}_2,\tilde{Z}_j)\Big\}.
\end{aligned}
\end{equation}

As for the $Z$ sfermion interferences:

\begin{equation}
\Gamma_{Z \tilde{f}_1} = (-1)^{\theta_i}(C_{1}^{Z \tilde{f}_1} I_{1}^{Z \tilde{f}_1} + C_{2}^{Z \tilde{f}_1} I_{2}^{Z \tilde{f}_1} + C_{3}^{Z \tilde{f}_1} I_{3}^{Z \tilde{f}_1} + C_{4}^{Z \tilde{f}_1} I_{4}^{Z \tilde{f}_1} + C_{5}^{Z \tilde{f}_1} I_{5}^{Z \tilde{f}_1} + C_{6}^{Z \tilde{f}_1} I_{6}^{Z \tilde{f}_1} + C_{7}^{Z \tilde{f}_1} I_{7}^{Z \tilde{f}_1} + C_{8}^{Z \tilde{f}_1} I_{8}^{Z \tilde{f}_1}),
\end{equation}

where
\begin{align}
C_{1}^{Z \tilde{f}_1} &= -4W_{ij}g \sin\theta_W [-{\alpha_{\tilde{f}_1}^{\tilde{Z}_i}}(\alpha_f-\beta_f){\beta_{\tilde{f}_1}^{\tilde{Z}_j}} + {\beta_{\tilde{f}_1}^{\tilde{Z}_i}}(\alpha_f+\beta_f){\alpha_{\tilde{f}_1}^{\tilde{Z}_j}}]m_{f}|m_{\tilde{Z}_i}|, \\
C_{2}^{Z \tilde{f}_1} &= -4(-1)^{\theta_i}(-1)^{\theta_j}W_{ij}g \sin\theta_W [-{\alpha_{\tilde{f}_1}^{\tilde{Z}_i}}(\alpha_f+\beta_f){\beta_{\tilde{f}_1}^{\tilde{Z}_j}} + {\beta_{\tilde{f}_1}^{\tilde{Z}_i}}(\alpha_f-\beta_f){\alpha_{\tilde{f}_1}^{\tilde{Z}_j}}]m_{f}|m_{\tilde{Z}_j}|, \\
C_{3}^{Z \tilde{f}_1} &= -4(-1)^{\theta_i}W_{ij}g \sin\theta_W [{\beta_{\tilde{f}_1}^{\tilde{Z}_i}}(\alpha_f + \beta_f){\beta_{\tilde{f}_1}^{\tilde{Z}_j}} - {\alpha_{\tilde{f}_1}^{\tilde{Z}_i}}(\alpha_f - \beta_f){\alpha_{\tilde{f}_1}^{\tilde{Z}_j}}], \\
C_{4}^{Z \tilde{f}_1} &= -8W_{ij}g \sin\theta_W [-{\alpha_{\tilde{f}_1}^{\tilde{Z}_i}}(\alpha_f + \beta_f){\beta_{\tilde{f}_1}^{\tilde{Z}_j}} + {\beta_{\tilde{f}_1}^{\tilde{Z}_i}}(\alpha_f - \beta_f){\alpha_{\tilde{f}_1}^{\tilde{Z}_j}}]|m_{\tilde{Z}_i}|m_{f}, \\
C_{5}^{Z \tilde{f}_1} &= -8(-1)^{\theta_i}(-1)^{\theta_j}W_{ij}g \sin\theta_W [-{\alpha_{\tilde{f}_1}^{\tilde{Z}_i}}(\alpha_f - \beta_f){\beta_{\tilde{f}_1}^{\tilde{Z}_j}} + {\beta_{\tilde{f}_1}^{\tilde{Z}_i}}(\alpha_f + \beta_f){\alpha_{\tilde{f}_1}^{\tilde{Z}_j}}]|m_{\tilde{Z}_j}|m_{f}, \\
C_{6}^{Z \tilde{f}_1} &= -4(-1)^{\theta_j}W_{ij}g \sin\theta_W [{\beta_{\tilde{f}_1}^{\tilde{Z}_i}}(\alpha_f + \beta_f){\beta_{\tilde{f}_1}^{\tilde{Z}_j}} - {\alpha_{\tilde{f}_1}^{\tilde{Z}_i}}(\alpha_f - \beta_f){\alpha_{\tilde{f}_1}^{\tilde{Z}_j}}]|m_{\tilde{Z}_i}||m_{\tilde{Z}_j}|, \\
C_{7}^{Z \tilde{f}_1} &= -4(-1)^{\theta_i}W_{ij}g \sin\theta_W [{\beta_{\tilde{f}_1}^{\tilde{Z}_i}}(\alpha_f - \beta_f){\beta_{\tilde{f}_1}^{\tilde{Z}_j}} - {\alpha_{\tilde{f}_1}^{\tilde{Z}_i}}(\alpha_f + \beta_f){\alpha_{\tilde{f}_1}^{\tilde{•}tilde{Z}_j}}]m_{f}^2, \\
C_{8}^{Z \tilde{f}_1} &= -16(-1)^{\theta_j}W_{ij}g \sin\theta_W [{\beta_{\tilde{f}_1}^{\tilde{Z}_i}}(\alpha_f - \beta_f){\beta_{\tilde{f}_1}^{\tilde{Z}_j}} - {\alpha_{\tilde{f}_1}^{\tilde{Z}_i}}(\alpha_f + \beta_f){\alpha_{\tilde{f}_1}^{\tilde{Z}_j}}]m_{f}^2 |m_{\tilde{Z}_i}||m_{\tilde{Z}_j}|.
\end{align}

The upper limit for the integrals here is $E_{upper} = {(m_{\tilde{Z}_i}^2 + m_{\tilde{Z}_j}^2 -4 m_{f}^2) \over 2|m_{\tilde{Z}_i}|}$. The argument of the logarithm in these integrals is as follows:

\begin{equation}
L = [|m_{\tilde{Z}_i}|(E_{Q} + Q^{'})- \mu^2]/[|m_{\tilde{Z}_i}|(E_{Q} - Q^{'}) - \mu^2],
\end{equation}

In these expressions, $E_{Q} = {s + m_{\tilde{Z}_i}^2 - m_{\tilde{Z}_j}^2 \over 2|m_{\tilde{Z}_i}|}$, $Q^{'} = \sqrt{E_{Q}^2 - s}\sqrt{1 - 4m_{f}^2/s}$ and $\mu^2 = s + m_{\tilde{f}_1}^2 - m_{\tilde{Z}_j}^2 - m_{q}^2$, where $s = m_{\tilde{Z}_i}^2 + m_{\tilde{Z}_j}^2 - 2|m_{\tilde{Z}_i}|E$. The necessary integrals are given by:

\begin{equation}
I_{1}^{Z \tilde{f}_1} = 2|m_{\tilde{Z}_i}|\int_{|m_{\tilde{Z}_j}|}^{E_{upper}} dE {1 \over s-m_{Z}^2}\left[-2|m_{\tilde{Z}_i}|\sqrt{1- 4m_{q}^2/s}\sqrt{E^2 - m_{\tilde{Z}_j}^2} - (m_{\tilde{f}_1}^2 - m_{f}^2 + m_{\tilde{Z}_j}^2 - 2|m_{\tilde{Z}_i}|E) \log L\right],
\end{equation}

\begin{equation}
I_{2}^{Z \tilde{f}_1} = 2|m_{\tilde{Z}_i}|\int_{|m_{\tilde{Z}_j}|}^{E_{upper}} dE {1 \over s-m_{Z}^2}\left[2|m_{\tilde{Z}_i}|\sqrt{1- 4m_{q}^2/s}\sqrt{E^2 - m_{\tilde{Z}_j}^2} + (m_{\tilde{f}_1}^2 + m_{\tilde{Z}_i}^2 - 2|m_{\tilde{Z}_i}|E - m_{f}^2 ) \log L\right]	,
\end{equation}

\begin{equation}
\begin{aligned}
I_{3}^{Z \tilde{f}_1} = & 2|m_{\tilde{Z}_i}|\int_{|m_{\tilde{Z}_j}|}^{E_{upper}} dE {1 \over s-m_{Z}^2}\Big[\{m_{\tilde{Z}_i}^2 + 2m_{f}^2 + m_{\tilde{Z}_j}^2 - {3 \over 2}m_{\tilde{f}_1}^2 - {1 \over 2}(m_{f}^2 + |m_{\tilde{Z}_i}|E \\ & + |m_{\tilde{Z}_i}|\sqrt{1-4m_{f}^2/s}\sqrt{E^2-m_{\tilde{Z}_j}^2}\ ) \} (m_{f}^2 + |m_{\tilde{Z}_i}|E + |m_{\tilde{Z}_i}|\sqrt{1-4m_{f}^2/s}\sqrt{E^2-m_{\tilde{Z}_j}^2} - m_{\tilde{f}_1}^2) \\ & - \{m_{\tilde{Z}_i}^2 + 2m_{f}^2 + m_{\tilde{Z}_j}^2 - {3 \over 2}m_{\tilde{f}_1}^2 - {1 \over 2}(m_{f}^2 + |m_{\tilde{Z}_i}|E - |m_{\tilde{Z}_i}|\sqrt{1-4m_{f}^2/s}\sqrt{E^2-m_{\tilde{Z}_j}^2}\ ) \\ & \times (m_{f}^2 + |m_{\tilde{Z}_i}|E - |m_{\tilde{Z}_i}|\sqrt{1-4m_{f}^2/s}\sqrt{E^2-m_{\tilde{Z}_j}^2}\ ) -m_{\tilde{f}_1}^2 \}\\ & + (m_{\tilde{Z}_i}^2 +m_{f}^2 - m_{\tilde{f}_1}^2)(m_{\tilde{f}_1}^2 - m_{f}^2 - m_{\tilde{Z}_j}^2) \log L\Big], 
\end{aligned}
\end{equation}

\begin{equation}
I_{4}^{Z \tilde{f}_1} = 2|m_{\tilde{Z}_i}|\int_{|m_{\tilde{Z}_j}|}^{E_{upper}} dE {1 \over s-m_{Z}^2}
\left[2|m_{\tilde{Z}_i}|\sqrt{1-4m_{f}^2/s}\sqrt{E^2 - m_{\tilde{Z}_j}^2} + (m_{\tilde{f}_1}^2 - m_{f}^2 -m_{\tilde{Z}_j}^2) \log L\right],
\end{equation}

\begin{equation}
I_{5}^{Z \tilde{f}_1} = 2|m_{\tilde{Z}_i}|\int_{|m_{\tilde{Z}_j}|}^{E_{upper}} dE {-1 \over s-m_{Z}^2}
\left[2|m_{\tilde{Z}_i}|\sqrt{1-4m_{f}^2/s}\sqrt{E^2 - m_{\tilde{Z}_j}^2} + (m_{\tilde{f}_1}^2 - m_{f}^2 -m_{\tilde{Z}_i}^2) \log L\right],
\end{equation}

\begin{equation}
I_{6}^{Z \tilde{f}_1} = 2|m_{\tilde{Z}_i}|\int_{|m_{\tilde{Z}_j}|}^{E_{upper}} dE {1 \over s-m_{Z}^2}(s - 2m_{f}^2)\log L,
\end{equation}

\begin{equation}
I_{7}^{Z \tilde{f}_1} = 2|m_{\tilde{Z}_i}|\int_{|m_{\tilde{Z}_j}|}^{E_{upper}} dE {1 \over s-m_{Z}^2} 2|m_{\tilde{Z}_i}| E \log L,
\end{equation}

\begin{equation}
I_{8}^{Z \tilde{f}_1} = 2|m_{\tilde{Z}_i}|\int_{|m_{\tilde{Z}_j}|}^{E_{upper}} dE {1 \over s-m_{Z}^2} \log L.
\end{equation}

$\Gamma_{Z \tilde{f}_2}$ is exactly the same as $\Gamma_{Z \tilde{f}_1}$ but with the change $\tilde{f}_1 \rightarrow \tilde{f}_2$ throughout to get the couplings $C_{1,\ldots,8}^{Z \tilde{f}_2}$ and the integrals $I_{1,\ldots,8}^{Z \tilde{f}_2}$.

\begin{equation}
\Gamma_{h \tilde{f}_1} = C_{1}^{h \tilde{f}_1} I_{1}^{h \tilde{f}_1} + C_{2}^{h \tilde{f}_1} I_{2}^{h \tilde{f}_1} + C_{3}^{h \tilde{f}_1} I_{3}^{h \tilde{f}_1} + C_{4}^{h \tilde{f}_1} I_{4}^{h \tilde{f}_1} + C_{5}^{h \tilde{f}_1} I_{5}^{h \tilde{f}_1} + C_{6}^{h \tilde{f}_1} I_{6}^{h \tilde{f}_1} + C_{7}^{h \tilde{f}_1} I_{7}^{h \tilde{f}_1} + C_{8}^{h \tilde{f}_1} I_{8}^{h \tilde{f}_1},
\end{equation}
where here the couplings are:
\begin{align}
C_{1}^{h \tilde{f}_1} &= -\frac{1}{2}(-1)^{\theta_i}(-1)^{\theta_j}(X_{ij}^h + X_{ji}^h) {f_q \over \sqrt{2}} t_{\alpha_h} ({\alpha_{\tilde{f}_1}^{\tilde{Z}_i}} {\beta_{\tilde{f}_1}^{\tilde{Z}_j}} + {\beta_{\tilde{f}_1}^{\tilde{Z}_i}}{\alpha_{\tilde{f}_1}^{\tilde{Z}_j}}), \\
C_{2}^{h \tilde{f}_1} &= -(-1)^{\theta_j}(X_{ij}^h + X_{ji}^h) {f_q \over \sqrt{2}} t_{\alpha_h} ({\beta_{\tilde{f}_1}^{\tilde{Z}_i}} {\beta_{\tilde{f}_1}^{\tilde{Z}_j}} + {\alpha_{\tilde{f}_1}^{\tilde{Z}_i}}{\alpha_{\tilde{f}_1}^{\tilde{Z}_j}})|m_{\tilde{Z}_i}| m_f, \\
C_{3}^{h \tilde{f}_1} &= (-1)^{\theta_i}(X_{ij}^h + X_{ji}^h) {f_q \over \sqrt{2}} t_{\alpha_h} ({\beta_{\tilde{f}_1}^{\tilde{Z}_i}} {\beta_{\tilde{f}_1}^{\tilde{Z}_j}} + {\alpha_{\tilde{f}_1}^{\tilde{Z}_i}}{\alpha_{\tilde{f}_1}^{\tilde{Z}_j}})|m_{\tilde{Z}_j}| m_f, \\
C_{4}^{h \tilde{f}_1} &= -(-1)^{\theta_j}(X_{ij}^h + X_{ji}^h) {f_q \over \sqrt{2}} t_{\alpha_h} (-{\beta_{\tilde{f}_1}^{\tilde{Z}_i}} {\beta_{\tilde{f}_1}^{\tilde{Z}_j}} - {\alpha_{\tilde{f}_1}^{\tilde{Z}_i}}{\alpha_{\tilde{f}_1}^{\tilde{Z}_j}})|m_{\tilde{Z}_i}| m_f, \\
C_{5}^{h \tilde{f}_1} &= -(-1)^{\theta_i}(X_{ij}^h + X_{ji}^h) {f_q \over \sqrt{2}} t_{\alpha_h} ({\beta_{\tilde{f}_1}^{\tilde{Z}_i}} {\beta_{\tilde{f}_1}^{\tilde{Z}_j}} + {\alpha_{\tilde{f}_1}^{\tilde{Z}_i}}{\alpha_{\tilde{f}_1}^{\tilde{Z}_j}})|m_{\tilde{Z}_j}| m_f, \\
C_{6}^{h \tilde{f}_1} &= (X_{ij}^h + X_{ji}^h) {f_q \over \sqrt{2}} t_{\alpha_h} (-{\alpha_{\tilde{f}_1}^{\tilde{Z}_i}} {\beta_{\tilde{f}_1}^{\tilde{Z}_j}} - {\beta_{\tilde{f}_1}^{\tilde{Z}_i}}{\alpha_{\tilde{f}_1}^{\tilde{Z}_j}})|m_{\tilde{Z}_i}||m_{\tilde{Z}_j}|, \\
C_{7}^{h \tilde{f}_1} &= (-1)^{\theta_i}(-1)^{\theta_j}(X_{ij}^h + X_{ji}^h) {f_q \over \sqrt{2}} t_{\alpha_h} ({\alpha_{\tilde{f}_1}^{\tilde{Z}_i}} {\beta_{\tilde{f}_1}^{\tilde{Z}_j}} + {\beta_{\tilde{f}_1}^{\tilde{Z}_i}}{\alpha_{\tilde{f}_1}^{\tilde{Z}_j}})m_{f}^2, \\
C_{8}^{h \tilde{f}_1} &= 2(X_{ij}^h + X_{ji}^h) {f_q \over \sqrt{2}} t_{\alpha_h} ({\alpha_{\tilde{f}_1}^{\tilde{Z}_i}} {\beta_{\tilde{f}_1}^{\tilde{Z}_j}} + {\beta_{\tilde{f}_1}^{\tilde{Z}_i}}{\alpha_{\tilde{f}_1}^{\tilde{Z}_j}})|m_{\tilde{Z}_i}|m_{f}^2 |m_{\tilde{Z}_j}|.
\end{align}
The necessary integrals are:
\begin{equation}
I_{1}^{h \tilde{f}_1} = 2|m_{\tilde{Z}_i}|\int_{|m_{\tilde{Z}_j}|}^{E_{upper}} dE {2 \over s-m_{h}^2} \left[2s|m_{\tilde{Z}_i}|\sqrt{E^2 - m_{\tilde{Z}_j}^2}\sqrt{1-4m_{f}^2/s} + \{m_{\tilde{f}_1}^2 s - m_{f}^2 (m_{\tilde{Z}_i}^2 + m_{\tilde{Z}_j}^2)\} \log L\right],
\end{equation}

\begin{equation}
I_{2}^{h \tilde{f}_1} = -2|m_{\tilde{Z}_i}|\int_{|m_{\tilde{Z}_j}|}^{E_{upper}} dE {1 \over s-m_{h}^2} \left[2|m_{\tilde{Z}_i}|\sqrt{E^2 - m_{\tilde{Z}_j}^2}\sqrt{1-4m_{f}^2/s} + (m_{\tilde{f}_1}^2 - m_{f}^2 + m_{\tilde{Z}_j}^2 - 2m_{\tilde{Z}_i}E) \log L\right],
\end{equation}

\begin{equation}
I_{3}^{h \tilde{f}_1} = 2|m_{\tilde{Z}_i}|\int_{|m_{\tilde{Z}_j}|}^{E_{upper}} dE {1 \over s-m_{h}^2} \left[2|m_{\tilde{Z}_i}|\sqrt{E^2 - m_{\tilde{Z}_j}^2}\sqrt{1-4m_{f}^2/s} + (m_{\tilde{f}_1}^2 - m_{f}^2 + m_{\tilde{Z}_i}^2 - 2m_{\tilde{Z}_i}E) \log L\right],
\end{equation}

\begin{equation}
I_{4}^{h \tilde{f}_1} = 2|m_{\tilde{Z}_i}|\int_{|m_{\tilde{Z}_j}|}^{E_{upper}} dE {1 \over s-m_{h}^2} \left[2|m_{\tilde{Z}_i}|\sqrt{E^2 - m_{\tilde{Z}_j}^2}\sqrt{1-4m_{f}^2/s} + (m_{\tilde{f}_1}^2 - m_{f}^2 + m_{\tilde{Z}_j}^2 ) \log L\right],
\end{equation}

\begin{equation}
I_{5}^{h \tilde{f}_1} = -2|m_{\tilde{Z}_i}|\int_{|m_{\tilde{Z}_j}|}^{E_{upper}} dE {1 \over s-m_{h}^2} \left[2|m_{\tilde{Z}_i}|\sqrt{E^2 - m_{\tilde{Z}_j}^2}\sqrt{1-4m_{f}^2/s} + (m_{\tilde{f}_1}^2 - m_{f}^2 + m_{\tilde{Z}_i}^2 ) \log L\right],
\end{equation}

\begin{equation}
I_{6}^{h \tilde{f}_1} = -2|m_{\tilde{Z}_i}|\int_{|m_{\tilde{Z}_j}|}^{E_{upper}} dE {1 \over s-m_{h}^2} (s - 2m_{f}^2) \log L,
\end{equation}

\begin{equation}
I_{7}^{h \tilde{f}_1} = -2|m_{\tilde{Z}_i}|\int_{|m_{\tilde{Z}_j}|}^{E_{upper}} dE {1 \over s-m_{h}^2} (2|m_{\tilde{Z}_i}|E) \log L,
\end{equation}

\begin{equation}
I_{8}^{h \tilde{f}_1} = -2|m_{\tilde{Z}_i}|\int_{|m_{\tilde{Z}_j}|}^{E_{upper}} dE {1 \over s-m_{h}^2} \log L.
\end{equation}

Note $\Gamma_{h \tilde{f}_2}$ is exactly the same as $\Gamma_{h \tilde{f}_1}$ but with the replacement $\tilde{f}_1 \rightarrow \tilde{f}_2$ throughout to get the couplings $C_{1,\ldots,8}^{h \tilde{f}_2}$ and the integrals $I_{1,\ldots,8}^{h \tilde{f}_2}$.
Similarly, one can obtain the $\Gamma_{H \tilde{f}_1}$ from $\Gamma_{h \tilde{f}_1}$ by replacing h by H throughout all the couplings, masses and integrals; therefore the changes $X_{ij}^h + X_{ji}^h \rightarrow X_{ij}^H + X_{ji}^H$ and $t_{\alpha_h} \rightarrow t_{\alpha_H}$ are made. One can then obtain $\Gamma_{H \tilde{f}_2}$ again by changing $\tilde{f}_1 \rightarrow \tilde{f}_2$ throughout the couplings, masses and integrals.
\begin{equation}
\Gamma_{A \tilde{f}_1} = C_{1}^{A \tilde{f}_1} I_{1}^{A \tilde{f}_1} + C_{2}^{A \tilde{f}_1} I_{2}^{A \tilde{f}_1} + C_{3}^{A \tilde{f}_1} I_{3}^{A \tilde{f}_1} + C_{4}^{A \tilde{f}_1} I_{4}^{A \tilde{f}_1} + C_{5}^{A \tilde{f}_1} I_{5}^{A \tilde{f}_1} + C_{6}^{A \tilde{f}_1} I_{6}^{A \tilde{f}_1} + C_{7}^{A \tilde{f}_1} I_{7}^{A \tilde{f}_1} + C_{8}^{A \tilde{f}_1} I_{8}^{A \tilde{f}_1},
\end{equation}
here we have:
\begin{align}
C_{1}^{A \tilde{f}_1} &= \frac{1}{2}(-1)^{\theta_i}(X_{ij}^A + X_{ji}^A) {A_q \over 2} ({\alpha_{\tilde{f}_1}^{\tilde{Z}_i}} {\beta_{\tilde{f}_1}^{\tilde{Z}_j}} + {\beta_{\tilde{f}_1}^{\tilde{Z}_i}}{\alpha_{\tilde{f}_1}^{\tilde{Z}_j}}), \\
C_{2}^{A \tilde{f}_1} &= -(X_{ij}^A + X_{ji}^A) {A_q \over 2} |m_{\tilde{Z}_i}|m_{f} ({\beta_{\tilde{f}_1}^{\tilde{Z}_i}} {\beta_{\tilde{f}_1}^{\tilde{Z}_j}} + {\alpha_{\tilde{f}_1}^{\tilde{Z}_i}}{\alpha_{\tilde{f}_1}^{\tilde{Z}_j}}), \\
C_{3}^{A \tilde{f}_1} &= (-1)^{\theta_i}(-1)^{\theta_j}(X_{ij}^A + X_{ji}^A) {A_q \over 2} |m_{\tilde{Z}_j}|m_{f} ({\beta_{\tilde{f}_1}^{\tilde{Z}_i}} {\beta_{\tilde{f}_1}^{\tilde{Z}_j}} + {\alpha_{\tilde{f}_1}^{\tilde{Z}_i}}{\alpha_{\tilde{f}_1}^{\tilde{Z}_j}}), \\
C_{4}^{A \tilde{f}_1} &= -(X_{ij}^A + X_{ji}^A) {A_q \over 2} |m_{\tilde{Z}_i}|m_{f} ({\beta_{\tilde{f}_1}^{\tilde{Z}_i}} {\beta_{\tilde{f}_1}^{\tilde{Z}_j}} + {\alpha_{\tilde{f}_1}^{\tilde{Z}_i}}{\alpha_{\tilde{f}_1}^{\tilde{Z}_j}}), \\
C_{5}^{A \tilde{f}_1} &= (-1)^{\theta_i}(-1)^{\theta_j}(X_{ij}^A + X_{ji}^A) {A_q \over 2} |m_{\tilde{Z}_j}|m_{f} ({\beta_{\tilde{f}_1}^{\tilde{Z}_i}} {\beta_{\tilde{f}_1}^{\tilde{Z}_j}} + {\alpha_{\tilde{f}_1}^{\tilde{Z}_i}}{\alpha_{\tilde{f}_1}^{\tilde{Z}_j}}), \\
C_{6}^{A \tilde{f}_1} &= -(-1)^{\theta_j}(X_{ij}^A + X_{ji}^A) {A_q \over 2} |m_{\tilde{Z}_i}||m_{\tilde{Z}_j}| ({\alpha_{\tilde{f}_1}^{\tilde{Z}_i}} {\beta_{\tilde{f}_1}^{\tilde{Z}_j}} + {\beta_{\tilde{f}_1}^{\tilde{Z}_i}}{\alpha_{\tilde{f}_1}^{\tilde{Z}_j}}), \\
C_{7}^{A \tilde{f}_1} &= (-1)^{\theta_i}(X_{ij}^A + X_{ji}^A) {A_q \over 2} m_{f}^2 ({\alpha_{\tilde{f}_1}^{\tilde{Z}_i}} {\beta_{\tilde{f}_1}^{\tilde{Z}_j}} + {\beta_{\tilde{f}_1}^{\tilde{Z}_i}}{\alpha_{\tilde{f}_1}^{\tilde{Z}_j}}), \\
C_{8}^{A \tilde{f}_1} &= -(-1)^{\theta_j}(X_{ij}^A + X_{ji}^A) {A_q} m_{f}^2 |m_{\tilde{Z}_i}||m_{\tilde{Z}_j}| ({\alpha_{\tilde{f}_1}^{\tilde{Z}_i}} {\beta_{\tilde{f}_1}^{\tilde{Z}_j}} + {\beta_{\tilde{f}_1}^{\tilde{Z}_i}}{\alpha_{\tilde{f}_1}^{\tilde{Z}_j}}).
\end{align}
The $I_{i}^{A \tilde{f}_1}$ are exactly as the $I_{i}^{h \tilde{f}_1}$ but with the change $m_{h} \rightarrow m_{A}$.
\begin{equation}
\Gamma_{Z A} = 2C_{1}^{Z A} I_{1}^{Z A} + 2C_{2}^{Z A} I_{2}^{Z A},
\end{equation}
where
\begin{align}
C_{1}^{Z A} &= -4(-1)^{\theta_i}(-1)^{\theta_j}W_{ij}(X_{ij}^A + X_{ji}^A)A_{q}g\sin\theta_W \beta_{f} |m_{\tilde{Z}_j}|m_{f}, \\
C_{2}^{Z A} &= 4W_{ij}(X_{ij}^A + X_{ji}^A)A_{q}g\sin\theta_W \beta_{f} |m_{\tilde{Z}_i}|m_{f}.
\end{align}
The integrals included here are:
\begin{equation}
I_{1}^{Z A} = 2|m_{\tilde{Z}_i}|\int_{|m_{\tilde{Z}_j}|}^{E_{upper}} d_{E} {1 \over (s-m_{z}^2)(s-m_{A}^2)}2|m_{\tilde{Z}_i}|\sqrt{E^2-m_{\tilde{Z}_j}^2} \sqrt{1-4m_{f}^2/s} \ \{m_{\tilde{Z}_i}^2 - |m_{\tilde{Z}_i}|E\},
\end{equation}
\begin{equation}
I_{2}^{Z A} = -2|m_{\tilde{Z}_i}|\int_{|m_{\tilde{Z}_j}|}^{E_{upper}} d_{E} {1 \over (s-m_{z}^2)(s-m_{A}^2)}2|m_{\tilde{Z}_i}|\sqrt{E^2-m_{\tilde{Z}_j}^2} \sqrt{1-4m_{f}^2/s} \ \{m_{\tilde{Z}_j}^2 - |m_{\tilde{Z}_i}|E\}.
\end{equation}
The goldstone contribution is:
\begin{equation}
\Gamma_{G} = 4 c_{G \tilde{Z}_i \tilde{Z}_j}^2 c_{G f f}^2\left[I_{4}^G + 2m_{f}^2 I_{3}^G - 2(-1)^{\theta_i}(-1)^{\theta_j}|m_{\tilde{Z}_i}||m_{\tilde{Z}_j}|I_{2}^G - 4(-1)^{\theta_i}(-1)^{\theta_j}m_{f}^2 |m_{\tilde{Z}_i}||m_{\tilde{Z}_j}|I_{1}^G\right].
\end{equation}
The goldstone interferences with fermions are given by:
\begin{equation}
\Gamma_{G \tilde{f}_1} = C_{1}^{G \tilde{f}_1} I_{1}^{G \tilde{f}_1} + C_{2}^{G \tilde{f}_1} I_{2}^{G \tilde{f}_1} + C_{3}^{G \tilde{f}_1} I_{3}^{G \tilde{f}_1} + C_{4}^{G \tilde{f}_1} I_{4}^{G \tilde{f}_1} + C_{5}^{G \tilde{f}_1} I_{5}^{G \tilde{f}_1} + C_{6}^{G \tilde{f}_1} I_{6}^{G \tilde{f}_1} + C_{7}^{G \tilde{f}_1} I_{7}^{G \tilde{f}_1} + C_{8}^{G \tilde{f}_1} I_{8}^{G \tilde{f}_1},
\end{equation}
where
\begin{align}
C_{1}^{G \tilde{f}_1} &= \frac{1}{2} c_{G \tilde{Z}_i \tilde{Z}_j} c_{G f f}({\alpha_{\tilde{f}_1}^{\tilde{Z}_i}}{\beta_{\tilde{f}_1}^{\tilde{Z}_j}} + {\beta_{\tilde{f}_1}^{\tilde{Z}_i}}{\alpha_{\tilde{f}_1}^{\tilde{Z}_j}}), \\
C_{2}^{G \tilde{f}_1} &= -(-1)^{\theta_i}m_{f}|m_{\tilde{Z}_i}| c_{G \tilde{Z}_i \tilde{Z}_j} c_{G f f}({\beta_{\tilde{f}_1}^{\tilde{Z}_i}}{\beta_{\tilde{f}_1}^{\tilde{Z}_j}} + {\alpha_{\tilde{f}_1}^{\tilde{Z}_i}}{\alpha_{\tilde{f}_1}^{\tilde{Z}_j}}), \\
C_{3}^{G \tilde{f}_1} &= (-1)^{\theta_j}m_{f}|m_{\tilde{Z}_j}| c_{G \tilde{Z}_i \tilde{Z}_j} c_{G f f}({\alpha_{\tilde{f}_1}^{\tilde{Z}_i}}{\alpha_{\tilde{f}_1}^{\tilde{Z}_j}} + {\beta_{\tilde{f}_1}^{\tilde{Z}_i}}{\beta_{\tilde{f}_1}^{\tilde{Z}_j}}), \\
C_{4}^{G \tilde{f}_1} &= -(-1)^{\theta_i}m_{f}|m_{\tilde{Z}_i}| c_{G \tilde{Z}_i \tilde{Z}_j} c_{G f f}({\alpha_{\tilde{f}_1}^{\tilde{Z}_i}}{\alpha_{\tilde{f}_1}^{\tilde{Z}_j}} + {\beta_{\tilde{f}_1}^{\tilde{Z}_i}}{\beta_{\tilde{f}_1}^{\tilde{Z}_j}}), \\
C_{5}^{G \tilde{f}_1} &= (-1)^{\theta_j}|m_{\tilde{Z}_i}|m_{f} c_{G \tilde{Z}_i \tilde{Z}_j} c_{G f f}({\alpha_{\tilde{f}_1}^{\tilde{Z}_i}}{\alpha_{\tilde{f}_1}^{\tilde{Z}_j}} + {\beta_{\tilde{f}_1}^{\tilde{Z}_i}}{\beta_{\tilde{f}_1}^{\tilde{Z}_j}}),  \\
C_{6}^{G \tilde{f}_1} &= -(-1)^{\theta_i}(-1)^{\theta_j}|m_{\tilde{Z}_i}||m_{\tilde{Z}_j}| c_{G \tilde{Z}_i \tilde{Z}_j} c_{G f f}({\alpha_{\tilde{f}_1}^{\tilde{Z}_i}}{\beta_{\tilde{f}_1}^{\tilde{Z}_j}} + {\beta_{\tilde{f}_1}^{\tilde{Z}_i}}{\alpha_{\tilde{f}_1}^{\tilde{Z}_j}}),  \\
C_{7}^{G \tilde{f}_1} &= m_{f}^2 c_{G \tilde{Z}_i \tilde{Z}_j} c_{G f f}({\alpha_{\tilde{f}_1}^{\tilde{Z}_i}}{\beta_{\tilde{f}_1}^{\tilde{Z}_j}} + {\beta_{\tilde{f}_1}^{\tilde{Z}_i}}{\alpha_{\tilde{f}_1}^{\tilde{Z}_j}}),  \\
C_{8}^{G \tilde{f}_1} &= -2(-1)^{\theta_i}(-1)^{\theta_j}m_{f}^2|m_{\tilde{Z}_i}||m_{\tilde{Z}_j}| c_{G \tilde{Z}_i \tilde{Z}_j} c_{G f f}({\alpha_{\tilde{f}_1}^{\tilde{Z}_i}}{\beta_{\tilde{f}_1}^{\tilde{Z}_j}} + {\beta_{\tilde{f}_1}^{\tilde{Z}_i}}{\alpha_{\tilde{f}_1}^{\tilde{Z}_j}}).  
\end{align}
The $I_{1,\ldots, 8}^{G \tilde{f}_1}$ integrals are the same as the $I_{1,\ldots,8}^{h \tilde{f}_1}$ but with the replacement $m_{h} \rightarrow m_{Z}$ as the mass of the goldstone is the $Z$ mass as it represents the longitudinal component of the $Z$ boson. Similar changes apply to the $I_{1,\ldots,8}^{G \tilde{f}_2}$, whilst in the couplings we apply the replacement $\tilde{f}_1 \rightarrow \tilde{f}_2$ throughout. The $Z$-goldstone interference contribution is:
\begin{equation}
\Gamma_{Z G} = 2C_{1}^{Z G}I_{1}^{Z G} + 2C_{2}^{Z G}I_{2}^{Z G},
\end{equation}
where
\begin{equation}
C_{1}^{Z G} = -8W_{ij}(-1)^{\theta_j}m_{f}|m_{\tilde{Z}_j}|c_{G f f} c_{G \tilde{Z}_i \tilde{Z}_j} g \sin\theta_W \beta_{f},
\end{equation}
\begin{equation}
C_{2}^{Z G} = 8W_{ij}(-1)^{\theta_i}m_{f}|m_{\tilde{Z}_i}|c_{G f f} c_{G \tilde{Z}_i \tilde{Z}_j} g \sin\theta_W \beta_{f}.
\end{equation}
The $I_{1,2}^{Z G}$ are the same as the $I_{1,2}^{Z A}$ but with the expected change $m_{A} \rightarrow m_{Z}$.
\begin{equation}
\Gamma_{G A} = C_{1}^{G A}C_{3}^{G A}I_{4}^{G A} - 2C_{1}^{GA}C_{4}^{GA}m_{f}^2I_{3}^{G A} + 2C_{2}^{G A} C_{3}^{G A}|m_{\tilde{Z}_i}||m_{\tilde{Z}_j}|I_{2}^{G A} - 4C_{2}^{G A}C_{4}^{G A}m_{f}^2|m_{\tilde{Z}_i}||m_{\tilde{Z}_j}|I_{1}^{G A},
\end{equation}
where here
\begin{equation}
C_{1}^{G A} = -2(-1)^{\theta_i}c_{G \tilde{Z}_i \tilde{Z}_j}(X_{ij}^A + X_{ji}^A),
\end{equation}
\begin{equation}
C_{2}^{G A} = 2(-1)^{\theta_j}c_{G \tilde{Z}_i \tilde{Z}_j}(X_{ij}^A + X_{ji}^A),
\end{equation}
\begin{equation}
C_{3}^{G A} = -A_{q} c_{G f f},
\end{equation}
\begin{equation}
C_{4}^{G A} = A_{q} c_{G f f}.
\end{equation}
The integrals are:
\begin{align}
I_{1}^{G A} &= 2|m_{\tilde{Z}_i}|\int_{|m_{\tilde{Z}_j}|}^{E_{max}}dE {2|m_{\tilde{Z}_i}|\sqrt{E^2 - m_{\tilde{Z}_j}^2}\sqrt{1-4m_{f}^2/s} \over (s-m_{Z}^2)(s-m_{A}^2)}, \\
I_{2}^{G A} &= 2|m_{\tilde{Z}_i}|\int_{|m_{\tilde{Z}_j}|}^{E_{max}}dE {2|m_{\tilde{Z}_i}| (s - 2m_{f}^2)\sqrt{E^2 - m_{\tilde{Z}_j}^2}\sqrt{1-4m_{f}^2/s} \over (s-m_{Z}^2)(s-m_{A}^2)}, \\
I_{3}^{G A} &= 2|m_{\tilde{Z}_i}|\int_{|m_{\tilde{Z}_j}|}^{E_{max}}dE {4|m_{\tilde{Z}_i}|^2 E \sqrt{E^2 - m_{\tilde{Z}_j}^2}\sqrt{1-4m_{f}^2/s} \over (s-m_{Z}^2)(s-m_{A}^2)}, \\
I_{4}^{G A} &= 2|m_{\tilde{Z}_i}|\int_{|m_{\tilde{Z}_j}|}^{E_{max}}dE {4|m_{\tilde{Z}_i}|^2 E (s-2m_{f}^2)\sqrt{E^2 - m_{\tilde{Z}_j}^2}\sqrt{1-4m_{f}^2/s} \over (s-m_{Z}^2)(s-m_{A}^2)}.
\end{align}
Now the list of the couplings used is:
\begin{equation}
W_{ij} = 0.25\sqrt{g^2 + g'^2}(N_{4i}N_{4j}- N_{3i}N_{3j}).
\end{equation}
The $X_{ij}^{\phi}$ couplings are:
\begin{align}
X_{ij}^h &= -\frac{1}{2}(-1)^{\theta_i}(-1)^{\theta_j}[-N_{3i}\sin\alpha - N_{4i} \cos\alpha](-gN_{2j}+g'N_{1j}), \\
X_{ij}^H &= -\frac{1}{2}(-1)^{\theta_i}(-1)^{\theta_j}[N_{3i}\cos\alpha - N_{4i} \sin\alpha](-gN_{2j}+g'N_{1j}), \\ 
X_{ij}^A &= \frac{1}{2}(-1)^{\theta_i}(-1)^{\theta_j}[N_{3i}\sin\beta - N_{4i} \cos\beta](-gN_{2j}+g'N_{1j}).
\end{align}
\begin{equation}
f_{q} = \begin{cases}
{gm_{q}^{run} \over \sqrt{2} m_{W} \sin\beta}, $ for $u$-type quarks,$ \\
{gm_{q}^{run} \over \sqrt{2} m_{W} \cos\beta}, $ for $d$-type quarks,$ \\
0, $ for neutrinos $ \nu$,$ \\
{gm_{l}^{run} \over \sqrt{2} m_{W} \cos\beta}, $ for charged leptons$. \\
\end{cases}
\end{equation}
\begin{equation}
A_{q} = \begin{cases}
{g m_{q}^{run} \over m_{W} \tan\beta}, $ for $u$-type quarks,$ \\
{g m_{q}^{run} \tan\beta \over m_{W}}, $ for $d$-type quarks,$ \\
{0}, $ for neutrinos $ \nu$,$ \\
{g m_{l}^{run} \tan\beta \over m_{W}}, $ for charged leptons$. 
\end{cases}
\end{equation}
\begin{equation}
t_{\alpha_h} = \begin{cases}
\cos\alpha, $ for $u$-type quarks,$ \\
-\sin\alpha, $ for $d$-type quarks,$ \\
\cos\alpha, $ for neutrinos $ \nu$,$ \\
-\sin\alpha, $ for charged leptons$. 
\end{cases}
\end{equation}
\begin{equation}
t_{\alpha_H} = \begin{cases}
\sin\alpha, $ for $u$-type quarks,$ \\
\cos\alpha, $ for $d$-type quarks,$ \\
\sin\alpha, $ for neutrinos $ \nu$,$ \\
\cos\alpha, $ for charged leptons$. 
\end{cases}
\end{equation}
\begin{equation}
\alpha_f = \begin{cases}
-{5g'p \over 12g} + {g \over 4g'}, $ for $u$-type quarks,$ \\
{g'p \over 12g} - {g \over 4g'}, $ for $d$-type quarks,$ \\
{1\over 4}({g' \over g} + {g \over g'}), $ for neutrinos $ \nu$,$ \\
{3 \over 4}{g' \over g} - {g \over 4 g'}, $ for charged leptons$. 
\end{cases}
\end{equation}
\begin{equation}
\beta_f = \begin{cases}
-{1 \over 4}({g' \over g} + {g \over g'}), $ for $u$-type quarks,$ \\
{1 \over 4}({g' \over g} + {g \over g'}), $ for $d$-type quarks,$ \\
-{1 \over 4}({g' \over g} + {g \over g'}), $ for neutrinos $ \nu$,$ \\
{1 \over 4}({g' \over g} + {g \over g'}), $ for charged leptons$. 
\end{cases}
\end{equation}
\begin{equation}
{\alpha_{\tilde{f}_1}^{\tilde{Z}_i}} = \begin{cases}
A_{\tilde{Z}_i}\cos\theta_q - f_{q}N_{4i}\sin\theta_q, $ for $u$-type quarks,$ \\
A_{\tilde{Z}_i}\cos\theta_q - f_{q}N_{3i}\sin\theta_q, $ for $d$-type quarks,$ \\
A_{\tilde{Z}_i}\cos\theta_q, $ for neutrinos $ \nu$,$ \\
A_{\tilde{Z}_i}\sin\theta_q + f_{q}N_{3i}\cos\theta_q, $ for charged leptons$. 
\end{cases}
\end{equation}
\begin{equation}
{\alpha_{\tilde{f}_2}^{\tilde{Z}_i}} = \begin{cases}
A_{\tilde{Z}_i}\sin\theta_q + f_{q}N_{4i}\cos\theta_q, $ for $u$-type quarks,$ \\
A_{\tilde{Z}_i}\sin\theta_q + f_{q}N_{3i}\cos\theta_q, $ for $d$-type quarks,$ \\
A_{\tilde{Z}_i}\sin\theta_q, $ for neutrinos $ \nu$,$ \\
-A_{\tilde{Z}_i}\cos\theta_q + f_{q}N_{3i}\sin\theta_q, $ for charged leptons$. 
\end{cases}
\end{equation}
\begin{equation}
{\beta_{\tilde{f}_1}^{\tilde{Z}_i}} = \begin{cases}
f_{q}N_{4i}\cos\theta_q + B_{\tilde{Z}_i}\sin\theta_q, $ for $u$-type quarks,$ \\
f_{q}N_{3i}\cos\theta_q + B_{\tilde{Z}_i}\sin\theta_q, $ for $d$-type quarks,$ \\
0, $ for neutrinos $ \nu$,$ \\
f_{q}N_{3i}\sin\theta_q - B_{\tilde{Z}_i}\cos\theta_q, $ for charged leptons$. 
\end{cases}
\end{equation}
\begin{equation}
{\beta_{\tilde{f}_2}^{\tilde{Z}_i}} = \begin{cases}
f_{q}N_{4i}\sin\theta_q - B_{\tilde{Z}_i}\cos\theta_q, $ for $u$-type quarks,$ \\
f_{q}N_{3i}\sin\theta_q - B_{\tilde{Z}_i}\cos\theta_q, $ for $d$-type quarks,$ \\
0, $ for neutrinos $ \nu$,$ \\
-f_{q}N_{3i}\cos\theta_q - B_{\tilde{Z}_i}\sin\theta_q, $ for charged leptons$. 
\end{cases}
\end{equation}
\begin{align}
{A_{\tilde{Z}_i}} &= \begin{cases}
-{g \over \sqrt{2}}N_{2i} - {g' \over 3 \sqrt{2}}N_{1i}, $ for $u$-type quarks,$ \\
{g \over \sqrt{2}}N_{2i} - {g' \over 3 \sqrt{2}}N_{1i}, $ for $d$-type quarks,$ \\
-{g \over \sqrt{2}}N_{2i} + {g' \over 3 \sqrt{2}}N_{1i}, $ for neutrinos $ \nu$,$ \\
{g \over \sqrt{2}}N_{2i} + {g' \over 3 \sqrt{2}}N_{1i}, $ for charged leptons$. 
\end{cases} \\
{B_{\tilde{Z}_i}} &= \begin{cases}
-{4 g' \over 3 \sqrt{2}}N_{1i}, $ for $u$-type quarks,$ \\
{2 g' \over 3 \sqrt{2}}N_{1i}, $ for $d$-type quarks,$ \\
0, $ for neutrinos $ \nu $,$\\
\sqrt{2}g' N_{1i}, $ for charged leptons$. 
\end{cases} \\
c_{G f f} &= \begin{cases}
{-f_{q} \sin\beta \over \sqrt{2}}, $ for $u$-type quarks,$ \\
{f_{q} \cos\beta \over \sqrt{2}}, $ for $d$-type quarks,$ \\
{0}, $ for neutrinos $ \nu$,$ \\
{f_{q} \cos\beta \over \sqrt{2}}, $ for charged leptons$. 
\end{cases}
\end{align}
\begin{equation}
c_{G \tilde{Z}_i \tilde{Z}_j} = \frac{1}{2}\left[(g^{'}N_{1i} - g N_{2i})(N_{3j} \cos\beta + N_{j4}\sin\beta) + (g^{'} N_{1j} - gN_{2j})(N_{3i}\cos\beta + N_{4i}\sin\beta)\right].
\end{equation}

\textbf{\underline{$\tilde{Z}_i \rightarrow \tilde{W}_j f' \bar{f}$}}

\begin{figure}
  \centering {\subfloat[$W$]{\includegraphics[scale = 0.25]{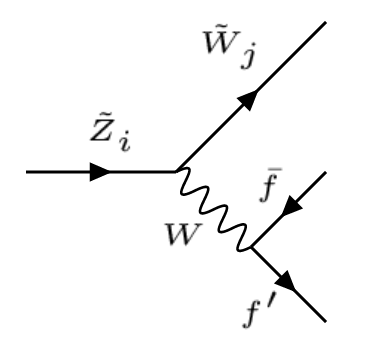}\label{fig:fa2}} \: \: \: \:\subfloat[$H^{\pm}$]{\includegraphics[scale = 0.25]{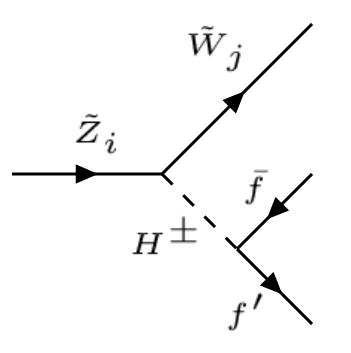}\label{fig:fb2}}}  \: \: \: \:
    \centering {\subfloat[$\tilde{f'}_k$]{\includegraphics[scale = 0.25]{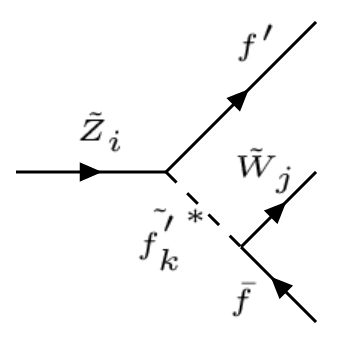}\label{fig:fc}} \: \: \: \: \subfloat[$\tilde{f}_k$]{\includegraphics[scale = 0.25]{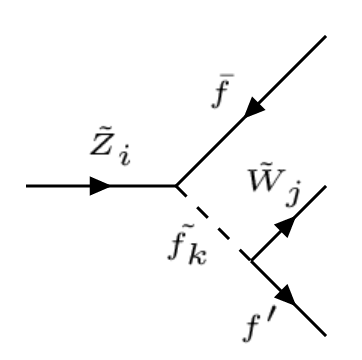}\label{fig:fd}}}
  \caption{$W$, $H^{\pm}$, $\tilde{f'}_k$, $\tilde{f}_k$ contributions to the $\tilde{Z}_i \rightarrow \tilde{W}_j f' \bar{f}$ decay. $i=1,2,3,4$, $j=1,2$, $k=1,2$. There are then also interferences between all these contributions.} \label{neutcharfeyn3}
\end{figure}

We turn now onto the 3 body decays of a neutralino into a chargino, fermion and anti-fermion. As for all the other 3 body modes included, this mode is only calculated if no 2 body modes are kinematically accessible. 
There are 4 main contributions to these decays, with $W$ boson, $H^{\pm}$, $\tilde{f'}_k$ and $\tilde{f}_k$ intermediates, the Feynman diagrams for these are shown in Figure~\ref{neutcharfeyn3}. Therefore there are nominally 6 squared contributions and 15 interferences; however, as the calculation is again done in Feynman gauge, the goldstone boson corresponding to the longitudinal components of the $W$ boson must be added, adding a further squared contribution and its 6 interferences.

For this decay mode, and the ``reverse'' decay mode $\tilde{W}_j \rightarrow
\tilde{Z}_i f' \bar{f}$, the formulae used are extracted from the {\tt sPHENO}
code, based on the work in references \cite{Baer:1998, TataBaer}. Note that
$f'$, $f$ are fermions with third components of weak isospin $1 \over 2$ and
$-{1 \over 2}$ respectively. A difference relative to these references is
that, following the formulae of {\tt sPHENO}, the expressions given do not
neglect $m_f$ in the Dirac algebra of the squared matrix element (whereas in
\cite{Baer:1998, TataBaer} it is neglected here, but of course included in the
phase space). As a result there is also $WH^{\pm}$ interference which is not
present if $m_f$ is neglected in the Dirac algebra. The possibilities of
positive and negative neutralino and chargino masses are included via
$(-1)^{\theta_{i}}$ and $(-1)^{\theta_{j}}$ factors. Similarly the fermion
Yukawa couplings are included and the formulae themselves allow for mixing of
the fermions. However in our program, mixing is only considered for the third
generation of sfermions and here this 3 body mode $\tilde{Z}_i \rightarrow
\tilde{W}_j t \bar{b}$ is not calculated as the 2 body modes $\tilde{Z}_i
\rightarrow W \tilde{W}_j$ and $\tilde{Z}_i \rightarrow h \tilde{W}_j$ are
then kinematically available and will dominate the branching ratios. 
The overall expression for the partial width is therefore given by:
\begin{equation} \label{neutchff}
\begin{aligned}
\Gamma = {N_{c} \over 512 \pi^3 |m_{Z_{i}}|^3}\Big[ & \Gamma_W + \Gamma_{\tilde{f}_1} + \Gamma_{\tilde{f}_2} + \Gamma_{\tilde{f'}_1} + \Gamma_{\tilde{f'}_2} -2\Gamma_{\tilde{f'}_{1} \tilde{f}_1} - 2\Gamma_{\tilde{f'}_{1}\tilde{f}_2} - 2\Gamma_{\tilde{f'}_{2}\tilde{f}_1} - 2\Gamma_{\tilde{f'}_{2}\tilde{f}_2} + 2\Gamma_{W H^{\pm}} \\ & + 2\Gamma_{W G} + \Gamma_{H^{\pm}} + \Gamma_{G} - 2\Gamma_{W \tilde{f'}_1} - 2\Gamma_{W \tilde{f'}_2} - 2\Gamma_{W \tilde{f}_1} - 2\Gamma_{W \tilde{f}_2} + 2\Gamma_{H^{\pm} G} - 2\Gamma_{G \tilde{f'}_1}  \\ & - 2\Gamma_{G \tilde{f'}_2} - 2\Gamma_{G \tilde{f}_1} - 2\Gamma_{G \tilde{f}_2} - 2\Gamma_{H^{\pm} \tilde{f'}_1} - 2\Gamma_{H^{\pm} \tilde{f'}_2} - 2\Gamma_{H^{\pm} \tilde{f}_1} - 2\Gamma_{H^{\pm} \tilde{f}_2} + 2\Gamma_{\tilde{f'}_1 \tilde{f'}_2} + 2\Gamma_{\tilde{f}_1 \tilde{f}_2} \Big].
\end{aligned}
\end{equation}
Note here $G$ refers to the goldstone contribution which is the longitudinal component of the $W$ and so has mass equal to the $W$ boson mass.
Here the following variables and couplings are used:
\begin{equation}
N_{c} = \begin{cases}  3, $ for $f'$ $\bar{f}$ quarks, $\\
					   1, $ for $f'$ $\bar{f}$ charged leptons or neutrinos$. \\
						\end{cases}
\end{equation}
There are several factors of (-1) depending on whether the neutralino or chargino have negative masses, and also there are factors of (-1) if the decay chargino $\rightarrow$ neutralino $f'$ $\bar{f}$ is being considered rather than neutralino $\rightarrow$ chargino $f'$ $\bar{f}$.
\begin{equation}
(-1)^{\theta_i} = \begin{cases}	1, $ for $ m_{\tilde{Z}_i} > 0$,$ \\
								-1, $ for $ m_{\tilde{Z}_i} < 0. \\
								\end{cases}
\end{equation}
\begin{equation}
(-1)^{\theta_j} = \begin{cases}	1, $ for $ m_{\tilde{W}_j} > 0$,$ \\
								-1, $ for $m_{\tilde{W}_j} < 0. \\
								\end{cases}
\end{equation}
\begin{equation}
(-1)^{\theta_c} = \begin{cases} 1, $ for neutralino decaying to chargino, $\\
								-1, $ for chargino decaying to neutralino. $ \\
								\end{cases}
\end{equation}
The following couplings are used:
\begin{equation}
f_{u} = {g m_{q'}^{run} \over \sqrt{2} \sin\beta m_{W}},
\end{equation}
\begin{equation}
f_{d} = {g m_{q}^{run} \over \sqrt{2} \cos\beta m_{W}}.
\end{equation}
For $\tilde{W}^{+}_{1}$, i.e.\ the lightest chargino ($j = 1$), and where $i$ is the index of the neutralino:
\begin{equation}
\mathcal{C}_{\tilde{W} \tilde{Z} W}^L = g\sin\theta_L N_{2i} + {\cos\theta_L N_{3i} \over \sqrt{2}},
\end{equation}
\begin{equation}
\mathcal{C}_{\tilde{W} \tilde{Z} W}^R = g\sin\theta_R N_{2i} - {\cos\theta_R N_{4i} \over \sqrt{2}},
\end{equation}
\begin{equation}
\mathcal{C}_{\tilde{W} \tilde{Z} H+}^L = (g\sin\theta_R N_{4i} + {\cos\theta_R \over \sqrt{2}})(g' N_{1i} + g N_{2i}),
\end{equation}
\begin{equation}
\mathcal{C}_{\tilde{W} \tilde{Z} H+}^R = (g\sin\theta_L N_{3i} - {\cos\theta_L) \over \sqrt{2}})(g' N_{1i} + g N_{2i}),
\end{equation}
\begin{align}
\alpha_{\tilde{f'}_1}^{\tilde{W}_1} &= -g \sin\theta_R \cos\theta_{q'} + f_{u}\cos\theta_{R}\sin\theta_{q'}, \\
\beta_{\tilde{f'}_1}^{\tilde{W}_1} &= -f_{d}\cos\theta_L \cos\theta_{q'} (-1)^{\theta_c}, \\
\alpha_{\tilde{f}_1}^{\tilde{W}_1} &= -g\sin\theta_L \cos\theta_q + f_{d}\cos\theta_L \sin\theta_q(-1)^{\theta_c}, \\
\beta_{\tilde{f}_1}^{\tilde{W}_1} &= -f_{u}\cos\theta_R \cos\theta_q, \\
\alpha_{\tilde{f'}_2}^{\tilde{W}_1} &= g\sin\theta_R \sin\theta_{q'}(-1)^{\theta_c} - f_{u}\cos\theta_R \cos\theta_{q'}, \\
\beta_{\tilde{f'}_2}^{\tilde{W}_1} &= -f_{d}\cos\theta_L \sin\theta_{q'}(-1)^{\theta_c}, \\
\alpha_{\tilde{f}_2}^{\tilde{W}_1} &= -f_{d}\cos\theta_{L}\cos\theta_{q} + g\sin\theta_{L}\sin\theta_{q}, \\
\beta_{\tilde{f}_2}^{\tilde{W}_1} &= -f_{u}\cos\theta_R\sin\theta_{q}.
\end{align}
Note that because of the conventions adopted, if the fermions considered are $\tau$ and $\nu_{\tau}$, so that the intermediates are $\tilde{\tau}_1$ and $\tilde{\tau}_2$, then the mixing angles in the formulae for this 3 body	 decay must be rotated so that one must take $\cos\theta_{\tau} \rightarrow \sin\theta_{\tau}$ and $\sin\theta_{\tau} \rightarrow -\cos\theta_{\tau}$ in the formulae listed for the $\tilde{Z}_i \rightarrow \tilde{W}_j  f' \bar{f}$ and for the reverse decay $\tilde{W}_j \rightarrow \tilde{Z}_i  f' \bar{f}$. Note that in this case where the fermions are $\tau$ and $\nu_{\tau}$, then $\theta_q$ would be the mixing angle for the $\tilde{\tau}$, whilst $\theta_{q'} = 0$ as the is no mixing for $\tilde{\nu_{\tau}}$.

For $\tilde{W}^{+}_{2}$, i.e.\ the heaviest chargino ($j = 2$), where $i$ is the index of the neutralino:
\begin{align}
\mathcal{C}_{\tilde{W} \tilde{Z} W}^L &= g\cos\theta_L N_{2i} - {\sin\theta_L N_{3i} \over \sqrt{2}}, \\
\mathcal{C}_{\tilde{W} \tilde{Z} W}^R &= g\cos\theta_R N_{2i} + {\sin\theta_R N_{4i} \over \sqrt{2}}, \\
\mathcal{C}_{\tilde{W} \tilde{Z} H+}^L &= \Big[g\cos\theta_R N_{4i} - {\sin\theta_R \over \sqrt{2}}\Big](g' N_{1i} + g N_{2i}), \\
\mathcal{C}_{\tilde{W} \tilde{Z} H+}^R &= \Big[g\cos\theta_L N_{3i} + {\sin\theta_L) \over \sqrt{2}}\Big](g' N_{1i} + g N_{2i}),
\end{align}
\begin{align}
\alpha_{\tilde{f'}_1}^{\tilde{W}_2} &= -g \cos\theta_R \cos\theta_{q'} - f_{u}\sin\theta_{R}\sin\theta_{q'}, \\
\beta_{\tilde{f'}_1}^{\tilde{W}_2} &= f_{d}\sin\theta_L \cos\theta_{q'}(-1)^{\theta_c}, \\
\alpha_{\tilde{f}_1}^{\tilde{W}_2} &= -g\cos\theta_L \cos\theta_q - f_{d}\sin\theta_L \sin\theta_q(-1)^{\theta_c}, \\
\beta_{\tilde{f}_1}^{\tilde{W}_2} &= f_{u}\sin\theta_R \cos\theta_q, \\
\alpha_{\tilde{f'}_2}^{\tilde{W}_2} &= g\cos\theta_R \sin\theta_{q'}(-1)^{\theta_c} + f_{u}\sin\theta_R \cos\theta_{q'}, \\
\beta_{\tilde{f'}_2}^{\tilde{W}_2} &= f_{d}\sin\theta_L \sin\theta_{q'}(-1)^{\theta_c}, \\
\alpha_{\tilde{f}_2}^{\tilde{W}_2} &= f_{d}\sin\theta_{L}\cos\theta_{q} + g\cos\theta_{L}\sin\theta_{q}, \\
\beta_{\tilde{f}_2}^{\tilde{W}_2} &= f_{u}\sin\theta_R\sin\theta_{q}.
\end{align}
There are also the neutralino couplings to the $f'$ $\bar{f}$ pair and these
depend upon whether they are quarks(q) or leptons(l) and whether they are
``$u$-type" or ``$d$-type" (i.e.\ third component of weak isospin being
$+\frac{1}{2}$ or $-\frac{1}{2}$ respectively).  
For quarks:
\begin{equation}
A_{\tilde{Z}_i}^{u} = -{g \over \sqrt{2}}N_{2i} - {g' \over 3 \sqrt{2}}N_{1i},
\end{equation}
\begin{equation}
B_{\tilde{Z}_i}^{u} = -{4 g' \over 3 \sqrt{2}}N_{1i},
\end{equation}
\begin{align}
\alpha_{\tilde{Z}_{i} \tilde{f}_1}^{u} &= -A_{\tilde{Z}_{i}}^{u}\cos\theta_{q'} (-1)^{\theta_j} (-1)^{\theta_i}(-1)^{\theta_c} - f_{u}N_{4i}\sin\theta_{q'}, \\
\beta_{\tilde{Z}_{i} \tilde{f}_1}^{u} &= f_{u}N_{4i}\cos\theta_{q'}(-1)^{\theta_c} - B_{\tilde{Z}_{i}}^{u}\sin\theta_{q'}, \\
\alpha_{\tilde{Z}_{i} \tilde{f}_2}^{u} &= -f_{u}N_{4i}\cos\theta_{q'}(-1)^{\theta_i} + A_{\tilde{Z}_{i}}^{u}\sin\theta_{q'}(-1)^{\theta_c}, \\
\beta_{\tilde{Z}_{i} \tilde{f}_2}^{u} &= B_{\tilde{Z}_{i}}^{u}\cos\theta_{q'}(-1)^{\theta_j}(-1)^{\theta_i}(-1)^{\theta_c} + f_{u}N_{4i}\sin\theta_{q'}.
\end{align}
\begin{equation}
A_{\tilde{Z}_{i}}^{d} = {g \over \sqrt{2}}N_{2i} - {g' \over 3 \sqrt{2}}N_{1i},
\end{equation}
\begin{equation}
B_{\tilde{Z}_{i}}^{d} = {2g' \over 3 \sqrt{2}}N_{1i},
\end{equation}
\begin{align}
\alpha_{\tilde{Z}_{i} \tilde{f}_1}^{d} &= -A_{\tilde{Z}_{i}}^{d}\cos\theta_{q}(-1)^{\theta_j}(-1)^{\theta_i}(-1)^{\theta_c} - f_{d}N_{4i}\sin\theta_{q}, \\
\beta_{\tilde{Z}_{i} \tilde{f}_1}^{d} &= f_{d}N_{3i}\cos\theta_{q} (-1)^{\theta_c} - B_{\tilde{Z}_{i}}^{d}\sin\theta_{q}(-1)^{\theta_i}, \\
\alpha_{\tilde{Z}_{i} \tilde{f}_2}^{d} &= f_{d}N_{3i}\cos\theta_{q}(-1)^{\theta_c} - (-1)^{\theta_c}(-1)^{\theta_i} A_{\tilde{Z}_{i}}^{d}\sin\theta_{q}, \\
\beta_{\tilde{Z}_{i} \tilde{f}_2}^{d} &= B_{\tilde{Z}_{i}}^{d}\cos\theta_{q}(-1)^{\theta_j}(-1)^{\theta_i}(-1)^{\theta_c} + f_{d}N_{4i}\sin\theta_q.
\end{align}
Again, remember for the case of $\tau$ and $\nu_{\tau}$ as $f$ and $f'$ respectively then one must take $\cos\theta_{\tau} \rightarrow \sin\theta_{\tau}$ and $\sin\theta_{\tau} \rightarrow -\cos\theta_{\tau}$ in the formulae listed for this decay mode. 

For leptons instead the neutralino couplings are:
\begin{equation}
A_{\tilde{Z}_i}^{u} = -{g \over \sqrt{2}}N_{2i}) + {g' \over \sqrt{2}}N_{1i},
\end{equation}
\begin{equation}
B_{\tilde{Z}_i}^{u} = 0,
\end{equation}
\begin{equation}
A_{\tilde{Z}_{i}}^{d} = {g \over \sqrt{2}}N_{2i} + {g' \over \sqrt{2}}N_{1i},
\end{equation}
\begin{equation}
B_{\tilde{Z}_{i}}^{d} = \sqrt{2} g' N_{1i}.
\end{equation}
The $\alpha$ and $\beta$ couplings are as before except $\alpha_{\tilde{Z}_{i} \tilde{f}_2}^{u} = 0$ and $\beta_{\tilde{Z}_{i} \tilde{f}_2}^{u} = 0$ as there are no RH sneutrinos.
Note in {\tt {\tt SOFTSUSY}} we use the same function for a neutralino decaying to a chargino as a chargino decaying to a neutralino, in general the changes required are $m_{\tilde{Z}_i} \leftrightarrow m_{\tilde{W}_j}$, $m_{f'} \leftrightarrow m_{f}$ and $m_{\tilde{f'}} \leftrightarrow m_{\tilde{f}}$, in some places there are further effects on the integrals or couplings, where this occurs it's listed in the following formulae.

Contribution by contribution, the couplings are:

\textbf{\underline{$\Gamma_{W}$}}

The upper limit of integration here is:
\begin{equation}
\mathcal{T} = {1 \over 2|m_{\tilde{Z}_{i}}|} (m_{\tilde{Z}_i}^2 + m_{\tilde{W}_j}^2 - m_{f}^2 - m_{f'}^2 - 2m_{f}m_{f'}),
\end{equation}
We also use $s$ and $\lambda$ given by:
\begin{equation}
s = m_{\tilde{Z}_{i}}^2 + m_{\tilde{Z}_j}^2 - 2|m_{\tilde{Z}_i}|E,
\end{equation}
\begin{equation}
\lambda = \sqrt{(s - (m_{f} + m_{f'})^2)(s - (m_{f} - m_{f'})^2)}.
\end{equation}
The necessary integrals are:	
\begin{equation}
\begin{aligned}
I_{W}^{1} = 2|m_{\tilde{Z}_i}| \int_{|m_{\tilde{W}_{j}}|}^{\mathcal{T}} dE & {2|m_{\tilde{Z}_i}| \over s} \lambda \sqrt{E^2 - m_{\tilde{W}_j}^2} \Big[-2s^4 + ((m_{\tilde{Z}_i}^2 + m_{\tilde{W}_j}^2 + m_{f}^2 + m_{f'}^2))s^3 \\ & + ((m_{\tilde{Z}_i}^2 - m_{\tilde{W}_j}^2)^2 + (m_{f}^2 - m_{f'}^2)^2 - 2(m_{\tilde{Z}_i}^2 + m_{\tilde{W}_j}^2)^2 (m_{f}^2 + m_{f'}^2)s^2 \\ & + ((m_{\tilde{Z}_i}^2 + m_{\tilde{W}_j}^2)(m_{f}^2 - m_{f'}^2)^2 + (m_{f}^2 + m_{f'}^2)(m_{\tilde{Z}_i}^2 - m_{\tilde{W}_j}^2)^2)s \\ & - 2(m_{\tilde{Z_{i}}}^2 - m_{\tilde{W}_j}^2)^2 (m_{f}^2 + m_{f'}^2)^2\Big]{1 \over 3s^2}{1 \over (s-m_{W}^2)^2},
\end{aligned}
\end{equation}
\begin{equation}
\begin{aligned}
I_{W}^{2} = 2|m_{\tilde{Z}_i}| \int_{|m_{\tilde{W}_{j}}|}^{\mathcal{T}} dE {2|m_{\tilde{Z}_i}| \over s} \lambda \sqrt{E^2 - m_{\tilde{W}_j}^2} (s - m_{f}^2 - m_{f'}^2){1 \over (s-m_{W}^2)^2}.
\end{aligned}
\end{equation}
Then
\begin{equation}
\begin{aligned}
\Gamma_W = & -8 \mathcal{C}_{\tilde{W} \tilde{Z} W}^L  \mathcal{C}_{\tilde{W} \tilde{Z} W}^R  {g^2 \over 2} |m_{\tilde{Z}_i}||m_{\tilde{W}_j}|I_{W}^2|(-1)^{\theta_i}(-1)^{\theta_j} + 2({\mathcal{C}_{\tilde{W} \tilde{Z} W}^L}^2 + {\mathcal{C}_{\tilde{W} \tilde{Z} W}^R}^2){g^2 \over 2} I_{W}^1.
\end{aligned}
\end{equation}

\textbf{\underline{$\Gamma_{H^{\pm}}$}}
\begin{equation}
\omega_{H^+ \tilde{W}^+ \tilde{Z}}^L = \mathcal{C}_{\tilde{W} \tilde{Z} H+}^L  \cos\beta,
\end{equation}
\begin{equation}
\omega_{H^+ \tilde{W}^+ \tilde{Z}}^R = \mathcal{C}_{\tilde{W} \tilde{Z} H+}^R  \sin\beta,
\end{equation}
\begin{equation}
\mathcal{C}_{H^+ f f'}^u = f_{u}\cos\beta,
\end{equation}
\begin{equation}
\mathcal{C}_{H^+ f f'}^d = f_{d}\sin\beta,
\end{equation}
The relevant combinations of these couplings for this contribution are:
\begin{align}
\mathcal{V}_{H+}^{(1)} &= {\omega_{H^+ \tilde{W}^+ \tilde{Z}}^L}^2 + {\omega_{H^+ \tilde{W}^+ \tilde{Z}}^R}^2, \\
\mathcal{V}_{H+}^{(2)} &= \omega_{H^+ \tilde{W}^+ \tilde{Z}}^L  \omega_{H^+ \tilde{W}^+ \tilde{Z}}^R  (-1)^{\theta_i}, \\
\mathcal{V}_{H+}^{(3)} &= {\mathcal{C}_{H^+ f f'}^u}^2 + {\mathcal{C}_{H^+ f f'}^d}^2, \\
\mathcal{V}_{H+}^{(4)} &= \mathcal{C}_{H^+ f f'}^u  \mathcal{C}_{H^+ f f'}^d.
\end{align}
The integrals are:
\begin{equation}  \label{IHpm1}
I_{H^{\pm}}^{1} = 2|m_{\tilde{Z}_i}| \int_{|m_{\tilde{W}_{j}}|}^{\mathcal{T}} dE {2|m_{\tilde{Z}_i}| \over s} \sqrt{(s-(m_{f'}+m_{f})^2)(s-(m_{f'}-m_{f})^2)} \sqrt{E^2 - m_{\tilde{W}_j}^2}{1 \over (s - m_{H^{\pm}}^2)^2},
\end{equation}
\begin{equation} \label{IHpm2}
\begin{aligned}
I_{H^{\pm}}^{2} = 2|m_{\tilde{Z}_i}| \int_{|m_{\tilde{W}_{j}}|}^{\mathcal{T}} dE \Big[ &{2|m_{\tilde{Z}_i}| \over s} \sqrt{(s-(m_{f'}+m_{f})^2)(s-(m_{f'}-m_{f})^2)} \sqrt{E^2 - m_{\tilde{W}_j}^2} \\ & \times (s-m_{f}^2 - m_{f'}^2){1 \over (s - m_{H^{\pm}}^2)^2}\Big],
\end{aligned}
\end{equation}
\begin{equation} \label{IHpm3}
\begin{aligned}
I_{H^{\pm}}^{3} = 2|m_{\tilde{Z}_i}| \int_{|m_{\tilde{W}_{j}}|}^{\mathcal{T}} dE \Big[ &{2|m_{\tilde{Z}_i}| \over s} \sqrt{(s-(m_{f'}+m_{f})^2)(s-(m_{f'}-m_{f})^2)} \sqrt{E^2 - m_{\tilde{W}_j}^2} \\ & \times 2|m_{\tilde{Z}_i}|E{1 \over (s - m_{H^{\pm}}^2)^2}\Big],
\end{aligned}
\end{equation}
\begin{equation} \label{IHpm4}
\begin{aligned}
I_{H^{\pm}}^{4} = 2|m_{\tilde{Z}_i}| \int_{|m_{\tilde{W}_{j}}|}^{\mathcal{T}} dE \Big[ &{2|m_{\tilde{Z}_i}| \over s} \sqrt{(s-(m_{f'}+m_{f})^2)(s-(m_{f'}-m_{f})^2)} \sqrt{E^2 - m_{\tilde{W}_j}^2} \\ & \times 2|m_{\tilde{Z}_i}|E (s-m_{f}^2 - m_{f'}^2){1 \over (s - m_{H^{\pm}}^2)^2}\Big]
\end{aligned}
\end{equation}
The overall contribution is then:
\begin{equation}
\begin{aligned}
\Gamma_{H^{\pm}} = & \mathcal{V}_{H+}^{(1)} \mathcal{V}_{H+}^{(3)} I_{H^{\pm}}^{4} - 4 \mathcal{V}_{H+}^{(1)} \mathcal{V}_{H+}^{(4)} I_{H^{\pm}}^3 m_{f}m_{f'} + 4 \mathcal{V}_{H+}^{(2)} \mathcal{V}_{H+}^{(3)} I_{H^{\pm}}^{2} |m_{\tilde{Z}_i}||m_{\tilde{W}_j}|(-1)^{\theta_j} \\ & - 16 \mathcal{V}_{H+}^{(2)} \mathcal{V}_{H+}^{(4)} I_{H^{\pm}}^{1} m_{f} m_{f'}|m_{\tilde{Z}_i}||m_{\tilde{W}_j}|(-1)^{\theta_j}.
\end{aligned}
\end{equation}

\textbf{\underline{$\Gamma_{G}$}}

Note here $G$ refers to the goldstone contribution which is the longitudinal component of the W and so has mass equal to the $W$ boson mass.
The couplings used are:
\begin{equation}
\omega_{G \tilde{W} \tilde{Z}}^L = \mathcal{C}_{\tilde{W} \tilde{Z} H+}^L \sin\beta,
\end{equation}
\begin{equation}
\omega_{G \tilde{W} \tilde{Z}}^R = -\mathcal{C}_{\tilde{W} \tilde{Z} H+}^R \cos\beta,
\end{equation}
\begin{equation}
\mathcal{C}_{G f f'}^u = f_{u}\sin\beta,
\end{equation}
\begin{equation}
\mathcal{C}_{G f f'}^d = -f_{d} \cos\beta,
\end{equation}
\begin{align}
\mathcal{V}_{G}^{(1)} &= {\omega_{G \tilde{W} \tilde{Z}}^L}^2 + {\omega_{G \tilde{W} \tilde{Z}}^R}^2, \\
\mathcal{V}_{G}^{(2)} &= \omega_{G \tilde{W} \tilde{Z}}^L  \omega_{G \tilde{W} \tilde{Z}}^R  (-1)^{\theta_i}, \\
\mathcal{V}_{G}^{(3)} &= {\mathcal{C}_{G f f'}^u}^2 + {\mathcal{C}_{G f f'}^d}^2, \\
\mathcal{V}_{G}^{(4)} &= {\mathcal{C}_{G f f'}^u}  \mathcal{C}_{G f f'}^d.
\end{align}
The integrals here $I_{G}^{1}$ etc are exactly the same as those for $H^{\pm}$ but with the change $m_{H^{\pm}} \rightarrow m_{W}$.
\begin{equation}
\begin{aligned}
\Gamma_{G} = & \mathcal{V}_{G}^{(1)} \mathcal{V}_{G}^{(3)} I_{G}^{4} - 4 \mathcal{V}_{G}^{(1)} \mathcal{V}_{G}^{(4)} I_{G}^3 m_{f} m_{f'} + 4 \mathcal{V}_{G}^{(2)} \mathcal{V}_{G}^{(3)} I_{G}^2 |m_{\tilde{Z}_i}||m_{\tilde{W}_j}|(-1)^{\theta_j} \\ & - 16 \mathcal{V}_{G}^{(2)} \mathcal{V}_{G}^{(4)} I_{G}^{1} m_{f} m_{f'} |m_{\tilde{Z}_i}||m_{\tilde{W}_j}|(-1)^{\theta_j}.
\end{aligned}
\end{equation}

\textbf{\underline{$\Gamma_{\tilde{f'}_1}$}}
\begin{align}
\mathcal{V}_{\tilde{f'}_1}^{(1)} &= {\alpha_{\tilde{Z}_i \tilde{f}_1}^{u}}^2 + {\beta_{\tilde{Z}_i \tilde{f}_1}^{u}}^2, \\
\mathcal{V}_{\tilde{f'}_1}^{(2)} &= -{\alpha_{\tilde{Z}_i \tilde{f}_1}^{u}}^2 {\beta_{\tilde{Z}_i \tilde{f}_1}^{u}}^2 (-1)^{\theta_i}, \\
\mathcal{V}_{\tilde{f'}_1}^{(3)} &= {\alpha_{\tilde{f'}_1}^{\tilde{W}}}^2 + {\beta_{\tilde{f'}_1}^{\tilde{W}}}^2, \\
\mathcal{V}_{\tilde{f'}_1}^{(4)} &= -\alpha_{\tilde{f'}_1}^{\tilde{W}} \beta_{\tilde{f'}_1}^{\tilde{W}}.
\end{align}
Now the integrals $I_{\tilde{f'}_1}^{1,2,3,4}$ are exactly as the $I_{H^{\pm}}^{1,2,3,4}$ integrals in \eqref{IHpm1} to \eqref{IHpm4} but with lower limit $m_{f'}$, upper limit of integration $E_{upper} = {1 \over 2|m_{\tilde{Z}_1}|}(m_{\tilde{Z}_i}^2 + m_{f'}^2 - m_{f}^2 - m_{\tilde{Z}_j}^2 -2 m_{f}|m_{\tilde{W}_j}|)$ and the replacements $m_{H^{\pm}} \rightarrow m_{\tilde{f'}_1}$, $|m_{\tilde{W}_j}| \rightarrow m_{f'}$ and $m_{f'} \rightarrow |m_{\tilde{W}_j}|$.
Then:
\begin{equation}
\begin{aligned}
\Gamma_{\tilde{f'}_1} = & \mathcal{V}_{\tilde{f'}_1}^{(1)} \mathcal{V}_{\tilde{f'}_1}^{(3)} I_{\tilde{f'}_1}^{4} - 4 \mathcal{V}_{\tilde{f'}_1}^{(1)} \mathcal{V}_{\tilde{f'}_1}^{(4)} m_{f}|m_{\tilde{W}_j}| I_{\tilde{f'}_1}^{3} + 4 \mathcal{V}_{\tilde{f'}_1}^{(2)} \mathcal{V}_{\tilde{f'}_1}^{(3)} |m_{\tilde{Z}_i}|m_{f'} I_{\tilde{f'}_1}^{2} \\ & - 16 \mathcal{V}_{\tilde{f'}_1}^{(2)} \mathcal{V}_{\tilde{f'}_1}^{(4)} |m_{\tilde{Z}_i}||m_{\tilde{W}_j}|m_{f'}m_{f} I_{\tilde{f'}_1}^{1}.
\end{aligned}
\end{equation}

\textbf{\underline{$\Gamma_{\tilde{f'}_2}$}}

Everything for $\Gamma_{\tilde{f'}_2}$ is exactly as for $\Gamma_{\tilde{f'}_1}$ but with the change $m_{\tilde{f'}_1} \rightarrow m_{\tilde{f'}_2}$ and the coupling combinations:
\begin{align}
\mathcal{V}_{\tilde{f'}_2}^{(1)} &= {\alpha_{\tilde{Z}_i \tilde{f}_2}^{u}}^2 + {\beta_{\tilde{Z}_i \tilde{f}_2}^{u}}^2, \\
\mathcal{V}_{\tilde{f'}_2}^{(2)} &= {\alpha_{\tilde{Z}_i \tilde{f}_2}^{u}}^2 {\beta_{\tilde{Z}_i \tilde{f}_2}^{u}}^2 , \\
\mathcal{V}_{\tilde{f'}_2}^{(3)} &= {\alpha_{\tilde{f'}_2}^{\tilde{W}}}^2 + {\beta_{\tilde{f'}_2}^{\tilde{W}}}^2, \\
\mathcal{V}_{\tilde{f'}_2}^{(4)} &= \alpha_{\tilde{f'}_2}^{\tilde{W}} \beta_{\tilde{f'}_2}^{\tilde{W}}.
\end{align}
The contribution is then:
\begin{equation}
\begin{aligned}
\Gamma_{\tilde{f'}_2} = & \mathcal{V}_{\tilde{f'}_2}^{(1)} \mathcal{V}_{\tilde{f'}_2}^{(3)} I_{\tilde{f'}_2}^{4} - 4 \mathcal{V}_{\tilde{f'}_2}^{(1)} \mathcal{V}_{\tilde{f'}_2}^{(4)} m_{f}|m_{\tilde{W}_j}| I_{\tilde{f'}_2}^{3}(-1)^{\theta_c} + 4 \mathcal{V}_{\tilde{f'}_2}^{(2)} \mathcal{V}_{\tilde{f'}_2}^{(3)} |m_{\tilde{Z}_i}|m_{f'} I_{\tilde{f'}_2}^{2} (-1)^{\theta_c} \\ & - 16 \mathcal{V}_{\tilde{f'}_2}^{(2)} \mathcal{V}_{\tilde{f'}_2}^{(4)} |m_{\tilde{Z}_i}||m_{\tilde{W}_j}|m_{f'}m_{f} I_{\tilde{f'}_2}^{1}.
\end{aligned}
\end{equation}

\textbf{\underline{$\Gamma_{\tilde{f}_1}$}}

The coupling combinations are now:
\begin{align}
\mathcal{V}_{\tilde{f}_1}^{(1)} &= {\alpha_{\tilde{Z}_i \tilde{f}_1}^{d}}^2 + {\beta_{\tilde{Z}_i \tilde{f}_1}^{d}}^2, \\
\mathcal{V}_{\tilde{f}_1}^{(2)} &= -(-1)^{\theta_i}{\alpha_{\tilde{Z}_i \tilde{f}_1}^{d}} {\beta_{\tilde{Z}_i \tilde{f}_1}^{d}}, \\
\mathcal{V}_{\tilde{f}_1}^{(3)} &= {\alpha_{\tilde{f}_1}^{\tilde{W}}}^2 + {\beta_{\tilde{f}_1}^{\tilde{W}}}^2, \\
\mathcal{V}_{\tilde{f}_1}^{(4)} &= -\alpha_{\tilde{f}_1}^{\tilde{W}} \beta_{\tilde{f}_1}^{\tilde{W}}(-1)^{\theta_j}.
\end{align}
The integrals are are exactly as the $I_{H^{\pm}}^{1,2,3,4}$ integrals in \eqref{IHpm1} to \eqref{IHpm4} but the lower limit is now $m_{f}$, the upper limit is $E_{upper2} = {1 \over 2|m_{\tilde{Z}_i}|}[m_{\tilde{Z}_i}^2 + m_{f}^2 - m_{\tilde{W}_j}^2 - m_{f'}^2 - 2m_{f'}|m_{\tilde{W}_j}|]$ and in general relative to the $H^{\pm}$ integrals we must make the changes $m_{H^{\pm}} \rightarrow m_{\tilde{f}_1}$, $|m_{\tilde{W}_j}| \rightarrow m_{f}$ and $m_f \rightarrow |m_{\tilde{W}_j}|$.
\begin{equation}
\begin{aligned}
\Gamma_{\tilde{f}_1} = & \mathcal{V}_{\tilde{f}_1}^{(1)} \mathcal{V}_{\tilde{f}_1}^{(3)} I_{\tilde{f}_1}^{4} - 4 \mathcal{V}_{\tilde{f}_1}^{(1)} \mathcal{V}_{\tilde{f}_1}^{(4)} m_{f'} |m_{\tilde{W}_j}| I_{\tilde{f}_1}^{3} (-1)^{\theta_c}(-1)^{\theta_j} + 4 \mathcal{V}_{\tilde{f}_1}^{(2)} \mathcal{V}_{\tilde{f}_1}^{(3)} |m_{\tilde{Z}_i}|m_{f} I_{\tilde{f}_1}^{2}(-1)^{\theta_c} \\ & - 16 \mathcal{V}_{\tilde{f}_1}^{(2)}  \mathcal{V}_{\tilde{f}_1}^{(4)} |m_{\tilde{Z}_i}||m_{\tilde{W}_j}|m_{f}m_{f'}(-1)^{\theta_j}I_{\tilde{f}_1}^{1}.
\end{aligned}
\end{equation}

\textbf{\underline{$\Gamma_{\tilde{f}_2}$}}

Nominally $\Gamma_{\tilde{f}_2}$ has the same expression as $\Gamma_{\tilde{f}_1}$ with the replacement $\tilde{f}_1 \rightarrow
\tilde{f}_2$, however differences in expressions for couplings mean we have
slight differences; 
the coupling combinations are now:
\begin{align}
\mathcal{V}_{\tilde{f}_2}^{(1)} &= {\alpha_{\tilde{Z}_i \tilde{f}_2}^{d}}^2 + {\beta_{\tilde{Z}_i \tilde{f}_2}^{d}}^2, \\
\mathcal{V}_{\tilde{f}_2}^{(2)} &= {\alpha_{\tilde{Z}_i \tilde{f}_2}^{d}} {\beta_{\tilde{Z}_i \tilde{f}_2}^{d}}, \\
\mathcal{V}_{\tilde{f}_2}^{(3)} &= {\alpha_{\tilde{f}_2}^{\tilde{W}}}^2 + {\beta_{\tilde{f}_2}^{\tilde{W}}}^2, \\
\mathcal{V}_{\tilde{f}_2}^{(4)} &= \alpha_{\tilde{f}_2}^{\tilde{W}} \beta_{\tilde{f}_2}^{\tilde{W}}(-1)^{\theta_c}.
\end{align}
Therefore the contribution is given by:
\begin{equation}
\begin{aligned}
\Gamma_{\tilde{f}_2} = & \mathcal{V}_{\tilde{f}_2}^{(1)} \mathcal{V}_{\tilde{f}_2}^{(3)} I_{\tilde{f}_2}^{4} - 4 \mathcal{V}_{\tilde{f}_2}^{(1)} \mathcal{V}_{\tilde{f}_2}^{(4)} m_{f'} |m_{\tilde{W}_j}| I_{\tilde{f}_2}^{3} (-1)^{\theta_c}+ 4 \mathcal{V}_{\tilde{f}_2}^{(2)} \mathcal{V}_{\tilde{f}_2}^{(3)} |m_{\tilde{Z}_i}|m_{f} I_{\tilde{f}_2}^{2}(-1)^{\theta_c} \\ & - 16 \mathcal{V}_{\tilde{f}_2}^{(2)}  \mathcal{V}_{\tilde{f}_2}^{(4)} |m_{\tilde{Z}_i}||m_{\tilde{W}_j}|m_{f}m_{f'}I_{\tilde{f}_2}^{1}.
\end{aligned}
\end{equation}

\textbf{\underline{$\Gamma_{\tilde{f'}_1 \tilde{f}_1}$}}

Here the coupling combinations differ significantly depending on which way around the decay is being considered, i.e.\ neutralino to chargino or chargino to neutralino. The following are fixed regardless of this:
\begin{align}
\mathcal{V}_{\tilde{f'}_1 \tilde{f}_1}^{(1)} &= \frac{1}{2}\left[\alpha_{\tilde{Z}_i \tilde{f}_1}^{u} \beta_{\tilde{Z}_i \tilde{f}_1}^{d} \beta_{\tilde{f'}_1}^{\tilde{W}}\alpha_{\tilde{f}_1}^{\tilde{W}}(-1)^{\theta_i} + \beta_{\tilde{Z}_i \tilde{f}_1}^{u} \alpha_{\tilde{Z}_i \tilde{f}_1}^{d} \alpha_{\tilde{f'}_1}^{\tilde{W}} \beta_{\tilde{f}_1}^{\tilde{W}}\right](-1)^{\theta_j}, \\
\mathcal{V}_{\tilde{f'}_1 \tilde{f}_1}^{(2)} &= -|m_{\tilde{Z}_i}||m_{\tilde{W}_j}|\left[(-1)^{\theta_i} \alpha_{\tilde{Z}_i \tilde{f}_1}^{u} \alpha_{\tilde{Z}_i \tilde{f}_1}^{d} \alpha_{\tilde{f'}_1}^{\tilde{W}} \alpha_{\tilde{f}_1}^{\tilde{W}} + \beta_{\tilde{Z}_i \tilde{f}_1}^{u} \beta_{\tilde{Z}_{i} \tilde{f}_1}^{d} \beta_{\tilde{f'}_1}^{\tilde{W}} \beta_{\tilde{f}_1}^{\tilde{W}}\right](-1)^{\theta_j}, \\
\mathcal{V}_{\tilde{f'}_1 \tilde{f}_1}^{(3)} &= -m_{f}m_{f'}\left[\beta_{\tilde{Z}_i \tilde{f}_1}^{u} \alpha_{\tilde{Z}_i \tilde{f}_1}^{d} \beta_{\tilde{f'}_1}^{\tilde{W}} \alpha_{\tilde{f}_1}^{\tilde{W}} + (-1)^{\theta_i} \alpha_{\tilde{Z}_i \tilde{f}_1}^{u} \beta_{\tilde{Z}_i \tilde{f}_1}^{d} \alpha_{\tilde{f}_1}^{\tilde{W}} \beta_{\tilde{f}_1}^{\tilde{W}}\right](-1)^{\theta_j}(-1)^{\theta_c}  , \\
\mathcal{V}_{\tilde{f'}_1 \tilde{f}_1}^{(8)} &= -2|m_{\tilde{Z}_i}||m_{\tilde{W}_j}|m_{f} m_{f'} \left[\beta_{\tilde{Z}_i \tilde{f}_1}^{u}\beta_{\tilde{Z}_i \tilde{f}_1}^{d} \alpha_{\tilde{f'}}^{\tilde{W}} \alpha_{\tilde{f}_1}^{\tilde{W}} + (-1)^{\theta_i}\alpha_{\tilde{Z}_i \tilde{f}_1}^{u} \alpha_{\tilde{Z}_i \tilde{f}_1}^{d} \beta_{\tilde{f'}}^{\tilde{W}} \beta_{\tilde{f}_1}^{\tilde{W}}\right] (-1)^{\theta_j}(-1)^{\theta_c}.
\end{align}

Meanwhile, if the decay is neutralino to chargino:
\begin{align}
\mathcal{V}_{\tilde{f'}_1 \tilde{f}_1}^{(4)} &= -|m_{\tilde{Z}_i}|m_{f}\left[(-1)^{\theta_i}\alpha_{\tilde{Z}_i \tilde{f}_1}^{u} \alpha_{\tilde{Z}_i \tilde{f}_1}^{d} \beta_{\tilde{f'}}^{\tilde{W}}\alpha_{\tilde{f}_1}^{\tilde{W}} +  \beta_{\tilde{Z}_i \tilde{f}_1}^{u} \beta_{\tilde{Z}_i \tilde{f}_1}^{d} \alpha_{\tilde{f'}}^{\tilde{W}} \beta_{\tilde{f}_1}^{\tilde{W}}\right], \\
\mathcal{V}_{\tilde{f'}_1 \tilde{f}_1}^{(5)} &= m_{f'} |m_{\tilde{W}_j}|\left[\beta_{\tilde{Z}_i \tilde{f}_1}^{u} \alpha_{\tilde{Z}_i \tilde{f}_1}^{d} \alpha_{\tilde{f'}_1}^{\tilde{W}} \alpha_{\tilde{f}_1}^{\tilde{W}} + (-1)^{\theta_i} \alpha_{\tilde{Z}_i \tilde{f}_1}^{u} \beta_{\tilde{Z}_i \tilde{f}_1}^{d} \beta_{\tilde{f'}_1}^{\tilde{W}} \beta_{\tilde{f}_1}^{\tilde{W}}\right](-1)^{\theta_j}, \\
\mathcal{V}_{\tilde{f'}_1 \tilde{f}_1}^{(6)} &= -|m_{\tilde{Z}_i}| m_{f'} \left[\beta_{\tilde{Z}_i \tilde{f}_1}^{u} \beta_{\tilde{Z}_i \tilde{f}_1}^{d} \beta_{\tilde{f'}}^{\tilde{W}} \alpha_{\tilde{f}_1}^{\tilde{W}} + (-1)^{\theta_i}  \alpha_{\tilde{Z}_i \tilde{f}_1}^{u} \alpha_{\tilde{Z}_i \tilde{f}_1}^{d} \alpha_{\tilde{f'}_1}^{\tilde{W}} \beta_{\tilde{f}_1}^{\tilde{W}}\right], \\
\mathcal{V}_{\tilde{f'}_1 \tilde{f}_1}^{(7)} &= |m_{\tilde{W}_j}|m_{f}\left[(-1)^{\theta_i} \alpha_{\tilde{Z}_i \tilde{f}_1}^{u} \beta_{\tilde{Z}_i \tilde{f}_1}^{d} \alpha_{\tilde{f'}_1}^{\tilde{W}} \alpha_{\tilde{f}_1}^{\tilde{W}} + \beta_{\tilde{Z}_i \tilde{f}_1}^{u} \alpha_{\tilde{Z}_i \tilde{f}_1}^{d} \beta_{\tilde{f'}_1}^{\tilde{W}} \beta_{\tilde{f}_1}^{\tilde{W}}\right](-1)^{\theta_j}.
\end{align}

Whilst if the decay is instead chargino to neutralino:
\begin{align}
\mathcal{V}_{\tilde{f'}_1 \tilde{f}_1}^{(4)} &= |m_{\tilde{Z}_i}| m_{f} \left[\alpha_{\tilde{f'}_1}^{\tilde{W}} \alpha_{\tilde{f}_1}^{\tilde{W}} \beta_{\tilde{Z}_i \tilde{f}_1}^{u} \alpha_{\tilde{Z}_i \tilde{f}_1}^{d} + \beta_{\tilde{f'}_1}^{\tilde{W}} \beta_{\tilde{f}_1}^{\tilde{W}} \alpha_{\tilde{Z}_i \tilde{f}_1}^{u} \beta_{\tilde{Z}_i \tilde{f}_1}^{d}\right](-1)^{\theta_j} ,\\
\mathcal{V}_{\tilde{f'}_1 \tilde{f}_1}^{(5)} &= -|m_{\tilde{W}_j}|m_{f'} \left[\alpha_{\tilde{f}_1}^{\tilde{W}} \beta_{\tilde{f'}_1}^{\tilde{W}} \alpha_{\tilde{Z}_i \tilde{f}_1}^{u} \alpha_{\tilde{Z}_i \tilde{f}_1}^{d}  + \alpha_{\tilde{f'}_1}^{\tilde{W}} \beta_{\tilde{f}_1}^{\tilde{W}} \beta_{\tilde{Z}_i \tilde{f}_1}^{u} \beta_{\tilde{Z}_i \tilde{f}_1}^{d}\right](-1)^{\theta_j} ,\\
\mathcal{V}_{\tilde{f'}_1 \tilde{f}_1}^{(6)} &= -|m_{\tilde{Z}_i}| m_{f'}\left[\beta_{\tilde{f'}_1}^{\tilde{W}}\beta_{\tilde{f}_1}^{\tilde{W}} \beta_{\tilde{Z}_i \tilde{f}_1}^{u} \alpha_{\tilde{Z}_i \tilde{f}_1}^{d}  + \alpha_{\tilde{f'}_1}^{\tilde{W}} \alpha_{\tilde{f}_1}^{\tilde{W}_1} \alpha_{\tilde{Z}_i \tilde{f}_1}^{u} \beta_{\tilde{Z}_i \tilde{f}_1}^{d}\right](-1)^{\theta_j}, \\
\mathcal{V}_{\tilde{f'}_1 \tilde{f}_1}^{(7)} &= |m_{\tilde{W}_j}|m_{f}\left[\alpha_{\tilde{f'}_1}^{\tilde{W}}\beta_{\tilde{f}_1}^{\tilde{W}} \alpha_{\tilde{Z}_i \tilde{f}_1}^{u} \alpha_{\tilde{Z}_i \tilde{f}_1}^{d} + \beta_{\tilde{f'}_1}^{\tilde{W}} \alpha_{\tilde{f}_1}^{\tilde{W}} \beta_{\tilde{Z}_i \tilde{f}_1}^{u} \beta_{\tilde{Z}_i \tilde{f}_1}^{d}\right](-1)^{\theta_j}.
\end{align}

We also need the following integrals, note $s = m_{\tilde{Z}_i}^2 + m_{f'}^2 - 2|m_{\tilde{Z}_i}|E$, $\lambda = \sqrt{(s-(m_{f}+m_{\tilde{W}})^2)(s-(m_{f}-m_{\tilde{W}})^2)}$, $A = \Big[m_{f}^2 + m_{\tilde{W}_j}^2 + 2|m_{\tilde{Z}_i}|E + (m_{\tilde{Z}_i}^2 - m_{f'}^2)(m_{\tilde{W}}^2 - m_{f}^2){1 \over s} + 2|m_{\tilde{Z}_i}| \lambda {1 \over s} \sqrt{E^2 - m_{f'}^2}  - 2m_{\tilde{f}_1}^2\Big]$ and $B = \Big[m_{f}^2 + m_{\tilde{W}_j}^2 + 2|m_{\tilde{Z}_i}|E + (m_{\tilde{Z}_i}^2 - m_{f'}^2)(m_{\tilde{W}}^2 - m_{f}^2){1 \over s} - 2|m_{\tilde{Z}_i}|\lambda {1 \over s} \sqrt{E^2 - m_{f'}^2} - 2m_{\tilde{f}_1}^2\Big]$:
\begin{equation}
\begin{aligned}
I_{\tilde{f'}_1 \tilde{f}_1}^{1} = 4|m_{\tilde{Z}_i}| \int_{m_{f'}}^{E_{upper}} dE {[2|m_{\tilde{Z}_i}| \lambda \sqrt{E^2 - m_{f'}^2}  + (m_{\tilde{f}_1}^2 s - m_{\tilde{Z}_i}^2 m_{\tilde{W}_j}^2 - m_{f}^2 m_{f'}^2) \log(A/B)] \over s - m_{\tilde{f'}_1}^2},
\end{aligned}
\end{equation}
\begin{equation}
\begin{aligned}
I_{\tilde{f'}_1 \tilde{f}_1}^{2} = -2|m_{\tilde{Z}_i}| \int_{m_{f'}}^{E_{upper}} dE {[2|m_{\tilde{Z}_i}| \lambda {1 \over s} \sqrt{E^2-m_{f'}^2}  + (m_{\tilde{f}_1}^2 - 2|m_{\tilde{Z}_i}E + m_{f'}^2 - m_{\tilde{W}_j}^2)\log(A/B)]  \over s - m_{\tilde{f'}_1}^2},
\end{aligned}
\end{equation}
\begin{equation}
\begin{aligned}
I_{\tilde{f'}_1 \tilde{f}_1}^{3} = 2|m_{\tilde{Z}_i}| \int_{m_{f'}}^{E_{upper}} dE {[2|m_{\tilde{Z}_i}|\lambda {1 \over s}\sqrt{E^2-m_{f'}^2}  + (m_{\tilde{f}_1}^2 - 2|m_{\tilde{Z}_i}E + m_{\tilde{Z}_i}^2 - m_{f}^2)\log(A/B)]  \over s - m_{\tilde{f'}_1}^2},
\end{aligned}
\end{equation}
\begin{equation}
\begin{aligned}
I_{\tilde{f'}_1 \tilde{f}_1}^{4} = 2|m_{\tilde{Z}_i}| \int_{m_{f'}}^{E_{upper}} dE {[2|m_{\tilde{Z}_i}|\lambda {1 \over s}\sqrt{E^2-m_{f'}^2} + (m_{\tilde{f}_1}^2 - m_{\tilde{W}_j}^2 - m_{f'}^2)\log(A/B)]  \over s - m_{\tilde{f'}_1}^2},
\end{aligned}
\end{equation}
\begin{equation}
\begin{aligned}
I_{\tilde{f'}_1 \tilde{f}_1}^{5} = 2|m_{\tilde{Z}_i}| \int_{m_{f'}}^{E_{upper}} dE {[2|m_{\tilde{Z}_i}| \lambda {1 \over s} \sqrt{E^2-m_{f'}^2}  + (m_{\tilde{f}_1}^2 - m_{\tilde{Z}_i}^2 - m_{f}^2)\log(A/B)]  \over s - m_{\tilde{f'}_1}^2},
\end{aligned}
\end{equation}
\begin{equation}
\begin{aligned}
I_{\tilde{f'}_1 \tilde{f}_1}^{6} = 2|m_{\tilde{Z}_i}| \int_{m_{f'}}^{E_{upper}} dE {\log(A/B) (s-m_{f}^2 - m_{\tilde{W}_j}^2)  \over s - m_{\tilde{f'}_1}^2},
\end{aligned}
\end{equation}
\begin{equation}
\begin{aligned}
I_{\tilde{f'}_1 \tilde{f}_1}^{7} = 2|m_{\tilde{Z}_i}| \int_{m_{f'}}^{E_{upper}} dE {\log(A/B) 2|m_{\tilde{Z}_i}|E  \over s - m_{\tilde{f'}_1}^2},
\end{aligned}
\end{equation}
\begin{equation}
\begin{aligned}
I_{\tilde{f'}_1 \tilde{f}_1}^{8} = 2|m_{\tilde{Z}_i}| \int_{m_{f'}}^{E_{upper}} dE {\log(A/B) \over s - m_{\tilde{f'}_1}^2}.
\end{aligned}
\end{equation}
The $\Gamma_{\tilde{f'}_1 \tilde{f}_1}$ contribution is then given by:
\begin{equation}
\begin{aligned}
\Gamma_{\tilde{f'}_1 \tilde{f}_1} = & \mathcal{V}_{\tilde{f'}_1 \tilde{f}_1}^{(1)} I_{\tilde{f'}_1 \tilde{f}_1}^{1} + \mathcal{V}_{\tilde{f'}_1 \tilde{f}_1}^{(2)} I_{\tilde{f'}_1 \tilde{f}_1}^{2} + \mathcal{V}_{\tilde{f'}_1 \tilde{f}_1}^{(3)} I_{\tilde{f'}_1 \tilde{f}_1}^{3} + \mathcal{V}_{\tilde{f'}_1 \tilde{f}_1}^{(4)} I_{\tilde{f'}_1 \tilde{f}_1}^{4} \\ & + \mathcal{V}_{\tilde{f'}_1 \tilde{f}_1}^{(5)} I_{\tilde{f'}_1 \tilde{f}_1}^{5} + \mathcal{V}_{\tilde{f'}_1 \tilde{f}_1}^{(6)} I_{\tilde{f'}_1 \tilde{f}_1}^{6} + \mathcal{V}_{\tilde{f'}_1 \tilde{f}_1}^{(7)} I_{\tilde{f'}_1 \tilde{f}_1}^{7} + \mathcal{V}_{\tilde{f'}_1 \tilde{f}_1}^{(8)} I_{\tilde{f'}_1 \tilde{f}_1}^{8}.
\end{aligned}
\end{equation}
The $\Gamma_{\tilde{f'}_1 \tilde{f}_2}$, $\Gamma_{\tilde{f'}_2 \tilde{f}_1}$ and $\Gamma_{\tilde{f'}_2 \tilde{f}_2}$ contributions follow analogously, they are given here as slight differences in the expressions for couplings complicated the expressions.

\textbf{\underline{$\Gamma_{\tilde{f'}_1 \tilde{f}_2}$}}

\begin{align}
\mathcal{V}_{\tilde{f'}_1 \tilde{f}_2}^{(1)} &= -\frac{1}{2}\left[\alpha_{\tilde{Z}_i \tilde{f}_1}^{u} \beta_{\tilde{Z}_i \tilde{f}_2}^{d} \beta_{\tilde{f'}_1}^{\tilde{W}}\alpha_{\tilde{f}_2}^{\tilde{W}} + \beta_{\tilde{Z}_i \tilde{f}_1}^{u} \alpha_{\tilde{Z}_i \tilde{f}_2}^{d} \alpha_{\tilde{f'}_1}^{\tilde{W}} \beta_{\tilde{f}_2}^{\tilde{W}}\right](-1)^{\theta_i}, \\
\mathcal{V}_{\tilde{f'}_1 \tilde{f}_2}^{(2)} &= |m_{\tilde{Z}_i}||m_{\tilde{W}_j}|\left[(-1)^{\theta_i} \alpha_{\tilde{Z}_i \tilde{f}_1}^{u} \alpha_{\tilde{Z}_i \tilde{f}_2}^{d} \alpha_{\tilde{f'}_1}^{\tilde{W}} \alpha_{\tilde{f}_2}^{\tilde{W}} + \beta_{\tilde{Z}_i \tilde{f}_1}^{u} \beta_{\tilde{Z}_{i} \tilde{f}_2}^{d} \beta_{\tilde{f'}_1}^{\tilde{W}} \beta_{\tilde{f}_2}^{\tilde{W}}\right], \\
\mathcal{V}_{\tilde{f'}_1 \tilde{f}_2}^{(3)} &= (-1)^{\theta_i} m_{f}m_{f'}\left[\beta_{\tilde{Z}_i \tilde{f}_1}^{u} \alpha_{\tilde{Z}_i \tilde{f}_2}^{d} \beta_{\tilde{f'}_1}^{\tilde{W}} \alpha_{\tilde{f}_2}^{\tilde{W}} - (-1)^{\theta_c} \alpha_{\tilde{Z}_i \tilde{f}_1}^{u} \beta_{\tilde{Z}_i \tilde{f}_2}^{d} \alpha_{\tilde{f}_1}^{\tilde{W}} \beta_{\tilde{f}_2}^{\tilde{W}}\right], \\
\mathcal{V}_{\tilde{f'}_1 \tilde{f}_2}^{(8)} &= 2 |m_{\tilde{Z}_i}||m_{\tilde{W}_j}|m_{f} m_{f'} \left[(-1)^{\theta_i}\beta_{\tilde{Z}_i \tilde{f}_1}^{u}\beta_{\tilde{Z}_i \tilde{f}_2}^{d} \alpha_{\tilde{f'}_1}^{\tilde{W}} \alpha_{\tilde{f}_2}^{\tilde{W}} + \alpha_{\tilde{Z}_i \tilde{f}_1}^{u} \alpha_{\tilde{Z}_i \tilde{f}_2}^{d} \beta_{\tilde{f'}_1}^{\tilde{W}} \beta_{\tilde{f}_2}^{\tilde{W}}\right].
\end{align}

If the decay is neutralino to chargino:
\begin{align}
\mathcal{V}_{\tilde{f'}_1 \tilde{f}_2}^{(4)} &= (-1)^{\theta_i} |m_{\tilde{Z}_i}|m_{f}\left[\alpha_{\tilde{Z}_i \tilde{f}_1}^{u} \alpha_{\tilde{Z}_i \tilde{f}_2}^{d} \beta_{\tilde{f'}}^{\tilde{W}}\alpha_{\tilde{f}_2}^{\tilde{W}} -  \beta_{\tilde{Z}_i \tilde{f}_1}^{u} \beta_{\tilde{Z}_i \tilde{f}_2}^{d} \alpha_{\tilde{f'}_1}^{\tilde{W}} \beta_{\tilde{f}_2}^{\tilde{W}}\right], \\
\mathcal{V}_{\tilde{f'}_1 \tilde{f}_2}^{(5)} &= m_{f'} |m_{\tilde{W}_j}|\left[(-1)^{\theta_i} \beta_{\tilde{Z}_i \tilde{f}_1}^{u} \alpha_{\tilde{Z}_i \tilde{f}_2}^{d} \alpha_{\tilde{f'}_1}^{\tilde{W}} \alpha_{\tilde{f}_2}^{\tilde{W}} + \alpha_{\tilde{Z}_i \tilde{f}_1}^{u} \beta_{\tilde{Z}_i \tilde{f}_2}^{d} \beta_{\tilde{f'}_1}^{\tilde{W}} \beta_{\tilde{f}_2}^{\tilde{W}}\right](-1)^{\theta_j}, \\
\mathcal{V}_{\tilde{f'}_1 \tilde{f}_2}^{(6)} &= -(-1)^{\theta_i} |m_{\tilde{Z}_i}| m_{f'} \left[-\beta_{\tilde{Z}_i \tilde{f}_1}^{u} \beta_{\tilde{Z}_i \tilde{f}_2}^{d} \beta_{\tilde{f'}_1}^{\tilde{W}} \alpha_{\tilde{f}_2}^{\tilde{W}} +  \alpha_{\tilde{Z}_i \tilde{f}_1}^{u} \alpha_{\tilde{Z}_i \tilde{f}_2}^{d} \alpha_{\tilde{f'}_1}^{\tilde{W}} \beta_{\tilde{f}_2}^{\tilde{W}}\right], \\
\mathcal{V}_{\tilde{f'}_1 \tilde{f}_2}^{(7)} &= |m_{\tilde{W}_j}|m_{f}\left[(-1)^{\theta_i} \alpha_{\tilde{Z}_i \tilde{f}_1}^{u} \beta_{\tilde{Z}_i \tilde{f}_2}^{d} \alpha_{\tilde{f'}_1}^{\tilde{W}} \alpha_{\tilde{f}_2}^{\tilde{W}} + \beta_{\tilde{Z}_i \tilde{f}_1}^{u} \alpha_{\tilde{Z}_i \tilde{f}_2}^{d} \beta_{\tilde{f'}_1}^{\tilde{W}} \beta_{\tilde{f}_2}^{\tilde{W}}\right](-1)^{\theta_j}.
\end{align}

Whilst if the decay is instead chargino to neutralino:
\begin{align}
\mathcal{V}_{\tilde{f'}_1 \tilde{f}_2}^{(4)} &= |m_{\tilde{Z}_i}| m_{f} \left[(-1)^{\theta_j}\alpha_{\tilde{f'}_1}^{\tilde{W}} \alpha_{\tilde{f}_2}^{\tilde{W}} \beta_{\tilde{Z}_i \tilde{f}_1}^{u} \alpha_{\tilde{Z}_i \tilde{f}_2}^{d} - \beta_{\tilde{f'}_1}^{\tilde{W}} \beta_{\tilde{f}_2}^{\tilde{W}} \alpha_{\tilde{Z}_i \tilde{f}_1}^{u} \beta_{\tilde{Z}_i \tilde{f}_2}^{d}\right] ,\\
\mathcal{V}_{\tilde{f'}_1 \tilde{f}_2}^{(5)} &= -|m_{\tilde{W}_j}|m_{f'} \left[\alpha_{\tilde{f}_2}^{\tilde{W}} \beta_{\tilde{f'}_1}^{\tilde{W}} \alpha_{\tilde{Z}_i \tilde{f}_1}^{u} \alpha_{\tilde{Z}_i \tilde{f}_2}^{d} -(-1)^{\theta_j} \alpha_{\tilde{f'}_1}^{\tilde{W}} \beta_{\tilde{f}_2}^{\tilde{W}} \beta_{\tilde{Z}_i \tilde{f}_1}^{u} \beta_{\tilde{Z}_i \tilde{f}_2}^{d}\right],\\
\mathcal{V}_{\tilde{f'}_1 \tilde{f}_2}^{(6)} &= |m_{\tilde{Z}_i}| m_{f'}\left[-\beta_{\tilde{f'}_1}^{\tilde{W}}\beta_{\tilde{f}_2}^{\tilde{W}} \beta_{\tilde{Z}_i \tilde{f}_1}^{u} \alpha_{\tilde{Z}_i \tilde{f}_2}^{d}  - \alpha_{\tilde{f'}_1}^{\tilde{W}} \alpha_{\tilde{f}_2}^{\tilde{W}_1} \alpha_{\tilde{Z}_i \tilde{f}_1}^{u} \beta_{\tilde{Z}_i \tilde{f}_2}^{d}\right], \\
\mathcal{V}_{\tilde{f'}_1 \tilde{f}_2}^{(7)} &= |m_{\tilde{W}_j}|m_{f}\left[-\alpha_{\tilde{f'}_1}^{\tilde{W}}\beta_{\tilde{f}_2}^{\tilde{W}} \alpha_{\tilde{Z}_i \tilde{f}_1}^{u} \alpha_{\tilde{Z}_i \tilde{f}_2}^{d} + (-1)^{\theta_j} \beta_{\tilde{f'}_1}^{\tilde{W}} \alpha_{\tilde{f}_2}^{\tilde{W}} \beta_{\tilde{Z}_i \tilde{f}_1}^{u} \beta_{\tilde{Z}_i \tilde{f}_2}^{d}\right].
\end{align}
The integrals are as in the $\tilde{f'}_1 \tilde{f}_1$ case, with the appropriate mass replacements. Similarly, $\Gamma_{\tilde{f'}_1 \tilde{f}_2}$ is just the product of each coupling combination $\mathcal{V}_{\tilde{f'}_1 \tilde{f}_2}^{(k)}$ with each corresponding integral $I_{\tilde{f'}_1 \tilde{f}_2}^{k}$.

\textbf{\underline{$\Gamma_{\tilde{f'}_2 \tilde{f}_1}$}}

\begin{align}
\mathcal{V}_{\tilde{f'}_2 \tilde{f}_1}^{(8)} &= |m_{\tilde{Z}_i}||m_{\tilde{W}_j}|m_{f} m_{f'} (-1)^{\theta_i} (-1)^{\theta_c} \left[\beta_{\tilde{Z}_i \tilde{f}_2}^{u}\beta_{\tilde{Z}_i \tilde{f}_1}^{d} \alpha_{\tilde{f'}_2}^{\tilde{W}} \alpha_{\tilde{f}_1}^{\tilde{W}} - \alpha_{\tilde{Z}_i \tilde{f}_2}^{u} \alpha_{\tilde{Z}_i \tilde{f}_1}^{d} \beta_{\tilde{f'}_2}^{\tilde{W}} \beta_{\tilde{f}_1}^{\tilde{W}}\right].
\end{align}
If the decay is neutralino to chargino:
\begin{align}
\mathcal{V}_{\tilde{f'}_2 \tilde{f}_1}^{(1)} &= \frac{1}{2}(-1)^{\theta_i}\left[-\alpha_{\tilde{Z}_i \tilde{f}_2}^{u} \beta_{\tilde{Z}_i \tilde{f}_1}^{d} \beta_{\tilde{f'}_2}^{\tilde{W}}\alpha_{\tilde{f}_1}^{\tilde{W}} + \beta_{\tilde{Z}_i \tilde{f}_2}^{u} \alpha_{\tilde{Z}_i \tilde{f}_1}^{d} \alpha_{\tilde{f'}_2}^{\tilde{W}} \beta_{\tilde{f}_1}^{\tilde{W}}\right], \\
\mathcal{V}_{\tilde{f'}_2 \tilde{f}_1}^{(2)} &= -|m_{\tilde{Z}_i}||m_{\tilde{W}_j}|\left[\alpha_{\tilde{Z}_i \tilde{f}_2}^{u} \alpha_{\tilde{Z}_i \tilde{f}_1}^{d} \alpha_{\tilde{f'}_2}^{\tilde{W}} \alpha_{\tilde{f}_1}^{\tilde{W}} - \beta_{\tilde{Z}_i \tilde{f}_2}^{u} \beta_{\tilde{Z}_{i} \tilde{f}_1}^{d} \beta_{\tilde{f'}_2}^{\tilde{W}} \beta_{\tilde{f}_1}^{\tilde{W}}\right], \\
\mathcal{V}_{\tilde{f'}_2 \tilde{f}_1}^{(3)} &= (-1)^{\theta_i} m_{f}m_{f'}\left[-\beta_{\tilde{Z}_i \tilde{f}_2}^{u} \alpha_{\tilde{Z}_i \tilde{f}_1}^{d} \beta_{\tilde{f'}_2}^{\tilde{W}} \alpha_{\tilde{f}_1}^{\tilde{W}} +  \alpha_{\tilde{Z}_i \tilde{f}_2}^{u} \beta_{\tilde{Z}_i \tilde{f}_1}^{d} \alpha_{\tilde{f}_2}^{\tilde{W}} \beta_{\tilde{f}_1}^{\tilde{W}}\right], \\
\mathcal{V}_{\tilde{f'}_2 \tilde{f}_1}^{(4)} &= |m_{\tilde{Z}_i}|m_{f}\left[\alpha_{\tilde{Z}_i \tilde{f}_2}^{u} \alpha_{\tilde{Z}_i \tilde{f}_1}^{d} \beta_{\tilde{f'}_2}^{\tilde{W}}\alpha_{\tilde{f}_1}^{\tilde{W}} + (-1)^{\theta_i} \beta_{\tilde{Z}_i \tilde{f}_2}^{u} \beta_{\tilde{Z}_i \tilde{f}_1}^{d} \alpha_{\tilde{f'}_2}^{\tilde{W}} \beta_{\tilde{f}_1}^{\tilde{W}}\right], \\
\mathcal{V}_{\tilde{f'}_2 \tilde{f}_1}^{(5)} &= (-1)^{\theta_i} m_{f'} |m_{\tilde{W}_j}|\left[ \beta_{\tilde{Z}_i \tilde{f}_2}^{u} \alpha_{\tilde{Z}_i \tilde{f}_1}^{d} \alpha_{\tilde{f'}_2}^{\tilde{W}} \alpha_{\tilde{f}_1}^{\tilde{W}} - \alpha_{\tilde{Z}_i \tilde{f}_2}^{u} \beta_{\tilde{Z}_i \tilde{f}_1}^{d} \beta_{\tilde{f'}_2}^{\tilde{W}} \beta_{\tilde{f}_1}^{\tilde{W}}\right](-1)^{\theta_j}, \\
\mathcal{V}_{\tilde{f'}_2 \tilde{f}_1}^{(6)} &= -(-1)^{\theta_i} |m_{\tilde{Z}_i}| m_{f'} \left[-\beta_{\tilde{Z}_i \tilde{f}_2}^{u} \beta_{\tilde{Z}_i \tilde{f}_1}^{d} \beta_{\tilde{f'}_2}^{\tilde{W}} \alpha_{\tilde{f}_1}^{\tilde{W}} + (-1)^{\theta_i} \alpha_{\tilde{Z}_i \tilde{f}_2}^{u} \alpha_{\tilde{Z}_i \tilde{f}_1}^{d} \alpha_{\tilde{f'}_2}^{\tilde{W}} \beta_{\tilde{f}_1}^{\tilde{W}}\right], \\
\mathcal{V}_{\tilde{f'}_2 \tilde{f}_1}^{(7)} &= -|m_{\tilde{W}_j}|m_{f}\left[ \alpha_{\tilde{Z}_i \tilde{f}_2}^{u} \beta_{\tilde{Z}_i \tilde{f}_1}^{d} \alpha_{\tilde{f'}_2}^{\tilde{W}} \alpha_{\tilde{f}_1}^{\tilde{W}} + (-1)^{\theta_i} \beta_{\tilde{Z}_i \tilde{f}_2}^{u} \alpha_{\tilde{Z}_i \tilde{f}_1}^{d} \beta_{\tilde{f'}_2}^{\tilde{W}} \beta_{\tilde{f}_1}^{\tilde{W}}\right](-1)^{\theta_j}. \\
\end{align}

Whilst if the decay is instead chargino to neutralino:
\begin{align}
\mathcal{V}_{\tilde{f'}_2 \tilde{f}_1}^{(1)} &= \frac{1}{2}\left[\alpha_{\tilde{Z}_i \tilde{f}_2}^{u} \beta_{\tilde{Z}_i \tilde{f}_1}^{d} \beta_{\tilde{f'}_2}^{\tilde{W}}\alpha_{\tilde{f}_1}^{\tilde{W}} + \beta_{\tilde{Z}_i \tilde{f}_2}^{u} \alpha_{\tilde{Z}_i \tilde{f}_1}^{d} \alpha_{\tilde{f'}_2}^{\tilde{W}} \beta_{\tilde{f}_1}^{\tilde{W}}\right](-1)^{\theta_i}, \\
\mathcal{V}_{\tilde{f'}_2 \tilde{f}_1}^{(2)} &= |m_{\tilde{Z}_i}||m_{\tilde{W}_j}|(-1)^{\theta_i}\left[ \alpha_{\tilde{Z}_i \tilde{f}_2}^{u} \alpha_{\tilde{Z}_i \tilde{f}_1}^{d} \alpha_{\tilde{f'}_2}^{\tilde{W}} \alpha_{\tilde{f}_1}^{\tilde{W}} + \beta_{\tilde{Z}_i \tilde{f}_2}^{u} \beta_{\tilde{Z}_{i} \tilde{f}_1}^{d} \beta_{\tilde{f'}_2}^{\tilde{W}} \beta_{\tilde{f}_1}^{\tilde{W}}\right], \\
\mathcal{V}_{\tilde{f'}_2 \tilde{f}_1}^{(3)} &= -m_{f}m_{f'}\left[\beta_{\tilde{Z}_i \tilde{f}_2}^{u} \alpha_{\tilde{Z}_i \tilde{f}_1}^{d} \beta_{\tilde{f'}_2}^{\tilde{W}} \alpha_{\tilde{f}_1}^{\tilde{W}} +  \alpha_{\tilde{Z}_i \tilde{f}_2}^{u} \beta_{\tilde{Z}_i \tilde{f}_1}^{d} \alpha_{\tilde{f}_2}^{\tilde{W}} \beta_{\tilde{f}_1}^{\tilde{W}}\right], \\
\mathcal{V}_{\tilde{f'}_2 \tilde{f}_1}^{(4)} &= -|m_{\tilde{Z}_i}| m_{f} \left[\alpha_{\tilde{f'}_2}^{\tilde{W}} \alpha_{\tilde{f}_1}^{\tilde{W}} \beta_{\tilde{Z}_i \tilde{f}_2}^{u} \alpha_{\tilde{Z}_i \tilde{f}_1}^{d} + \beta_{\tilde{f'}_2}^{\tilde{W}} \beta_{\tilde{f}_1}^{\tilde{W}} \alpha_{\tilde{Z}_i \tilde{f}_2}^{u} \beta_{\tilde{Z}_i \tilde{f}_1}^{d}\right] ,\\
\mathcal{V}_{\tilde{f'}_2 \tilde{f}_1}^{(5)} &= |m_{\tilde{W}_j}|m_{f'} \left[\alpha_{\tilde{f}_1}^{\tilde{W}} \beta_{\tilde{f'}_2}^{\tilde{W}} \alpha_{\tilde{Z}_i \tilde{f}_2}^{u} \alpha_{\tilde{Z}_i \tilde{f}_1}^{d} + \alpha_{\tilde{f'}_2}^{\tilde{W}} \beta_{\tilde{f}_1}^{\tilde{W}} \beta_{\tilde{Z}_i \tilde{f}_2}^{u} \beta_{\tilde{Z}_i \tilde{f}_1}^{d}\right],\\
\mathcal{V}_{\tilde{f'}_2 \tilde{f}_1}^{(6)} &= |m_{\tilde{Z}_i}| m_{f'}\left[\beta_{\tilde{f'}_2}^{\tilde{W}}\beta_{\tilde{f}_1}^{\tilde{W}} \beta_{\tilde{Z}_i \tilde{f}_2}^{u} \alpha_{\tilde{Z}_i \tilde{f}_1}^{d} + \alpha_{\tilde{f'}_2}^{\tilde{W}} \alpha_{\tilde{f}_1}^{\tilde{W}_1} \alpha_{\tilde{Z}_i \tilde{f}_2}^{u} \beta_{\tilde{Z}_i \tilde{f}_1}^{d}\right], \\
\mathcal{V}_{\tilde{f'}_2 \tilde{f}_1}^{(7)} &= -|m_{\tilde{W}_j}|m_{f}\left[\alpha_{\tilde{f'}_2}^{\tilde{W}}\beta_{\tilde{f}_1}^{\tilde{W}} \alpha_{\tilde{Z}_i \tilde{f}_2}^{u} \alpha_{\tilde{Z}_i \tilde{f}_1}^{d} + \beta_{\tilde{f'}_2}^{\tilde{W}} \alpha_{\tilde{f}_1}^{\tilde{W}} \beta_{\tilde{Z}_i \tilde{f}_2}^{u} \beta_{\tilde{Z}_i \tilde{f}_1}^{d}\right].
\end{align}
Again, the integrals are as in the $\tilde{f'}_1 \tilde{f}_1$ case with the obvious mass replacements. $\Gamma_{\tilde{f'}_2 \tilde{f}_1}$ is just the product of each coupling combination $\mathcal{V}_{\tilde{f'}_2 \tilde{f}_1}^{(k)}$ with each corresponding integral $I_{\tilde{f'}_2 \tilde{f}_1}^{k}$.

\textbf{\underline{$\Gamma_{\tilde{f'}_2 \tilde{f}_2}$}}

\begin{align}
\mathcal{V}_{\tilde{f'}_2 \tilde{f}_2}^{(8)} &= -|m_{\tilde{Z}_i}||m_{\tilde{W}_j}|m_{f} m_{f'} (-1)^{\theta_i} (-1)^{\theta_j} \left[\beta_{\tilde{Z}_i \tilde{f}_2}^{u}\beta_{\tilde{Z}_i \tilde{f}_2}^{d} \alpha_{\tilde{f'}_2}^{\tilde{W}} \alpha_{\tilde{f}_2}^{\tilde{W}} - \alpha_{\tilde{Z}_i \tilde{f}_2}^{u} \alpha_{\tilde{Z}_i \tilde{f}_2}^{d} \beta_{\tilde{f'}_2}^{\tilde{W}} \beta_{\tilde{f}_2}^{\tilde{W}}\right].
\end{align}

If the decay is neutralino to chargino:
\begin{align}
\mathcal{V}_{\tilde{f'}_2 \tilde{f}_2}^{(1)} &= \frac{1}{2}(-1)^{\theta_j}\left[-\alpha_{\tilde{Z}_i \tilde{f}_2}^{u} \beta_{\tilde{Z}_i \tilde{f}_2}^{d} \beta_{\tilde{f'}_2}^{\tilde{W}}\alpha_{\tilde{f}_2}^{\tilde{W}} + (-1)^{\theta_i}\beta_{\tilde{Z}_i \tilde{f}_2}^{u} \alpha_{\tilde{Z}_i \tilde{f}_2}^{d} \alpha_{\tilde{f'}_2}^{\tilde{W}} \beta_{\tilde{f}_2}^{\tilde{W}}\right], \\
\mathcal{V}_{\tilde{f'}_2 \tilde{f}_2}^{(2)} &= |m_{\tilde{Z}_i}||m_{\tilde{W}_j}|(-1)^{\theta_j}\left[\alpha_{\tilde{Z}_i \tilde{f}_2}^{u} \alpha_{\tilde{Z}_i \tilde{f}_2}^{d} \alpha_{\tilde{f'}_2}^{\tilde{W}} \alpha_{\tilde{f}_2}^{\tilde{W}} - \beta_{\tilde{Z}_i \tilde{f}_2}^{u} \beta_{\tilde{Z}_{i} \tilde{f}_2}^{d} \beta_{\tilde{f'}_2}^{\tilde{W}} \beta_{\tilde{f}_2}^{\tilde{W}}\right], \\
\mathcal{V}_{\tilde{f'}_2 \tilde{f}_2}^{(3)} &= -m_{f}m_{f'}\left[-(-1)^{\theta_i}\beta_{\tilde{Z}_i \tilde{f}_2}^{u} \alpha_{\tilde{Z}_i \tilde{f}_2}^{d} \beta_{\tilde{f'}_2}^{\tilde{W}} \alpha_{\tilde{f}_2}^{\tilde{W}} +  \alpha_{\tilde{Z}_i \tilde{f}_2}^{u} \beta_{\tilde{Z}_i \tilde{f}_2}^{d} \alpha_{\tilde{f}_2}^{\tilde{W}} \beta_{\tilde{f}_2}^{\tilde{W}}\right], \\
\mathcal{V}_{\tilde{f'}_2 \tilde{f}_2}^{(4)} &=-|m_{\tilde{Z}_i}|m_{f}(-1)^{\theta_j}\left[\alpha_{\tilde{Z}_i \tilde{f}_2}^{u} \alpha_{\tilde{Z}_i \tilde{f}_2}^{d} \beta_{\tilde{f'}_2}^{\tilde{W}}\alpha_{\tilde{f}_2}^{\tilde{W}} - (-1)^{\theta_i} \beta_{\tilde{Z}_i \tilde{f}_2}^{u} \beta_{\tilde{Z}_i \tilde{f}_2}^{d} \alpha_{\tilde{f'}_2}^{\tilde{W}} \beta_{\tilde{f}_2}^{\tilde{W}}\right], \\
\mathcal{V}_{\tilde{f'}_2 \tilde{f}_2}^{(5)} &= -m_{f'} |m_{\tilde{W}_j}|\left[(-1)^{\theta_i}(-1)^{\theta_j} \beta_{\tilde{Z}_i \tilde{f}_2}^{u} \alpha_{\tilde{Z}_i \tilde{f}_2}^{d} \alpha_{\tilde{f'}_2}^{\tilde{W}} \alpha_{\tilde{f}_2}^{\tilde{W}} + \alpha_{\tilde{Z}_i \tilde{f}_2}^{u} \beta_{\tilde{Z}_i \tilde{f}_2}^{d} \beta_{\tilde{f'}_2}^{\tilde{W}} \beta_{\tilde{f}_2}^{\tilde{W}}\right], \\
\mathcal{V}_{\tilde{f'}_2 \tilde{f}_2}^{(6)} &= |m_{\tilde{Z}_i}| m_{f'} \left[(-1)^{\theta_i}\beta_{\tilde{Z}_i \tilde{f}_2}^{u} \beta_{\tilde{Z}_i \tilde{f}_2}^{d} \beta_{\tilde{f'}_2}^{\tilde{W}} \alpha_{\tilde{f}_2}^{\tilde{W}} - \alpha_{\tilde{Z}_i \tilde{f}_2}^{u} \alpha_{\tilde{Z}_i \tilde{f}_2}^{d} \alpha_{\tilde{f'}_2}^{\tilde{W}} \beta_{\tilde{f}_2}^{\tilde{W}}\right], \\
\mathcal{V}_{\tilde{f'}_2 \tilde{f}_2}^{(7)} &= |m_{\tilde{W}_j}|m_{f}\left[ \alpha_{\tilde{Z}_i \tilde{f}_2}^{u} \beta_{\tilde{Z}_i \tilde{f}_2}^{d} \alpha_{\tilde{f'}_2}^{\tilde{W}} \alpha_{\tilde{f}_2}^{\tilde{W}} -  \beta_{\tilde{Z}_i \tilde{f}_2}^{u} \alpha_{\tilde{Z}_i \tilde{f}_2}^{d} \beta_{\tilde{f'}_2}^{\tilde{W}} \beta_{\tilde{f}_2}^{\tilde{W}}\right](-1)^{\theta_j}.
\end{align}

Whilst if the decay is instead chargino to neutralino:
\begin{align}
\mathcal{V}_{\tilde{f'}_2 \tilde{f}_2}^{(1)} &=-\frac{1}{2}\left[(-1)^{\theta_j}\alpha_{\tilde{Z}_i \tilde{f}_2}^{u} \beta_{\tilde{Z}_i \tilde{f}_2}^{d} \beta_{\tilde{f'}_2}^{\tilde{W}}\alpha_{\tilde{f}_2}^{\tilde{W}} - \beta_{\tilde{Z}_i \tilde{f}_2}^{u} \alpha_{\tilde{Z}_i \tilde{f}_2}^{d} \alpha_{\tilde{f'}_2}^{\tilde{W}} \beta_{\tilde{f}_2}^{\tilde{W}}\right], \\
\mathcal{V}_{\tilde{f'}_2 \tilde{f}_2}^{(2)} &= -|m_{\tilde{Z}_i}||m_{\tilde{W}_j}(-1)^{\theta_i}|\left[(-1)^{\theta_j} \alpha_{\tilde{Z}_i \tilde{f}_2}^{u} \alpha_{\tilde{Z}_i \tilde{f}_2}^{d} \alpha_{\tilde{f'}_2}^{\tilde{W}} \alpha_{\tilde{f}_2}^{\tilde{W}} - \beta_{\tilde{Z}_i \tilde{f}_2}^{u} \beta_{\tilde{Z}_{i} \tilde{f}_2}^{d} \beta_{\tilde{f'}_2}^{\tilde{W}} \beta_{\tilde{f}_2}^{\tilde{W}}\right], \\
\mathcal{V}_{\tilde{f'}_2 \tilde{f}_2}^{(3)} &= m_{f}m_{f'}\left[-(-1)^{\theta_j}\beta_{\tilde{Z}_i \tilde{f}_2}^{u} \alpha_{\tilde{Z}_i \tilde{f}_2}^{d} \beta_{\tilde{f'}_2}^{\tilde{W}} \alpha_{\tilde{f}_2}^{\tilde{W}} -  \alpha_{\tilde{Z}_i \tilde{f}_2}^{u} \beta_{\tilde{Z}_i \tilde{f}_2}^{d} \alpha_{\tilde{f}_2}^{\tilde{W}} \beta_{\tilde{f}_2}^{\tilde{W}}\right], \\
\mathcal{V}_{\tilde{f'}_2 \tilde{f}_2}^{(4)} &= |m_{\tilde{Z}_i}| m_{f} (-1)^{\theta_j} \left[\alpha_{\tilde{f'}_2}^{\tilde{W}} \alpha_{\tilde{f}_2}^{\tilde{W}} \beta_{\tilde{Z}_i \tilde{f}_2}^{u} \alpha_{\tilde{Z}_i \tilde{f}_2}^{d} + \beta_{\tilde{f'}_2}^{\tilde{W}} \beta_{\tilde{f}_2}^{\tilde{W}} \alpha_{\tilde{Z}_i \tilde{f}_2}^{u} \beta_{\tilde{Z}_i \tilde{f}_2}^{d}\right] ,\\
\mathcal{V}_{\tilde{f'}_2 \tilde{f}_2}^{(5)} &= |m_{\tilde{W}_j}|m_{f'} \left[\alpha_{\tilde{f}_2}^{\tilde{W}} \beta_{\tilde{f'}_2}^{\tilde{W}} \alpha_{\tilde{Z}_i \tilde{f}_2}^{u} \alpha_{\tilde{Z}_i \tilde{f}_2}^{d} - \alpha_{\tilde{f'}_2}^{\tilde{W}} \beta_{\tilde{f}_2}^{\tilde{W}} \beta_{\tilde{Z}_i \tilde{f}_2}^{u} \beta_{\tilde{Z}_i \tilde{f}_2}^{d}\right],\\
\mathcal{V}_{\tilde{f'}_2 \tilde{f}_2}^{(6)} &= (-1)^{\theta_j}|m_{\tilde{Z}_i}| m_{f'}\left[\beta_{\tilde{f'}_2}^{\tilde{W}}\beta_{\tilde{f}_2}^{\tilde{W}} \beta_{\tilde{Z}_i \tilde{f}_2}^{u} \alpha_{\tilde{Z}_i \tilde{f}_2}^{d}  + \alpha_{\tilde{f'}_2}^{\tilde{W}} \alpha_{\tilde{f}_2}^{\tilde{W}_1} \alpha_{\tilde{Z}_i \tilde{f}_2}^{u} \beta_{\tilde{Z}_i \tilde{f}_2}^{d}\right], \\
\mathcal{V}_{\tilde{f'}_2 \tilde{f}_2}^{(7)} &= |m_{\tilde{W}_j}|m_{f}\left[-\alpha_{\tilde{f'}_2}^{\tilde{W}}\beta_{\tilde{f}_2}^{\tilde{W}} \alpha_{\tilde{Z}_i \tilde{f}_2}^{u} \alpha_{\tilde{Z}_i \tilde{f}_2}^{d} + \beta_{\tilde{f'}_2}^{\tilde{W}} \alpha_{\tilde{f}_2}^{\tilde{W}} \beta_{\tilde{Z}_i \tilde{f}_2}^{u} \beta_{\tilde{Z}_i \tilde{f}_2}^{d}\right].
\end{align}

The integrals are again as in the $\tilde{f'}_1 \tilde{f}_1$ case with the obvious mass replacements and $\Gamma_{\tilde{f'}_2 \tilde{f}_2}$ is just the product of each coupling combination $\mathcal{V}_{\tilde{f'}_2 \tilde{f}_2}^{(k)}$ with each corresponding integral $I_{\tilde{f'}_2 \tilde{f}_2}^{k}$.

\textbf{\underline{$\Gamma_{W H^{\pm}}$}}

For this contribution the relevant coupling combinations are:
\begin{equation}
\begin{aligned}
\mathcal{V}_{W H^{\pm}}^{(1)} = & -(\mathcal{C}_{\tilde{W} \tilde{Z} W}^R \omega_{H^+ \tilde{W}^+ \tilde{Z}}^R + \mathcal{C}_{\tilde{W} \tilde{Z} W}^L \omega_{H^+ \tilde{W}^+ \tilde{Z}}^L){g \over \sqrt{2}} \mathcal{C}_{H^+ f f'}^u |m_{\tilde{W}_j}|m_{f'}(-1)^{\theta_c},
\end{aligned}
\end{equation}
\begin{equation}
\begin{aligned}
\mathcal{V}_{W H^{\pm}}^{(2)} = & (\mathcal{C}_{\tilde{W} \tilde{Z} W}^L \omega_{H^+ \tilde{W}^+ \tilde{Z}}^R + \mathcal{C}_{\tilde{W} \tilde{Z} W}^R \omega_{H^+ \tilde{W}^+ \tilde{Z}}^L){g \over \sqrt{2}} \mathcal{C}_{H^+ f f'}^d |m_{\tilde{Z}_i}|m_{f} (-1)^{\theta_i}(-1)^{\theta_j}(-1)^{\theta_c},
\end{aligned}
\end{equation}
\begin{equation}
\begin{aligned}
\mathcal{V}_{W H^{\pm}}^{(3)} = & (\mathcal{C}_{\tilde{W} \tilde{Z} W}^R \omega_{H^+ \tilde{W}^+ \tilde{Z}}^R + \mathcal{C}_{\tilde{W} \tilde{Z} W}^L  \omega_{H^+ \tilde{W}^+ \tilde{Z}}^L){g \over \sqrt{2}} \mathcal{C}_{H^+ f f'}^d  |m_{\tilde{W}_j}|m_{f} (-1)^{\theta_c},
\end{aligned}
\end{equation}
\begin{equation}
\begin{aligned}
\mathcal{V}_{W H^{\pm}}^{(4)} = & -(\mathcal{C}_{\tilde{W} \tilde{Z} W}^L  \omega_{H^+ \tilde{W}^+ \tilde{Z}}^R + \mathcal{C}_{\tilde{W} \tilde{Z} W}^R  \omega_{H^+ \tilde{W}^+ \tilde{Z}}^L) {g \over \sqrt{2}} \mathcal{C}_{H^+ f f'}^u |m_{\tilde{Z}_i}| m_{f'}(-1)^{\theta_i} (-1)^{\theta_j} (-1)^{\theta_c}.
\end{aligned}
\end{equation}

The integrals are as follows, with upper limit $E_{upper3} = {1 \over 2 |m_{\tilde{Z}_i}|} (m_{\tilde{Z}_i}^2 + m_{\tilde{W}_j}^2 - m_{f}^2 - m_{f'}^2 - 2m_{f}m_{f'})$, $s = m_{\tilde{Z}_i}^2 + m_{\tilde{W}_j}^2 - 2|m_{\tilde{Z}_i}|E$, $\lambda = \sqrt{(s-(m_{f} + m_{f'})^2)(s-(m_{f}-m_{f'})^2)}$, $\mathcal{A} = 2|m_{\tilde{Z}_i}E + m_{f}^2 + m_{f'}^2 - (m_{\tilde{Z}_i}^2 - m_{\tilde{W}_j}^2)(m_{f}^2-m_{f'}^2)/s$ and $\mathcal{B} = {2|m_{\tilde{Z}_i}| \over s} \lambda \sqrt{E^2 - m_{\tilde{W}_j}^2} $:
\begin{align}
I_{W H^{\pm}}^{1} &= 2|m_{\tilde{Z}_i}| \int_{|m_{\tilde{W}_j}|}^{E_{upper3}} dE {-\frac{1}{2}\mathcal{A}\mathcal{B} + (m_{\tilde{Z}_i}^2 + m_{f}^2)\mathcal{B} \over (s-m_{W}^2)(s-m_{H^{\pm}}^2)}, \\
I_{W H^{\pm}}^{2} &= 2|m_{\tilde{Z}_i}| \int_{|m_{\tilde{W}_j}|}^{E_{upper3}} dE {\frac{1}{2}\mathcal{A}\mathcal{B} - (m_{\tilde{W}_j}^2 + m_{f'}^2)\mathcal{B} \over (s-m_{W}^2)(s-m_{H^{\pm}}^2)}, \\
I_{W H^{\pm}}^{3} &= 2|m_{\tilde{Z}_i}| \int_{|m_{\tilde{W}_j}|}^{E_{upper3}} dE {-\frac{1}{2}\mathcal{A}\mathcal{B} + (m_{\tilde{Z}_i}^2 -2|m_{\tilde{Z}_i}|E	 - m_{f}^2)\mathcal{B} \over (s-m_{W}^2)(s-m_{H^{\pm}}^2)}, \\
I_{W H^{\pm}}^{4} &= 2|m_{\tilde{Z}_i}| \int_{|m_{\tilde{W}_j}|}^{E_{upper3}} dE {-\frac{1}{2}\mathcal{A}\mathcal{B} - (m_{\tilde{Z}_j}^2 -2|m_{\tilde{Z}_i}| - m_{f'}^2)\mathcal{B} \over (s-m_{W}^2)(s-m_{H^{\pm}}^2)},
\end{align}
So overall:
\begin{equation}
\Gamma_{W H^{\pm}} = \mathcal{V}_{W H^{\pm}}^{(1)} I_{W H^{\pm}}^{1}(-1)^{\theta_c}(-1)^{\theta_j} + \mathcal{V}_{W H^{\pm}}^{(2)} I_{W H^{\pm}}^{2}(-1)^{\theta_c}(-1)^{\theta_j} + \mathcal{V}_{W H^{\pm}}^{(3)} I_{W H^{\pm}}^{3} + \mathcal{V}_{W H^{\pm}}^{(4)} I_{W H^{\pm}}^{4}.
\end{equation}

\textbf{\underline{$\Gamma_{W G}$}}

Here everything is as above but in the coupling combinations we must make the appropriate replacements $\omega_{H^+ \tilde{W}^+ \tilde{Z}}^{L/R} \rightarrow \omega_{G \tilde{W} \tilde{Z}}^{L/R}$, whilst in the integrals we make the change $m_{H^{\pm}} \rightarrow m_{goldstone} = m_{W}$. However because of subtle differences in the definitions of the couplings, the overall contribution here is given by:
\begin{equation}
\Gamma_{W G} = \mathcal{V}_{W G}^{(1)} I_{W G}^{1} + \mathcal{V}_{W G}^{(2)} I_{W G}^{2} + \mathcal{V}_{W G}^{(3)} I_{W G}^{3} + \mathcal{V}_{W G}^{(4)} I_{W G}^{4}.
\end{equation}

\textbf{\underline{$\Gamma_{W \tilde{f'}_1}$}}

The coupling combinations are:
\begin{align}
\mathcal{V}_{W \tilde{f'}_1}^{(1)} &= -2 \mathcal{C}_{\tilde{W} \tilde{Z} W}^L \alpha_{\tilde{Z}_i \tilde{f}_1}^{u} {g \over \sqrt{2}} \beta_{\tilde{f'}_1}^{\tilde{W}_j}|m_{\tilde{Z}_i}|m_{f}(-1)^{\theta_j}, \\
\mathcal{V}_{W \tilde{f'}_1}^{(2)} &= -2 \mathcal{C}_{\tilde{W} \tilde{Z} W}^L \beta_{\tilde{Z}_i \tilde{f}_1} {g \over \sqrt{2}} \alpha_{\tilde{f}_1}^{\tilde{W}_j} m_{f'}|m_{\tilde{W}_j}|(-1)^{\theta_c}, \\
\mathcal{V}_{W \tilde{f'}_1}^{(3)} &= 2 \mathcal{C}_{\tilde{W} \tilde{Z} W}^R \alpha_{\tilde{Z}_i \tilde{f}_1}^{u} {g \over \sqrt{2}} \alpha_{\tilde{f'}_1}^{\tilde{W}_j}(-1)^{\theta_i}(-1)^{\theta_j}(-1)^{\theta_c}, \\
\mathcal{V}_{W \tilde{f'}_1}^{(4)} &= (-1)^{\theta_i}4 \mathcal{C}_{\tilde{W} \tilde{Z} W}^R \beta_{\tilde{Z}_i \tilde{f}_1}^{u} {g \over \sqrt{2}} \alpha_{\tilde{f'}_1}^{\tilde{W}_j}|m_{\tilde{Z}_i}|m_{f'}(-1)^{\theta_c}, \\
\mathcal{V}_{W \tilde{f'}_1}^{(5)} &= 4 \mathcal{C}_{\tilde{W} \tilde{Z} W}^R \alpha_{\tilde{Z}_i \tilde{f}_1}^{u} {g \over \sqrt{2}} \beta_{\tilde{f'}_1}^{\tilde{W}_j}m_{f}|m_{\tilde{W}_j}|(-1)^{\theta_i}, \\
\mathcal{V}_{W \tilde{f'}_1}^{(6)} &= -2 \mathcal{C}_{\tilde{W} \tilde{Z} W}^L \alpha_{\tilde{Z}_i \tilde{f}_1}^{u} {g \over \sqrt{2}} \alpha_{\tilde{f'}_1}^{\tilde{W}_j}|m_{\tilde{Z}_i}||m_{\tilde{W}_j}|(-1)^{\theta_c}, \\
\mathcal{V}_{W \tilde{f'}_1}^{(7)} &= -2 \mathcal{C}_{\tilde{W} \tilde{Z} W}^L \beta_{\tilde{Z}_i \tilde{f}_1}^{u} {g \over \sqrt{2}} \beta_{\tilde{f'}_1}^{\tilde{W}_f} m_{f'}m_{f}, \\ 
\mathcal{V}_{W \tilde{f'}_1}^{(8)} &= 8 \mathcal{C}_{\tilde{W} \tilde{Z} W}^R \beta_{\tilde{Z}_i \tilde{f}_1}^{u} {g \over \sqrt{2}} \beta_{\tilde{f'}_1}^{\tilde{W}_1} |m_{\tilde{Z}_i}|m_{f'} m_{f}|m_{\tilde{W}_j}|(-1)^{\theta_i}(-1)^{\theta_j}. \\
\end{align}
The integrals are as follows with $s = m_{\tilde{Z}_i}^2 + m_{\tilde{W}_j}^2 - 2|m_{\tilde{Z}_i}|E$, $\lambda = \sqrt{(s-(m_{f'}+m_{f})^2)(s-(m_{f'}-m_{f})^2)}$, $A = m_{f'}^2 + m_{f}^2 + 2|m_{\tilde{Z}_i}|E + (m_{\tilde{Z}_i}^2 - m_{\tilde{W}_j}^2)(m_{f}^2 - m_{f'}^2)/s$, $B = 2|m_{\tilde{Z}_i}| \lambda \sqrt{E^2-m_{\tilde{W}_j}^2} $, $Z = {A + B - 2m_{\tilde{f'}_1}^2 \over A - B - 2m_{\tilde{f'}_1}^2}$:
\begin{equation}
I_{W \tilde{f'}_1}^{1} = -2|m_{\tilde{Z}_i}| \int_{|m_{\tilde{W}_j}|}^{E_{upper3}} dE {B + (m_{\tilde{f'}_1}^2 + m_{\tilde{W}_j}^2 - 2|m_{\tilde{Z}_i}|E - m_{f}^2)\log(Z) \over s-m_{W}^2} ,
\end{equation}
\begin{equation}
I_{W \tilde{f'}_1}^{2} = 2|m_{\tilde{Z}_i}| \int_{|m_{\tilde{W}_j}|}^{E_{upper3}} dE {B + (m_{\tilde{f'}_1}^2 + m_{\tilde{Z}_i}^2 - 2|m_{\tilde{Z}_i}|E - m_{f'}^2)\log(Z) \over s - m_{W}^2},
\end{equation}
\begin{equation}
\begin{aligned}
I_{W \tilde{f'}_1}^{3} = 2|m_{\tilde{Z}_i}| \int_{|m_{\tilde{W}_j}|}^{E_{upper3}} dE & \Big[\{m_{\tilde{Z}_i}^2 + m_{f}^2 + m_{f'}^2 + m_{\tilde{W}_j}^2 - 1.5 m_{\tilde{f'}_1}^2 - 0.25(A+B)\}(\frac{1}{2}(A+B) - m_{\tilde{f'}_1}^2 \\ & - (m_{\tilde{Z}_i}^2 + m_{f}^2 + m_{f'}^2 +m_{\tilde{W}_j}^2 - 1.5m_{\tilde{f'}_1}^2 - 0.25(A-B))(\frac{1}{2}(A-B) - m_{\tilde{f'}_1}^2) \\ & + (m_{\tilde{Z}_i}^2 + m_{f'}^2 - m_{\tilde{f'}_1}^2)(m_{\tilde{f'}_1}^2-m_{f}^2 - m_{\tilde{W}_j}^2)\log(Z)\Big]{1 \over s - m_{W}^2},
\end{aligned}
\end{equation}
\begin{equation}
I_{W \tilde{f'}_1}^{4} = 2|m_{\tilde{Z}_i}| \int_{|m_{\tilde{W}_j}|}^{E_{upper3}} dE {B + (m_{\tilde{f'}}^2 - m_{f}^2 - m_{\tilde{W}_j}^2)\log(Z) \over s - m_{W}^2},
\end{equation}
\begin{equation}
I_{W \tilde{f'}_1}^{5} = -2|m_{\tilde{Z}_i}| \int_{|m_{\tilde{W}_j}|}^{E_{upper3}} dE {B + (m_{\tilde{f'}_1}^2 - m_{\tilde{Z}_i}^2 - m_{f'}^2)\log(Z) \over s - m_{W}^2},
\end{equation}
\begin{equation}
I_{W \tilde{f'}_1}^{6} = 2|m_{\tilde{Z}_i}| \int_{|m_{\tilde{W}_j}|}^{E_{upper3}} dE {(s-m_{f'}^2 - m_{f}^2)\log(Z) \over s- m_{W}^2},
\end{equation}
\begin{equation}
I_{W \tilde{f'}_1}^{7} = 2|m_{\tilde{Z}_i}| \int_{|m_{\tilde{W}_j}|}^{E_{upper3}} dE {2|m_{\tilde{Z}_i}|E \log(Z) \over s-m_{W}^2},
\end{equation}
\begin{equation}
I_{W \tilde{f'}_1}^{8} = 2|m_{\tilde{Z}_i}| \int_{|m_{\tilde{W}_j}|}^{E_{upper3}} dE {\log(Z) \over s-m_{W}^2}.
\end{equation}
Therefore the overall contribution is:
\begin{equation} 
\begin{aligned}
\Gamma_{W \tilde{f'}_1} = & \mathcal{V}_{W \tilde{f'}_1}^{(1)} I_{W \tilde{f'}_1}^{1} + \mathcal{V}_{W \tilde{f'}_1}^{(2)} I_{W \tilde{f'}_1}^{2} + \mathcal{V}_{W \tilde{f'}_1}^{(3)} I_{W \tilde{f'}_1}^{3} + \mathcal{V}_{W \tilde{f'}_1}^{(4)} I_{W \tilde{f'}_1}^{4} \\ & + \mathcal{V}_{W \tilde{f'}_1}^{(5)} I_{W \tilde{f'}_1}^{5} + \mathcal{V}_{W \tilde{f'}_1}^{(6)} I_{W \tilde{f'}_1}^{6} + \mathcal{V}_{W \tilde{f'}_1}^{(7)} I_{W \tilde{f'}_1}^{7} + \mathcal{V}_{W \tilde{f'}_1}^{(8)} I_{W \tilde{f'}_1}^{8}.
\end{aligned}
\end{equation}

\textbf{\underline{$\Gamma_{W \tilde{f'}_2}$}}

The coupling combinations here are:
\begin{align}
\mathcal{V}_{W \tilde{f'}_2}^{(1)} &= 2 \mathcal{C}_{\tilde{W} \tilde{Z} W}^L \alpha_{\tilde{Z}_i \tilde{f}_2}^{u} {g \over \sqrt{2}} \beta_{\tilde{f'}_2}^{\tilde{W}_j} |m_{\tilde{Z}_i}| m_{f}(-1)^{\theta_j}(-1)^{\theta_c}, \\
\mathcal{V}_{W \tilde{f'}_2}^{(2)} &= 2 \mathcal{C}_{\tilde{W} \tilde{Z} W}^L \beta_{\tilde{Z}_i \tilde{f}_2}^{u} {g \over \sqrt{2}} \alpha_{\tilde{f'}_2}^{\tilde{W}_j} m_{f'}|m_{\tilde{W}_j}|(-1)^{\theta_i}(-1)^{\theta_j}, \\
\mathcal{V}_{W \tilde{f'}_2}^{(3)} &= 2 \mathcal{C}_{\tilde{W} \tilde{Z} W}^R \alpha_{\tilde{Z}_i \tilde{f}_2}^{u} {g \over \sqrt{2}} \alpha_{\tilde{f'}_2}^{\tilde{W}_j}(-1)^{\theta_i}, \\
\mathcal{V}_{W \tilde{f'}_2}^{(4)} &= -4 \mathcal{C}_{\tilde{W} \tilde{Z} W}^R \beta_{\tilde{Z}_i \tilde{f}_2}^{u} {g \over \sqrt{2}} \alpha_{\tilde{f'}_2}^{\tilde{W}_j}|m_{\tilde{Z}_i}|m_{f'}, \\
\mathcal{V}_{W \tilde{f'}_2}^{(5)} &= -4 \mathcal{C}_{\tilde{W} \tilde{Z} W}^R \alpha_{\tilde{Z}_i \tilde{f}_2}^{u} {g \over \sqrt{2}} \beta_{\tilde{f'}_2}^{\tilde{W}_j}m_{f}|m_{\tilde{W}_j}|(-1)^{\theta_i}(-1)^{\theta_c}, \\
\mathcal{V}_{W \tilde{f'}_2}^{(6)} &= -2 \mathcal{C}_{\tilde{W} \tilde{Z} W}^L \alpha_{\tilde{Z}_i \tilde{f}_2}^{u} {g \over \sqrt{2}} \alpha_{\tilde{f'}_2}^{\tilde{W}_j} |m_{\tilde{Z}_i}||m_{\tilde{W}_j}|(-1)^{\theta_j}, \\
\mathcal{V}_{W \tilde{f'}_2}^{(7)} &= -2 \mathcal{C}_{\tilde{W} \tilde{Z} W}^L \beta_{\tilde{Z}_i \tilde{f}_2}^{u} {g \over \sqrt{2}} \beta_{\tilde{f'}_2}^{\tilde{W}_j} m_{f'} m_{f} (-1)^{\theta_i}(-1)^{\theta_j} (-1)^{\theta_c}, \\
\mathcal{V}_{W \tilde{f'}_2}^{(8)} &= 8 \mathcal{C}_{\tilde{W} \tilde{Z} W}^R \beta_{\tilde{Z}_i \tilde{f}_2}^{u} {g \over \sqrt{2}} \beta_{\tilde{f'}_2}^{\tilde{W}_j} |m_{\tilde{Z}_i}| m_{f'} m_{f}|m_{\tilde{W}_j}|(-1)^{\theta_c}.
\end{align}
The integrals here are exactly as for $W \tilde{f'}_1$ with the change $m_{\tilde{f'}_1} \rightarrow m_{\tilde{f'}_2}$. As above $\Gamma_{W \tilde{f'}_2}$ is then the sum of the products of coupling combinations, $\mathcal{V}^{(i)}$ and integrals, $I_{W \tilde{f'}_2}^{i}$.

\textbf{\underline{$\Gamma_{W \tilde{f}_1}$}}

Here the coupling combinations required are:
\begin{align}
\mathcal{V}_{W \tilde{f}_1}^{(6)} &= -2 \mathcal{C}_{\tilde{W} \tilde{Z} W}^R \alpha_{\tilde{Z}_i \tilde{f}_1}^{d} {g \over \sqrt{2}} \alpha_{\tilde{f}_1}^{\tilde{W}_j}|m_{\tilde{Z}_i}||m_{\tilde{W}_j}|(-1)^{\theta_c}, \\
\mathcal{V}_{W \tilde{f}_1}^{(7)} &= -2 \mathcal{C}_{\tilde{W} \tilde{Z} W}^R \beta_{\tilde{Z}_i \tilde{f}_1}^{d} {g \over \sqrt{2}} \beta_{\tilde{f}_1}^{\tilde{W}_j}m_{f}m_{f'}, \\
\mathcal{V}_{W \tilde{f}_1}^{(8)} &= 8 \mathcal{C}_{\tilde{W} \tilde{Z} W}^L \beta_{\tilde{Z}_i \tilde{f}_1}^{d} {g \over \sqrt{2}} \beta_{\tilde{f}_1}^{\tilde{W}_j} |m_{\tilde{Z}_i}|m_{f} m_{f'}|m_{\tilde{W}_j}|(-1)^{\theta_i}(-1)^{\theta_j}.
\end{align}

The other coupling combinations depend upon if it is a neutralino decaying into a chargino or a chargino decaying into a neutralino, for a neutralino decaying:
\begin{align}
\mathcal{V}_{W \tilde{f}_1}^{(1)} &= -2 \mathcal{C}_{\tilde{W} \tilde{Z} W}^R \alpha_{\tilde{Z}_i \tilde{f}_1}^{d} {g \over \sqrt{2}} \beta_{\tilde{f}_1}^{\tilde{W}_j}|m_{\tilde{Z}_i}| m_{f'}, \\
\mathcal{V}_{W \tilde{f}_1}^{(2)} &= -2 \mathcal{C}_{\tilde{W} \tilde{Z} W}^R \beta_{\tilde{Z}_i \tilde{f}_1}^{d} {g \over \sqrt{2}} \alpha_{\tilde{f}_1}^{\tilde{W}_j} m_{f} |m_{\tilde{W}_j}|(-1)^{\theta_j}, \\
\mathcal{V}_{W \tilde{f}_1}^{(3)} &= 2 \mathcal{C}_{\tilde{W} \tilde{Z} W}^R \alpha_{\tilde{Z}_i \tilde{f}_1}^{d} {g \over \sqrt{2}} \alpha_{\tilde{f}_1}^{\tilde{W}_j}, \\
\mathcal{V}_{W \tilde{f}_1}^{(4)} &= 4 \mathcal{C}_{\tilde{W} \tilde{Z} W}^L \beta_{\tilde{Z}_i \tilde{f}_1}^{d} {g \over \sqrt{2}} \alpha_{\tilde{f}_1}^{\tilde{W}_j}|m_{\tilde{Z}_i}|m_{f}(-1)^{\theta_i}, \\
\mathcal{V}_{W \tilde{f}_1}^{(5)} &= 4 \mathcal{C}_{\tilde{W} \tilde{Z} W}^L \alpha_{\tilde{Z}_i \tilde{f}_1}^{d} {g \over \sqrt{2}} \beta_{\tilde{f}_1}^{\tilde{W}_j} m_{f'}|m_{\tilde{W}_j}|(-1)^{\theta_i}(-1)^{\theta_j},
\end{align}

whilst if it's a chargino decaying:
\begin{align}
\mathcal{V}_{W \tilde{f}_1}^{(1)} &= 2 \mathcal{C}_{\tilde{W} \tilde{Z} W}^R \beta_{\tilde{Z}_i \tilde{f}_1}^{d} \alpha_{\tilde{f}_1}^{\tilde{W}_j}{g \over \sqrt{2}} |m_{\tilde{Z}_i}| m_{f'}(-1)^{\theta_i}, \\
\mathcal{V}_{W \tilde{f}_1}^{(2)} &= -2 \mathcal{C}_{\tilde{W} \tilde{Z} W}^R \alpha_{\tilde{Z}_i \tilde{f}_1}^{d} \beta_{\tilde{f}_1}^{\tilde{W}_j}{g \over \sqrt{2}} m_{f} |m_{\tilde{W}_j}|, \\
\mathcal{V}_{W \tilde{f}_1}^{(3)} &= -2 \mathcal{C}_{\tilde{W} \tilde{Z} W}^L \alpha_{\tilde{Z}_i \tilde{f}_1}^{d} \alpha_{\tilde{f}_1}^{\tilde{W}_j}{g \over \sqrt{2}} (-1)^{\theta_i}(-1)^{\theta_j}, \\
\mathcal{V}_{W \tilde{f}_1}^{(4)} &= 4 \mathcal{C}_{\tilde{W} \tilde{Z} W}^L \alpha_{\tilde{Z}_i \tilde{f}_1}^{d} \beta_{\tilde{f}_1}^{\tilde{W}_j} {g \over \sqrt{2}} |m_{\tilde{Z}_i}|m_{f}(-1)^{\theta_i}(-1)^{\theta_j}, \\
\mathcal{V}_{W \tilde{f}_1}^{(5)} &= -4 \mathcal{C}_{\tilde{W} \tilde{Z} W}^L \beta_{\tilde{Z}_i \tilde{f}_1}^{d} \alpha_{\tilde{f}_1}^{\tilde{W}_j}{g \over \sqrt{2}}m_{f'}|m_{\tilde{W}_j}|(-1)^{\theta_j}.
\end{align}
Then the integrals are exactly as for $W \tilde{f'}_1$ but with the changes $m_{f'} \leftrightarrow m_{f}$, $m_{\tilde{f'}_1} \rightarrow m_{\tilde{f}_1}$. $\Gamma_{W \tilde{f}_1}$ is, as above, just the sum of the products of coupling combinations and corresponding integrals.

\textbf{\underline{$\Gamma_{W \tilde{f}_2}$}}

The coupling combinations now are:
\begin{align}
\mathcal{V}_{W \tilde{f}_2}^{(6)} &= 2 \mathcal{C}_{\tilde{W} \tilde{Z} W}^R \alpha_{\tilde{Z}_i \tilde{f}_2}^{d} {g \over \sqrt{2}} \alpha_{\tilde{f}_2}^{\tilde{W}_j} |m_{\tilde{Z}_i}||m_{\tilde{W}_j}|(-1)^{\theta_j}(-1)^{\theta_c}, \\
\mathcal{V}_{W \tilde{f}_2}^{(7)} &= -2 \mathcal{C}_{\tilde{W} \tilde{Z} W}^R \beta_{\tilde{Z}_i \tilde{f}_2}^{d} {g \over \sqrt{2}} \beta_{\tilde{f}_2}^{\tilde{W}_j}m_{f}m_{f'}(-1)^{\theta_c}, \\
\mathcal{V}_{W \tilde{f}_2}^{(8)} &= 8 \mathcal{C}_{\tilde{W} \tilde{Z} W}^L \beta_{\tilde{Z}_i \tilde{f}_2}^{d} {g \over \sqrt{2}} \beta_{\tilde{f}_2} |m_{\tilde{Z}_i}|m_{f}m_{f'}|m_{\tilde{W}_j}|(-1)^{\theta_i}(-1)^{\theta_j}(-1)^{\theta_c}.
\end{align}

Again, here some of the coupling combinations depend upon which way around the decay occurs, i.e.\ neutralino to chargino or chargino to neutralino, for neutralino decaying:
\begin{align}
\mathcal{V}_{W \tilde{f}_2}^{(1)} &= -2 \mathcal{C}_{\tilde{W} \tilde{Z} W}^R \alpha_{\tilde{Z}_i \tilde{f}_2}^{d} {g \over \sqrt{2}} \beta_{\tilde{f}_2}^{\tilde{W}_j}|m_{\tilde{Z}_i}|m_{f'}, \\
\mathcal{V}_{W \tilde{f}_2}^{(2)} &= 2 \mathcal{C}_{\tilde{W} \tilde{Z} W}^R \beta_{\tilde{Z}_i \tilde{f}_2}^{d} {g \over \sqrt{2}} \alpha_{\tilde{f}_2}^{\tilde{W}_j} m_{f} |m_{\tilde{W}_j}|(-1)^{\theta_j}, \\
\mathcal{V}_{W \tilde{f}_2}^{(3)} &= -2 \mathcal{C}_{\tilde{W} \tilde{Z} W}^R  \alpha_{\tilde{Z}_i \tilde{f}_2}^{d} {g \over \sqrt{2}} \alpha_{\tilde{f}_2}^{\tilde{W}_j}, \\
\mathcal{V}_{W \tilde{f}_2}^{(4)} &= -4 \mathcal{C}_{\tilde{W} \tilde{Z} W}^L \beta_{\tilde{Z}_i \tilde{f}_2}^{d} {g \over \sqrt{2}} \alpha_{\tilde{f}_2}^{\tilde{W}_j}|m_{\tilde{Z}_i}|m_{f}(-1)^{\theta_i}, \\
\mathcal{V}_{W \tilde{f}_2}^{(5)} &= 4 \mathcal{C}_{\tilde{W} \tilde{Z} W}^L \alpha_{\tilde{Z}_i \tilde{f}_2}^{d} {g \over \sqrt{2}} \beta_{\tilde{f}_2}^{\tilde{W}_j}m_{f'}|m_{\tilde{W}_j}|(-1)^{\theta_i}(-1)^{\theta_j},
\end{align}

whilst if it's a chargino decaying:
\begin{align}
\mathcal{V}_{W \tilde{f}_2}^{(1)} &= -2 \mathcal{C}_{\tilde{W} \tilde{Z} W}^R \beta_{\tilde{Z}_i \tilde{f}_2}^{d} \alpha_{\tilde{f}_2}^{\tilde{W}_j} {g \over \sqrt{2}} |m_{\tilde{Z}_i}|m_{f'}(-1)^{\theta_i}(-1)^{\theta_j}(-1)^{\theta_c}, \\
\mathcal{V}_{W \tilde{f}_2}^{(2)} &= 2 \mathcal{C}_{\tilde{W} \tilde{Z} W}^R \alpha_{\tilde{Z}_i \tilde{f}_2}^{d} \beta_{\tilde{f}_2}^{\theta{W}_j} {g \over \sqrt{2}} m_{f} |m_{\tilde{W}_j}|(-1)^{\theta_c}, \\
\mathcal{V}_{W \tilde{f}_2}^{(3)} &= -2 \mathcal{C}_{\tilde{W} \tilde{Z} W}^L \alpha_{\tilde{Z}_i \tilde{f}_2}^{d} \alpha_{\tilde{f}_2}^{\tilde{W}_j} {g \over \sqrt{2}} (-1)^{\theta_i} (-1)^{\theta_c}, \\
\mathcal{V}_{W \tilde{f}_2}^{(4)} &= -4 \mathcal{C}_{\tilde{W} \tilde{Z} W}^L \alpha_{\tilde{Z}_i \tilde{f}_2}^{d} \beta_{\tilde{f}_2}^{\tilde{W}_j} {g \over \sqrt{2}} |m_{\tilde{Z}_i}| m_{f} (-1)^{\theta_i}(-1)^{\theta_j}(-1)^{\theta_c}, \\
\mathcal{V}_{W \tilde{f}_2}^{(5)} &= 4 \mathcal{C}_{\tilde{W} \tilde{Z} W}^L \beta_{\tilde{Z}_i \tilde{f}_2}^{d} \alpha_{\tilde{f}_2}^{\tilde{W}_j}{g \over \sqrt{2}} m_{f'}|m_{\tilde{W}_j}| (-1)^{\theta_c}.
\end{align}

Then the integrals, and indeed the overall expression for $\Gamma_{W \tilde{f}_2}$, are just like that for $W \tilde{f}_1$ but with the expected replacement $m_{\tilde{f}_1} \rightarrow m_{\tilde{f}_2}$.

\textbf{\underline{$\Gamma_{H^{\pm} G}$}}

The coupling combinations are:
\begin{align}
\mathcal{V}_{H^{\pm} G}^{(1)} &= \omega_{G \tilde{W} \tilde{Z}}^L  \omega_{H^+ \tilde{W}^+ \tilde{Z}}^L + \omega_{G \tilde{W} \tilde{Z}}^R  \omega_{H^+ \tilde{W}^+ \tilde{Z}}^R, \\
\mathcal{V}_{H^{\pm} G}^{(2)} &= (\omega_{G \tilde{W} \tilde{Z}}^R  \omega_{H^+ \tilde{W}^+ \tilde{Z}}^L + \omega_{G \tilde{W} \tilde{Z}}^L  \omega_{H^+ \tilde{W}^+ \tilde{Z}}^R)(-1)^{\theta_i}(-1)^{\theta_j}, \\
\mathcal{V}_{H^{\pm} G}^{(3)} &= \mathcal{C}_{G f f'}^u  \mathcal{C}_{H^+ f f'}^u + \mathcal{C}_{G f f'}^d  \mathcal{C}_{H^+ f f'}^d, \\
\mathcal{V}_{H^{\pm} G}^{(4)} &= \mathcal{C}_{G f f'}^d  \mathcal{C}_{H^+ f f'}^u + \mathcal{C}_{G f f'}^u  \mathcal{C}_{H^+ f f'}^d,
\end{align}
and the integrals with $s = m_{\tilde{Z}_i}^2 + m_{\tilde{W}_j}^2 - 2|m_{\tilde{Z}_i}|E$, $\lambda = \sqrt{(s-(m_{f}+m_{f'})^2)(s-(m_{f}-m_{f'})^2)}$ are:
\begin{equation}
I_{H^{\pm} G}^{1} = 2|m_{\tilde{Z}_i}| \int_{|m_{\tilde{W}_j}|}^{E_{upper3}} dE  {\lambda \sqrt{E^2 - m_{\tilde{W}_j}^2}   \over s(s-m_{W}^2)(s-m_{H^{\pm}}^2)},
\end{equation}
\begin{equation}
I_{H^{\pm} G}^{2} = 2|m_{\tilde{Z}_i}| \int_{|m_{\tilde{W}_j}|}^{E_{upper3}} dE {2|m_{\tilde{Z}_i}| \lambda \sqrt{E^2-m_{\tilde{W}_j}^2} (s- m_{f}^2 - m_{f'}^2) \over s(s-m_{W}^2)(s-m_{H^{\pm}}^2)},
\end{equation}
\begin{equation}
I_{H^{\pm} G}^{3} = 2|m_{\tilde{Z}_i}| \int_{|m_{\tilde{W}_j}|}^{E_{upper3}} dE {2|m_{\tilde{Z}_i}| \lambda \sqrt{E^2-m_{\tilde{W}_j}^2} 2|m_{\tilde{Z}_i}|E \over s(s-m_{W}^2)(s-m_{H^{\pm}}^2)},
\end{equation}
\begin{equation}
I_{H^{\pm} G}^{4} = 2|m_{\tilde{Z}_i}| \int_{|m_{\tilde{W}_j}|}^{E_{upper3}} dE {2|m_{\tilde{Z}_i}| \lambda \sqrt{E^2-m_{\tilde{W}_j}^2} (s- m_{f}^2 - m_{f'}^2) 2|m_{\tilde{Z}_i}|E \over s(s-m_{W}^2)(s-m_{H^{\pm}}^2)}.
\end{equation}
The overall contribution is then:
\begin{equation}
\begin{aligned}
\Gamma_{H^{\pm} G} = & \mathcal{V}_{H^{\pm} G}^{(1)}  \mathcal{V}_{H^{\pm} G}^{(3)}  I_{H^{\pm} G}^{4} - 2  \mathcal{V}_{H^{\pm} G}^{(1)}  \mathcal{V}_{H^{\pm} G}^{(4)} m_{f} m_{f'} I_{H^{\pm} G}^{3} + 2  \mathcal{V}_{H^{\pm} G}^{(2)}  \mathcal{V}_{H^{\pm} G}^{(3)}  |m_{\tilde{Z}_i}||m_{\tilde{W}_j}| I_{H^{\pm} G}^{2} \\ & - 4  \mathcal{V}_{H^{\pm} G}^{(2)}  \mathcal{V}_{H^{\pm} G}^{(4)} |m_{\tilde{Z}_i}||m_{\tilde{W}_j}|m_{f} m_{f'} I_{H^{\pm} G}^{1} (-1)^{\theta_j}.
\end{aligned}
\end{equation}

\textbf{\underline{$\Gamma_{G \tilde{f'}_1}$}}

Here the required coupling combinations are dependent on whether it's
neutralino to chargino or chargino to neutralino. For a neutralino decaying:
\begin{align}
\mathcal{V}_{G \tilde{f'}_1}^{(1)} &= \frac{1}{2}(\omega_{G \tilde{W} \tilde{Z}}^R   \alpha_{\tilde{Z}_i \tilde{f}_1}^{u}  \mathcal{C}_{G f f'}^d  \beta_{\tilde{f'}_1}^{\tilde{W}_j} +  \omega_{G \tilde{W} \tilde{Z}}^L  \beta_{\tilde{Z}_i \tilde{f}_1}^{u}  \mathcal{C}_{G f f'}^u  \alpha_{\tilde{f'}_1}^{\tilde{W}_j}), \\
\mathcal{V}_{G \tilde{f'}_1}^{(2)} &= -(\omega_{G \tilde{W} \tilde{Z}}^R   \beta_{\tilde{Z}_i \tilde{f}_1}^{u}  \mathcal{C}_{G f f'}^u  \beta_{\tilde{f'}_1}^{\tilde{W}_j} +  \omega_{G \tilde{W} \tilde{Z}}^L  \alpha_{\tilde{Z}_i \tilde{f}_1}^{u}  \mathcal{C}_{G f f'}^d  \alpha_{\tilde{f'}_1}^{\tilde{W}_j})m_{f}|m_{\tilde{Z}_i}|, \\
\mathcal{V}_{G \tilde{f'}_1}^{(3)} &= (\omega_{G \tilde{W} \tilde{Z}}^L  \beta_{\tilde{Z}_i \tilde{f}_1}^{u}  \mathcal{C}_{G f f'}^d  \beta_{\tilde{f'}_1}^{\tilde{W}_j} +  \omega_{G \tilde{W} \tilde{Z}}^R  \alpha_{\tilde{Z}_i \tilde{f}_1}^{u}  \mathcal{C}_{G f f'}^u  \alpha_{\tilde{f'}_1}^{\tilde{W}_j})(-1)^{\theta_i} m_{f'}|m_{\tilde{W}_j}|, \\
\mathcal{V}_{G \tilde{f'}_1}^{(4)} &= (\omega_{G \tilde{W} \tilde{Z}}^R   \beta_{\tilde{Z}_i \tilde{f}_1}^{u}  \mathcal{C}_{G f f'}^d  \beta_{\tilde{f'}_1}^{\tilde{W}_j} +  \omega_{G \tilde{W} \tilde{Z}}^L  \alpha_{\tilde{Z}_i \tilde{f}_1}^{u}  \mathcal{C}_{G f f'}^u  \alpha_{\tilde{f'}_1}^{\tilde{W}_j})m_{f'}|m_{\tilde{Z}_i}|, \\
\mathcal{V}_{G \tilde{f'}_1}^{(5)} &= (\omega_{G \tilde{W} \tilde{Z}}^L \beta_{\tilde{Z}_i \tilde{f}_1}^{u}  \mathcal{C}_{G f f'}^u  \beta_{\tilde{f'}_1}^{\tilde{W}_j} -  (-1)^{\theta_i} \omega_{G \tilde{W} \tilde{Z}}^R  \alpha_{\tilde{Z}_i \tilde{f}_1}^{u}  \mathcal{C}_{G f f'}^d  \alpha_{\tilde{f'}_1}^{\tilde{W}_j})m_{f} |m_{\tilde{W}_j}|,\\
\mathcal{V}_{G \tilde{f'}_1}^{(6)} &= (-1)^{\theta_i}(\omega_{G \tilde{W} \tilde{Z}}^L   \alpha_{\tilde{Z}_i \tilde{f}_1}^{u}  \mathcal{C}_{G f f'}^d  \beta_{\tilde{f'}_1}^{\tilde{W}_j} + \omega_{G \tilde{W} \tilde{Z}}^R  \beta_{\tilde{Z}_i \tilde{f}_1}^{u}  \mathcal{C}_{G f f'}^u  \alpha_{\tilde{f'}_1}^{\tilde{W}_j})|m_{\tilde{Z}_i}||m_{\tilde{W}_j}| ,\\
\mathcal{V}_{G \tilde{f'}_1}^{(7)} &= -(\omega_{G \tilde{W} \tilde{Z}}^R   \alpha_{\tilde{Z}_i \tilde{f}_1}^{u}  \mathcal{C}_{G f f'}^u  \beta_{\tilde{f'}_1}^{\tilde{W}_j} +  \omega_{G \tilde{W} \tilde{Z}}^L  \beta_{\tilde{Z}_i \tilde{f}_1}^{u}  \mathcal{C}_{G f f'}^d  \alpha_{\tilde{f'}_1}^{\tilde{W}_j})m_{f'}m_{f}, \\
\mathcal{V}_{G \tilde{f'}_1}^{(8)} &= 2(\omega_{G \tilde{W} \tilde{Z}}^L   \alpha_{\tilde{Z}_i \tilde{f}_1}^{u}  \mathcal{C}_{G f f'}^u  \beta_{\tilde{f'}_1}^{\tilde{W}_j} + (-1)^{\theta_i}\omega_{G \tilde{W} \tilde{Z}}^R  \beta_{\tilde{Z}_i \tilde{f}_1}^{u}  \mathcal{C}_{G f f'}^d  \alpha_{\tilde{f'}_1}^{\tilde{W}_j})|m_{\tilde{Z}_i}|m_{f'}m_{f}|m_{\tilde{W}_j}|.
\end{align}

For a chargino decaying:
\begin{align}
\mathcal{V}_{G \tilde{f'}_1}^{(1)} &= -\frac{1}{2}(\omega_{G \tilde{W} \tilde{Z}}^L   \alpha_{\tilde{Z}_i \tilde{f}_1}^{u}  \mathcal{C}_{G f f'}^u  \beta_{\tilde{f'}_1}^{\tilde{W}_j} +  (-1)^{\theta_f}\omega_{G \tilde{W} \tilde{Z}}^R  \beta_{\tilde{Z}_i \tilde{f}_1}^{u}  \mathcal{C}_{G f f'}^d  \alpha_{\tilde{f'}_1}^{\tilde{W}_j}), \\
\mathcal{V}_{G \tilde{f'}_1}^{(2)} &= (\omega_{G \tilde{W} \tilde{Z}}^L  \alpha_{\tilde{Z}_i \tilde{f}_1}^{u}  \mathcal{C}_{G f f'}^d  \alpha_{\tilde{f'}_1}^{\tilde{W}_j} +  (-1)^{\theta_j}\omega_{G \tilde{W} \tilde{Z}}^R  \beta_{\tilde{Z}_i \tilde{f}_1}^{u}  \mathcal{C}_{G f f'}^u  \beta_{\tilde{f'}_1}^{\tilde{W}_j})m_{f'}|m_{\tilde{Z}_i}|, \\
\mathcal{V}_{G \tilde{f'}_1}^{(3)} &= -(\omega_{G \tilde{W} \tilde{Z}}^R  \alpha_{\tilde{Z}_i \tilde{f}_1}^{u}  \mathcal{C}_{G f f'}^u  \alpha_{\tilde{f'}_1}^{\tilde{W}_j} +  (-1)^{\theta_j}\omega_{G \tilde{W} \tilde{Z}}^L  \beta_{\tilde{Z}_i \tilde{f}_1}^{u}  \mathcal{C}_{G f f'}^d  \beta_{\tilde{f'}_1}^{\tilde{W}_j})(-1)^{\theta_j} m_{f}|m_{\tilde{W}_j}|, \\
\mathcal{V}_{G \tilde{f'}_1}^{(4)} &= -(\omega_{G \tilde{W} \tilde{Z}}^L   \alpha_{\tilde{Z}_i \tilde{f}_1}^{u}  \mathcal{C}_{G f f'}^u  \alpha_{\tilde{f'}_1}^{\tilde{W}_j} +  (-1)^{\theta_j}\omega_{G \tilde{W} \tilde{Z}}^R  \beta_{\tilde{Z}_i \tilde{f}_1}^{u}  \mathcal{C}_{G f f'}^d  \beta_{\tilde{f'}_1}^{\tilde{W}_j})m_{f}|m_{\tilde{Z}_i}|, \\
\mathcal{V}_{G \tilde{f'}_1}^{(5)} &= ((-1)^{\theta_j}\omega_{G \tilde{W} \tilde{Z}}^R \alpha_{\tilde{Z}_i \tilde{f}_1}^{u}  \mathcal{C}_{G f f'}^d  \alpha_{\tilde{f'}_1}^{\tilde{W}_j} + \omega_{G \tilde{W} \tilde{Z}}^L  \beta_{\tilde{Z}_i \tilde{f}_1}^{u}  \mathcal{C}_{G f f'}^u  \beta_{\tilde{f'}_1}^{\tilde{W}_j})m_{f'} |m_{\tilde{W}_j}|,\\
\mathcal{V}_{G \tilde{f'}_1}^{(6)} &= -((-1)^{\theta_j}\omega_{G \tilde{W} \tilde{Z}}^R \alpha_{\tilde{Z}_i \tilde{f}_1}^{u}  \mathcal{C}_{G f f'}^u  \beta_{\tilde{f'}_1}^{\tilde{W}_j} + \omega_{G \tilde{W} \tilde{Z}}^L  \beta_{\tilde{Z}_i \tilde{f}_1}^{u}  \mathcal{C}_{G f f'}^d  \alpha_{\tilde{f'}_1}^{\tilde{W}_j})|m_{\tilde{Z}_i}||m_{\tilde{W}_j}| ,\\
\mathcal{V}_{G \tilde{f'}_1}^{(7)} &= (\omega_{G \tilde{W} \tilde{Z}}^L \alpha_{\tilde{Z}_i \tilde{f}_1}^{u}  \mathcal{C}_{G f f'}^d  \beta_{\tilde{f'}_1}^{\tilde{W}_j} + (-1)^{\theta_j}\omega_{G \tilde{W} \tilde{Z}}^R  \beta_{\tilde{Z}_i \tilde{f}_1}^{u} \mathcal{C}_{G f f'}^u \alpha_{\tilde{f'}_1}^{\tilde{W}_j})m_{f'}m_{f}, \\
\mathcal{V}_{G \tilde{f'}_1}^{(8)} &= 2((-1)^{\theta_j}\omega_{G \tilde{W} \tilde{Z}}^R  \alpha_{\tilde{Z}_i \tilde{f}_1}^{u}  \mathcal{C}_{G f f'}^d  \beta_{\tilde{f'}_1}^{\tilde{W}_j} + \omega_{G \tilde{W} \tilde{Z}}^L  \beta_{\tilde{Z}_i \tilde{f}_1}^{u}  \mathcal{C}_{G f f'}^u  \alpha_{\tilde{f'}_1}^{\tilde{W}_j})|m_{\tilde{Z}_i}|m_{f'}m_{f}|m_{\tilde{W}_j}|.
\end{align}
The integrals necessary are, for neutralino decaying with \\ $s = m_{\tilde{Z}_i}^2 + m_{\tilde{W}_j}^2 -2|m_{\tilde{Z}_i}|E$,  $\lambda = \sqrt{(s-(m_{f'}+m_{f})^2)(s-(m_{f'}-m_{f})^2)}$, $A = m_{f}^2 + m_{f'}^2 +2|m_{\tilde{Z}_i}|E + (m_{\tilde{Z}_i}^2 - m_{\tilde{W}_j}^2)(m_{f}^2 - m_{f'}^2)/s$, $B = 2|m_{\tilde{Z}_i}|\lambda/s \sqrt{E^2 - m_{\tilde{W}_j}^2} $, $Z = {(\frac{1}{2}(A+B) - m_{\tilde{f'}_1}^2) \over (\frac{1}{2}(A-B) - m_{\tilde{f'}_1}^2)}$, given by:
\begin{align}
I_{G \tilde{f'}_1}^{1} &= 2|m_{\tilde{Z}_i}| \int_{|m_{\tilde{W}_j}|}^{E_{upper3}} dE {2[sB + (m_{\tilde{f'}_1}^2 s - m_{\tilde{Z}_i}^2 m_{f}^2 - m_{f'}^2 m_{\tilde{W}_j}^2)\log(Z)] \over s - m_{W}^2}, \\
I_{G \tilde{f'}_1}^{2} &= -2|m_{\tilde{Z}_i}| \int_{|m_{\tilde{W}_j}|}^{E_{upper3}} dE {[B + (m_{\tilde{f'}_1}^2 + m_{\tilde{W}_j}^2 - 2|m_{\tilde{Z}_i}|E - m_{f}^2)\log(Z)] \over s - m_{W}^2}, \\
I_{G \tilde{f'}_1}^{3} &= 2|m_{\tilde{Z}_i}| \int_{|m_{\tilde{W}_j}|}^{E_{upper3}} dE {[B + (m_{\tilde{f'}_1}^2 + m_{\tilde{Z}_i}^2 - 2|m_{\tilde{Z}_i}|E - m_{f'}^2)\log(Z)] \over s - m_{W}^2}, \\
I_{G \tilde{f'}_1}^{4} &= 2|m_{\tilde{Z}_i}| \int_{|m_{\tilde{W}_j}|}^{E_{upper3}} dE {[B + (m_{\tilde{f'}_1}^2 - m_{f}^2 - m_{\tilde{W}_j}^2)\log(Z)] \over s - m_{W}^2}, \\
I_{G \tilde{f'}_1}^{5} &= -2|m_{\tilde{Z}_i}| \int_{|m_{\tilde{W}_j}|}^{E_{upper3}} dE {[B + (m_{\tilde{f'}_1}^2 - m_{\tilde{Z}_i}^2 - m_{f'}^2)\log(Z)] \over s - m_{W}^2}, \\
I_{G \tilde{f'}_1}^{6} &= 2|m_{\tilde{Z}_i}| \int_{|m_{\tilde{W}_j}|}^{E_{upper3}} dE {(s-m_{f'}^2-m_{f}^2)\log(Z) \over s - m_{W}^2}, \\
I_{G \tilde{f'}_1}^{7} &= 2|m_{\tilde{Z}_i}| \int_{|m_{\tilde{W}_j}|}^{E_{upper3}} dE {2|m_{\tilde{Z}_i}|E \log(Z) \over s - m_{W}^2}, \\
I_{G \tilde{f'}_1}^{8} &= 2|m_{\tilde{Z}_i}| \int_{|m_{\tilde{W}_j}|}^{E_{upper3}} dE {\log(Z) \over s - m_{W}^2}.
\end{align}
For a chargino decaying the integrals have the same expressions but one must
swap integrals 2 and 4 and integrals 3 and 5. The overall contribution is the product of each coupling combination with the corresponding integral:
\begin{equation}
\begin{aligned}
\Gamma_{G \tilde{f'}_1} = & \mathcal{V}_{G \tilde{f'}_1}^{(1)} I_{G \tilde{f'}_1}^{1} + \mathcal{V}_{G \tilde{f'}_1}^{(2)} I_{G \tilde{f'}_1}^{2} + \mathcal{V}_{G \tilde{f'}_1}^{(3)} I_{G \tilde{f'}_1}^{3} + \mathcal{V}_{G \tilde{f'}_1}^{(4)} I_{G \tilde{f'}_1}^{4} \\ & + \mathcal{V}_{G \tilde{f'}_1}^{(5)} I_{G \tilde{f'}_1}^{5} + \mathcal{V}_{G \tilde{f'}_1}^{(6)} I_{G \tilde{f'}_1}^{6} + \mathcal{V}_{G \tilde{f'}_1}^{(7)} I_{G \tilde{f'}_1}^{7} + \mathcal{V}_{G \tilde{f'}_1}^{(8)} I_{G \tilde{f'}_1}^{8}.
\end{aligned}
\end{equation}

\textbf{\underline{$\Gamma_{G \tilde{f'}_2}$}}

Again here the coupling combinations depend upon if we are considering neutralino to chargino or chargino to neutralino, for neutralino decaying:
\begin{align}
\mathcal{V}_{G \tilde{f'}_2}^{(1)} &= -\frac{1}{2}(-1)^{\theta_i}[-(-1)^{\theta_i} \omega_{G \tilde{W} \tilde{Z}}^R \alpha_{\tilde{Z}_i \tilde{f}_2}^{u} \mathcal{C}_{G f f'}^d \beta_{\tilde{f'}_2}^{\tilde{W}_j}  + \omega_{G \tilde{W} \tilde{Z}}^L \beta_{\tilde{Z}_i \tilde{f}_2}^{u} \mathcal{C}_{G f f'}^u  \alpha_{\tilde{f'}_2}^{\tilde{W}_j}], \\
\mathcal{V}_{G \tilde{f'}_2}^{(2)} &= (-1)^{\theta_i}(\omega_{G \tilde{W} \tilde{Z}}^R \beta_{\tilde{Z}_i \tilde{f}_2}^{u}  \mathcal{C}_{G f f'}^u \beta_{\tilde{f'}_2}^{\tilde{W}_j} - \omega_{G \tilde{W} \tilde{Z}}^L \alpha_{\tilde{Z}_i \tilde{f}_2}^{u} \mathcal{C}_{G f f'}^d \alpha_{\tilde{f'}_2}^{\tilde{W}_j})|m_{\tilde{Z}_i}|m_{f}, \\
\mathcal{V}_{G \tilde{f'}_2}^{(3)} &= -(\omega_{G \tilde{W} \tilde{Z}}^L \beta_{\tilde{Z}_i \tilde{f}_2}^{u} \mathcal{C}_{G f f'}^d \beta_{\tilde{f'}_2}^{\tilde{W}_j} - \omega_{G \tilde{W} \tilde{Z}}^R \alpha_{\tilde{Z}_i \tilde{f}_2}^{u} \mathcal{C}_{G f f'}^u \alpha_{\tilde{f'}_2}^{\tilde{W}_j})m_{f'}|m_{\tilde{W}_j}|(-1)^{\theta_j}, \\
\mathcal{V}_{G \tilde{f'}_2}^{(4)} &= -(\omega_{G \tilde{W} \tilde{Z}}^R \beta_{\tilde{Z}_i \tilde{f}_2}^{u} \mathcal{C}_{G f f'}^d \beta_{\tilde{f'}_2}^{\tilde{W}_j} - (-1)^{\theta_i} \omega_{G \tilde{W} \tilde{Z}}^L \alpha_{\tilde{Z}_i \tilde{f}_2}^{u} \mathcal{C}_{G f f'}^u \alpha_{\tilde{f'}_2}^{\tilde{W}_j})|m_{\tilde{Z}_i}|m_{f'}, \\
\mathcal{V}_{G \tilde{f'}_2}^{(5)} &= (\omega_{G \tilde{W} \tilde{Z}}^L \beta_{\tilde{Z}_i \tilde{f}_2}^{u} \mathcal{C}_{G f f'}^u \beta_{\tilde{f'}_2}^{\tilde{W}_j} + \omega_{G \tilde{W} \tilde{Z}}^R \alpha_{\tilde{Z}_i \tilde{f}_2}^{u} \mathcal{C}_{G f f'}^d \alpha_{\tilde{f'}_2}^{\tilde{W}_j})m_{f}|m_{\tilde{W}_j}|(-1)^{\theta_j}, \\
\mathcal{V}_{G \tilde{f'}_1}^{(6)} &= [\omega_{G \tilde{W} \tilde{Z}}^L \alpha_{\tilde{Z}_i \tilde{f}_2}^{u} \mathcal{C}_{G f f'}^d \beta_{\tilde{f'}_2}^{\tilde{W}_j} - (-1)^{\theta_i} \omega_{G \tilde{W} \tilde{Z}}^R \beta_{\tilde{Z}_i \tilde{f}_2}^{u} \mathcal{C}_{G f f'}^u \alpha_{\tilde{f'}_2}^{\tilde{W}_j}]|m_{\tilde{Z}_i}||m_{\tilde{W}_j}|, \\
\mathcal{V}_{G \tilde{f'}_2}^{(7)} &= [-(-1)^{\theta_i} \omega_{G \tilde{W} \tilde{Z}}^R \alpha_{\tilde{Z}_i \tilde{f}_2}^{u} \mathcal{C}_{G f f'}^u \beta_{\tilde{f'}_2}^{\tilde{W}_j} + \omega_{G \tilde{W} \tilde{Z}}^L \beta_{\tilde{Z}_i \tilde{f}_2}^{u} \mathcal{C}_{G f f'}^d \alpha_{\tilde{f'}_2}^{\tilde{W}_j}]m_{f'}m_{f}, \\
\mathcal{V}_{G \tilde{f'}_2}^{(8)} &= 2[-\omega_{G \tilde{W} \tilde{Z}}^L \alpha_{\tilde{Z}_i \tilde{f}_2}^{u} \mathcal{C}_{G f f'}^u \beta_{\tilde{f'}_2}^{\tilde{W}_j} + (-1)^{\theta_i} \omega_{G \tilde{W} \tilde{Z}}^R \beta_{\tilde{Z}_i \tilde{f}_2}^{u} \mathcal{C}_{G f f'}^d \alpha_{\tilde{f'}_2}^{\tilde{W}_j}]|m_{\tilde{Z}_i}|m_{f'}m_{f}|m_{\tilde{W}_j}|,
\end{align}

whilst if it's a chargino decaying into a neutralino:
\begin{align}
\mathcal{V}_{G \tilde{f'}_2}^{(1)} &= \frac{1}{2}[\omega_{G \tilde{W} \tilde{Z}}^L \beta_{\tilde{f'}_2}^{\tilde{W}_j} \mathcal{C}_{G f f'}^u \alpha_{\tilde{Z}_i \tilde{f}_2}^{u} (-1)^{\theta_j} + \omega_{G \tilde{W} \tilde{Z}}^R \alpha_{\tilde{f'}_2}^{\tilde{W}_j} \mathcal{C}_{G f f'}^d \beta_{\tilde{Z}_i \tilde{f}_2}^{u} ], \\
\mathcal{V}_{G \tilde{f'}_2}^{(2)} &= -((-1)^{\theta_j}\omega_{G \tilde{W} \tilde{Z}}^L \alpha_{\tilde{f'}_2}^{\tilde{W}_j} \mathcal{C}_{G f f'}^d \alpha_{\tilde{Z}_i \tilde{f}_2}^{u} +  \omega_{G \tilde{W} \tilde{Z}}^R \beta_{\tilde{f'}_2}^{\tilde{W}_j} \mathcal{C}_{G f f'}^u \beta_{\tilde{Z}_i \tilde{f}_2}^{u})|m_{\tilde{Z}_i}|m_{f'}, \\
\mathcal{V}_{G \tilde{f'}_2}^{(3)} &= (\omega_{G \tilde{W} \tilde{Z}}^R \alpha_{\tilde{f'}_2}^{\tilde{W}_j} \mathcal{C}_{G f f'}^u \alpha_{\tilde{Z}_i \tilde{f}_2}^{u} + (-1)^{\theta_j}\omega_{G \tilde{W} \tilde{Z}}^L \beta_{\tilde{f'}_2}^{\tilde{W}_j} \mathcal{C}_{G f f'}^d \beta_{\tilde{Z}_i \tilde{f}_2}^{u})m_{f}|m_{\tilde{W}_j}| ,\\
\mathcal{V}_{G \tilde{f'}_2}^{(4)} &= ((-1)^{\theta_j}\omega_{G \tilde{W} \tilde{Z}}^L \alpha_{\tilde{f'}_2}^{\tilde{W}_j} \mathcal{C}_{G f f'}^u \alpha_{\tilde{Z}_i \tilde{f}_2}^{u} + \omega_{G \tilde{W} \tilde{Z}}^R \beta_{\tilde{f'}_2}^{\tilde{W}_j} \mathcal{C}_{G f f'}^d \beta_{\tilde{Z}_i \tilde{f}_2}^{u} ) |m_{\tilde{Z}_i}| m_{f}, \\
\mathcal{V}_{G \tilde{f'}_2}^{(5)} &= -(\omega_{G \tilde{W} \tilde{Z}}^R \alpha_{\tilde{f'}_2}^{\tilde{W}_j} \mathcal{C}_{G f f'}^d \alpha_{\tilde{Z}_i \tilde{f}_2}^{u} + (-1)^{\theta_j} \omega_{G \tilde{W} \tilde{Z}}^L \beta_{\tilde{f'}_2}^{\tilde{W}_j} \mathcal{C}_{G f f'}^u \beta_{\tilde{Z}_i \tilde{f}_2}^{u})m_{f'}|m_{\tilde{W}_j}|, \\
\mathcal{V}_{G \tilde{f'}_2}^{(6)} &= [\omega_{G \tilde{W} \tilde{Z}}^R \beta_{\tilde{f'}_2}^{\tilde{W}_j} \mathcal{C}_{G f f'}^u \alpha_{\tilde{Z}_i \tilde{f}_2}^{u} + (-1)^{\theta_j}\omega_{G \tilde{W} \tilde{Z}}^L \alpha_{\tilde{f'}_2}^{\tilde{W}_j} \mathcal{C}_{G f f'}^d \beta_{\tilde{Z}_i \tilde{f}_2}^{u}]|m_{\tilde{W}_j}||m_{\tilde{Z}_i}|, \\
\mathcal{V}_{G \tilde{f'}_2}^{(7)} &= -[(-1)^{\theta_j}\omega_{G \tilde{W} \tilde{Z}}^L \beta_{\tilde{f'}_2}^{\tilde{W}_j} \mathcal{C}_{G f f'}^d \alpha_{\tilde{Z}_i \tilde{f}_2}^{u} + \omega_{G \tilde{W} \tilde{Z}}^R \alpha_{\tilde{f'}_2}^{\tilde{W}_j} \mathcal{C}_{G f f'}^u \beta_{\tilde{Z}_i \tilde{f}_2}^{u}]m_{f}m_{f'}, \\
\mathcal{V}_{G \tilde{f'}_2}^{(8)} &= -2[\omega_{G \tilde{W} \tilde{Z}}^R \beta_{\tilde{f'}_2}^{\tilde{W}_j} \mathcal{C}_{G f f'}^d \alpha_{\tilde{Z}_i \tilde{f}_2}^{u} + (-1)^{\theta_j}\omega_{G \tilde{W} \tilde{Z}}^L \alpha_{\tilde{f'}_2}^{\tilde{W}_j} \mathcal{C}_{G f f'}^u \beta_{\tilde{Z}_i \tilde{f}_2}^{u}]m_{f}m_{f'}|m_{\tilde{W}_j}|m_{\tilde{Z}_i}|.
\end{align}
The integrals are exactly as for $G \tilde{f'}_1$ but with the change $m_{\tilde{f'}_1} \rightarrow m_{\tilde{f'}_2}$, and similar changes produce the overall expression for $\Gamma_{G \tilde{f'}_2}$.

\textbf{\underline{$\Gamma_{H^{\pm} \tilde{f'}_1}$}}

Here the couplings required are dependent again on which particle is initial state and which final state, for the neutralino as the decaying (i.e.\ initial state) particle:
\begin{align}
\mathcal{V}_{H^{\pm} \tilde{f'}_1}^{(1)} &= -\frac{1}{2}[(-1)^{\theta_i}\omega_{H^+ \tilde{W}^+ \tilde{Z}}^R \alpha_{\tilde{Z}_i \tilde{f}_1}^{u} \mathcal{C}_{H^+ f f'}^d \beta_{\tilde{f'}_1}^{\tilde{W}_j} + \omega_{H^+ \tilde{W}^+ \tilde{Z}}^L \beta_{\tilde{Z}_i \tilde{f}_1}^{u} \mathcal{C}_{H^+ f f'}^u \alpha_{\tilde{f'}_1}^{\tilde{W}_j}], \\
\mathcal{V}_{H^{\pm} \tilde{f'}_1}^{(2)} &= -(\omega_{H^+ \tilde{W}^+ \tilde{Z}}^R \beta_{\tilde{Z}_i \tilde{f}_1}^{u} \mathcal{C}_{H^+ f f'}^u \beta_{\tilde{f'}_1}^{\tilde{W}_j} + \omega_{H^+ \tilde{W}^+ \tilde{Z}}^L \alpha_{\tilde{Z}_i \tilde{f}_1}^{u} \mathcal{C}_{H^+ f f'}^d \alpha_{\tilde{f'}_1}^{\tilde{W}_j})|m_{\tilde{Z}_i}|m_{f}, \\
\mathcal{V}_{H^{\pm} \tilde{f'}_1}^{(3)} &= (\omega_{H^+ \tilde{W}^+ \tilde{Z}}^L \beta_{\tilde{Z}_i \tilde{f}_1}^{u} \mathcal{C}_{H^+ f f'}^d \beta_{\tilde{f'}_1}^{\tilde{W}_j} + \omega_{H^+ \tilde{W}^+ \tilde{Z}}^R \alpha_{\tilde{Z}_i \tilde{f}_1}^{u} \mathcal{C}_{H^+ f f'}^u \alpha_{\tilde{f'}_1}^{\tilde{W}_j})m_{f'}|m_{\tilde{W}_j}|(-1)^{\theta_i}, \\
\mathcal{V}_{H^{\pm} \tilde{f'}_1}^{(4)} &= -[(-1)^{\theta_i}\omega_{H^+ \tilde{W}^+ \tilde{Z}}^R \beta_{\tilde{Z}_i \tilde{f}_1}^{u} \mathcal{C}_{H^+ f f'}^d \beta_{\tilde{f'}_1}^{\tilde{W}_j} - \omega_{H^+ \tilde{W}^+ \tilde{Z}}^L \alpha_{\tilde{Z}_i \tilde{f}_1}^{u} \mathcal{C}_{H^+ f f'}^u \alpha_{\tilde{f'}_1}^{\tilde{W}_j}]|m_{\tilde{Z}_i}|m_{f'}, \\
\mathcal{V}_{H^{\pm} \tilde{f'}_1}^{(5)} &= -(\omega_{H^+ \tilde{W}^+ \tilde{Z}}^L \beta_{\tilde{Z}_i \tilde{f}_1}^{u} \mathcal{C}_{H^+ f f'}^u \beta_{\tilde{f'}_1}^{\tilde{W}_1} + \omega_{H^+ \tilde{W}^+ \tilde{Z}}^R \alpha_{\tilde{Z}_i \tilde{f}_1}^{u} \mathcal{C}_{H^+ f f'}^d \alpha_{\tilde{f'}_1}^{\tilde{W}_j})m_{f}|m_{\tilde{W}_j}|(-1)^{\theta_i}, \\
\mathcal{V}_{H^{\pm} \tilde{f'}_1}^{(6)} &= (-1)^{\theta_i}[\omega_{H^+ \tilde{W}^+ \tilde{Z}}^L \alpha_{\tilde{Z}_i \tilde{f}_1}^{u} \mathcal{C}_{H^+ f f'}^d \beta_{\tilde{f'}_1}^{\tilde{W}_j} + \omega_{H^+ \tilde{W}^+ \tilde{Z}}^R \beta_{\tilde{Z}_i \tilde{f}_1}^{u} \mathcal{C}_{H^+ f f'}^u \alpha_{\tilde{f'}_1}^{\tilde{W}_j}]|m_{\tilde{Z}_i}||m_{\tilde{W}_j}|, \\
\mathcal{V}_{H^{\pm} \tilde{f'}_1}^{(7)} &= -(\omega_{H^+ \tilde{W}^+ \tilde{Z}}^R \alpha_{\tilde{Z}_i \tilde{f}_1}^{u} \mathcal{C}_{H^+ f f'}^u \beta_{\tilde{f'}_1}^{\tilde{W}_j} + \omega_{H^+ \tilde{W}^+ \tilde{Z}}^L  \beta_{\tilde{Z}_i \tilde{f}_1}^{u} \mathcal{C}_{H^+ f f'}^d \alpha_{\tilde{f'}_1}^{\tilde{W}_j})m_{f'}m_{f},
\\
\mathcal{V}_{H^{\pm} \tilde{f'}_1}^{(8)} &= -2(-1)^{\theta_i}[\omega_{H^+ \tilde{W}^+ \tilde{Z}}^L \alpha_{\tilde{Z}_i \tilde{f}_1}^{u} \mathcal{C}_{H^+ f f'}^u \beta_{\tilde{f'}_1}^{\tilde{W}_j} + \omega_{H^+ \tilde{W}^+ \tilde{Z}}^R \beta_{\tilde{Z}_i \tilde{f}_1}^{u} \mathcal{C}_{H^+ f f'}^d \alpha_{\tilde{f'}_1}^{\tilde{W}_j}]|m_{\tilde{Z}_i}||m_{\tilde{W}_j}|m_{f'}m_{f}.
\end{align}

If the initial state is a chargino:
\begin{align}
\mathcal{V}_{H^{\pm} \tilde{f'}_1}^{(1)} &= \frac{1}{2}(-1)^{\theta_j}[\omega_{H^+ \tilde{W}^+ \tilde{Z}}^L \beta_{\tilde{f'}_1}^{\tilde{W}_j} \mathcal{C}_{H^+ f f'}^u \alpha_{\tilde{Z}_i \tilde{f}_1}^{u} + (-1)^{\theta_j}\omega_{H^+ \tilde{W}^+ \tilde{Z}}^R \alpha_{\tilde{f'}_1}^{\tilde{W}_j} \mathcal{C}_{H^+ f f'}^d \beta_{\tilde{Z}_i \tilde{f}_1}^{u}], \\
\mathcal{V}_{H^{\pm} \tilde{f'}_1}^{(2)} &= [\omega_{H^+ \tilde{W}^+ \tilde{Z}}^L \alpha_{\tilde{f'}_1}^{\tilde{W}_j} \mathcal{C}_{H^+ f f'}^d \alpha_{\tilde{Z}_i \tilde{f}_1}^{u} + (-1)^{\theta_j} \omega_{H^+ \tilde{W}^+ \tilde{Z}}^R \beta_{\tilde{f'}_1}^{\tilde{W}_j} \mathcal{C}_{H^+ f f'}^u \beta_{\tilde{Z}_i \tilde{f}_1}^{u}]|m_{\tilde{Z}_i}|m_{f'}, \\
\mathcal{V}_{H^{\pm} \tilde{f'}_1}^{(3)} &= -[(-1)^{\theta_j}\omega_{H^+ \tilde{W}^+ \tilde{Z}}^R \alpha_{\tilde{f'}_1}^{\tilde{W}_j} \mathcal{C}_{H^+ f f'}^u \alpha_{\tilde{Z}_i \tilde{f}_1}^{u} + \omega_{H^+ \tilde{W}^+ \tilde{Z}}^L \beta_{\tilde{f'}_1}^{\tilde{W}_j} \mathcal{C}_{H^+ f f'}^d \beta_{\tilde{Z}_i \tilde{f}_1}^{u} ]m_{f}|m_{\tilde{W}_j}|, \\
\mathcal{V}_{H^{\pm} \tilde{f'}_1}^{(4)} &= -[\omega_{H^+ \tilde{W}^+ \tilde{Z}}^L \alpha_{\tilde{f'}_1}^{\tilde{W}_j} \mathcal{C}_{H^+ f f'}^u \alpha_{\tilde{Z}_i \tilde{f}_1}^{u} + (-1)^{\theta_j}\omega_{H^+ \tilde{W}^+ \tilde{Z}}^R \beta_{\tilde{f'}_1}^{\tilde{W}_j} \mathcal{C}_{H^+ f f'}^d \beta_{\tilde{Z}_i \tilde{f}_1}^{u}]|m_{\tilde{Z}_i}|m_{f}, \\
\mathcal{V}_{H^{\pm} \tilde{f'}_1}^{(5)} &= [(-1)^{\theta_j}\omega_{H^+ \tilde{W}^+ \tilde{Z}}^R \alpha_{\tilde{f'}_1}^{\tilde{W}_j} \mathcal{C}_{H^+ f f'}^d \alpha_{\tilde{Z}_i \tilde{f}_1}^{u} + \omega_{H^+ \tilde{W}^+ \tilde{Z}}^L \beta_{\tilde{f'}_1}^{\tilde{W}_j} \mathcal{C}_{H^+ f f'}^u \beta_{\tilde{Z}_i \tilde{f}_1}^{u}]m_{f'}|m_{\tilde{W}_j}|, \\
\mathcal{V}_{H^{\pm} \tilde{f'}_1}^{(6)} &= [\omega_{H^+ \tilde{W}^+ \tilde{Z}}^R \beta_{\tilde{f'}_1}^{\tilde{W}_j} \mathcal{C}_{H^+ f f'}^u \alpha_{\tilde{Z}_i \tilde{f}_1}^{u} + (-1)^{\theta_j}\omega_{H^+ \tilde{W}^+ \tilde{Z}}^L \alpha_{\tilde{f'}_1}^{\tilde{W}_j} \mathcal{C}_{H^+ f f'}^d \beta_{\tilde{Z}_i \tilde{f}_1}^{u}]|m_{\tilde{W}_j}||m_{\tilde{Z}_i}|, \\
\mathcal{V}_{H^{\pm} \tilde{f'}_1}^{(7)} &= -[(-1)^{\theta_j}\omega_{H^+ \tilde{W}^+ \tilde{Z}}^L \beta_{\tilde{f'}_1}^{\tilde{W}_j}  \mathcal{C}_{H^+ f f'}^d \alpha_{\tilde{Z}_i \tilde{f}_1}^{u} +  \omega_{H^+ \tilde{W}^+ \tilde{Z}}^R  \alpha_{\tilde{f'}_1}^{\tilde{W}_j} \mathcal{C}_{H^+ f f'}^u \beta_{\tilde{Z}_i \tilde{f}_1}^{u}]m_{f}m_{f'}, \\
\mathcal{V}_{H^{\pm} \tilde{f'}_1}^{(8)} &= -2[\omega_{H^+ \tilde{W}^+ \tilde{Z}}^R \beta_{\tilde{f'}_1}^{\tilde{W}_j} \mathcal{C}_{H^+ f f'}^d \alpha_{\tilde{Z}_i \tilde{f}_1}^{u} + (-1)^{\theta_j}\omega_{H^+ \tilde{W}^+ \tilde{Z}}^L \alpha_{\tilde{f'}_1}^{\tilde{W}_j} \mathcal{C}_{H^+ f f'}^u  \beta_{\tilde{Z}_i \tilde{f}_1}^{u}]m_{f}m_{f'}|m_{\tilde{W}_j}||m_{\tilde{Z}_i}|.
\end{align}

The integrals required are exactly as in the $G \tilde{f'}_1$ but with the expected change $m_{W} \rightarrow m_{H^{\pm}}$. 
$\Gamma_{H^{\pm} \tilde{f'}_1}$ is then given exactly as $\Gamma_{G \tilde{f'}_1}$.

\textbf{\underline{$\Gamma_{H^{\pm} \tilde{f'}_2}$}}

The coupling combinations now are, if it's a neutralino decaying:
\begin{align}
\mathcal{V}_{H^{\pm} \tilde{f'}_2}^{(1)} &= -\frac{1}{2}[(-1)^{\theta_i}\omega_{H^+ \tilde{W}^+ \tilde{Z}}^R \alpha_{\tilde{Z}_i \tilde{f}_2}^{u} \mathcal{C}_{H^+ f f'}^d \beta_{\tilde{f'}_2}^{\tilde{W}_j} + \omega_{H^+ \tilde{W}^+ \tilde{Z}}^L \beta_{\tilde{Z}_i \tilde{f}_2}^{u} \mathcal{C}_{H^+ f f'}^u \alpha_{\tilde{f'}_2}^{\tilde{W}_j}], \\
\mathcal{V}_{H^{\pm} \tilde{f'}_2}^{(2)} &= (\omega_{H^+ \tilde{W}^+ \tilde{Z}}^R \beta_{\tilde{Z}_i \tilde{f}_2}^{u}  \mathcal{C}_{H^+ f f'}^u \beta_{\tilde{f'}_2}^{\tilde{W}_j} - \omega_{H^+ \tilde{W}^+ \tilde{Z}}^L \alpha_{\tilde{Z}_i \tilde{f}_2}^{u} \mathcal{C}_{H^+ f f'}^d \alpha_{\tilde{f'}_2}^{\tilde{W}_j})m_{f}|m_{\tilde{Z}_i}|(-1)^{\theta_i}, \\
\mathcal{V}_{H^{\pm} \tilde{f'}_2}^{(3)} &= -[(-1)^{\theta_i}\omega_{H^+ \tilde{W}^+ \tilde{Z}}^L \beta_{\tilde{Z}_i \tilde{f}_2}^{u} \mathcal{C}_{H^+ f f'}^d \beta_{\tilde{f'}_2}^{\tilde{W}_j} +  \omega_{H^+ \tilde{W}^+ \tilde{Z}}^R \alpha_{\tilde{Z}_i \tilde{f}_2}^{u} \mathcal{C}_{H^+ f f'}^u \alpha_{\tilde{f'}_2}^{\tilde{W}_j}]m_{f'}|m_{\tilde{W}_j}|, \\
\mathcal{V}_{H^{\pm} \tilde{f'}_2}^{(4)} &= -[(-1)^{\theta_i}\omega_{H^+ \tilde{W}^+ \tilde{Z}}^R \beta_{\tilde{Z}_i \tilde{f}_2}^{u} \mathcal{C}_{H^+ f f'}^d \beta_{\tilde{f'}_2}^{\tilde{W}_j} + \omega_{H^+ \tilde{W}^+ \tilde{Z}}^L \alpha_{\tilde{Z}_i \tilde{f}_2}^{u} \mathcal{C}_{H^+ f f'}^u \alpha_{\tilde{f'}_2}^{\tilde{W}_j}]|m_{\tilde{Z}_i}|m_{f'}, \\
\mathcal{V}_{H^{\pm} \tilde{f'}_2}^{(5)} &= (\omega_{H^+ \tilde{W}^+ \tilde{Z}}^L \beta_{\tilde{Z}_i \tilde{f}_2}^{u} \mathcal{C}_{H^+ f f'}^u \beta_{\tilde{f'}_2}^{\tilde{W}_j} - \omega_{H^+ \tilde{W}^+ \tilde{Z}}^R \alpha_{\tilde{Z}_i \tilde{f}_2}^{u} \mathcal{C}_{H^+ f f'}^d \alpha_{\tilde{f'}_2}^{\tilde{W}_j})m_{f}|m_{\tilde{W}_j}|(-1)^{\theta_j}, \\
\mathcal{V}_{H^{\pm} \tilde{f'}_2}^{(6)} &= [ \omega_{H^+ \tilde{W}^+ \tilde{Z}}^L \alpha_{\tilde{Z}_i \tilde{f}_2}^{u} \mathcal{C}_{H^+ f f'}^d \beta_{\tilde{f'}_2}^{\tilde{W}_j} - (-1)^{\theta_i}\omega_{H^+ \tilde{W}^+ \tilde{Z}}^R \beta_{\tilde{Z}_i \tilde{f}_2}^{u} \mathcal{C}_{H^+ f f'}^u \alpha_{\tilde{f'}_2}^{\tilde{W}_j}]|m_{\tilde{Z}_i}||m_{\tilde{W}_j}|, \\
\mathcal{V}_{H^{\pm} \tilde{f'}_2}^{(7)} &= [-(-1)^{\theta_i}\omega_{H^+ \tilde{W}^+ \tilde{Z}}^R \alpha_{\tilde{Z}_i \tilde{f}_2}^{u} \mathcal{C}_{H^+ f f'}^u \beta_{\tilde{f'}}^{\tilde{W}_j} + \omega_{H^+ \tilde{W}^+ \tilde{Z}}^L \beta_{\tilde{Z}_i \tilde{f}_2}^{u} \mathcal{C}_{H^+ f f'}^d \alpha_{\tilde{f'}_2}^{\tilde{W}_j}]m_{f'}m_{f}, \\
\mathcal{V}_{H^{\pm} \tilde{f'}_2}^{(8)} &= 2(-1)^{\theta_i}[-\omega_{H^+ \tilde{W}^+ \tilde{Z}}^L \alpha_{\tilde{Z}_i \tilde{f}_2}^{u} \mathcal{C}_{H^+ f f'}^u \beta_{\tilde{f'}_2}^{\tilde{W}_j} + \omega_{H^+ \tilde{W}^+ \tilde{Z}}^R \beta_{\tilde{Z}_i \tilde{f}_2}^{u} \mathcal{C}_{H^+ f f'}^d \alpha_{\tilde{f'}_2}^{\tilde{W}_j}]|m_{\tilde{Z}_i}||m_{\tilde{W}_j}|m_{f'}m_{f}.
\end{align}

Whilst if it's a chargino decaying:
\begin{align}
\mathcal{V}_{H^{\pm} \tilde{f'}_2}^{(1)} &	= \frac{1}{2}[(-1)^{\theta_j}\omega_{H^+ \tilde{W}^+ \tilde{Z}}^L \beta_{\tilde{f'}_2}^{\tilde{W}_j} \mathcal{C}_{H^+ f f'}^u \alpha_{\tilde{Z}_i \tilde{f}_2}^{u} + \omega_{H^+ \tilde{W}^+ \tilde{Z}}^R \alpha_{\tilde{f'}_2}^{\tilde{W}_j} \mathcal{C}_{H^+ f f'}^d \beta_{\tilde{Z}_i \tilde{f}_2}^{u}], \\
\mathcal{V}_{H^{\pm} \tilde{f'}_2}^{(2)} &= -[(-1)^{\theta_j}\omega_{H^+ \tilde{W}^+ \tilde{Z}}^L \alpha_{\tilde{f'}_2}^{\tilde{W}_j} \mathcal{C}_{H^+ f f'}^d \alpha_{\tilde{Z}_i \tilde{f}_2}^{u} + \omega_{H^+ \tilde{W}^+ \tilde{Z}}^R \beta_{\tilde{f'}_2}^{\tilde{W}_j} \mathcal{C}_{H^+ f f'}^u \beta_{\tilde{Z}_i \tilde{f}_2}^{u}]|m_{\tilde{Z}_i}|m_{f'}, \\
\mathcal{V}_{H^{\pm} \tilde{f'}_2}^{(3)} &= (\omega_{H^+ \tilde{W}^+ \tilde{Z}}^R \alpha_{\tilde{f'}_2}^{\tilde{W}_j} \mathcal{C}_{H^+ f f'}^u \alpha_{\tilde{Z}_i \tilde{f}_2}^{u} +  (-1)^{\theta_j} \omega_{H^+ \tilde{W}^+ \tilde{Z}}^L \beta_{\tilde{f'}_2}^{\tilde{W}_j} \mathcal{C}_{H^+ f f'}^d \beta_{\tilde{Z}_i \tilde{f}_2}^{u})m_{f}|m_{\tilde{W}_j}|, \\
\mathcal{V}_{H^{\pm} \tilde{f'}_2}^{(4)} &= [(-1)^{\theta_j}\omega_{H^+ \tilde{W}^+ \tilde{Z}}^L \alpha_{\tilde{f'}_2}^{\tilde{W}_j} \mathcal{C}_{H^+ f f'}^u \alpha_{\tilde{Z}_i \tilde{f}_2}^{u} + \omega_{H^+ \tilde{W}^+ \tilde{Z}}^R \beta_{\tilde{f'}_2}^{\tilde{W}_j} \mathcal{C}_{H^+ f f'}^d \beta_{\tilde{Z}_i \tilde{f}_2}^{u}]|m_{\tilde{Z}_i}|m_{f}, \\
\mathcal{V}_{H^{\pm} \tilde{f'}_2}^{(5)} &= -[\omega_{H^+ \tilde{W}^+ \tilde{Z}}^R \alpha_{\tilde{f'}_2}^{\tilde{W}_j} \mathcal{C}_{H^+ f f'}^d \alpha_{\tilde{Z}_i \tilde{f}_2}^{u} + (-1)^{\theta_j} \omega_{H^+ \tilde{W}^+ \tilde{Z}}^L \beta_{\tilde{f'}_2}^{\tilde{W}_j} \mathcal{C}_{H^+ f f'}^u \beta_{\tilde{Z}_i \tilde{f}_2}^{u}]m_{f'}|m_{\tilde{W}_j}|, \\
\mathcal{V}_{H^{\pm} \tilde{f'}_2}^{(6)} &= (\omega_{H^+ \tilde{W}^+ \tilde{Z}}^R \beta_{\tilde{f'}_2}^{\tilde{W}_j} \mathcal{C}_{H^+ f f'}^u \alpha_{\tilde{Z}_i \tilde{f}_2}^{u} + (-1)^{\theta_j} \omega_{H^+ \tilde{W}^+ \tilde{Z}}^L \alpha_{\tilde{f'}_2}^{\tilde{W}_j} \mathcal{C}_{H^+ f f'}^d \beta_{\tilde{Z}_i \tilde{f}_2}^{u})|m_{\tilde{W}_j}||m_{\tilde{Z}_i}|, \\
\mathcal{V}_{H^{\pm} \tilde{f'}_2}^{(7)} &= -[(-1)^{\theta_j}\omega_{H^+ \tilde{W}^+ \tilde{Z}}^L \beta_{\tilde{f'}_2}^{\tilde{W}_j} \mathcal{C}_{H^+ f f'}^d \alpha_{\tilde{Z}_i \tilde{f}_2}^{u} + \omega_{H^+ \tilde{W}^+ \tilde{Z}}^R \alpha_{\tilde{f'}_2}^{\tilde{W}_j} \mathcal{C}_{H^+ f f'}^u \beta_{\tilde{Z}_i \tilde{f}_2}^{u}]m_{f}m_{f'}, \\
\mathcal{V}_{H^{\pm} \tilde{f'}_2}^{(8)} &= -2[\omega_{H^+ \tilde{W}^+ \tilde{Z}}^R \beta_{\tilde{f'}_2}^{\tilde{W}_j}  \mathcal{C}_{H^+ f f'}^d \alpha_{\tilde{Z}_i \tilde{f}_2}^{u} + (-1)^{\theta_j}\omega_{H^+ \tilde{W}^+ \tilde{Z}}^L \alpha_{\tilde{f'}_2}^{\tilde{W}_j} \mathcal{C}_{H^+ f f'}^u \beta_{\tilde{Z}_i \tilde{f}_2}^{u}]m_{f}m_{f'}|m_{\tilde{W}_j}||m_{\tilde{Z}_i}|.
\end{align}

The required integrals are as for $G \tilde{f'}_2$ but with with the mass change $m_W \rightarrow m_{H^{\pm}}$. Similarly, $\Gamma_{H^{\pm} \tilde{f'}_2}$ is given analogously.

\textbf{\underline{$\Gamma_{G \tilde{f}_1}$}}

The coupling combinations here again depend upon if it's a neutralino decaying or chargino decaying, if it's a neutralino decaying:
\begin{align}
\mathcal{V}_{G \tilde{f}_1}^{(1)} &= -\frac{1}{2}[\omega_{G \tilde{W} \tilde{Z}}^R \beta_{\tilde{f}_1 \tilde{Z}_i}^{d} \mathcal{C}_{G f f'}^d \alpha_{\tilde{f}_1}^{\tilde{W}_j} + (-1)^{\theta_i} \omega_{G \tilde{W} \tilde{Z}}^L \alpha_{\tilde{f}_1 \tilde{Z}_i}^{d} \mathcal{C}_{G f f'}^u \beta_{\tilde{f}_1}^{\tilde{W}_j}], \\
\mathcal{V}_{G \tilde{f}_1}^{(2)} &= -(\omega_{G \tilde{W} \tilde{Z}}^R \alpha_{\tilde{f}_1 \tilde{Z}_i}^{d} \mathcal{C}_{G f f'}^u \alpha_{\tilde{f}_1}^{\tilde{W}_j} + (-1)^{\theta_i} \omega_{G \tilde{W} \tilde{Z}}^L \beta_{\tilde{f}_1 \tilde{Z}_i}^{d} \mathcal{C}_{G f f'}^d \beta_{\tilde{f}_1}^{\tilde{W}_j})|m_{\tilde{Z}_i}|m_{f'}, \\
\mathcal{V}_{G \tilde{f}_1}^{(3)} &= [(-1)^{\theta_i} \omega_{G \tilde{W} \tilde{Z}}^L \alpha_{\tilde{f}_1 \tilde{Z}_i}^{d}  \mathcal{C}_{G f f'}^d \alpha_{\tilde{f}_1}^{\tilde{W}_j} + \omega_{G \tilde{W} \tilde{Z}}^R \beta_{\tilde{f}_1 \tilde{Z}_i}^{d} \mathcal{C}_{G f f'}^u \beta_{\tilde{f}_1}^{\tilde{W}_j}]m_{f}|m_{\tilde{W}_j}|, \\
\mathcal{V}_{G \tilde{f}_1}^{(4)} &= (\omega_{G \tilde{W} \tilde{Z}}^R \alpha_{\tilde{f}_1 \tilde{Z}_i}^{d} \mathcal{C}_{G f f'}^d \alpha_{\tilde{f}_1}^{\tilde{W}_j} - (-1)^{\theta_i}\omega_{G \tilde{W} \tilde{Z}}^L \beta_{\tilde{f}_1 \tilde{Z}_i}^{d} \mathcal{C}_{G f f'}^u \beta_{\tilde{f}_1}^{\tilde{W}_j})|m_{\tilde{Z}_i}|m_{f}, \\
\mathcal{V}_{G \tilde{f}_1}^{(5)} &= -[(-1)^{\theta_i}\omega_{G \tilde{W} \tilde{Z}}^L \alpha_{\tilde{f}_1 \tilde{Z}_i}^{d} \mathcal{C}_{G f f'}^u \alpha_{\tilde{f}_1}^{\tilde{W}_j} + \omega_{G \tilde{W} \tilde{Z}}^R \beta_{\tilde{f}_1 \tilde{Z}_i}^{d} \mathcal{C}_{G f f'}^d \beta_{\tilde{f}_1}^{\tilde{W}_j}]m_{f'}|m_{\tilde{W}_j}|, \\
\mathcal{V}_{G \tilde{f}_1}^{(6)} &= -[(-1)^{\theta_i} \omega_{G \tilde{W} \tilde{Z}}^L \beta_{\tilde{f}_1 \tilde{Z}_i}^{d} \mathcal{C}_{G f f'}^d \alpha_{\tilde{f}_1}^{\tilde{W}_j} - \omega_{G \tilde{W} \tilde{Z}}^R \alpha_{\tilde{f}_1 \tilde{Z}_i}^{d } \mathcal{C}_{G f f'}^u \beta_{\tilde{f}_1}^{\tilde{W}_j}]|m_{\tilde{Z}_i}||m_{\tilde{W}_j}|, \\
\mathcal{V}_{G \tilde{f}_1}^{(7)} &= -(\omega_{G \tilde{W} \tilde{Z}}^R \beta_{\tilde{f}_1 \tilde{Z}_i}^{d} \mathcal{C}_{G f f'}^u \alpha_{\tilde{f}_1}^{\tilde{W}_j} - (-1)^{\theta_i}\omega_{G \tilde{W} \tilde{Z}}^L \alpha_{\tilde{f}_1 \tilde{Z}_i}^{d} \mathcal{C}_{G f f'}^d \beta_{\tilde{f}_1}^{\tilde{W}_j})m_{f'}m_{f}, \\
\mathcal{V}_{G \tilde{f}_1}^{(8)} &= -2[(-1)^{\theta_i}\omega_{G \tilde{W} \tilde{Z}}^L \beta_{\tilde{f}_1 \tilde{Z}_i}^{d} \mathcal{C}_{G f f'}^u \alpha_{\tilde{f}_1}^{\tilde{W}_j} + \omega_{G \tilde{W} \tilde{Z}}^R \alpha_{\tilde{f}_1 \tilde{Z}_i}^{d} \mathcal{C}_{G f f'}^d \beta_{\tilde{f}_1}^{\tilde{W}_j}]|m_{\tilde{Z}_i}||m_{\tilde{W}_j}|m_{f'}m_{f}.
\end{align}

On the other hand, if it's instead a chargino decaying to neutralino:
\begin{align}
\mathcal{V}_{G \tilde{f}_1}^{(1)} &= \frac{1}{2}[(-1)^{\theta_j}\omega_{G \tilde{W} \tilde{Z}}^L \beta_{\tilde{f}_1}^{\tilde{W}_j} \mathcal{C}_{G f f'}^u \alpha_{\tilde{f}_1 \tilde{Z}_{i}}^{d} + \omega_{G \tilde{W} \tilde{Z}}^R \alpha_{\tilde{f}_1}^{\tilde{W}_j} \mathcal{C}_{G f f'}^d \beta_{\tilde{f}_1 \tilde{Z}_{i}}^{d}], \\
\mathcal{V}_{G \tilde{f}_1}^{(2)} &= [(-1)^{\theta_j}\omega_{G \tilde{W} \tilde{Z}}^L \alpha_{\tilde{f}_1}^{\tilde{W}_j} \mathcal{C}_{G f f'}^d \alpha_{\tilde{f}_1 \tilde{Z}_{i}}^{d} + \omega_{G \tilde{W} \tilde{Z}}^R \beta_{\tilde{f}_1}^{\tilde{W}_j} \mathcal{C}_{G f f'}^u \beta_{\tilde{f}_1 \tilde{Z}_{i}}^{d}]|m_{\tilde{Z}_i}|m_{f}, \\
\mathcal{V}_{G \tilde{f}_1}^{(3)} &= -(\omega_{G \tilde{W} \tilde{Z}}^R \alpha_{\tilde{f}_1}^{\tilde{W}_j} \mathcal{C}_{G f f'}^u \alpha_{\tilde{f}_1 \tilde{Z}_i}^{d} + (-1)^{\theta_j} \omega_{G \tilde{W} \tilde{Z}}^L \beta_{\tilde{f}_1}^{\tilde{W}_j} \mathcal{C}_{G f f'}^d \beta_{\tilde{f}_1 \tilde{Z}_{i}}^{d})m_{f'}|m_{\tilde{W}_j}|, \\
\mathcal{V}_{G \tilde{f}_1}^{(4)} &= -[(-1)^{\theta_j}\omega_{G \tilde{W} \tilde{Z}}^L \alpha_{\tilde{f}_1}^{\tilde{W}_j} \mathcal{C}_{G f f'}^u  \alpha_{\tilde{f}_1 \tilde{Z}_i}^{d} + \omega_{G \tilde{W} \tilde{Z}}^R \beta_{\tilde{f}_1}^{\tilde{W}_j} \mathcal{C}_{G f f'}^d \beta_{\tilde{f}_1 \tilde{Z}_{i}}^{d}]|m_{\tilde{Z}_i}|m_{f'}, \\
\mathcal{V}_{G \tilde{f}_1}^{(5)} &= (\omega_{G \tilde{W} \tilde{Z}}^R \alpha_{\tilde{f}_1}^{\tilde{W}_j} \mathcal{C}_{G f f'}^d + (-1)^{\theta_j}\omega_{G \tilde{W} \tilde{Z}}^L \beta_{\tilde{f}_1}^{\tilde{W}_j} \mathcal{C}_{G f f'}^u \beta_{\tilde{f}_1 \tilde{Z}_{i}}^{d})m_{f}|m_{\tilde{W}_j}| ,\\
\mathcal{V}_{G \tilde{f}_1}^{(6)} &= (\omega_{G \tilde{W} \tilde{Z}}^R \beta_{\tilde{f}_1}^{\tilde{W}_j} \mathcal{C}_{G f f'}^u \alpha_{\tilde{f}_1 \tilde{Z}_i}^{d} + (-1)^{\theta_j} \omega_{G \tilde{W} \tilde{Z}}^L \alpha_{\tilde{f}_1}^{\tilde{W}_j} \mathcal{C}_{G f f'}^d \beta_{\tilde{f}_1 \tilde{Z}_{i}}^{d})|m_{\tilde{W}_j}||m_{\tilde{Z}_i}|, \\
\mathcal{V}_{G \tilde{f}_1}^{(7)} &= -[(-1)^{\theta_j}\omega_{G \tilde{W} \tilde{Z}}^L \beta_{\tilde{f}_1}^{\tilde{W}_j} \mathcal{C}_{G f f'}^d \alpha_{\tilde{f}_1 \tilde{Z}_i}^{d} + \omega_{G \tilde{W} \tilde{Z}}^R \alpha_{\tilde{f}_1}^{\tilde{W}_j} \mathcal{C}_{G f f'}^u \beta_{\tilde{f}_1 \tilde{Z}_{i}}^{d}]m_{f}m_{f'}, \\
\mathcal{V}_{G \tilde{f}_1}^{(8)} &= -2[\omega_{G \tilde{W} \tilde{Z}}^R \beta_{\tilde{f}_1}^{\tilde{W}_j} \mathcal{C}_{G f f'}^d \alpha_{\tilde{f}_1 \tilde{Z}_i}^{d} + (-1)^{\theta_j} \omega_{G \tilde{W} \tilde{Z}}^L \alpha_{\tilde{f}_1}^{\tilde{W}_j} \mathcal{C}_{G f f'}^u \beta_{\tilde{f}_1 \tilde{Z}_{i}}^{d}]m_{f}m_{f'}|m_{\tilde{W}_j}||m_{\tilde{Z}_i}|.
\end{align}
The integrals are exactly as for $G \tilde{f'}_1$ except you must swap $m_{f'}
\leftrightarrow m_{f}$ and $m_{\tilde{f'}_1} \leftrightarrow
m_{\tilde{f}_1}$. As for $G \tilde{f'}_1$, for the case of a chargino decaying
to a neutralino you must relabel integrals such that integrals 2 and 4 are
swapped as are integrals 3 and 5. 
The $\Gamma_{G \tilde{f}_1}$ is then given analogously to $\Gamma_{G \tilde{f'}_1}$ as the sum of the products of the coupling combinations with corresponding integrals.

\textbf{\underline{$\Gamma_{G \tilde{f}_2}$}}

Here the coupling combinations are, if a neutralino is decaying:
\begin{align}
\mathcal{V}_{G \tilde{f}_2}^{(1)} &= -\frac{1}{2}[\omega_{G \tilde{W} \tilde{Z}}^R \beta_{\tilde{f}_2 \tilde{Z}_i}^{d} \mathcal{C}_{G f f'}^d \alpha_{\tilde{f}_2}^{\tilde{W}_j} + (-1)^{\theta_i}\omega_{G \tilde{W} \tilde{Z}}^L  \alpha_{\tilde{f}_2 \tilde{Z}_i}^{d} \mathcal{C}_{G f f'}^u \beta_{\tilde{f}_2}^{\tilde{W}_j}], \\
\mathcal{V}_{G \tilde{f}_2}^{(2)} &= [\omega_{G \tilde{W} \tilde{Z}}^R \alpha_{\tilde{f}_2 \tilde{Z}_i}^{d} \mathcal{C}_{G f f'}^u \alpha_{\tilde{f}_2}^{\tilde{W}_j} + (-1)^{\theta_i}\omega_{G \tilde{W} \tilde{Z}}^L \beta_{\tilde{f}_2 \tilde{Z}_i}^{d} \mathcal{C}_{G f f'}^d \beta_{\tilde{f}_2}^{\tilde{W}_j}]|m_{\tilde{Z}_i}|m_{f'}, \\
\mathcal{V}_{G \tilde{f}_2}^{(3)} &= -((-1)^{\theta_i}\omega_{G \tilde{W} \tilde{Z}}^L \alpha_{\tilde{f}_2 \tilde{Z}_i}^{d} \mathcal{C}_{G f f'}^d \alpha_{\tilde{f}_2}^{\tilde{W}_j} +  \omega_{G \tilde{W} \tilde{Z}}^R \beta_{\tilde{f}_2 \tilde{Z}_i}^{d} \mathcal{C}_{G f f'}^u \beta_{\tilde{f}_2}^{\tilde{W}_j})m_{f}|m_{\tilde{W}_j}|, \\
\mathcal{V}_{G \tilde{f}_2}^{(4)} &= -[\omega_{G \tilde{W} \tilde{Z}}^R \alpha_{\tilde{f}_2 \tilde{Z}_i}^{d} \mathcal{C}_{G f f'}^d \alpha_{\tilde{f}_2}^{\tilde{W}_j} + (-1)^{\theta_i}\omega_{G \tilde{W} \tilde{Z}}^L \beta_{\tilde{f}_2 \tilde{Z}_i}^{d} \mathcal{C}_{G f f'}^u \beta_{\tilde{f}_2}^{\tilde{W}_j}]|m_{\tilde{Z}_i}|m_{f}, \\
\mathcal{V}_{G \tilde{f}_2}^{(5)} &= [(-1)^{\theta_i}\omega_{G \tilde{W} \tilde{Z}}^L \alpha_{\tilde{f}_2 \tilde{Z}_i}^{d} \mathcal{C}_{G f f'}^u \alpha_{\tilde{f}_2}^{\tilde{W}_j} + \omega_{G \tilde{W} \tilde{Z}}^R \beta_{\tilde{f}_2 \tilde{Z}_i}^{d} \mathcal{C}_{G f f'}^d \beta_{\tilde{f}_2}^{\tilde{W}_j}]m_{f'}|m_{\tilde{W}_j}|, \\
\mathcal{V}_{G \tilde{f}_2}^{(6)} &= -((-1)^{\theta_i}\omega_{G \tilde{W} \tilde{Z}}^L \beta_{\tilde{f}_2 \tilde{Z}_i}^{d} \mathcal{C}_{G f f'}^d \alpha_{\tilde{f}_2}^{\tilde{W}_j} + \omega_{G \tilde{W} \tilde{Z}}^R \alpha_{\tilde{f}_2 \tilde{Z}_i}^{d} \mathcal{C}_{G f f'}^u \beta_{\tilde{f}_2}^{\tilde{W}_j})|m_{\tilde{Z}_i}||m_{\tilde{W}_j}|(-1)^{\theta_j}, \\
\mathcal{V}_{G \tilde{f}_2}^{(7)} &= [\omega_{G \tilde{W} \tilde{Z}}^R \beta_{\tilde{f}_2 \tilde{Z}_i}^{d} \mathcal{C}_{G f f'}^u \alpha_{\tilde{f}_2}^{\tilde{W}_j} + (-1)^{\theta_i}\omega_{G \tilde{W} \tilde{Z}}^L \alpha_{\tilde{f}_2 \tilde{Z}_i}^{d} \mathcal{C}_{G f f'}^d \beta_{\tilde{f}_2}^{\tilde{W}_j}]m_{f'}m_{f} ,\\
\mathcal{V}_{G \tilde{f}_2}^{(8)} &= 2[(-1)^{\theta_i}\omega_{G \tilde{W} \tilde{Z}}^L \beta_{\tilde{f}_2 \tilde{Z}_i}^{d} \mathcal{C}_{G f f'}^u \alpha_{\tilde{f}_2}^{\tilde{W}_j} + \omega_{G \tilde{W} \tilde{Z}}^R \alpha_{\tilde{f}_2 \tilde{Z}_i}^{d} \mathcal{C}_{G f f'}^d \beta_{\tilde{f}_2}^{\tilde{W}_j}]|m_{\tilde{Z}_i}|m_{f'}m_{f}|m_{\tilde{W}_j}|(-1)^{\theta_j}.
\end{align}

If it's a chargino decaying:
\begin{align}
\mathcal{V}_{G \tilde{f}_2}^{(1)} &= -\frac{1}{2}(-1)^{\theta_j}[\omega_{G \tilde{W} \tilde{Z}}^L \beta_{\tilde{f}_2}^{\tilde{W}_j}\mathcal{C}_{G f f'}^u \alpha_{\tilde{f}_2 \tilde{Z}_i}^{d} - \omega_{G \tilde{W} \tilde{Z}}^R \alpha_{\tilde{f}_2}^{\tilde{W}_j} \mathcal{C}_{G f f'}^d \beta_{\tilde{f}_2 \tilde{Z}_i}^{d}],\\
\mathcal{V}_{G \tilde{f}_2}^{(2)} &= -[\omega_{G \tilde{W} \tilde{Z}}^L \alpha_{\tilde{f}_2}^{\tilde{W}_j} \mathcal{C}_{G f f'}^d \alpha_{\tilde{f}_2 \tilde{Z}_i}^{d} - (-1)^{\theta_j} \omega_{G \tilde{W} \tilde{Z}}^R \beta_{\tilde{f}_2}^{\tilde{W}_j} \mathcal{C}_{G f f'}^u \beta_{\tilde{f}_2 \tilde{Z}_i}^{d}]|m_{\tilde{Z}_i}|m_{f}, \\
\mathcal{V}_{G \tilde{f}_2}^{(3)} &= [(-1)^{\theta_j}\omega_{G \tilde{W} \tilde{Z}}^R \alpha_{\tilde{f}_2}^{\tilde{W}_j} \mathcal{C}_{G f f'}^u \alpha_{\tilde{f}_2 \tilde{Z}_i}^{d} - \omega_{G \tilde{W} \tilde{Z}}^L \beta_{\tilde{f}_2}^{\tilde{W}_j} \mathcal{C}_{G f f'}^d \beta_{\tilde{f}_2 \tilde{Z}_i}^{d}]m_{f'}|m_{\tilde{W}_j}| ,\\
\mathcal{V}_{G \tilde{f}_2}^{(4)} &= (\omega_{G \tilde{W} \tilde{Z}}^L \alpha_{\tilde{f}_2}^{\tilde{W}_j} \mathcal{C}_{G f f'}^u \alpha_{\tilde{f}_2 \tilde{Z}_i}^{d} + (-1)^{\theta_j} \omega_{G \tilde{W} \tilde{Z}}^R \beta_{\tilde{f}_2}^{\tilde{W}_j} \mathcal{C}_{G f f'}^d \beta_{\tilde{f}_2 \tilde{Z}_i}^{d})|m_{\tilde{Z}_i}|m_{f'}, \\
\mathcal{V}_{G \tilde{f}_2}^{(5)} &= -[(-1)^{\theta_j}\omega_{G \tilde{W} \tilde{Z}}^R \alpha_{\tilde{f}_2}^{\tilde{W}_j} \mathcal{C}_{G f f'}^d \alpha_{\tilde{f}_2 \tilde{Z}_i}^{d} + \omega_{G \tilde{W} \tilde{Z}}^L \beta_{\tilde{f}_2}^{\tilde{W}_j} \mathcal{C}_{G f f'}^u \beta_{\tilde{f}_2 \tilde{Z}_i}^{d}]m_{f}|m_{\tilde{W}_j}|, \\
\mathcal{V}_{G \tilde{f}_2}^{(6)} &= [\omega_{G \tilde{W} \tilde{Z}}^R \beta_{\tilde{f}_2}^{\tilde{W}_j} \mathcal{C}_{G f f'}^u \alpha_{\tilde{f}_2 \tilde{Z}_i}^{d} + \omega_{G \tilde{W} \tilde{Z}}^L \alpha_{\tilde{f}_2}^{\tilde{W}_j} \mathcal{C}_{G f f'}^d \beta_{\tilde{f}_2 \tilde{Z}_i}^{d}]|m_{\tilde{W}_j}||m_{\tilde{Z}_i}| ,\\
\mathcal{V}_{G \tilde{f}_2}^{(7)} &= -(-1)^{\theta_j}[\omega_{G \tilde{W} \tilde{Z}}^L \beta_{\tilde{f}_2}^{\tilde{W}_j} \mathcal{C}_{G f f'}^d \alpha_{\tilde{f}_2 \tilde{Z}_i}^{d} + \omega_{G \tilde{W} \tilde{Z}}^R \alpha_{\tilde{f}_2}^{\tilde{W}_j} \mathcal{C}_{G f f'}^u \beta_{\tilde{f}_2 \tilde{Z}_i}^{d}]m_{f}m_{f'}, \\
\mathcal{V}_{G \tilde{f}_2}^{(8)} &= 2[(-1)^{\theta_j}\omega_{G \tilde{W} \tilde{Z}}^R \beta_{\tilde{f}_2}^{\tilde{W}_j} \mathcal{C}_{G f f'}^d \alpha_{\tilde{f}_2 \tilde{Z}_i}^{d} + \omega_{G \tilde{W} \tilde{Z}}^L \alpha_{\tilde{f}_2}^{\tilde{W}_j} \mathcal{C}_{G f f'}^u \beta_{\tilde{f}_2 \tilde{Z}_i}^{d}]m_{f}m_{f'}|m_{\tilde{W}_j}||m_{\tilde{Z}_i}|.
\end{align}
The integrals are exactly as for $G \tilde{f'}_2$ except you must swap $m_{f'} \leftrightarrow m_{f}$ and $m_{\tilde{f'}_2} \leftrightarrow m_{\tilde{f}_2}$. As for $G \tilde{f'}_2$, for the case of a chargino decaying to a neutralino you must relabel integrals such that integrals 2 and 4 are swapped as are integrals 3 and 5.
The $\Gamma_{G \tilde{f}_2}$ is then given analogously to $\Gamma_{G \tilde{f'}_2}$ as the sum of the products of the coupling combinations with corresponding integrals.

\textbf{\underline{$\Gamma_{H^{\pm} \tilde{f}_1}$}}

Again couplings depend upon which direction the decay occurs, if it is a neutralino decaying the couplings are:
\begin{align}
\mathcal{V}_{H^{\pm} \tilde{f}_1}^{(1)} &= \frac{1}{2}[\omega_{H^+ \tilde{W}^+ \tilde{Z}}^R \beta_{\tilde{Z}_i \tilde{f}_1}^{d} \mathcal{C}_{H^+ f f'}^d \alpha_{\tilde{f}_1}^{\tilde{W}_j} + \omega_{H^+ \tilde{W}^+ \tilde{Z}}^L \alpha_{\tilde{Z}_i \tilde{f}_1}^{d} \mathcal{C}_{H^+ f f'}^u \beta_{\tilde{f}_1}^{\tilde{W}_j}], \\
\mathcal{V}_{H^{\pm} \tilde{f}_1}^{(2)} &= -[\omega_{H^+ \tilde{W}^+ \tilde{Z}}^R \alpha_{\tilde{Z}_i \tilde{f}_1}^{d} \mathcal{C}_{H^+ f f'}^u \alpha_{\tilde{f}_1}^{\tilde{W}_j} + \omega_{H^+ \tilde{W}^+ \tilde{Z}}^L \beta_{\tilde{Z}_i \tilde{f}_1}^{d} \mathcal{C}_{H^+ f f'}^d \beta_{\tilde{f}_1}^{\tilde{W}_j}]|m_{\tilde{Z}_i}|m_{f'}, \\
\mathcal{V}_{H^{\pm} \tilde{f}_1}^{(3)} &= (\omega_{H^+ \tilde{W}^+ \tilde{Z}}^L \alpha_{\tilde{Z}_i \tilde{f}_1}^{d}  \mathcal{C}_{H^+ f f'}^d \alpha_{\tilde{f}_1}^{\tilde{W}_j} + \omega_{H^+ \tilde{W}^+ \tilde{Z}}^R \beta_{\tilde{Z}_i \tilde{f}_1}^{d} \mathcal{C}_{H^+ f f'}^u \beta_{\tilde{f}_1}^{\tilde{W}_j})m_{f}|m_{\tilde{W}_j}|(-1)^{\theta_i}, \\
\mathcal{V}_{H^{\pm} \tilde{f}_1}^{(4)} &= (\omega_{H^+ \tilde{W}^+ \tilde{Z}}^R \alpha_{\tilde{Z}_i \tilde{f}_1}^{d} \mathcal{C}_{H^+ f f'}^d \alpha_{\tilde{f}_1}^{\tilde{W}_j} + \omega_{H^+ \tilde{W}^+ \tilde{Z}}^L \beta_{\tilde{Z}_i \tilde{f}_1}^{d} \mathcal{C}_{H^+ f f'}^u \beta_{\tilde{f}_1}^{\tilde{W}_j})|m_{\tilde{Z}_i}|m_{f}, \\
\mathcal{V}_{H^{\pm} \tilde{f}_1}^{(5)} &= -[(-1)^{\theta_i}\omega_{H^+ \tilde{W}^+ \tilde{Z}}^L \alpha_{\tilde{Z}_i \tilde{f}_1}^{d} \mathcal{C}_{H^+ f f'}^u \alpha_{\tilde{f}_1}^{\tilde{W}_j} + \omega_{H^+ \tilde{W}^+ \tilde{Z}}^R \beta_{\tilde{Z}_i \tilde{f}_1}^{d} \mathcal{C}_{H^+ f f'}^d \beta_{\tilde{f}_1}^{\tilde{W}_j}]m_{f'}|m_{\tilde{W}_j}|, \\
\mathcal{V}_{H^{\pm} \tilde{f}_1}^{(6)} &= [(-1)^{\theta_i} \omega_{H^+ \tilde{W}^+ \tilde{Z}}^L \beta_{\tilde{Z}_i \tilde{f}_1}^{d} \mathcal{C}_{H^+ f f'}^d \alpha_{\tilde{f}_1}^{\tilde{W}_j} - \omega_{H^+ \tilde{W}^+ \tilde{Z}}^R \alpha_{\tilde{Z}_i \tilde{f}_1}^{d} \mathcal{C}_{H^+ f f'}^u \beta_{\tilde{f}_1}^{\tilde{W}_j}]|m_{\tilde{Z}_i}||m_{\tilde{W}_j}|, \\
\mathcal{V}_{H^{\pm} \tilde{f}_1}^{(7)} &= -[\omega_{H^+ \tilde{W}^+ \tilde{Z}}^R \beta_{\tilde{Z}_i \tilde{f}_1}^{d} \mathcal{C}_{H^+ f f'}^u \alpha_{\tilde{f}_1}^{\tilde{W}_j} - (-1)^{\theta_i} \omega_{H^+ \tilde{W}^+ \tilde{Z}}^L \alpha_{\tilde{Z}_i \tilde{f}_1}^{d} \mathcal{C}_{H^+ f f'}^d \beta_{\tilde{f}_1}^{\tilde{W}_j}]m_{f'}m_{f}, \\
\mathcal{V}_{H^{\pm} \tilde{f}_1}^{(8)} &= -2[(-1)^{\theta_i} \omega_{H^+ \tilde{W}^+ \tilde{Z}}^L \beta_{\tilde{Z}_i \tilde{f}_1}^{d} \mathcal{C}_{H^+ f f'}^u \alpha_{\tilde{f}_1}^{\tilde{W}_j} - \omega_{H^+ \tilde{W}^+ \tilde{Z}}^R \alpha_{\tilde{Z}_i \tilde{f}_1}^{d} \mathcal{C}_{H^+ f f'}^d \beta_{\tilde{f}_1}^{\tilde{W}_j}]|m_{\tilde{Z}_i}||m_{\tilde{W}_j}|m_{f'}m_{f}.
\end{align}

Whilst if it's a chargino decaying:
\begin{align}
\mathcal{V}_{H^{\pm} \tilde{f}_1}^{(1)} &= \frac{1}{2}[(-1)^{\theta_j} \omega_{H^+ \tilde{W}^+ \tilde{Z}}^L \beta_{\tilde{f}_1}^{\tilde{W}_j} \mathcal{C}_{H^+ f f'}^u \alpha_{\tilde{Z}_i \tilde{f}_1}^{d} + \omega_{H^+ \tilde{W}^+ \tilde{Z}}^R \alpha_{\tilde{f}_1}^{\tilde{W}_j} \mathcal{C}_{H^+ f f'}^d \beta_{\tilde{Z}_i \tilde{f}_1}^{d}], \\
\mathcal{V}_{H^{\pm} \tilde{f}_1}^{(2)} &= [(-1)^{\theta_j}\omega_{H^+ \tilde{W}^+ \tilde{Z}}^L \alpha_{\tilde{f}_1}^{\tilde{W}_j} \mathcal{C}_{H^+ f f'}^d \alpha_{\tilde{Z}_i \tilde{f}_1}^{d} + \omega_{H^+ \tilde{W}^+ \tilde{Z}}^R \beta_{\tilde{f}_1}^{\tilde{W}_j} \mathcal{C}_{H^+ f f'}^u \beta_{\tilde{Z}_i \tilde{f}_1}^{d}]|m_{\tilde{Z}_i}|m_{f}, \\
\mathcal{V}_{H^{\pm} \tilde{f}_1}^{(3)} &= -[\omega_{H^+ \tilde{W}^+ \tilde{Z}}^R \alpha_{\tilde{f}_1}^{\tilde{W}_j} \mathcal{C}_{H^+ f f'}^u \alpha_{\tilde{Z}_i \tilde{f}_1}^{d} + (-1)^{\theta_j} \omega_{H^+ \tilde{W}^+ \tilde{Z}}^L \beta_{\tilde{f}_1}^{\tilde{W}_j} \mathcal{C}_{H^+ f f'}^d \beta_{\tilde{Z}_i \tilde{f}_1}^{d}]m_{f'}|m_{\tilde{W}_j}|, \\
\mathcal{V}_{H^{\pm} \tilde{f}_1}^{(4)} &= -[(-1)^{\theta_j} \omega_{H^+ \tilde{W}^+ \tilde{Z}}^L \alpha_{\tilde{f}_1}^{\tilde{W}_j} \mathcal{C}_{H^+ f f'}^u \alpha_{\tilde{Z}_i \tilde{f}_1}^{d} + \omega_{H^+ \tilde{W}^+ \tilde{Z}}^R \beta_{\tilde{f}_1}^{\tilde{W}_j} \mathcal{C}_{H^+ f f'}^d \beta_{\tilde{Z}_i \tilde{f}_1}^{d}]|m_{\tilde{Z}_i}|m_{f'}, \\
\mathcal{V}_{H^{\pm} \tilde{f}_1}^{(5)} &= (\omega_{H^+ \tilde{W}^+ \tilde{Z}}^R \alpha_{\tilde{f}_1}^{\tilde{W}_j} \mathcal{C}_{H^+ f f'}^d \alpha_{\tilde{Z}_i \tilde{f}_1}^{d} + (-1)^{\theta_j} \omega_{H^+ \tilde{W}^+ \tilde{Z}}^L  \beta_{\tilde{f}_1}^{\tilde{W}_j} \mathcal{C}_{H^+ f f'}^u \beta_{\tilde{Z}_i \tilde{f}_1}^{d})m_{f}|m_{\tilde{W}_j}|, \\
\mathcal{V}_{H^{\pm} \tilde{f}_1}^{(6)} &= (\omega_{H^+ \tilde{W}^+ \tilde{Z}}^R \beta_{\tilde{f}_1}^{\tilde{W}_j} \mathcal{C}_{H^+ f f'}^u \alpha_{\tilde{Z}_i \tilde{f}_1}^{d} + (-1)^{\theta_j} \omega_{H^+ \tilde{W}^+ \tilde{Z}}^L \alpha_{\tilde{f}_1}^{\tilde{W}_j} \mathcal{C}_{H^+ f f'}^d \beta_{\tilde{Z}_i \tilde{f}_1}^{d})|m_{\tilde{W}_j}||m_{\tilde{Z}_i}|, \\
\mathcal{V}_{H^{\pm} \tilde{f}_1}^{(7)} &= -[(-1)^{\theta_j} \omega_{H^+ \tilde{W}^+ \tilde{Z}}^L \beta_{\tilde{f}_1}^{\tilde{W}_j} \mathcal{C}_{H^+ f f'}^d \alpha_{\tilde{Z}_i \tilde{f}_1}^{d} + \omega_{H^+ \tilde{W}^+ \tilde{Z}}^R \alpha_{\tilde{f}_1}^{\tilde{W}_j} \mathcal{C}_{H^+ f f'}^u \beta_{\tilde{Z}_i \tilde{f}_1}^{d}]m_{f}m_{f'}, \\
\mathcal{V}_{H^{\pm} \tilde{f}_1}^{(8)} &= -2[\omega_{H^+ \tilde{W}^+ \tilde{Z}}^R \beta_{\tilde{f}_1}^{\tilde{W}_j} \mathcal{C}_{H^+ f f'}^d \alpha_{\tilde{Z}_i \tilde{f}_1}^{d} + (-1)^{\theta_j} \omega_{H^+ \tilde{W}^+ \tilde{Z}}^L \alpha_{\tilde{f}_1}^{\tilde{W}_j} \mathcal{C}_{H^+ f f'}^u \beta_{\tilde{Z}_i \tilde{f}_1}^{d}]m_{f}m_{f'}|m_{\tilde{W}_j}||m_{\tilde{Z}_i}|.
\end{align}
The integrals and overall contribution are given exactly as for $H^{\pm} \tilde{f'}_1$ but $m_{\tilde{f'}_1} \rightarrow m_{\tilde{f}_1}$. Remember the integrals' labelling depends on whether it's a neutralino decaying (to a chargino) or a chargino decaying (to a neutralino).

\textbf{\underline{$\Gamma_{H^{\pm} \tilde{f}_2}$}}

The coupling combinations again depend upon which way around the decay is occurring, for a neutralino decaying the coupling combinations are:
\begin{align}
\mathcal{V}_{H^{\pm} \tilde{f}_2}^{(1)} &= -\frac{1}{2}[\omega_{H^+ \tilde{W}^+ \tilde{Z}}^R \beta_{\tilde{f}_2 \tilde{Z}_i}^{d} \mathcal{C}_{H^+ f f'}^d \alpha_{\tilde{f}_2}^{\tilde{W}_j} + \omega_{H^+ \tilde{W}^+ \tilde{Z}}^L \alpha_{\tilde{f}_2 \tilde{Z}_i}^{d} \mathcal{C}_{H^+ f f'}^u \beta_{\tilde{f}_2}^{\tilde{W}_j}], \\
\mathcal{V}_{H^{\pm} \tilde{f}_2}^{(2)} &= (\omega_{H^+ \tilde{W}^+ \tilde{Z}}^R \alpha_{\tilde{f}_2 \tilde{Z}_i}^{d} \mathcal{C}_{H^+ f f'}^u \alpha_{\tilde{f}_2}^{\tilde{W}_j} + \omega_{H^+ \tilde{W}^+ \tilde{Z}}^L \beta_{\tilde{f}_2 \tilde{Z}_i}^{d} \mathcal{C}_{H^+ f f'}^d \beta_{\tilde{f}_2}^{\tilde{W}_j}]|m_{\tilde{Z}_i}|m_{f'}, \\
\mathcal{V}_{H^{\pm} \tilde{f}_2}^{(3)} &= -[\omega_{H^+ \tilde{W}^+ \tilde{Z}}^LL \alpha_{\tilde{f}_2 \tilde{Z}_i}^{d} \mathcal{C}_{H^+ f f'}^d \alpha_{\tilde{f}_2}^{\tilde{W}_j} + \omega_{H^+ \tilde{W}^+ \tilde{Z}}^R \beta_{\tilde{f}_2 \tilde{Z}_i}^{d} \mathcal{C}_{H^+ f f'}^u \beta_{\tilde{f}_2}^{\tilde{W}_j}]m_{f}|m_{\tilde{W}_j}|(-1)^{\theta_i}, \\
\mathcal{V}_{H^{\pm} \tilde{f}_2}^{(4)} &= -[\omega_{H^+ \tilde{W}^+ \tilde{Z}}^R \alpha_{\tilde{f}_2 \tilde{Z}_i}^{d} \mathcal{C}_{H^+ f f'}^d \alpha_{\tilde{f}_2}^{\tilde{W}_j} + \omega_{H^+ \tilde{W}^+ \tilde{Z}}^L \beta_{\tilde{f}_2 \tilde{Z}_i}^{d} \mathcal{C}_{H^+ f f'}^u \beta_{\tilde{f}_2}^{\tilde{W}_j}]|m_{\tilde{Z}_i}|m_{f}, \\
\mathcal{V}_{H^{\pm} \tilde{f}_2}^{(5)} &= [(-1)^{\theta_i}\omega_{H^+ \tilde{W}^+ \tilde{Z}}^L \alpha_{\tilde{f}_2 \tilde{Z}_i}^{d} \mathcal{C}_{H^+ f f'}^u \alpha_{\tilde{f}_2}^{\tilde{W}_j} + \omega_{H^+ \tilde{W}^+ \tilde{Z}}^R \beta_{\tilde{f}_2 \tilde{Z}_i}^{d} \mathcal{C}_{H^+ f f'}^d \beta_{\tilde{f}_2}^{\tilde{W}_j}]m_{f'}|m_{\tilde{W}_j}|, \\
\mathcal{V}_{H^{\pm} \tilde{f}_2}^{(6)} &= -[\omega_{H^+ \tilde{W}^+ \tilde{Z}}^L \beta_{\tilde{f}_2 \tilde{Z}_i}^{d} \mathcal{C}_{H^+ f f'}^d \alpha_{\tilde{f}_2}^{\tilde{W}_j} + \omega_{H^+ \tilde{W}^+ \tilde{Z}}^R \alpha_{\tilde{f}_2 \tilde{Z}_i}^{d} \mathcal{C}_{H^+ f f'}^u \beta_{\tilde{f}_2}^{\tilde{W}_j}]|m_{\tilde{Z}_i}||m_{\tilde{W}_j}|(-1)^{\theta_i}(-1)^{\theta_j}, \\
\mathcal{V}_{H^{\pm} \tilde{f}_2}^{(7)} &= [\omega_{H^+ \tilde{W}^+ \tilde{Z}}^R \beta_{\tilde{f}_2 \tilde{Z}_i}^{d} \mathcal{C}_{H^+ f f'}^u \alpha_{\tilde{f}_2}^{\tilde{W}_j} + \omega_{H^+ \tilde{W}^+ \tilde{Z}}^L \alpha_{\tilde{f}_2 \tilde{Z}_i}^{d} \mathcal{C}_{H^+ f f'}^d \beta_{\tilde{f}_2}^{\tilde{W}_j}]m_{f'}m_{f}, \\
\mathcal{V}_{H^{\pm} \tilde{f}_2}^{(8)} &= 2[(-1)^{\theta_i}\omega_{H^+ \tilde{W}^+ \tilde{Z}}^L  \beta_{\tilde{f}_2 \tilde{Z}_i}^{d} \mathcal{C}_{H^+ f f'}^u \alpha_{\tilde{f}_2}^{\tilde{W}_j} +  \omega_{H^+ \tilde{W}^+ \tilde{Z}}^R \alpha_{\tilde{f}_2 \tilde{Z}_i}^{d} \mathcal{C}_{H^+ f f'}^d \beta_{\tilde{f}_2}^{\tilde{W}_j}]|m_{\tilde{Z}_i}|m_{f'}m_{f}|m_{\tilde{W}_j}|(-1)^{\theta_j}.
\end{align}

If it's a chargino decaying:
\begin{align}
\mathcal{V}_{H^{\pm} \tilde{f}_2}^{(1)} &= \frac{1}{2}[\omega_{H^+ \tilde{W}^+ \tilde{Z}}^L \beta_{\tilde{f}_2}^{\tilde{W}_j} \mathcal{C}_{H^+ f f'}^u \alpha_{\tilde{f}_2 \tilde{Z}_i}^{d} + (-1)^{\theta_j} \omega_{H^+ \tilde{W}^+ \tilde{Z}}^R \alpha_{\tilde{f}_2}^{\tilde{W}_j} \mathcal{C}_{H^+ f f'}^d \beta_{\tilde{f}_2 \tilde{Z}_i}^{d}], \\
\mathcal{V}_{H^{\pm} \tilde{f}_2}^{(2)} &= -[\omega_{H^+ \tilde{W}^+ \tilde{Z}}^L \alpha_{\tilde{f}_2}^{\tilde{W}_j} \mathcal{C}_{H^+ f f'}^d \alpha_{\tilde{f}_2 \tilde{Z}_i}^{d} - (-1)^{\theta_j} \omega_{H^+ \tilde{W}^+ \tilde{Z}}^R \beta_{\tilde{f}_2}^{\tilde{W}_j} \mathcal{C}_{H^+ f f'}^u \beta_{\tilde{f}_2 \tilde{Z}_i}^{d}]|m_{\tilde{Z}_i}| m_{f}], \\
\mathcal{V}_{H^{\pm} \tilde{f}_2}^{(3)} &= [(-1)^{\theta_j}\omega_{H^+ \tilde{W}^+ \tilde{Z}}^R \alpha_{\tilde{f}_2}^{\tilde{W}_j} \mathcal{C}_{H^+ f f'}^u \alpha_{\tilde{f}_2 \tilde{Z}_i}^{d} -  \omega_{H^+ \tilde{W}^+ \tilde{Z}}^L \beta_{\tilde{f}_2}^{\tilde{W}_j} \mathcal{C}_{H^+ f f'}^d \beta_{\tilde{f}_2 \tilde{Z}_i}^{d}]m_{f'}|m_{\tilde{W}_j}|, \\
\mathcal{V}_{H^{\pm} \tilde{f}_2}^{(4)} &= (\omega_{H^+ \tilde{W}^+ \tilde{Z}}^L \alpha_{\tilde{f}_2}^{\tilde{W}_j} \mathcal{C}_{H^+ f f'}^u \alpha_{\tilde{f}_2 \tilde{Z}_i}^{d} - (-1)^{\theta_j} \omega_{H^+ \tilde{W}^+ \tilde{Z}}^R \beta_{\tilde{f}_2}^{\tilde{W}_j} \mathcal{C}_{H^+ f f'}^d \beta_{\tilde{f}_2 \tilde{Z}_i}^{d})|m_{\tilde{Z}_i}|m_{f'}, \\
\mathcal{V}_{H^{\pm} \tilde{f}_2}^{(5)} &= -[(-1)^{\theta_j}\omega_{H^+ \tilde{W}^+ \tilde{Z}}^R \alpha_{\tilde{f}_2}^{\tilde{W}_j} \mathcal{C}_{H^+ f f'}^d \alpha_{\tilde{f}_2 \tilde{Z}_i}^{d} + \omega_{H^+ \tilde{W}^+ \tilde{Z}}^L \beta_{\tilde{f}_2}^{\tilde{W}_j} \mathcal{C}_{H^+ f f'}^u \beta_{\tilde{f}_2 \tilde{Z}_i}^{d}]m_{f}|m_{\tilde{W}_j}|, \\
\mathcal{V}_{H^{\pm} \tilde{f}_2}^{(6)} &= (\omega_{H^+ \tilde{W}^+ \tilde{Z}}^R \beta_{\tilde{f}_2}^{\tilde{W}_j} \mathcal{C}_{H^+ f f'}^u \alpha_{\tilde{f}_2 \tilde{Z}_i}^{d} +  \omega_{H^+ \tilde{W}^+ \tilde{Z}}^L \alpha_{\tilde{f}_2}^{\tilde{W}_j} \mathcal{C}_{H^+ f f'}^d \beta_{\tilde{f}_2 \tilde{Z}_i}^{d})|m_{\tilde{W}_j}||m_{\tilde{Z}_i}|, \\
\mathcal{V}_{H^{\pm} \tilde{f}_2}^{(7)} &= -(-1)^{\theta_j}[\omega_{H^+ \tilde{W}^+ \tilde{Z}}^L \beta_{\tilde{f}_2}^{\tilde{W}_j} \mathcal{C}_{H^+ f f'}^d \alpha_{\tilde{f}_2 \tilde{Z}_i}^{d} + \omega_{H^+ \tilde{W}^+ \tilde{Z}}^R \alpha_{\tilde{f}_2}^{\tilde{W}_j} \mathcal{C}_{H^+ f f'}^u \beta_{\tilde{f}_2 \tilde{Z}_i}^{d}]m_{f}m_{f'}, \\
\mathcal{V}_{H^{\pm} \tilde{f}_2}^{(8)} &= 2[(-1)^{\theta_j}\omega_{H^+ \tilde{W}^+ \tilde{Z}}^R \beta_{\tilde{f}_2}^{\tilde{W}_j} \mathcal{C}_{H^+ f f'}^d \alpha_{\tilde{f}_2 \tilde{Z}_i}^{d} - \omega_{H^+ \tilde{W}^+ \tilde{Z}}^L \alpha_{\tilde{f}_2}^{\tilde{W}_j} \mathcal{C}_{H^+ f f'}^u \beta_{\tilde{f}_2 \tilde{Z}_i}^{d}]m_{f}m_{f'}|m_{\tilde{W}_j}||m_{\tilde{Z}_i}|.
\end{align}
The integrals and overall contribution are given exactly as for $H^{\pm} \tilde{f'}_2$ but $m_{\tilde{f'}_2} \rightarrow m_{\tilde{f}_2}$. Remember the integrals labelling depends on whether it's a neutralino decaying (to a chargino) or a chargino decaying (to a neutralino).

\textbf{\underline{$\Gamma_{\tilde{f'}_1 \tilde{f'}_2}$}}

Here the coupling combinations for the interference of the two positively charged sfermions depend upon whether the decay is neutralino to chargino or chargino to neutralino. For neutralino to chargino:
\begin{align}
\mathcal{V}_{\tilde{f'}_2 \tilde{f'}_2}^{(1)} &= (\beta_{\tilde{f}_1 \tilde{Z}_i}^{u} \beta_{\tilde{f}_2 \tilde{Z}_i}^{u} +  \alpha_{\tilde{f}_1 \tilde{Z}_i}^{u}\alpha_{\tilde{f}_2 \tilde{Z}_i}^{u})(-1)^{\theta_i}	, \\
\mathcal{V}_{\tilde{f'}_2 \tilde{f'}_2}^{(2)} &= (\alpha_{\tilde{f}_1 \tilde{Z}_i}^{u} \beta_{\tilde{f}_2 \tilde{Z}_i}^{u} +  \beta_{\tilde{f}_1 \tilde{Z}_i}^{u}\alpha_{\tilde{f}_2 \tilde{Z}_i}^{u})(-1)^{\theta_i}, \\
\mathcal{V}_{\tilde{f'}_2 \tilde{f'}_2}^{(3)} &= (-\alpha_{\tilde{f'}_1}^{\tilde{W}_j} \alpha_{\tilde{f'}_2}^{\tilde{W}_j} + \beta_{\tilde{f'}_1}^{\tilde{W}_j} \beta_{\tilde{f'}_2}^{\tilde{W}_j})(-1)^{\theta_i}, \\
\mathcal{V}_{\tilde{f'}_2 \tilde{f'}_2}^{(4)} &= (\beta_{\tilde{f'}_1}^{\tilde{W}_j}\alpha_{\tilde{f'}_2}^{\tilde{W}_j} - \alpha_{\tilde{f'}_1}^{\tilde{W}_j} \beta_{\tilde{f'}_2}^{\tilde{W}_j})(-1)^{\theta_i}(-1)^{\theta_j},
\end{align} 
whilst for chargino to neutralino:
\begin{align}
\mathcal{V}_{\tilde{f'}_2 \tilde{f'}_2}^{(1)} &= -\beta_{\tilde{f'}_1}^{\tilde{W}_j}\beta_{\tilde{f'}_2}^{\tilde{W}_j} + \alpha_{\tilde{f'}_1}^{\tilde{W}_j} \alpha_{\tilde{f'}_2}^{\tilde{W}_j}(-1)^{\theta_j}, \\
\mathcal{V}_{\tilde{f'}_2 \tilde{f'}_2}^{(2)} &= (-1)^{\theta_j} \alpha_{\tilde{f'}_1}^{\tilde{W}_j}\beta_{\tilde{f'}_2}^{\tilde{W}_j} - \alpha_{\tilde{f'}_2}^{\tilde{W}_j}\beta_{\tilde{f'}_2}^{\tilde{W}_j}, \\
\mathcal{V}_{\tilde{f'}_2 \tilde{f'}_2}^{(3)} &= (-\alpha_{\tilde{f}_1 \tilde{Z}_i}^{u}\alpha_{\tilde{f}_2 \tilde{Z}_i}^{u} + (-1)^{\theta_j}\beta_{\tilde{f}_1 \tilde{Z}_i}^{u} \beta_{\tilde{f}_2 \tilde{Z}_i}^{u}), \\
\mathcal{V}_{\tilde{f'}_2 \tilde{f'}_2}^{(4)} &= (-\alpha_{\tilde{f}_1 \tilde{Z}_i}^{u}\beta_{\tilde{f}_2 \tilde{Z}_i}^{u} + (-1)^{\theta_j}\alpha_{\tilde{f}_2 \tilde{Z}_i}^{u} \beta_{\tilde{f}_1 \tilde{Z}_i}^{u}).
\end{align}
The integrals are as follows with $s = m_{\tilde{Z}_i}^2 + m_{f'}^2 - 2|m_{\tilde{Z}_i}|E$ and $\lambda = \sqrt{(s-(m_{f}+m_{\tilde{W}_j})^2)(s-(m_{f}-m_{\tilde{W}_j})^2}$:
\begin{equation}
I_{\tilde{f'}_1 \tilde{f'}_2}^{1} = 2|m_{\tilde{Z}_i}| \int_{m_{f'}}^{E_{upper}} 2|m_{\tilde{Z}_i}|\sqrt{E^2-m_{f'}^2} {\lambda \over s(s-m_{\tilde{f'}_1}^2)(s-m_{\tilde{f'}_2})^2)},
\end{equation}
\begin{equation}
I_{\tilde{f'}_1 \tilde{f'}_2}^{2} = 2|m_{\tilde{Z}_i}| \int_{m_{f'}}^{E_{upper}} 2|m_{\tilde{Z}_i}|\sqrt{E^2-m_{f'}^2} {(s-m_{f}^2-m_{\tilde{W}_j}^2)\lambda \over s(s-m_{\tilde{f'}_1}^2)(s-m_{\tilde{f'}_2})^2)},
\end{equation}
\begin{equation}
I_{\tilde{f'}_1 \tilde{f'}_2}^{3} = 2|m_{\tilde{Z}_i}| \int_{m_{f'}}^{E_{upper}} 2|m_{\tilde{Z}_i}|\sqrt{E^2-m_{f'}^2} {2|m_{\tilde{Z}_i}|E \lambda \over s(s-m_{\tilde{f'}_1}^2)(s-m_{\tilde{f'}_2})^2)},
\end{equation}
\begin{equation}
I_{\tilde{f'}_1 \tilde{f'}_2}^{4} = 2|m_{\tilde{Z}_i}| \int_{m_{f'}}^{E_{upper}} 2|m_{\tilde{Z}_i}|\sqrt{E^2-m_{f'}^2}{(s-m_{f}^2-m_{\tilde{W}_j}^2)2|m_{\tilde{Z}_i}|E \lambda \over s(s-m_{\tilde{f'}_1}^2)(s-m_{\tilde{f'}_2})^2)}.
\end{equation}
Now if it's instead a chargino decaying, as described before, swap the chargino and neutralino masses throughout, but also here you must relabel the integrals $I_{\tilde{f'}_1 \tilde{f'}_2}^{2} \leftrightarrow I_{\tilde{f'}_1 \tilde{f'}_2}^{3}$. Also for a chargino decaying one must interchange $m_f$ and $m_{f'}$. For a neutralino decaying:
\begin{equation}
\begin{aligned}
\Gamma_{\tilde{f'}_1 \tilde{f'}_2} = & -4 \mathcal{V}_{\tilde{f'}_2 \tilde{f'}_2}^{(2)} \mathcal{V}_{\tilde{f'}_2 \tilde{f'}_2}^{(4)} |m_{\tilde{Z}_i}||m_{\tilde{W}_j}|m_{f'}m_{f} I_{\tilde{f'}_1 \tilde{f'}_2}^{1} + 2 \mathcal{V}_{\tilde{f'}_2 \tilde{f'}_2}^{(2)} \mathcal{V}_{\tilde{f'}_2 \tilde{f'}_2}^{(3)}|m_{\tilde{Z}_i}|m_{f'} I_{\tilde{f'}_1 \tilde{f'}_2}^{2} \\ & - 2 \mathcal{V}_{\tilde{f'}_2 \tilde{f'}_2}^{(1)} \mathcal{V}_{\tilde{f'}_2 \tilde{f'}_2}^{(4)} m_{f}|m_{\tilde{W}_j}| I_{\tilde{f'}_1 \tilde{f'}_2}^{3} + \mathcal{V}_{\tilde{f'}_2 \tilde{f'}_2}^{(1)} \mathcal{V}_{\tilde{f'}_2 \tilde{f'}_2}^{(3)} I_{\tilde{f'}_1 \tilde{f'}_2}^{4}.
\end{aligned}
\end{equation}

\textbf{\underline{$\Gamma_{\tilde{f}_1 \tilde{f}_2}$}}

For this interference contribution the coupling combinations used are as follows:
\begin{align}
\mathcal{V}_{\tilde{f}_1 \tilde{f}_2}^{(1)} &= -[(-1)^{\theta_c}\alpha_{\tilde{Z}_i \tilde{f}_1}^{d}\alpha_{\tilde{Z}_i \tilde{f}_2}^{d} + \beta_{\tilde{Z}_i \tilde{f}_1}^{d} \beta_{\tilde{Z}_i \tilde{f}_2}^{d}](-1)^{\theta_j}, \\
\mathcal{V}_{\tilde{f}_1 \tilde{f}_2}^{(2)} &= [(-1)^{\theta_i}\beta_{\tilde{Z}_i \tilde{f}_1}^{d} \alpha_{\tilde{Z}_i \tilde{f}_2}^{d} - \alpha_{\tilde{Z}_i \tilde{f}_1}^{d} \beta_{\tilde{Z}_i \tilde{f}_2}^{d}](-1)^{\theta_j}(-1)^{\theta_c}, \\
\mathcal{V}_{\tilde{f}_1 \tilde{f}_2}^{(3)} &= [\beta_{\tilde{f}_1}^{\tilde{W}} \beta_{\tilde{f}_2}^{\tilde{W}} - \alpha_{\tilde{f}_1}^{\tilde{W}} \alpha_{\tilde{f}_2}^{\tilde{W}}](-1)^{\theta_i}, \\
\mathcal{V}_{\tilde{f}_1 \tilde{f}_2}^{(4)} &= (\alpha_{\tilde{f}_1}^{\tilde{W}}\beta_{\tilde{f}_2}^{\tilde{W}} - \beta_{\tilde{f}_1}^{\tilde{W}} \alpha_{\tilde{f}_2}^{\tilde{W}})(-1)^{\theta_c}.
\end{align}
The integrals are as follows, now with $s = m_{\tilde{Z}_i}^2 + m_{f}^2 - 2|m_{\tilde{Z}_i}|E$ and $\lambda = \sqrt{(s - (m_{f'}-m_{\tilde{W}_j})^2)(s - (m_{f'}+m_{\tilde{W}_j})^2)}$:
\begin{equation}
I_{\tilde{f}_1 \tilde{f}_2}^{1} = 2|m_{\tilde{Z}_i}| \int_{m_{f}}^{E_{upper2}} dE {2|m_{\tilde{Z}_i}| \over s} \lambda \sqrt{E^2 - m_{f}^2} {1 \over (s-m_{\tilde{f}_1}^2)(s-m_{\tilde{f}_2}^2)},
\end{equation}
\begin{equation}
I_{\tilde{f}_1 \tilde{f}_2}^{2} = 2|m_{\tilde{Z}_i}| \int_{m_{f}}^{E_{upper2}} dE {2|m_{\tilde{Z}_i}| \over s} \lambda \sqrt{E^2 - m_{f}^2} \ {s- m_{\tilde{W}_j}^2 - m_{f'}^2 \over (s-m_{\tilde{f}_1}^2)(s-m_{\tilde{f}_2}^2)},
\end{equation}
\begin{equation}
I_{\tilde{f}_1 \tilde{f}_2}^{3} = 2|m_{\tilde{Z}_i}| \int_{m_{f}}^{E_{upper2}} dE {2|m_{\tilde{Z}_i}| \over s} \lambda \sqrt{E^2 - m_{f}^2} \ {2|m_{\tilde{Z}_i}| E \over (s-m_{\tilde{f}_1}^2)(s-m_{\tilde{f}_2}^2)},
\end{equation}
\begin{equation}
I_{\tilde{f}_1 \tilde{f}_2}^{4} = 2|m_{\tilde{Z}_i}| \int_{m_{f}}^{E_{upper2}} dE {2|m_{\tilde{Z}_i}| \over s} \lambda \sqrt{E^2 - m_{f}^2} \ {2|m_{\tilde{Z}_i}| E (s- m_{\tilde{W}_j}^2 - m_{f'}^2) \over (s-m_{\tilde{f}_1}^2)(s-m_{\tilde{f}_2}^2)},
\end{equation}
The expression for the overall contribution as a product of these couplings and integrals is:
\begin{equation} 
\begin{aligned}
\Gamma_{\tilde{f}_1 \tilde{f}_2} = - \Big[ & (-1)^{\theta_i}(-1)^{\theta_c}\mathcal{V}_{\tilde{f}_1 \tilde{f}_2}^{(1)} \mathcal{V}_{\tilde{f}_1 \tilde{f}_2}^{(3)} I_{\tilde{f}_1 \tilde{f}_2}^{4} + 2(-1)^{\theta_i}\mathcal{V}_{\tilde{f}_1 \tilde{f}_2}^{(1)} \mathcal{V}_{\tilde{f}_1 \tilde{f}_2}^{(4)} m_{f} |m_{\tilde{W}_j}| I_{\tilde{f}_1 \tilde{f}_2}^{2} \\ & + 2(-1)^{\theta_c} \mathcal{V}_{\tilde{f}_1 \tilde{f}_2}^{(2)} \mathcal{V}_{\tilde{f}_1 \tilde{f}_2}^{(3)} |m_{\tilde{Z}_i}| m_{f'} I_{\tilde{f}_1 \tilde{f}_2}^{3} + 4 \mathcal{V}_{\tilde{f}_1 \tilde{f}_2}^{(2)} \mathcal{V}_{\tilde{f}_1 \tilde{f}_2}^{(4)} |m_{\tilde{Z}_i}||m_{\tilde{W}_j}|m_{f} m_{f'} I_{\tilde{f}_1 \tilde{f}_2}^{1}) \Big].
\end{aligned}
\end{equation}

It should be noted the partial widths of the decays $\tilde{Z}_i \rightarrow \tilde{W}_j f' \bar{f}$ and the ``reverse'' decay $\tilde{W}_j \rightarrow \tilde{Z}_i f' \bar{f}$ may show strong dependence on the quark masses taken for kinematic masses and for the running masses (which is used to set Yukawa couplings), of course depending on the details of the exact spectrum considered. These mass choice effects, along with the fact that {\tt sPHENO} allows (small) mixing in the first two generations of sfermions whereas it is neglected in {\tt SOFTSUSY}, can cause larger differences between {\tt SOFTSUSY} and {\tt sPHENO} of around $25\%$. If the same mass choices are made and {\tt sPHENO}'s small mixing angles inserted by hand into the {\tt SOFTSUSY} code then these differences are reduced to around $10\%$.

\subsection{Chargino $1 \rightarrow 3$ Decays}

\textbf{\underline{$\tilde{W}_j \rightarrow \tilde{Z}_i f' \bar{f}$}}

It is detailed above in the formulae for $\tilde{Z}_i \rightarrow \tilde{W}_j f' \bar{f}$ how to adapt the formula for the chargino decaying into the neutralino rather than the neutralino decaying into the chargino.

\section{Decays to Gravitinos} \label{appendix:gravitinos}

In certain SUSY-breaking scenarios, particularly gauge-mediated SUSY breaking (GMSB), the gravitino can be very light and therefore may be the Lightest Supersymmetric Particle (LSP). Moreover, the gravitino has longitudinal components from the goldstino which couple much more strongly than gravitational strength, this therefore provides interactions relevant to collider phenomenology, resulting in gravitino-SUSY-SM couplings that affect collider signatures, when the gravitino is the LSP\@. Consequently Next-to-Lightest Supersymmetric Particle (NLSP) decays to gravitino LSPs may be of interest and are included within our decay calculator program {\tt SOFTSUSY}.
The following decay modes are relevant when the initial SUSY particle is the NLSP:
\begin{enumerate}
\item $\tilde{g} \rightarrow g \tilde{G}$
\item $\tilde{q}_{i} \rightarrow q \tilde{G}$
\item $\tilde{l} \rightarrow l \tilde{G}$
\item $\tilde{Z}_{i} \rightarrow \gamma \tilde{G}$
\item $\tilde{Z}_{i} \rightarrow Z \tilde{G}$
\item $\tilde{Z}_{i} \rightarrow \phi \tilde{G}$
\end{enumerate}
Decays of Higgs bosons to the gravitino are not relevant as there are always other decays available to the Higgs (whether $h$ or $A$) which dominate its branching ratio, for example even a Higgs boson were the NLSP then decays to $\gamma \gamma$ would be available and occur much more quickly than the Planck suppressed decays to gravitinos.
The formulae used were rederived but nonetheless are as provided in \cite{TataBaer}:
\begin{equation} \label{gluggrav}
\Gamma(\tilde{g} \rightarrow g \tilde{G}) = {m_{\tilde{g}}^5 \over 48 \pi m_{\tilde{G}}^2 {M_{P}^{red}}^2}.
\end{equation}
\begin{equation} \label{sqqgrav}
\Gamma(\tilde{q} \rightarrow q \tilde{G}) = {(m_{\tilde{q}}^2 - m_{q}^2)^4 \over 48 \pi m_{\tilde{q}}^3 m_{\tilde{G}}^2 {M_{P}^{red}}^2}.
\end{equation}
\begin{equation} \label{neutgamgrav}
\Gamma(\tilde{Z}_i \rightarrow \gamma \tilde{G}) = {|m_{\tilde{Z}_i}|^5 \over 48 \pi m_{\tilde{G}}^2 {M_{P}^{red}}^2}[N_{1i}\cos\theta_W + N_{2i}\sin\theta_W]^2.
\end{equation}
\begin{equation} \label{neutZgrav}
\Gamma(\tilde{Z}_i \rightarrow Z \tilde{G}) = {(m_{\tilde{Z}_i}^2 - m_{Z}^2)^4 \over 96 \pi m_{\tilde{G}}^2 {M_{P}^{red}}^2 |m_{\tilde{Z}_i}|^3}\left[2(N_{1i}\sin\theta_W - N_{2i}\cos\theta_W)^2 + (N_{4i}\sin\beta-N_{3i}\cos\beta)^2\right].
\end{equation}
\begin{equation} \label{neutphigrav}
\Gamma(\tilde{Z}_i \rightarrow \phi \tilde{G}) = {(m_{\tilde{Z}_i}^2 - m_{\phi}^2)^4 \over 96 \pi m_{\tilde{G}}^2 {M_{P}^{red}}^2 |m_{\tilde{Z}_i}|^3}\mathcal{C}_{h/H/A}^2,
\end{equation}
where
\begin{equation}
\mathcal{C}_{h/H/A} = \begin{cases} N_{4i}\cos\alpha - N_{3i}\sin\alpha $, for h, $ \\
									N_{4i}\sin\alpha + N_{3i}\cos\alpha $, for H, $ \\
									N_{4i}\cos\beta + N_{3i} \sin\beta $, for A. $
						\end{cases}
\end{equation}
Note that $M_{P}^{red} = {M_{P} \over \sqrt{8\pi}} \approx 2.4 \times 10^{18} $ GeV.

\section{NMSSM Decays} \label{appendix:NMSSMdec}
Only 2 body decays have been included in the NMSSM, this does however include
the important loop decays of the neutral Higgs bosons to $\gamma\gamma$,
$Z\gamma$ and $gg$, as well as QCD corrections to the neutral Higgs decays to
$q\bar{q}$ and $gg$. We do however include the decays $\phi \rightarrow WW^*
\rightarrow Wf'\bar{f}$ and $\phi \rightarrow ZZ^* \rightarrow
Zf\bar{f}$.\footnote{The decays $\phi \rightarrow WW^* \rightarrow Wf'\bar{f}$
  and $\phi \rightarrow ZZ^* \rightarrow Zf\bar{f}$, whilst strictly being 3
  body, are classified as having NDA (Number of Daughters) of 2 according to
  SLHA conventions.} 
Note that throughout $S(A,B)$ is the now 3x3 CP even Higgs mixing matrix and $P(A,B)$ is the  $2\times3$ CP odd Higgs mixing matrix (with goldstone excluded). The additional NMSSM variables include $\lambda$ (distinct from the $\tilde{\lambda}(A,B,C)$ used for kinematic part of decay widths given above), $\kappa$ and $\mu_{eff} = {\lambda \langle S \rangle \over \sqrt{2}}$. The conventions used for the NMSSM decay formulae were detailed earlier in section~\ref{NMSSMconventions}, with differences with respect to the {\tt SOFTSUSY} NMSSM manual \cite{Allanach:2013kza} noted. The conventions are those of {\tt NMSSMTools}~\cite{Ellwanger:2004xm,Ellwanger:2006rn,Ellwanger:2012dd,Ellwanger:2006ch}, against which the formulae were checked for consistency and which provided a useful guide.

The NMSSM simply involves the addition of a gauge singlet chiral superfield to the MSSM, therefore the NMSSM has an additional neutralino, additional CP even neutral Higgs and additional CP odd neutral Higgs, therefore any decays not involving the extended neutralino or extended Higgs sectors are as in the MSSM.

\subsection{CP Even Higgs Decays}

First the decay to a fermion and anti-fermion, with no QCD corrections (QCD corrected formulae given later in~\ref{appendix:QCDcorrdec}).
\begin{equation} \label{hiqqNMSSM}
\Gamma (h_i \rightarrow f \bar{f}) = {N_c G_F m_{h_i} \over \sqrt{2} 4 \pi} m_{q}^2 \Big(1-4{m_{q}^2 \over m_{h_i}^2}\Big)^{3 \over 2} \mathcal{A}_{h_i ff}^{NMSSM},
\end{equation}
where
\begin{equation}
\mathcal{A}_{h_i ff}^{NMSSM} = \begin{cases} {S(i,1) \over \sin\beta} $, for `u'-type fermions, $\\
											 {S(i,2) \over \cos\beta} $, for `d'-type fermions. $\\
								\end{cases}
\end{equation}

For squarks of the same handedness of the first two generations (so no mixing and negligible quark masses) the decay widths for the CP even Higgs i (i=1,2,3 are mass ordered CP even neutral Higgs bosons) are
\begin{equation} \label{hqLRqLRNMSSM}
\Gamma( h_{i} \rightarrow \tilde{q}_{L/R} \tilde{q}_{L/R} ) = {N_{c} \over 16\pi m_{h_i}} \tilde{\lambda}^{1 \over 2}(m_{{h}_i},m_{\tilde{q}_{L/R}},m_{\tilde{q}_{L/R}})\mathcal{C}^2,
\end{equation}
where
\begin{equation}
\mathcal{C} = \begin{cases} g\Big(m_{W}(\frac{1}{2}-{\tan^2 \theta_W \over 6})[\sin\beta S(i,1) - \cos\beta S(i,2)] - {m_{q}^2 S(i,1) \over m_{W} \sin\beta}\Big) $ , for up-type LH squarks,$ \\
							g\Big(m_{W}(-\frac{1}{2}-{\tan^2 \theta_W \over 6})[\sin\beta S(i,1) - \cos\beta S(i,2)] - {m_{q}^2 S(i,2) \over m_{W} \cos\beta}\Big) $ , for down-type LH squarks,$ \\
							g\Big({2 \over 3}m_{W}\tan^2 \theta_W[\sin\beta S(i,1) - \cos\beta S(i,2)] - {m_{q}^2 S(i,1) \over m_{W} \sin\beta}\Big) $ , for up-type RH squarks,$ \\
							g\Big({-1 \over 3}m_{W}\tan^2 \theta_W[\sin\beta S(i,1) - \cos\beta S(i,2)] - {m_{q}^2 S(i,2) \over m_{W} \cos\beta}\Big) $ , for down-type RH squarks$. \\
							\end{cases}
\end{equation}
For sleptons the same formulae apply but without the factor of 3 in the pre-factor from $N_{c}$ and with the coupling $\mathcal{C}$ now (note ``down type" here means the charged sleptons (i.e.\ sleptons of the electron, muon, tau)):

\begin{equation}
\mathcal{C} = \begin{cases} g\Big(m_{W}(\frac{1}{2}+{\tan^2 \theta_W \over 2})[\sin\beta S(i,1) - \cos\beta S(i,2)]\Big) $ , for sneutrinos (i.e.\ equivalent of up-type LH),$ \\
							g\Big(m_{W}(\frac{1}{2}-{\tan^2 \theta_W \over 6})[\sin\beta S(i,1) - \cos\beta S(i,2)] - {m_{q}^2 S(i,2) \over m_{W} \cos\beta}\Big) $ , for down-type LH sleptons, $ \\
							g\Big(m_{W}(\tan^2 \theta_W[\sin\beta S(i,1) - \cos\beta S(i,2)] - {m_{q}^2 S(i,1) \over m_{W} \cos\beta}\Big) $ , for down-type RH sleptons$. \\
							\end{cases}
\end{equation}
For squarks of opposite handedness:
\begin{equation} \label{hqLRqRLNMSSM}
\Gamma( h_{i} \rightarrow \tilde{q}_{L/R} \tilde{q}_{R/L} ) = {N_{c} \over 16\pi m_{h_i}} \tilde{\lambda}^{1 \over 2}(m_{{h}_i},m_{\tilde{q}_L},m_{\tilde{q}_R})\mathcal{D}^2,
\end{equation}
where
\begin{equation}
\mathcal{D} = \begin{cases} {g m_{q} \over 2 m_{W} \sin\beta} [A_{q} S(i,1) - \mu_{eff}S(i,2) - \lambda {\sqrt{2} m_{W} \cos\beta \over g S(i,3)}]  $ , for up-type squarks of different handedness,$ \\
							{g m_{q} \over 2 m_{W} \cos\beta} [A_{q} S(i,2) - \mu_{eff}S(i,1) - \lambda {\sqrt{2} m_{W} \sin\beta \over g S(i,3)}]  $ , for down-type squarks of different handedness$. \\
							\end{cases}
\end{equation}
For the decay to two sleptons with different handedness then we can use the same formulae as for two squarks of opposite handedness above but dividing by 3 to account for the fact sleptons aren't coloured (i.e.\ no factor $N_{c}$). Note as only LH sneutrinos exist, only decays to charged sleptons of opposite handedness are possible, i.e.\ only ``down-type sleptons" of different handedness.
For third generation squarks and sleptons the formulae are more complicated ($j = 1,2$ indicates $\tilde{t}_1$ and $\tilde{t}_2$):
\begin{equation} \label{hststNMSSM}
\Gamma( h_{i} \rightarrow \tilde{t}_{j} \tilde{t}_{j} ) = {3 \over 16\pi m_{h_i}} \tilde{\lambda}^{1 \over 2}(m_{{h}_i},m_{\tilde{t}_j},m_{\tilde{t}_j})\mathcal{C}_{t_{j}t_{j}}^2,
\end{equation}
where
\begin{equation}
\begin{aligned}
\mathcal{C}_{t_{1}t_{1}} = & \cos^2\theta_t \sqrt{2}\Big[h_{u}^2 \langle h_{1} \rangle S(i,1) + ({{g'}^{2} \over 12} - {g^2 \over 4})\{\langle h_{1} \rangle S(i,1) - \langle h_{2} \rangle S(i,2)\}\Big] \\ & + \sin^2 \theta_t \sqrt{2} \Big[h_{u}^2 \langle h_{1} \rangle S(i,1) - {{g'}^2 \over 3}\{\langle h_{1} \rangle S(i,1) - \langle h_{2} \rangle S(i,2)\}\Big] \\ & + 2\cos\theta_t \sin\theta_t {h_{u} \over \sqrt{2}}\Big(A_t S(i,1) - \mu_{eff} S(i,2) - \lambda \langle h_{2} \rangle S(i,3)\Big).
\end{aligned}
\end{equation}
This was for $j=1$ i.e.\ for $\tilde{t}_1 \tilde{t}_1$. For $j = 2$, make the replacements $\cos\theta_t \rightarrow -\sin\theta_t$ and $\sin\theta_t \rightarrow \cos\theta_t$.

For different stops:
\begin{equation} \label{hst1st2NMSSM}
\Gamma( h_{i} \rightarrow \tilde{t}_{1} \tilde{t}_{2} ) = {3 \over 16\pi m_{h_i}} \tilde{\lambda}^{1 \over 2}(m_{{h}_i},m_{\tilde{t}_j},m_{\tilde{t}_j})\mathcal{C}_{t_{1}t_{2}}^2,
\end{equation}
where
\begin{equation}
\begin{aligned}
\mathcal{C}_{t_{1}t_{2}} = & \cos\theta_t \sin\theta_t \Big[\sqrt{2} \Big(h_{u}^2 \langle h_{1} \rangle S(i,1) - {{g'}^2 \over 3}\{\langle h_{1} \rangle S(i,1) - \langle h_{2} \rangle S(i,2)\}\Big) \\ & - \sqrt{2}\Big(h_{u}^2 \langle h_{1} \rangle S(i,1)  + ({{g'}^2 \over 12} - {g^2 \over 4})\{\langle h_{1} \rangle S(i,1) - \langle h_{2} \rangle S(i,2)\}\Big)\Big] \\ & + (\cos^2 \theta_t - \sin^2 \theta_t) {h_{u} \over \sqrt{2}} \Big(A_{t} S(i,1) - \mu_{eff} S(i,2) - \lambda \langle h_{2} \rangle S(i,3)\Big).
\end{aligned}
\end{equation}
Note $\langle h_{1} \rangle = {\sqrt{2} m_{W} \sin\beta \over g}$ and $\langle h_{2} \rangle = {\sqrt{2} m_{W} \cos\beta \over g}$, whilst $h_{u} = {m_{t}^{run} \over \langle h_{1} \rangle}$. \label{vevsh}
For decays to sbottoms the decay formulae are the same (with the expected mass changes) except:
\begin{equation}
\begin{aligned}
\mathcal{C}_{b_{1}b_{1}} = & \cos^2\theta_b \sqrt{2}\Big[h_{d}^2 \langle h_{2} \rangle S(i,2) + ({{g'}^{2} \over 12} + {g^2 \over 4})\{\langle h_{1} \rangle S(i,1) - \langle h_{2} \rangle S(i,2)\}\Big] \\ & + \sin^2 \theta_b \sqrt{2} \Big[h_{d}^2 \langle h_{2} \rangle S(i,2) + {{g'}^2 \over 6}\{\langle h_{1} \rangle S(i,1) - \langle h_{2} \rangle S(i,2)\}\Big] \\ & + 2\cos\theta_b \sin\theta_b {h_{d} \over \sqrt{2}}\Big(A_b S(i,2) - \mu_{eff} S(i,1) - \lambda \langle h_{1} \rangle S(i,3)\Big),
\end{aligned}
\end{equation}
and
\begin{equation}
\begin{aligned}
\mathcal{C}_{b_{1}b_{2}} = & \cos\theta_b \sin\theta_b \Big[\sqrt{2} \Big(h_{d}^2 \langle h_{2} \rangle S(i,2) + {{g'}^2 \over 6}\{\langle h_{1} \rangle S(i,1) - \langle h_{2} \rangle S(i,2)\}\Big) \\ & - \sqrt{2}\Big(h_{d}^2 \langle h_{2} \rangle S(i,2)  + ({{g'}^2 \over 12} + {g^2 \over 4})\{\langle h_{1} \rangle S(i,1) - \langle h_{2} \rangle S(i,2)\}\Big)\Big] \\ & + (\cos^2 \theta_b - \sin^2 \theta_b) {h_{d} \over \sqrt{2}} \Big(A_{b} S(i,2) - \mu_{eff} S(i,1) - \lambda \langle h_{1} \rangle S(i,3)\Big).
\end{aligned}
\end{equation}
Note $h_{d} = {m_{b}^{run} \over \langle h_{2} \rangle}$. Again to get the $\tilde{b}_2 \tilde{b}_2$ coupling from the $\tilde{b}_1 \tilde{b}_1$, make the changes $\cos\theta_b \rightarrow -\sin\theta_b$ and $\sin\theta_b \rightarrow \cos\theta_b$.
For staus:
\begin{equation} \label{hstaustauNMSSM}
\Gamma( h_{i} \rightarrow \tilde{\tau}_{j} \tilde{\tau}_{k} ) = {1 \over 16\pi m_{h_i}} \tilde{\lambda}^{1 \over 2}(m_{{h}_i},m_{\tilde{\tau}_j},m_{\tilde{\tau}_k})\mathcal{C}_{\tau_{j}\tau_{k}}^2.
\end{equation}
\begin{equation}
\begin{aligned}
\mathcal{C}_{\tau_{1}\tau_{1}} = & \sin^2\theta_\tau \sqrt{2}\Big[h_{\tau}^2 \langle h_{2} \rangle S(i,2) + (-{{g'}^{2} \over 4} + {g^2 \over 4})\{\langle h_{1} \rangle S(i,1) - \langle h_{2} \rangle S(i,2)\}\Big] \\ & + \cos^2 \theta_\tau \sqrt{2} \Big[h_{\tau}^2 \langle h_{2} \rangle S(i,2) + {{g'}^2 \over 2}\{\langle h_{1} \rangle S(i,1) - \langle h_{2} \rangle S(i,2)\}\Big] \\ & + 2\cos\theta_\tau \sin\theta_\tau {h_{\tau} \over \sqrt{2}}\Big(A_{\tau} S(i,2) - \mu_{eff} S(i,1) - \lambda \langle h_{1} \rangle S(i,3)\Big),
\end{aligned}
\end{equation}
\begin{equation}
\begin{aligned}
\mathcal{C}_{\tau_{2}\tau_{2}} = & \cos^2\theta_\tau \sqrt{2}\Big[h_{\tau}^2 \langle h_{2} \rangle S(i,2) + (-{{g'}^{2} \over 4} + {g^2 \over 4})\{\langle h_{1} \rangle S(i,1) - \langle h_{2} \rangle S(i,2)\}\Big] \\ & + \sin^2 \theta_\tau \sqrt{2} \Big[h_{\tau}^2 \langle h_{2} \rangle S(i,2) + {{g'}^2 \over 2}\{\langle h_{1} \rangle S(i,1) - \langle h_{2} \rangle S(i,2)\}\Big] \\ & - 2\cos\theta_\tau \sin\theta_\tau {h_{\tau} \over \sqrt{2}}\Big(A_{\tau} S(i,2) - \mu_{eff} S(i,1) - \lambda \langle h_{1} \rangle S(i,3)\Big), 
\end{aligned}
\end{equation}
and 
\begin{equation}
\begin{aligned}
\mathcal{C}_{\tau_{1}\tau_{2}} = & -\cos\theta_\tau \sin\theta_\tau \Big[\sqrt{2} \Big(h_{\tau}^2 \langle h_{2} \rangle S(i,2) + {{g'}^2 \over 2}\{\langle h_{1} \rangle S(i,1) - \langle h_{2} \rangle S(i,2)\}\Big) \\ & - \sqrt{2}\Big(h_{\tau}^2 \langle h_{2} \rangle S(i,2)  + (-{{g'}^2 \over 4} + {g^2 \over 4})\{\langle h_{1} \rangle S(i,1) - \langle h_{2} \rangle S(i,2)\}\Big)\Big] \\ & + (\sin^2 \theta_\tau - \cos^2 \theta_\tau) {h_{\tau} \over \sqrt{2}} \Big(A_{\tau} S(i,2) - \mu_{eff} S(i,1) - \lambda \langle h_{1} \rangle S(i,3)\Big).
\end{aligned}
\end{equation}
Now the decays to charginos, first of all decays to the same chargino:
\begin{equation} \label{hchchNMSSM}
\Gamma( h_{i} \rightarrow \tilde{W}_j \tilde{W}_j) = {m_{h_i} \over 8\pi} \tilde{\lambda}^{3 \over 2}(m_{h_i},m_{\tilde{W}_j},m_{\tilde{W}_j})\mathcal{F}_{jj}^2,
\end{equation}
where
\begin{equation}
\mathcal{F}_{jj} = \begin{cases}
					{\lambda \over \sqrt{2}}S(i,3)\cos\theta_L \cos\theta_R + {g \over \sqrt{2}}[S(i,1)\sin\theta_L \cos\theta_R + S(i,2)\cos\theta_L \sin\theta_R] $, for $j=1$, $ \\
					{\lambda \over \sqrt{2}}S(i,3)\sin\theta_L \sin\theta_R - {g \over \sqrt{2}}[S(i,1)\cos\theta_L \sin\theta_R + S(i,2)\sin\theta_L \cos\theta_R] $, for $j=2$. $ \\
					\end{cases}
\end{equation}
For decays to different charginos:
\begin{equation} \label{hch1ch2NMSSM}
\Gamma( h_{i} \rightarrow \tilde{W}_1 \tilde{W}_2) = {m_{h_i} \over 16\pi} \tilde{\lambda}^{1 \over 2}(m_{h_i},m_{\tilde{W}_1},m_{\tilde{W}_2})\left[(c_{1}^2 + c_{2}^2){1 \over m_{h_i}^2}(m_{h_i}^2- m_{\tilde{W}_1}^2 - m_{\tilde{W}_2}^2) - 4c_{1}c_{2}{m_{\tilde{W}_1}m_{\tilde{W}_2} \over m_{h_i}^2}\right] ,
\end{equation}
where
\begin{equation}
c_{1} = {\lambda \over \sqrt{2}}S(i,3)\cos\theta_L \sin\theta_R + {g \over \sqrt{2}}\left(S(i,1)\sin\theta_L \sin\theta_R - S(i,2)\cos\theta_L \cos\theta_R\right),
\end{equation}
\begin{equation}
c_{2} = {\lambda \over \sqrt{2}}S(i,3)\sin\theta_L \cos\theta_R - {g \over \sqrt{2}}\left(S(i,1)\cos\theta_L \cos\theta_R - S(i,2)\sin\theta_L \sin\theta_R\right).
\end{equation}
Now the decays to neutralinos:
\begin{equation} \label{hneutneutNMSSM}
\Gamma(h_{i} \rightarrow \tilde{Z}_j \tilde{Z}_k) = {\alpha_{jk} m_{h_i} \over 16 \pi} (1 - ({m_{\tilde{Z}_j} + m_{\tilde{Z}_k} \over m_{h_i}})^2) \tilde{\lambda}^{1 \over 2}(m_{h_i},m_{\tilde{Z}_j},m_{\tilde{Z}_k}) \mathcal{G}_{ijk}^2,
\end{equation}
where
\begin{equation}
\begin{aligned}
\mathcal{G}_{ijk} = & {\lambda \over \sqrt{2}}\Big[S(i,1)(N_{3j}N_{5k} + N_{5j}N_{3k}) + S(i,2)(N_{4j}N_{5k} + N_{5j}N_{4k}) \\ & + S(i,3)(N_{3j}N_{4k} + N_{4j}N_{3k})\Big]  - \sqrt{2}\kappa S(i,3)N_{5j}N_{5k} \\ & + {g' \over 2}\Big[-S(i,1)(N_{1j}N_{4k} + N_{1k}N_{4j}) \\ & + S(i,2)(N_{1j}N_{3k}+N_{3j}N_{1k})\Big] + {g \over 2}\Big[S(i,1)(N_{2j}N_{4k} + N_{4j}N_{2k}) \\ & - S(i,2)(N_{2j}N_{3k}+N_{3j}N_{2k})\Big].
\end{aligned}
\end{equation}
Here $N_{ab}$ is the neutralino mixing matrix which is now $5\times5$ as the singlino mixes with the four original neutralinos. The neutralinos here are in order of increasing mass. The conventions for the NMSSM were detailed previously in section~\ref{NMSSMconventions}. Note the $\alpha_{jk}$ is 2 if $j\neq k$ and 1 if $j=k$ in order to account for indistinguishability of particles.

The neutral Higgs decays to CP odd neutral Higgs bosons are given by:
\begin{equation} \label{hAANMSSM}
\Gamma(h_{i} \rightarrow A_{j} A_{k}) = {1 \over 16 \pi m_{h_i}} \tilde{\lambda}^{1 \over 2}(m_{h_i},m_{A_j},m_{A_k}) \mathcal{Q}_{jk}^2,
\end{equation}
where 
\begin{equation}
\begin{aligned}
\mathcal{Q}_{jk} = & {g^2 + {g'}^2 \over 4\sqrt{2}}\Big[\langle h_{1} \rangle \{ \mathcal{C}(i,j,k,1,1,1) - \mathcal{C}(i,j,k,1,2,2) \} + \langle h_{2} \rangle \{ \mathcal{C}(i,j,k,2,2,2) - \mathcal{C} (i,j,k,2,1,1)\}\Big] \\ & + {\lambda A_{\lambda} \over \sqrt{2}}\{ \mathcal{C}(i,j,k,1,2,3) +\mathcal{C}(i,j,k,2,1,3) + \mathcal{C}(i,j,k,3,1,2)\} - {\kappa A_{\kappa} \over \sqrt{2}}\mathcal{C}(i,j,k,3,3,3) \\ & + {\lambda^2 \over \sqrt{2}}\Big[\langle h_{1} \rangle \{\mathcal{C}(i,j,k,1,2,2) + \mathcal{C}(i,j,k,1,3,3)\} + \langle h_{2} \rangle \{ \mathcal{C}(i,j,k,2,1,1) + \mathcal{C}(i,j,k,2,3,3)\} \\ & + {\mu_{eff} \over \lambda}\{\mathcal{C}(i,j,k,3,1,1) + \mathcal{C}(i,j,k,3,2,2)\}\Big] + {\kappa^2 \sqrt{2} \mu_{eff} \over \lambda} \mathcal{C}(i,j,k,3,3,3) \\ & + {\lambda \kappa \over \sqrt{2}}\Big[\langle h_{1} \rangle \{\mathcal{C}(i,j,k,2,3,3) - 2\mathcal{C}(i,j,k,3,2,3)\} + \langle h_{2} \rangle \{\mathcal{C}(i,j,k,1,3,3) - 2\mathcal{C}(i,j,k,3,1,3)\} \\ & + 2 {\mu_{eff} \over \lambda}\{\mathcal{C}(i,j,k,3,1,2) - \mathcal{C}(i,j,k,1,2,3) - \mathcal{C}(i,j,k,2,1,3)\}\Big].
\end{aligned}
\end{equation}
The $\mathcal{C}(i,j,k,x,y,z)$ is the same coupling which appears later in \eqref{CAA2hi}.
\begin{equation} \label{hAZNMSSM}
\Gamma(h_i \rightarrow A_j Z) = {(g^2 + {g'}^2) m_{h_i}^3 \over 64\pi m_{Z}^2}\tilde{\lambda}^{3 \over 2}(m_{h_i},m_{A_j},m_Z) \mathcal{R}_{ij}^2,
\end{equation}
where
\begin{equation}
\mathcal{R}_{ij} = (S(i,1)\cos\beta - S(i,2)\sin\beta)(P(j,1)\cos\beta - P(j,2)\sin\beta).
\end{equation}
\begin{equation} \label{hHpmHpmNMSSM}
\Gamma(h_i \rightarrow H^+ H^-) = {1 \over 16 \pi m_{h_i}} \tilde{\lambda}^{1 \over 2}(m_{h_i},m_{H^{\pm}},m_{H^{\pm}}) \mathcal{S}_{i}^2,
\end{equation}
where
\begin{equation}
\begin{aligned}
\mathcal{S}_i = & {\lambda \mu_{eff} \over \sqrt{2}}\Big[2S(i,3)\cos^2 \beta + 2S(i,3) \sin^2 \beta\Big] - {\lambda^2 m_{W} \sin\beta \over g}(2S(i,2) + 2S(i,1))\cos\beta \sin\beta \\ & + \mu_{eff} \kappa 2\sqrt{2}S(i,3)\cos\beta \sin\beta  + \lambda A_{\lambda} \sqrt{2} S(i,3)\cos\beta \sin\beta \\ & + {{g'}^2 \over 4}{m_{W} \over g}\Big[\sin\beta (2S(i,1)\cos^2 \beta - 2S(i,1)\sin^2 \beta) + \cos\beta(2S(i,2)\sin^2 \beta - 2S(i,2) \cos^2 \beta)\Big] \\ & + {g m_{W} \over 4}\Big[\sin\beta(2S(i,1) \sin^2 \beta - 2S(i,2)\cos^2 \beta + 4S(i,2)\sin\beta\cos\beta) \\ & + \cos\beta(2S(i,2) \cos^2 \beta + 2S(i,2)\sin^2 \beta + 4S(i,1)\sin\beta \cos\beta)\Big].
\end{aligned}
\end{equation}
\begin{equation} \label{hWHpmNMSSM}
\Gamma(h_{i} \rightarrow W^{\pm} H^{\pm}) = {G_{F} m_{h_i}^3 \over 8 \pi}[S(i,1)\cos\beta - S(i,2)\sin\beta]^2 \tilde{\lambda}^{3 \over 2} (m_{h_i},m_{W^{\pm}},m_{H^{\pm}}).
\end{equation}
In this equation a factor of 2 has been included as it could be either $W^+ H^-$ or $W^- H^+$.
\begin{equation} \label{hHH3NMSSM}
\Gamma(h_i \rightarrow h_j h_k) = {\alpha_{jk} \over 32 \pi m_{h_i}} \tilde{\lambda}^{1 \over 2}(m_{h_{i}},m_{h_{j}}, m_{h_{k}}) [\mathcal{C}_{h_{i}h_{j}h_{k}}^{NMSSM}]^2 ,
\end{equation}
where again $\alpha_{jk} = 1$ if $j=k$ and $2$ otherwise. Also remember that $\mu_{eff} = {\lambda \langle s \rangle \over \sqrt{2}}$, $\langle h_1 \rangle = {\sqrt{2}m_W \sin\beta \over g}$ and $\langle h_2 \rangle = {\sqrt{2}m_W \cos\beta \over g}$, where $\lambda$ is a parameter of the NMSSM which couples the singlino to higgsinos.
Here $\mathcal{C}_{h_{i}h_{j}h_{k}}^{NMSSM}$ is given by:
\begin{equation}
\begin{aligned}
\mathcal{C}_{h_{i}h_{j}h_{k}}^{NMSSM} = & {g^2 + {g'}^2 \over 4\sqrt{2}}\Big[ \langle h_1 \rangle [S_{ijk}(1,1,1) - S_{ijk}(1,2,2)] + \langle h_2 \rangle [S_{ijk}(2,2,2) - S_{ijk}(2,1,1)]\Big] - {\lambda A_{\lambda} \over \sqrt{2}} S_{ijk}(1,2,3) + {\kappa A_{\kappa} \over 3 \sqrt{2}}S_{ijk}(3,3,3) \\ & + {\lambda^2 \over \sqrt{2}}\Big[ \langle h_1 \rangle [S_{ijk}(1,2,2) + S_{ijk}(1,3,3)] + \langle h_2 \rangle [S_{ijk}(2,1,1) + S_{ijk}(2,3,3)] + {\mu_{eff} \over \lambda}[S_{ijk}(3,1,1)+S_{ijk}(3,2,2)]\Big]\\ & + \kappa^2 \sqrt{2} {\mu_{eff} \over \lambda} S_{ijk}(3,3,3) - {\lambda \kappa \over \sqrt{2}}[\langle h_1 \rangle S_{ijk}(3,2,3) + \langle h_2 \rangle S_{ijk}(3,1,3) + 2{\mu_{eff} \over \lambda}S_{ijk}(1,2,3)],
\end{aligned}
\end{equation}
where $S_{ijk}(x,y,z)$ is just the symmetric combination of triples of $S$ matrix elements with each of $i,j,k$ with each of $x,y,z$, i.e.:
\begin{equation}
\begin{aligned}
S_{ijk}(x,y,z) = & S(i,x)S(j,y)S(k,z) + S(i,x)S(k,y)S(j,z) + S(j,x)S(i,y)S(k,z) \\ & + S(j,x)S(i,z)S(k,y) + S(k,x)S(i,y)S(j,z) + S(k,x)S(j,y)S(i,z).
\end{aligned}
\end{equation}
Decays to two vector bosons are complicated by the consideration of whether the Higgs boson mass is greater than twice the mass of the vector boson, just as they were complicated in the MSSM\@. Included in {\tt SOFTSUSY} for the NMSSM are the cases both where the Higgs has mass $m_{h/H/H3} > 2m_V $, and so decays to two on-shell vector bosons, and also the case where the Higgs has mass $m_V < m_{h/H/H3} < 2m_V$ so that it may only undergo a decay to one on-shell vector boson and one off-shell vector boson, which then decays into a fermion anti-fermion pair, i.e. $h/H/H3 \rightarrow WW^* \rightarrow Wf'\bar{f}$ and $h/H/H3 \rightarrow ZZ^* \rightarrow Zf\bar{f}$, exactly as were included for the MSSM\@.

As in the MSSM, first consider $m_V < m_{h/H/H3} < 2m_V$. The only difference compared with the MSSM formulae is in the couplings $c_{h/H/H3VV}$.
\begin{align} \label{hZZstarNMSSM}
\Gamma(h/H/H3 \rightarrow ZZ^*) &= {G_{F}^2 m_{h/H/H3} m_{W}^4 c_{h/H/H3VV}^2 \over 64 \pi^3 \cos^4\theta_W}
F(\epsilon_{Z}) \left[7 - {40 \over 3} \sin^2 \theta_W + {160 \over 9} \sin^4 \theta_W\right], \\ 
\Gamma(h/H/H3 \rightarrow WW^*) &= {3G_{F}^2 m_{W}^4 m_{h/H/H3} c_{h/H/H3VV}^2 \over 16 \pi^3}  F(\epsilon_{W}) , \label{hWWstarNMSSM}
\end{align}
where here
\begin{equation}
\epsilon_{V} = {m_V \over m_{h/H/H3}},
\end{equation}
\begin{align}
c_{hVV} &= S(1,1)\sin\beta + S(1,2)\cos\beta, \\
c_{HVV} &= S(2,1)\sin\beta + S(2,2)\cos\beta, \\
c_{H3VV} &= S(3,1)\sin\beta + S(3,2)\cos\beta,  
\end{align}
and as before
\begin{equation}
\begin{aligned}
F(\epsilon_{V}) = {3(1-8\epsilon_{V}^2 + 20\epsilon_{V}^4) \over \sqrt{4\epsilon_{V}^2-1}}\cos^{-1} \Big[{3\epsilon_{V}^2-1 \over 2\epsilon_{V}^3}\Big] - (1-\epsilon_{V}^2)\Big({47 \over 2}\epsilon_{V}^2 -{13 \over 2} + {1 \over \epsilon_{V}^2}\Big) - 3(1-6\epsilon_{V}^2+4\epsilon_{V}^4)\log(\epsilon_{V}).
\end{aligned}
\end{equation}
If however $m_{h/H/H3} > 2m_{V}$ then the decay to two on-shell vector bosons occurs instead and the formulae are then:
\begin{equation} \label{hWWNMSSM}
\Gamma(h/H/H3 \rightarrow WW) = {G_{F} m_{h/H/H3}^3 \over  8\pi \sqrt{2}} \tilde{\lambda}^{1 \over 2}(m_{h/H/H3},m_{W},m_{W})(1-r^2 + {3 \over 4}r^4)c_{h/H/H3WW}^2,
\end{equation}
\begin{equation} \label{hZZNMSSM}
\Gamma(h/H/H3 \rightarrow ZZ) = {G_{F} m_{h/H/H3}^3 \over 16\pi \sqrt{2}} \tilde{\lambda}^{1 \over 2}(m_{h/H/H3},m_{Z},m_{Z})(1-r^2 + {3 \over 4}r^4)c_{h/H/H3ZZ}^2.
\end{equation}
Remember $r = 2{m_{V} \over m_{h/H/H3}}$.

Now onto the loop decays of the neutral Higgs bosons in the NMSSM:
\begin{equation} \label{hgamgamNMSSM}
\Gamma(h_i \rightarrow \gamma \gamma) = {G_{F} \over \sqrt{2}} {m_{h_i}^3 \alpha_{em}^2 (m_{h_i}) \over 32 \pi^2} |M_{\gamma \gamma}|^2,
\end{equation}
where 
\begin{equation}
\begin{aligned}
|M_{\gamma \gamma}|^2 = & \Big[I_{t}^{r} + I_{b}^{r} + I_{c}^{r} + I_{\tau}^{r} + I_{\tilde{W}_{1}}^{r} + I_{\tilde{W}_2}^{r} + I_{W}^{r} + I_{H^{\pm}}^{r} + I_{\tilde{c}_L}^{r} + I_{\tilde{c}_R}^{r} + I_{\tilde{s}_L}^{r} + I_{\tilde{s}_R}^{r} + I_{\tilde{\mu}_L}^{r} + I_{\tilde{\mu}_R}^{r} + I_{\tilde{t}_1}^{r} + I_{\tilde{t}_2}^{r} \\ & + I_{\tilde{b}_1}^{r} + I_{\tilde{b}_2}^{r} + I_{\tilde{\tau}_1}^{r} + I_{\tilde{\tau}_2}^{r}\Big]^2 + \Big[I_{t}^{i} + I_{b}^{i} + I_{c}^{i} + I_{\tau}^{i} + I_{\tilde{W}_{1}}^{i} + I_{\tilde{W}_2}^{i} + I_{W}^{i} + I_{H^{\pm}}^{i} + I_{\tilde{c}_L}^{i} + I_{\tilde{c}_R}^{i} + I_{\tilde{s}_L}^{i} \\ & + I_{\tilde{s}_R}^{i} + I_{\tilde{\mu}_L}^{i} + I_{\tilde{\mu}_R}^{i} + I_{\tilde{t}_1}^{i} + I_{\tilde{t}_2}^{i} + I_{\tilde{b}_1}^{i} + I_{\tilde{b}_2}^{i} + I_{\tilde{\tau}_1}^{i} + I_{\tilde{\tau}_2}^{i}\Big]^2.
\end{aligned}
\end{equation}
The $I_{a}^{r/i}$ are given below and are the real (r) and imaginary (i) parts.
\begin{equation}
I_{a} = c_{a}k_{a},
\end{equation}
where the $c_{a}$ is the coupling for particle $a$ and the $k_{a}$ is the kinetic part for particle $a$. The coupling is real whilst the kinetic part may be complex, it is from the kinetic part therefore that we get real and imaginary contributions. The couplings of the various loop particles are:
\begin{equation}
c_{t} = {4 \over 3}{S(i,1) \over \sin\beta},
\end{equation}
\begin{equation}
c_{c} = {4 \over 3}{S(i,1) \over \sin\beta},
\end{equation}
\begin{equation}
c_{b} = {1 \over 3}{S(i,2) \over \cos\beta},
\end{equation}
\begin{equation}
c_{\tau} = {S(i,2) \over \cos\beta},
\end{equation}
\begin{equation}
c_{W} = S(i,1)\sin\beta + S(i,2)\cos\beta,
\end{equation}
\begin{equation}
c_{\tilde{W}_1} = {1 \over \sqrt{\sqrt{2} G_{F}} m_{\tilde{W}_1}}{\lambda \over \sqrt{2}}S(i,3)\cos\theta_L \cos\theta_R + {g \over \sqrt{2}}\left[S(i,1)\sin\theta_L \cos\theta_R + S(i,2) \cos\theta_L \sin\theta_R\right],
\end{equation}
\begin{equation}
c_{\tilde{W}_2} = {1 \over \sqrt{\sqrt{2} G_{F}} m_{\tilde{W}_2}}{\lambda \over \sqrt{2}}S(i,3)\sin\theta_L \sin\theta_R - {g \over \sqrt{2}}\left[S(i,1)\cos\theta_L \sin\theta_R + S(i,2) \sin\theta_L \cos\theta_R\right],
\end{equation}
\begin{align}
c_{\tilde{c}_L} &= {4 \over 3} {2 m_{W} \over g m_{\tilde{c}_L}^2} [{{g'}^2 \over 12} + {g^2 \over 4}]{2 m_{W} \over g}(\sin\beta S(i,1) - \cos\beta S(i,2)),  \\
c_{\tilde{c}_R} &= {4 \over 3} {2 m_{W} \over g m_{\tilde{c}_R}^2} {{g'}^2 \over 6}{2 m_{W} \over g}(\sin\beta S(i,1) - \cos\beta S(i,2)), \\
c_{\tilde{s}_L} &= {1 \over 3} {2 m_{W} \over g m_{\tilde{s}_L}^2} [{{g'}^2 \over 12} + {g^2 \over 4}]{2 m_{W} \over g}(\sin\beta S(i,1) - \cos\beta S(i,2)),  \\
c_{\tilde{s}_R} &= {1 \over 3} {2 m_{W} \over g m_{\tilde{s}_R}^2} {{g'}^2 \over 6}{2 m_{W} \over g}(\sin\beta S(i,1) - \cos\beta S(i,2)), \\
c_{\tilde{\mu}_L} &= {2 m_{W} \over g m_{\tilde{\mu}_L}^2} [{-{g'}^2 \over 4} + {g^2 \over 4}]{2 m_{W} \over g}(\sin\beta S(i,1) - \cos\beta S(i,2)),  \\
c_{\tilde{\mu}_R} &= {2 m_{W} \over g m_{\tilde{\mu}_R}^2} {{g'}^2 \over 2}{2 m_{W} \over g}(\sin\beta S(i,1) - \cos\beta S(i,2)),
\end{align}
\begin{equation}
\begin{aligned}
c_{\tilde{t}_1} = & {1 \over 2\sqrt{\sqrt{2}G_{F}} m_{\tilde{t}_1}^2}\Big\{\cos^2 \theta_t \sqrt{2}\Big[f_{t}^2 \sqrt{2} {m_{W} \sin\beta \over g} S(i,1) + ({{g'}^2 \over 12} - {g^2 \over 4}) \{{\sqrt{2} m_{W} \sin\beta \over g} S(i,1)  \\ & -{\sqrt{2} m_{W} \cos\beta \over g}S(i,2)\}\Big] + \sin^2 \theta_t \sqrt{2} \Big[ f_{t}^2 {\sqrt{2} m_{W} \sin\beta \over g}S(i,1) \\ & - {{g'}^2 \over 3}\{{\sqrt{2} m_{W} \sin\beta \over g}S(i,1) - {\sqrt{2} m_{W} \cos\beta \over g} S(i,2)\}\Big] \\ & + 2\sin\theta_t \cos\theta_t {f_{t} \over \sqrt{2}}\Big[A_{t}S(i,1) - \mu_{eff}S(i,2) - \lambda {\sqrt{2} m_{W} \cos\beta \over g}S(i,3)\Big]\Big\},
\end{aligned}
\end{equation}
\begin{equation}
\begin{aligned}
c_{\tilde{t}_2} = & {1 \over 2\sqrt{\sqrt{2}G_{F}} m_{\tilde{t}_2}^2}\Big\{\sin^2 \theta_t \sqrt{2}\Big[f_{t}^2 \sqrt{2} {m_{W} \sin\beta \over g} S(i,1) + ({{g'}^2 \over 12} - {g^2 \over 4}) \{{\sqrt{2} m_{W} \sin\beta \over g} S(i,1) \\ & -{\sqrt{2} m_{W} \cos\beta \over g}S(i,2)\}\Big] + \cos^2 \theta_t \sqrt{2} \Big[ f_{t}^2 {\sqrt{2} m_{W} \sin\beta \over g}S(i,1) \\ & - {{g'}^2 \over 3}\{{\sqrt{2} m_{W} \sin\beta \over g}S(i,1) - {\sqrt{2} m_{W} \cos\beta \over g} S(i,2)\}\Big] \\ & - 2\sin\theta_t \cos\theta_t {f_{t} \over \sqrt{2}}\Big[A_{t}S(i,1) - \mu_{eff}S(i,2) - \lambda {\sqrt{2} m_{W} \cos\beta \over g}S(i,3)\Big]\Big\},
\end{aligned}
\end{equation}
\begin{equation}
\begin{aligned}
c_{\tilde{b}_1} = & {1 \over 2\sqrt{\sqrt{2}G_{F}} m_{\tilde{b}_1}^2}\Big\{\cos^2 \theta_b \sqrt{2}\Big[f_{b}^2 \sqrt{2} {m_{W} \cos\beta \over g} S(i,2) + ({{g'}^2 \over 12} + {g^2 \over 4}) \{{\sqrt{2} m_{W} \sin\beta \over g} S(i,1) \\ & -{\sqrt{2} m_{W} \cos\beta \over g}S(i,2)\}\Big] + \sin^2 \theta_b \sqrt{2} \Big[ f_{b}^2 {\sqrt{2} m_{W} \cos\beta \over g}S(i,2) + {{g'}^2 \over 6}\{{\sqrt{2} m_{W} \sin\beta \over g}S(i,1) \\ & - {\sqrt{2} m_{W} \cos\beta \over g} S(i,2)\}\Big] + 2\sin\theta_b \cos\theta_b {f_{b} \over \sqrt{2}}\Big[A_{b}S(i,2) - \mu_{eff}S(i,1) - \lambda {\sqrt{2} m_{W} \sin\beta \over g}S(i,3)\Big]\Big\},
\end{aligned}
\end{equation}
\begin{equation}
\begin{aligned}
c_{\tilde{b}_2} = & {1 \over 2\sqrt{\sqrt{2}G_{F}} m_{\tilde{b}_2}^2}\Big\{\sin^2 \theta_b \sqrt{2}\Big[f_{b}^2 \sqrt{2} {m_{W} \cos\beta \over g} S(i,2) + ({{g'}^2 \over 12} + {g^2 \over 4}) \{{\sqrt{2} m_{W} \sin\beta \over g} S(i,1) \\ & -{\sqrt{2} m_{W} \cos\beta \over g}S(i,2)\}\Big] + \cos^2 \theta_b \sqrt{2} \Big[ f_{b}^2 {\sqrt{2} m_{W} \cos\beta \over g}S(i,2) + {{g'}^2 \over 6}\{{\sqrt{2} m_{W} \sin\beta \over g}S(i,1) \\ & - {\sqrt{2} m_{W} \cos\beta \over g} S(i,2)\}\Big] - 2\sin\theta_b \cos\theta_b {f_{b} \over \sqrt{2}}\Big[A_{b}S(i,2) - \mu_{eff}S(i,1) - \lambda {\sqrt{2} m_{W} \sin\beta \over g}S(i,3)\Big]\Big\},
\end{aligned}
\end{equation}
\begin{equation}
\begin{aligned}
c_{\tilde{\tau}_1} = & {1 \over 2\sqrt{\sqrt{2}G_{F}} m_{\tilde{\tau}_1}^2}\Big\{\cos^2 \theta_{\tau} \sqrt{2}\Big[f_{\tau}^2 \sqrt{2} {m_{W} \cos\beta \over g} S(i,2) + (-{{g'}^2 \over 4} + {g^2 \over 4}) \{{\sqrt{2} m_{W} \sin\beta \over g} S(i,1) \\ & -{\sqrt{2} m_{W} \cos\beta \over g}S(i,2)\}\Big] + \sin^2 \theta_{\tau} \sqrt{2} \Big[ f_{\tau}^2 {\sqrt{2} m_{W} \cos\beta \over g}S(i,2) + {{g'}^2 \over 2}\{{\sqrt{2} m_{W} \sin\beta \over g}S(i,1) \\ & - {\sqrt{2} m_{W} \cos\beta \over g} S(i,2)\}\Big] + 2\sin\theta_{\tau} \cos\theta_{\tau} {f_{\tau} \over \sqrt{2}}\Big[A_{\tau}S(i,2) - \mu_{eff}S(i,1) - \lambda {\sqrt{2} m_{W} \sin\beta \over g}S(i,3)\Big]\Big\},
\end{aligned}
\end{equation}
\begin{equation}
\begin{aligned}
c_{\tilde{\tau}_2} = & {1 \over 2\sqrt{\sqrt{2}G_{F}} m_{\tilde{\tau}_2}^2}\Big\{\sin^2 \theta_{\tau} \sqrt{2}\Big[f_{\tau}^2 \sqrt{2} {m_{W} \cos\beta \over g} S(i,2) + (-{{g'}^2 \over 4} + {g^2 \over 4}) \{{\sqrt{2} m_{W} \sin\beta \over g} S(i,1) \\ & -{\sqrt{2} m_{W} \cos\beta \over g}S(i,2)\}\Big] + \cos^2 \theta_{\tau} \sqrt{2} \Big[ f_{\tau}^2 {\sqrt{2} m_{W} \cos\beta \over g}S(i,2) + {{g'}^2 \over 2}\{{\sqrt{2} m_{W} \sin\beta \over g}S(i,1) \\ & - {\sqrt{2} m_{W} \cos\beta \over g} S(i,2)\}\Big] - 2\sin\theta_{\tau} \cos\theta_{\tau} {f_{\tau} \over \sqrt{2}}\Big[A_{\tau}S(i,2) - \mu_{eff}S(i,1) - \lambda {\sqrt{2} m_{W} \sin\beta \over g}S(i,3)\Big]\Big\},
\end{aligned}
\end{equation}
\begin{equation}
\begin{aligned}
c_{H^{\pm}} = & {\lambda \mu_{eff} \over \sqrt{2}}\Big[2S(i,3)\cos^2 \beta + 2S(i,3)\sin^2 \beta\Big] - \lambda^2 {m_{W} \sin\beta \over g}2S(i,2)\cos\beta\sin\beta  \\ & - {m_{W} \sin\beta \over g}2S(i,1)\cos\beta\sin\beta + \mu_{eff} \kappa 2\sqrt{2} S(i,3)\cos\beta\sin\beta + {\lambda A_{\lambda} \over \sqrt{2}} 2S(i,3)\cos\beta\sin\beta \\ & + {{g'}^2 \over 4}{m_{W} \over g}\Big[\sin\beta \{2S(i,1)\cos^2 \beta - 2S(i,1) \sin^2 \beta\}  + \cos\beta\{2S(i,2)\sin^2 \beta - 2S(i,2) \cos^2 \beta\}\Big] \\ & + {g m_{W} \over 4}\Big[\sin\beta \{2S(i,1) \cos^2 \beta + 2S(i,1) \sin^2 \beta + 4S(i,2)\sin\beta\cos\beta\} \\ & + \cos\beta\{2S(i,2)\cos^2  \beta + 2S(i,2) \sin^2 \beta + 4S(i,1)\sin\beta\cos\beta\}\Big],
\end{aligned}
\end{equation}
where we remember that
\begin{equation}
f_{t} = {g m_{t} \over \sqrt{2} m_{W} \sin\beta}, \quad \quad \\
f_{b} = {g m_{b} \over \sqrt{2} m_{W} \cos\beta}, \quad \quad \\
f_{\tau} = {g m_{\tau} \over \sqrt{2} m_{W} \cos\beta}. \\
\end{equation}
The kinetic parts meanwhile are as follows, they depend upon the $f({\tau})$ given in Eq.~\eqref{ftau}.

For fermions (spin ${1 \over 2}$), i.e.\ the quarks and the charginos:
\begin{equation}
k_{a} = 2 \tau_{a} [1-\tau_{a} f({\tau_{a}})].
\end{equation}

For scalars (e.g.\ sfermions and $H^{\pm}$):
\begin{equation}
k_{a} = \tau_{a} (\tau_{a} f({\tau_{a}) - 1}).
\end{equation}

For spin 1 (i.e\ $W^\pm$ bosons):
\begin{equation}
k_{a} = -[2 + 3\tau_{a} + 3\tau_{a}(2-\tau_{a})f({\tau_{a}})].
\end{equation}
That's all the information needed for $h_{i} \rightarrow \gamma \gamma$.

\begin{equation} \label{hZgamNMSSM}
\Gamma (h_{i} \rightarrow Z \gamma) = {G_{F} \over \sqrt{2}} {m_{h_i}^3 \alpha_{em}^2 (m_{h_i}) \over 64 \pi^3} \Big(1 - \big({m_{Z} \over m_{h_i}}\big)^2\Big)^3 |M_{Z \gamma}|^2,
\end{equation}
where
\begin{equation}
|M_{Z \gamma}|^2 = \Big(\mathcal{I}_{t}^{r} + \mathcal{I}_{b}^{r} + \mathcal{I}_{c}^{r} + \mathcal{I}_{\tilde{W}_1}^{r} + \mathcal{I}_{\tilde{W}_2}^{r} + \mathcal{I}_{W}^{r} + \mathcal{I}_{H^{\pm}}^{r}\Big)^2 + \Big(\mathcal{I}_{t}^{i} + \mathcal{I}_{b}^{i} + \mathcal{I}_{c}^{i} + \mathcal{I}_{\tilde{W}_1}^{i} + \mathcal{I}_{\tilde{W}_2}^{i} + \mathcal{I}_{W}^{i} + \mathcal{I}_{H^{\pm}}^{i}\Big)^2.
\end{equation}
Again $\mathcal{I}_{a}^{r/i}$ are the real and imaginary parts of each contribution and $\mathcal{I}_{a} = c_{a} k_{a}$, where now the $c_{a}$ and $k_{a}$ are different to above as this is now for decays to $Z \gamma$.
Now the couplings are:
\begin{equation}
c_{t} = -2(1-{8 \over 3}\sin^2 \theta_W){1 \over \sin\theta_W \cos\theta_W}{S(i,1) \over \sin\beta},
\end{equation}
\begin{equation}
c_{c} = -2(1-{8 \over 3}\sin^2 \theta_W){1 \over \sin\theta_W \cos\theta_W}{S(i,1) \over \sin\beta},
\end{equation}
\begin{equation}
c_{b} = (-1+{4 \over 3}\sin^2 \theta_W){1 \over \sin\theta_W \cos\theta_W}{S(i,2) \over \cos\beta},
\end{equation}
\begin{equation}
c_{W} = -{g \over g'} [S(i,1)\sin\beta + S(i,2)\cos\beta],
\end{equation}
\begin{equation}
\begin{aligned}
c_{\tilde{W}_1} = & {4 m_{W} \over m_{\tilde{W}_1} g \sin\theta_W \cos\theta_W}\Big[ {\lambda \over \sqrt{2}} S(i,3) \cos\theta_L \cos\theta_R + {g \over \sqrt{2}}\{S(i,1)\sin\theta_L \cos\theta_R + S(i,2)\cos\theta_L \sin\theta_R\}\Big] \\ & \times [-\sin^2 \theta_R - {1 \over 2} \cos^2 \theta_R + 2 \sin^2 \theta_W - \sin^2 \theta_L - {1 \over 2}\cos^2 \theta_L],
\end{aligned}
\end{equation}
\begin{equation}
\begin{aligned}
c_{\tilde{W}_2} = & {4 m_{W} \over m_{\tilde{W}_2} g \sin\theta_W \cos\theta_W}\Big[ {\lambda \over \sqrt{2}} S(i,3) \sin\theta_L \sin\theta_R - {g \over \sqrt{2}}\{S(i,1)\cos\theta_L \sin\theta_R + S(i,2)\sin\theta_L \cos\theta_R\}\Big]\\ & \times [-\cos^2 \theta_R - {1 \over 2} \sin^2 \theta_R + 2 \sin^2 \theta_W - \cos^2 \theta_L - {1 \over 2}\sin^2 \theta_L],
\end{aligned}
\end{equation}
\begin{equation}
\begin{aligned}
c_{H^{\pm}} = & (1 - 2\sin^2 \theta_W){1 \over 2 \sin\theta_W \cos\theta_W  m_{H^{\pm}}^2} {1 \over \sqrt{\sqrt{2} G_{F}}} \Big\{ {\lambda \mu_{eff} \over \sqrt{2}}[2S(i,3)\cos^2 \beta + 2 S(i,3) \sin^2 \beta] \\ & - {\lambda^2 m_{W} \sin\beta \over g}2S(i,2)\cos\beta \sin\beta - {\lambda^2 m_{W} \sin\beta \over g} 2S(i,1)\cos\beta\sin\beta + \mu_{eff}\kappa 2 \sqrt{2} S(i,3)\cos\beta\sin\beta \\ & + \lambda A_{\lambda} \sqrt{2} S(i,3)\cos\beta \sin\beta + {{g'}^2 \over 4}{m_{W} \over g}[\sin\beta(2S(i,1)\cos^2 \beta - 2S(i,1) \sin^2 \beta] + \cos\beta[2S(i,2)\sin^2 \beta \\ & - 2 S(i,2) \cos^2 \beta] + {g m_{W} \over 4} \Big(\sin\beta [2S(i,1)\cos^2 \beta + 2S(i,1)\sin^2 \beta + 4S(i,2)\sin\beta \cos\beta] \\ & + \cos\beta[2S(i,2)\cos^2 \beta + 2S(i,2) \sin^2 \beta + 4S(i,1)\sin\beta \cos\beta]\Big) \Big\}.
\end{aligned}
\end{equation}
Now the kinetic parts depend upon both $f({\tau_{a}})$ in Eq. \eqref{ftau}, $g({\tau_{a}})$ in Eq. \eqref{gtau} and $f({\tau_{aZ}})$, $g({\tau_{aZ}})$ where $\tau_{aZ} = 4({m_{a} \over m_{Z}})^2$ cf $\tau_{a} = 4({m_{a} \over m_{h_i}})^2$.

For the spin ${1 \over 2}$ particles (quarks or charginos):
\begin{equation}
k_{a} =  {\tau_{a} \tau_{aZ} \over 2(\tau_{a} - \tau_{aZ})} +  {(\tau_{a} \tau_{aZ})^2 \over 2(\tau_{a} - \tau_{aZ})^2 } [f({\tau_{a}}) - f({\tau_{aZ}})] +  {\tau_{a}^2 \tau_{aZ} \over (\tau_{a} - \tau_{aZ})^2}[g({\tau_{a}}) - g({\tau_{aZ}})] +  {\tau_{a} \tau_{aZ} \over 2(\tau_{a} - \tau_{aZ})}[f({\tau_{a}}) - f({\tau_{aZ}})].
\end{equation}
For scalars (charged Higgs bosons):
\begin{equation}
k_{a} = {\tau_{a} \tau_{aZ}  \over 2 ({\tau_{a}} - {\tau_{aZ}})} + {(\tau_{a} \tau_{aZ})^2 \over 2(\tau_{a} - \tau_{aZ})^2}[f({\tau_{a}})-f({\tau_{aZ}})] + {\tau_{a}^2 \tau_{aZ} \over (\tau_{a} - \tau_{aZ})^2}[g({\tau_{a}}) - g({\tau_{aZ}})].
\end{equation}
For the spin 1 bosons (W bosons):
\begin{equation}
\begin{aligned}
k_{a} = & -4(3-\tan^2 \theta_W){\tau_{a}\tau_{aZ} \over 2 (\tau_{a} - \tau_{aZ})}[f({\tau_{a}}) - f({\tau_{aZ}})] + \{(1 + {2 \over \tau_{a}})\tan^2 \theta_W - (5 + {2 \over \tau_{a}})\}\Big[{\tau_{a} \tau_{aZ} \over 2(\tau_{a} - \tau_{aZ})} \\ & + {(\tau_{a} \tau_{aZ})^2 \over 2 (\tau_{a}-\tau_{aZ})^2}[f({\tau_{a}}) - f({\tau_{aZ}})] + {\tau_{a}^2 \tau_{aZ} \over (\tau_{a} - \tau_{aZ})^2}[g({\tau_{a}}) - g({\tau_{aZ}})]\Big]. 	
\end{aligned}	
\end{equation}
That's all the information required for $h_{i} \rightarrow Z \gamma$.

Next consider gluon gluon:
\begin{equation} \label{hggNMSSM}
\Gamma(h_{i} \rightarrow gg) = {G_{F} m_{h_i}^3 \alpha_{s}^2 (m_{h_i}) \over 64 \sqrt{2} \pi^3} |M_{gg}|^2,
\end{equation}
where
\begin{equation}
\begin{aligned}
|M_{gg}|^2 = & [J_{t}^{r} + J_{b}^{r} + J_{c}^{r} + J_{\tilde{c}_L}^{r} + J_{\tilde{c}_R}^{r} + J_{\tilde{s}_L}^{r} + J_{\tilde{s}_R}^{r} + J_{\tilde{t}_1}^{r} + J_{\tilde{t}_2}^{r} + J_{\tilde{b}_1}^{r} + J_{\tilde{b}_2}^{r}]^2 \\ & + [J_{t}^{i} + J_{b}^{i} + J_{c}^{i} + J_{\tilde{c}_L}^{i} + J_{\tilde{c}_R}^{i} + J_{\tilde{s}_L}^{i} + J_{\tilde{s}_R}^{i} + J_{\tilde{t}_1}^{i} + J_{\tilde{t}_2}^{i} + J_{\tilde{b}_1}^{i} + J_{\tilde{b}_2}^{i}]^2.
\end{aligned}
\end{equation}
As usual each of the $J_a$ are given by $J_{a} = c_{a}k_{a}$, where the coupling parts $c_{a}$ and kinetic parts $k_{a}$ are given below:
\begin{align}
c_{t} &= {S(i,1) \over \sin\beta} \quad \quad c_{c} = {S(i,1) \over \sin\beta} \quad \quad c_{b} = {S(i,2) \over \cos\beta}, \\
c_{\tilde{c}_L} &= {2 m_{W} \over g m_{\tilde{c}_L}^2}\Big({{g'}^2 \over 12} + {g^2 \over 4}\Big){2 m_{W} \over g}[\sin\beta S(i,1) - \cos\beta S(i,2)], \\
c_{\tilde{c}_R} &= {2 m_{W} \over g m_{\tilde{c}_R}^2}{{g'}^2 \over 6}{2 m_{W} \over g}[\sin\beta S(i,1) - \cos\beta S(i,2)], \\
c_{\tilde{s}_L} &= {2 m_{W} \over g m_{\tilde{s}_L}^2}\Big({{g'}^2 \over 12} + {g^2 \over 4}\Big){2 m_{W} \over g}[\sin\beta S(i,1) - \cos\beta S(i,2)], \\
c_{\tilde{s}_R} &= {2 m_{W} \over g m_{\tilde{s}_R}^2}{{g'}^2 \over 6}{2 m_{W} \over g}[\sin\beta S(i,1) - \cos\beta S(i,2)], \\
\end{align}
\begin{equation}
\begin{aligned}
c_{\tilde{t}_1} = & {m_{W} \over g m_{\tilde{t}_1}^2}\Big[\cos^2 \theta_t \sqrt{2}\Big(f_{t}^2\sqrt{2} {m_{W} \sin\beta \over g}S(i,1) + ({{g'}^2 \over 12} - {g^2 \over 4})\{\sqrt{2} {m_{W} \over g}(S(i,1)\sin\beta - \cos\beta S(i,2))\}\Big) \\ & + \sin^2 \theta_t \sqrt{2} \Big(f_{t}^2 \sqrt{2} {m_{W} \sin\beta \over g}S(i,1) - {{g'}^2 \over 3}\{\sqrt{2} {m_{W} \over g}(\sin\beta S(i,1) - \cos\beta S(i,2))\}\Big) \\ & + 2\sin\theta_t \cos\theta_t {f_{t} \over \sqrt{2}}\Big(A_{t}S(i,1) - \mu_{eff}S(i,2) - \lambda\sqrt{2}{m_{W} \cos\beta \over g}S(i,3)\Big)\Big],
\end{aligned}
\end{equation}
\begin{equation}
\begin{aligned}
c_{\tilde{t}_2} = & {m_{W} \over g m_{\tilde{t}_2}^2}\Big[\sin^2 \theta_t \sqrt{2}\Big(f_{t}^2\sqrt{2} {m_{W} \sin\beta \over g}S(i,1) + ({{g'}^2 \over 12} - {g^2 \over 4})\{\sqrt{2} {m_{W} \over g}(S(i,1)\sin\beta - \cos\beta S(i,2))\}\Big) \\ & + \cos^2 \theta_t \sqrt{2} \Big(f_{t}^2 \sqrt{2} {m_{W} \sin\beta \over g}S(i,1) - {{g'}^2 \over 3}\{\sqrt{2} {m_{W} \over g}(\sin\beta S(i,1) - \cos\beta S(i,2))\}\Big) \\ & - 2\sin\theta_t \cos\theta_t {f_{t} \over \sqrt{2}}\Big(A_{t}S(i,1) - \mu_{eff}S(i,2) - \lambda\sqrt{2}{m_{W} \cos\beta \over g}S(i,3)\Big)\Big],
\end{aligned}
\end{equation}
\begin{equation}
\begin{aligned}
c_{\tilde{b}_1} = & {m_{W} \over g m_{\tilde{b}_1}^2}\Big[\cos^2 \theta_b \sqrt{2}\Big(f_{b}^2\sqrt{2} {m_{W} \cos\beta \over g}S(i,2) + ({{g'}^2 \over 12} + {g^2 \over 4})\{\sqrt{2} {m_{W} \over g}(S(i,1)\sin\beta - \cos\beta S(i,2))\}\Big) \\ & + \sin^2 \theta_b \sqrt{2} \Big(f_{b}^2 \sqrt{2} {m_{W} \cos\beta \over g}S(i,2) + {{g'}^2 \over 6}\{\sqrt{2} {m_{W} \over g}(\sin\beta S(i,1) - \cos\beta S(i,2))\}\Big) \\ & + 2\sin\theta_b \cos\theta_b {f_{b} \over \sqrt{2}}\Big(A_{b}S(i,2) - \mu_{eff}S(i,1) - \lambda\sqrt{2}{m_{W} \sin\beta \over g}S(i,3)\Big)\Big],
\end{aligned}
\end{equation}
\begin{equation}
\begin{aligned}
c_{\tilde{b}_2} = & {m_{W} \over g m_{\tilde{b}_2}^2}\Big[\sin^2 \theta_B \sqrt{2}\Big(f_{b}^2\sqrt{2} {m_{W} \cos\beta \over g}S(i,2) + ({{g'}^2 \over 12} + {g^2 \over 4})\{\sqrt{2} {m_{W} \over g}(S(i,1)\sin\beta - \cos\beta S(i,2))\}\Big) \\ & + \cos^2 \theta_b \sqrt{2} \Big(f_{b}^2 \sqrt{2} {m_{W} \cos\beta \over g}S(i,2) + {{g'}^2 \over 6}\{\sqrt{2} {m_{W} \over g}(\sin\beta S(i,1) - \cos\beta S(i,2))\}\Big) \\ & - 2\sin\theta_b \cos\theta_b {f_{b} \over \sqrt{2}}\Big(A_{b}S(i,2) - \mu_{eff}S(i,1) - \lambda\sqrt{2}{m_{W} \sin\beta \over g}S(i,3)\Big)\Big].
\end{aligned}
\end{equation}
The kinetic parts are then exactly as in the $h_{i} \rightarrow \gamma \gamma$ case.

\subsection{CP Odd Higgs Decays}

First of all decays to a quark and an anti-quark:
\begin{equation} \label{AqqNMSSM}
\Gamma(A_{i} \rightarrow q \bar{q}) = {3 G_{F} \over 4 \pi \sqrt {2}} m_{q}^2 m_{A_i} \sqrt{1- {4m_{q}^2 \over m_{A_i}^2}} \mathcal{A},
\end{equation}
where 
\begin{equation}
\mathcal{A} = \begin{cases}
				[{S(i,1) \over \sin\beta}]^2 $, for up-type quarks (u,c,t), $ \\
				[{S(i,2) \over \cos\beta}]^2 $, for down-type quarks (d,s,b)$.	\\
				\end{cases}
\end{equation}			
Use the same formulae for decays to leptons but divide by 3 as the 3 in the pre-factor is $N_c$.
Now decays to sfermions, because of CP conservation, decays can only go to sfermions of different handedness.
\begin{equation} \label{AqLqRNMSSM}
\Gamma(A_{i} \rightarrow \tilde{q}_L \tilde{q}_R^*) = {1 \over 16 \pi m_{A_i}} \tilde{\lambda}^{1 \over 2}(m_{A_i}, m_{\tilde{q}_L}, m_{\tilde{q}_R}) \mathcal{C}_{A_{i}\tilde{q}\tilde{q}}^2,
\end{equation}
where 
\begin{equation}
\mathcal{C}_{A_{i}\tilde{q}\tilde{q}} = \begin{cases}
										{f_{q} \over \sqrt{2}} [A_{\tilde{q}} P(i,1) + \mu_{eff}P(i,2) + \lambda \sqrt{2} {m_{W} \cos\beta \over g} P(i,3)] $, for $u$-type squarks,$ \\
										{f_{q} \over \sqrt{2}} [A_{\tilde{q}} P(i,2) + \mu_{eff}P(i,1) + \lambda \sqrt{2} {m_{W} \sin\beta \over g} P(i,3)] $, for $d$-type squarks$. \\
										\end{cases}
\end{equation}
Where you must remember that the expression for $f_{q}$ differs for up type and down type quarks, for example see \eqref{ftfb}.

Note, \eqref{AqLqRNMSSM} holds even for third generation squarks; as we see in the MSSM, in the NMSSM the decays of CP odd Higgs bosons to squarks are independent of the sfermion mixing angles.
The formulae for the decays to squarks also hold for decays to sleptons, but again one must divide by 3.

Decays to neutralinos:
\begin{equation} \label{AneutneutNMSSM}
\Gamma(A_i \rightarrow \tilde{Z}_j \tilde{Z}_k) = {m_{A_i} \over 16 \pi} \Big[1 - ({m_{\tilde{Z}_j} - m_{\tilde{Z}_k} \over m_{A_i}})^2\Big] \tilde{\lambda}^{1 \over 2}(m_{A_i},m_{\tilde{Z}_j}, m_{\tilde{Z}_k}) \alpha_{ij} \mathcal{C}_{A_i \tilde{Z}_j \tilde{Z}_k}^2,
\end{equation}
where, as for the CP even decays to neutralinos, the $\alpha_{jk}$ factor accounts for indistinguishability and so is 1 if $j=k$ (i.e.\ decay to two of the same neutralinos) and 2 if $j\neq k$ (i.e.\ decays to two different neutralino mass eigenstates).
The coupling is given by:
\begin{equation}
\begin{aligned}
\mathcal{C}_{A_i \tilde{Z}_j \tilde{Z}_k} = & {\lambda \over \sqrt{2}}\Big[P(i,1)(N_{3j}N_{5k} + N_{5j}N_{3k}) + P(i,2)(N_{4j}N_{5k} + N_{5j}N_{4k}) \\ & + P(i,3)(N_{3j}N_{4k} + N_{4j}N_{3k})\Big] - \sqrt{2}\kappa P(i,3)N_{5j}N_{5k} \\ & - \tan\theta_W {g \over 2}\Big[-P(i,1)(N_{1j}N_{4k} + N_{4j}N_{1k}) + P(i,2)(N_{1j}N_{3k} \\ & + N_{3j}N_{1k})\Big] - {g \over 2}\Big[P(i,1)(N_{2j}N_{4k} + N_{4j}N_{2k}) - P(i,2)(N_{2j}N_{3k} \\ & + N_{3j}N_{2k})\Big].
\end{aligned}
\end{equation}
Decays to charginos, first consider decays to the two of the same chargino:
\begin{equation} \label{AchchNMSSM}
\Gamma(A_i \rightarrow \tilde{W_{j}} \tilde{W_{j}}) = {m_{A_i} \over 8 \pi} \tilde{\lambda}^{1 \over 2}(m_{A_i},m_{\tilde{W}_j}, m_{\tilde{W}_j}) S^2,
\end{equation}
here
\begin{equation}
S = \begin{cases}
	{\lambda \over \sqrt{2}}P(i,3)\cos\theta_L \cos\theta_R - {g \over \sqrt{2}}[P(i,1)\sin\theta_L \cos \theta_R + P(i,2) \cos\theta_L \sin\theta_R] $, for $j=1, \\
	{\lambda \over \sqrt{2}}P(i,3)\sin\theta_L \sin\theta_R + {g \over \sqrt{2}}[P(i,1)\cos\theta_L \sin \theta_R + P(i,2) \sin\theta_L \cos\theta_R] $, for $j=2. \\
	\end{cases}
\end{equation}
Meanwhile for decays to different charginos:
\begin{equation} \label{Ach1ch2NMSSM}
\Gamma(A_i \rightarrow \tilde{W}_1 \tilde{W}_2) = {m_{A_i} \over 8 \pi} \tilde{\lambda}^{1 \over 2}(m_{A_i},m_{\tilde{W}_1},m_{\tilde{W}_2})\Big[(c_{1}^2 + c_{2}^2){1 \over m_{A_i}^2}(m_{A_i}^2 - m_{\tilde{W}_1}^2 - m_{\tilde{W}_2}^2) + 4c_{1}c_{2}{m_{\tilde{W}_1}m_{\tilde{W}_2} \over m_{A_i}^2}\Big],
\end{equation}

Here,
\begin{equation}
c_{1} = {\lambda \over \sqrt{2}}P(i,3)\cos\theta_L \sin\theta_R - {g \over \sqrt{2}}[P(i,1)\sin\theta_L \sin\theta_R - P(i,2)\cos\theta_L \cos\theta_R],
\end{equation}
\begin{equation}
c_{2} = {\lambda \over \sqrt{2}}P(i,3)\sin\theta_L \cos\theta_R + {g \over \sqrt{2}}[P(i,1)\cos\theta_L \cos\theta_R - P(i,2)\sin\theta_L \sin\theta_R].
\end{equation}
Decays to CP even neutral Higgs bosons and a $Z$ boson:
\begin{equation} \label{AhZNMSSM}
\Gamma(A_i \rightarrow h_j Z) = {G_{F} m_{A_i}^3 \over 8 \pi \sqrt{2}} \tilde{\lambda}^{3 \over 2}(m_{A_i},m_{h_j},m_{Z}) \mathcal{C}_{A_i h_j Z}^2,
\end{equation}
where
\begin{equation}
\mathcal{C}_{A_i h_j Z} = \begin{cases}
							[S(1,1)\cos\beta - S(1,2)\sin\beta]\cos\theta_A $, for $i=j=1$,$ \\
							[S(1,1)\cos\beta - S(1,2)\sin\beta]\sin\theta_A $, for $i=2$, $j=1$,$ \\
							[S(2,1)\cos\beta - S(2,2)\sin\beta]\cos\theta_A $, for $i=1$, $j=2$,$ \\
							[S(2,1)\cos\beta - S(2,2)\sin\beta]\sin\theta_A $, for $i=j=2$,$ \\
							[S(3,1)\cos\beta - S(3,2)\sin\beta]\cos\theta_A $, for $i=1$, $j=3$,$ \\
							[S(3,1)\cos\beta - S(3,2)\sin\beta]\sin\theta_A $, for $i=2$, $j=3. \\
							\end{cases}
\end{equation}
The decay of a CP odd neutral Higgs to a charged Higgs and a $W$ boson in the NMSSM is given by:
\begin{equation} \label{AHpmWNMSSM}
\Gamma(A \rightarrow H^{\pm} W^{\pm}) = {G_{F} m_{A}^3 \over 8 \pi \sqrt{2}} \tilde{\lambda}^{3 \over 2}(m_{A},m_{H^{\pm}},m_{W}) \cos^2 \theta_A,
\end{equation}
for $A_2$ undergoing the same decay transform $\cos\theta_A \rightarrow \sin\theta_A$ and $m_{A} \rightarrow m_{A_2}$.
\begin{equation} \label{A2AhiNMSSM}
\Gamma(A_2 \rightarrow A h_i) = {1 \over 16 \pi m_{h_i}} \tilde{\lambda}^{1 \over 2}(m_{A_2},m_{A},m_{h_i})[\mathcal{C}_{AA2h_i}^{NMSSM}]^2,
\end{equation}
where the coupling $\mathcal{C}_{AA2h_i}^{NMSSM}$ is:
\begin{equation}
\begin{aligned}
\mathcal{C}_{AA2h_i}^{NMSSM} = & {g^2 + {g'}^2 \over 4 \sqrt{2}} \Big[ \langle h_1 \rangle [C(i,1,2,1,1,1) -C(i,1,2,1,2,2)] + \langle h_2 \rangle [C(i,1,2,2,2,2) - C(i,1,2,2,1,1)] \Big] \\ & + {\lambda A_{\lambda} \over \sqrt{2}}  [ C(i,1,2,1,2,3) + C(i,1,2,2,1,3) + C(i,1,2,3,1,2)] - {\kappa A_{\kappa} \over \sqrt{2}} C(i,1,2,3,3,3) \\ & + {\lambda^2 \over \sqrt{2}} \Big [  \langle h_1 \rangle[C(i,1,2,1,2,2) + C(i,1,2,1,3,3)] + \langle h_2 \rangle [C(i,1,2,2,1,1) + C(i,1,2,2,3,3)] \\ & + \mu_{eff} \lambda [C(i,1,2,3,1,1) + C(i,1,2,3,2,2)] \Big]  + {\kappa^2 \sqrt{2} \mu_{eff} \over \lambda}C(i,1,2,3,3,3) \\ & + {\lambda \kappa \over \sqrt{2}}\Big[ \langle h_1 \rangle [C(i,1,2,2,3,3) - 2C(i,1,2,3,2,3)] + \langle h_2 \rangle [ C(i,1,2,1,3,3) - 2 C(i,1,2,3,1,3)] \\ & + 2{\mu_{eff} \over \lambda} [C(i,1,2,3,1,2)-C(i,1,2,1,2,3)-C(i,1,2,2,1,3)]\Big],
\end{aligned}
\end{equation}
where here $C(i,1,2,x,y,z)$ is notation for
\begin{equation} \label{CAA2hi}
C(i,1,2,x,y,z) = S(i,x)[P(1,y)P(2,z) + P(1,z)P(2,y)].
\end{equation}
Now the loop decays of the CP odd Higgs bosons. First consider decays to $\gamma\gamma$:
\begin{equation} \label{AgamgamNMSSM}
\Gamma(A_i \rightarrow \gamma \gamma) = {G_{F} m_{A_i}^3 \alpha_{em}^2 (m_{A_i})  \over 32 \pi^3 \sqrt{2}} |M_{A_i \gamma \gamma}|^2,
\end{equation}
where
\begin{equation}
|M_{A_i \gamma \gamma}|^2 = (\mathcal{J}_{t}^{r} + \mathcal{J}_{b}^{r} + \mathcal{J}_{c}^{r} + \mathcal{J}_{\tau}^{r} + \mathcal{J}_{\tilde{W}_1}^{r} + \mathcal{J}_{\tilde{W}_2}^{r})^2 + (\mathcal{J}_{t}^{i} + \mathcal{J}_{b}^{i} + \mathcal{J}_{c}^{i} + \mathcal{J}_{\tau}^{i} + \mathcal{J}_{\tilde{W}_1}^{i} + \mathcal{J}_{\tilde{W}_2}^{i})^2.
\end{equation}
The $\mathcal{J}_{a}^{r/i}$ here are the real and imaginary parts respectively of $c_{a}k_{a}$ where the $c_{a}$ and $k_{a}$ for this decay mode are given below:
\begin{equation}
c_{t} = {4 \over 3}{P(i,1) \over \sin\beta}, \quad \quad\\
c_{b} = {1 \over 3}{P(i,2) \over \cos\beta}, \quad \quad\\
c_{c} = {4 \over 3}{P(i,1) \over \sin\beta}, \quad \quad\\
c_{\tau} = {P(i,2) \over \cos\beta} ,\\
\end{equation}
\begin{equation}
c_{\tilde{W}_1} = {2 m_{W} \over g m_{\tilde{W}_1}} [{\lambda \over \sqrt{2}} P(i,3) \cos\theta_L \cos\theta_R - {g \over \sqrt{2}}(P(i,1)\sin\theta_L \cos\theta_R + P(i,2) \cos\theta_L \sin\theta_R)],
\end{equation}
\begin{equation}
c_{\tilde{W}_2} = {2 m_{W} \over g m_{\tilde{W}_2}} [{\lambda \over \sqrt{2}} P(i,3) \sin\theta_L \sin\theta_R + {g \over \sqrt{2}}(P(i,1)\cos\theta_L \sin\theta_R + P(i,2) \sin\theta_L \cos\theta_R)].
\end{equation}
Meanwhile the kinetic parts are, for the quarks or the charginos (as both are spin ${1 \over 2}$):
\begin{equation}
k_{a} = \tau_{a}f({\tau_{a}}).
\end{equation}
The next loop decay is to $Z\gamma$:
\begin{equation} \label{AZgamNMSSM}
\Gamma(A_i \rightarrow Z \gamma) = {G_{F} \over \sqrt{2}} {m_{A_i}^3 \alpha_{em}^2(m_{A_i}) \over 64 \pi^3} \Big(1 - ({m_{Z} \over m_{A_i}})^2 \Big) |M_{A_i Z \gamma}|^2,
\end{equation}
where
\begin{equation}
|M_{A_i Z \gamma}|^2 = (\mathcal{K}_{t}^{r} + \mathcal{K}_{b}^{r} +  \mathcal{K}_{c}^{r} +  \mathcal{K}_{\tilde{W}_1}^{r} + \mathcal{K}_{\tilde{W}_2}^{r})^2 + (\mathcal{K}_{t}^{i} + \mathcal{K}_{b}^{i} +  \mathcal{K}_{c}^{i} +  \mathcal{K}_{\tilde{W}_1}^{i} + \mathcal{K}_{\tilde{W}_2}^{i})^2.
\end{equation}
As usual each $\mathcal{K}_{a}^{r/i}$ is the real/imaginary part of $c_{a}k_{a}$ where the $c_{a}$ and $k_{a}$ for this mode are given below:
\begin{equation}
c_{t} = -2(1-{8 \over 3}\sin^2 \theta_W){1 \over \sin\theta_W \cos\theta_W}{P(i,1) \over \sin\beta}, \quad \quad \\
c_{c} = -2(1-{8 \over 3}\sin^2 \theta_W){1 \over \sin\theta_W \cos\theta_W}{P(i,1) \over \sin\beta}, \\
\end{equation}
\begin{equation}
c_{b} = (-1+{4 \over 3}\sin^2 \theta_W){1 \over \sin\theta_W \cos\theta_W}{P(i,2) \over \cos\beta} ,
\end{equation}
\begin{equation}
\begin{aligned}
c_{\tilde{W}_1} = & 4{m_{W} \over m_{\tilde{W}_1} g \sin\theta_W \cos\theta_W}\Big[-\sin^2 \theta_R - {1 \over 2}\cos^2 \theta_R - \sin^2 \theta_L - {1 \over 2}\cos^2 \theta_L + 2\sin^2 \theta_W\Big] \\ & \times \Big({\lambda \over \sqrt{2}}P(i,3)\cos\theta_L \cos\theta_R - {g \over \sqrt{2}}[P(i,1)\sin\theta_L \cos\theta_R + P(i,2)\cos\theta_L \sin\theta_R]\Big),
\end{aligned}
\end{equation}
\begin{equation}
\begin{aligned}
c_{\tilde{W}_2} = & 4{m_{W} \over m_{\tilde{W}_2} g \sin\theta_W \cos\theta_W}\Big[-\cos^2 \theta_R - {1 \over 2}\sin^2 \theta_R - \cos^2 \theta_L - {1 \over 2}\sin^2 \theta_L + 2\sin^2 \theta_W\Big] \\ & \times \Big({\lambda \over \sqrt{2}}P(i,3)\sin\theta_L \sin\theta_R + {g \over \sqrt{2}}[P(i,1)\cos\theta_L \sin\theta_R + P(i,2)\sin\theta_L \cos\theta_R]\Big),
\end{aligned}
\end{equation}
The kinetic parts are all of the following form, where remember $\tau_{aZ} = 4({m_{a} \over m_{Z}})^2$ cf $\tau_{a} = 4({m_{a} \over m_{\phi}})^2$:
\begin{equation}
k_{a} = {\tau_{a}\tau_{aZ} \over 2(\tau_{a} - \tau_{aZ})}[f(\tau_{a})-f(\tau_{aZ})].
\end{equation}
Finally, for the loop decay to $gg$:
\begin{equation} \label{AggNMSSM}
\Gamma(A_i \rightarrow g g) = {G_{F} \over \sqrt{2}} {m_{A_i}^3 \alpha_s^2(m_{A_i}) \over 16 \pi^3} |M_{A_i gg}|^2,
\end{equation}
where 
\begin{equation}
|M_{A_i gg}|^2 = (\mathcal{R}_{t}^{r} + \mathcal{R}_{b}^{r} + \mathcal{R}_{c}^{r})^2 + (\mathcal{R}_{t}^{i} + \mathcal{R}_{b}^{i} + \mathcal{R}_{c}^{i})^2.
\end{equation}
Again the $\mathcal{R}_{a}^{r/i}$ are the real and imaginary parts of $c_{a}k_{a}$, where for this mode they are:
\begin{equation}
c_{t} = {P(i,1) \over \sin\beta}, \quad \quad \\
c_{b} = {P(i,2) \over \cos\beta}, \quad \quad \\
c_{c} = {P(i,1) \over \sin\beta}. \\
\end{equation}
The kinetic parts are just:
\begin{equation}
k_{a} = \tau_{a} f({\tau_a}).
\end{equation}

\subsection{Decays into Higgs Bosons}
\begin{equation} \label{b2b1hNMSSM}
\Gamma(\tilde{b}_2 \rightarrow \tilde{b}_1 h_{i}) = {1 \over 16 \pi m_{\tilde{b}_2}}\tilde{\lambda}^{1 \over 2}(m_{\tilde{b}_2}, m_{\tilde{b}_1}, m_{h_i})\Big[\cos\theta_b \sin\theta_b (c_R - c_L) + (\cos^2 \theta_b - \sin^2 \theta_b) c_{LR}\Big]^2,
\end{equation}
where
\begin{align}
c_{L} &= -\sqrt{2}\left[f_{b}^2 \langle h_2 \rangle S(i,2) + ({{g'}^2 \over 12} + {g^2 \over 4})[\langle h_1 \rangle S(i,1) - \langle h_{2} \rangle S(i,2)]\right], \\
c_{R} &= -\sqrt{2}\left[f_{b}^2 \langle h_2 \rangle S(i,2) + ({{g'}^2 \over 6})[\langle h_1 \rangle S(i,1) - \langle h_{2} \rangle S(i,2)]\right], \\
c_{LR} &= -{f_{b} \over \sqrt{2}}\left[A_{b}S(i,2) - \mu_{eff}S(i,1)-\lambda \langle h_{1} \rangle S(i,3)\right].
\end{align}
\begin{equation} \label{t2t1hNMSSM}
\Gamma(\tilde{t}_2 \rightarrow \tilde{t}_1 h_{i}) = {1 \over 16 \pi m_{\tilde{t}_2}}\tilde{\lambda}^{1 \over 2}(m_{\tilde{t}_2}, m_{\tilde{t}_1}, m_{h_i})\Big[\cos\theta_t \sin\theta_t (c_R - c_L) + (\cos^2 \theta_t - \sin^2 \theta_t) c_{LR}\Big]^2,
\end{equation}
where
\begin{align}
c_{L} &= -\sqrt{2}\left[f_{t}^2 \langle h_1 \rangle S(i,1) + ({{g'}^2 \over 12} - {g^2 \over 4})[\langle h_1 \rangle S(i,1) - \langle h_{2} \rangle S(i,2)]\right], \\
c_{R} &= -\sqrt{2}\left[f_{t}^2 \langle h_1 \rangle S(i,1) - ({{g'}^2 \over 3})[\langle h_1 \rangle S(i,1) - \langle h_{2} \rangle S(i,2)]\right], \\
c_{LR} &= -{f_{t} \over \sqrt{2}}\left[A_{t}S(i,1) - \mu_{eff}S(i,2)-\lambda \langle h_{2} \rangle S(i,3)\right].
\end{align}
\begin{equation} \label{stau2stau1hNMSSM}
\Gamma(\tilde{\tau}_2 \rightarrow \tilde{\tau}_1 h_{i}) = {1 \over 16 \pi m_{\tilde{\tau}_2}}\tilde{\lambda}^{1 \over 2}(m_{\tilde{\tau}_2}, m_{\tilde{\tau}_1}, m_{h_i})\Big[\cos\theta_{\tau} \sin\theta_{\tau} (c_R - c_L) + (\cos^2 \theta_{\tau} - \sin^2 \theta_{\tau}) c_{LR}\Big]^2,
\end{equation}
where
\begin{align}
c_{L} &= -\sqrt{2}\left[f_{\tau}^2 \langle h_2 \rangle S(i,2) + ({-{g'}^2 \over 4} + {g^2 \over 4})[\langle h_1 \rangle S(i,1) - \langle h_{2} \rangle S(i,2)]\right], \\
c_{R} &= -\sqrt{2}\left[f_{\tau}^2 \langle h_2 \rangle S(i,2) + ({{g'}^2 \over 2})[\langle h_1 \rangle S(i,1) - \langle h_{2} \rangle S(i,2)]\right], \\
c_{LR} &= -{f_{\tau} \over \sqrt{2}}\left[A_{\tau}S(i,2) - \mu_{eff}S(i,1)-\lambda \langle h_{1} \rangle S(i,3)\right].
\end{align}
\begin{equation} \label{sb2sb1ANMSSM}
\Gamma(\tilde{b}_2 \rightarrow \tilde{b}_1 A_{i}) = {1 \over 16 \pi m_{\tilde{b}_2}} \tilde{\lambda}^{1 \over 2}(m_{\tilde{b}_2},m_{\tilde{b}_1},m_{\tilde{A}_i})[\cos^2 \theta_b - \sin^2 \theta_b]^2 A_{LR}^2,
\end{equation}
where
\begin{equation}
A_{LR} = {f_{b} \over \sqrt{2}} \left[A_{b}P(i,2) + \mu_{eff} P(i,1) + \lambda \langle h_{1} \rangle P(i,3)\right].
\end{equation}
For $\tilde{\tau}_2 \rightarrow \tilde{\tau}_1 A_{i}$ the formulae are the same except the changes $\theta_{b} \rightarrow \theta_{\tau}$, $m_b \rightarrow m_{\tau}$ and now $A_{LR}$ is given by:
\begin{equation}
A_{LR} = {f_{\tau} \over \sqrt{2}}\left[A_{\tau}P(i,2) + \mu_{eff}P(i,1) + \lambda \langle h_{1} \rangle P(i,3)\right].
\end{equation}
For $\tilde{t}_2 \rightarrow \tilde{t}_1 A_{i}$ the formulae are the same except the changes $\theta_b \rightarrow \theta_t$, $m_b \rightarrow m_t$ and now $A_{LR}$ is given by:
\begin{equation}
A_{LR} = {f_{t} \over \sqrt{2}}\left[A_{t}P(i,1) + \mu_{eff}P(i,2) + \lambda \langle h_{2} \rangle P(i,3)\right].
\end{equation}
The chargino decays to lighter charginos and a CP even neutral Higgs:
\begin{equation} \label{ch2ch1hNMSSM}
\Gamma(\tilde{W}_2 \rightarrow \tilde{W}_1 h_{i}) = {1 \over 32 \pi |m_{\tilde{W}_2}|} \tilde{\lambda}^{1/2}
(m_{\tilde{W}_2},m_{\tilde{W}_1},m_{h_i})\left[(c_{1}^2 + c_{2}^2)(m_{\tilde{W}_1}^2 + m_{\tilde{W}_2}^2 - m_{h_i}^2) + 4c_{1} c_{2} m_{\tilde{W}_1} m_{\tilde{W}_2}\right]^2,
\end{equation}
here the $c_{1}$ and $c_{2}$ are:
\begin{equation}
c_{1} = {\lambda \over \sqrt{2}}S(i,3)\cos\theta_L \sin\theta_R + {g \over \sqrt{2}}[S(i,1)\sin\theta_L \sin\theta_R - S(i,2) \cos\theta_L \cos\theta_R],
\end{equation}
\begin{equation}
c_{2} = {\lambda \over \sqrt{2}}S(i,3)\sin\theta_L \cos\theta_R - {g \over \sqrt{2}}[S(i,1)\cos\theta_L \cos\theta_R - S(i,2) \sin\theta_L \sin\theta_R].
\end{equation}
The chargino decays to lighter charginos and a CP odd neutral Higgs:
\begin{equation} \label{ch2ch1ANMSSM}
\Gamma(\tilde{W}_2 \rightarrow \tilde{W}_1 A_{i}) = {1 \over 32 \pi |m_{\tilde{W}_2}|} \tilde{\lambda}^{1/2}(m_{\tilde{W}_2},m_{\tilde{W}_1},m_{A_i})\left[(C_{1}^2 + C_{2}^2)(m_{\tilde{W}_1}^2 + m_{\tilde{W}_2}^2 - m_{h_i}^2) + 4C_{1} C_{2} m_{\tilde{W}_1} m_{\tilde{W}_2}\right]^2,
\end{equation}
here the $C_{1}$ and $C_{2}$ are:
\begin{equation}
C_{1} = {\lambda \over \sqrt{2}}P(i,3)\cos\theta_L \sin\theta_R - {g \over \sqrt{2}}[P(i,1)\sin\theta_L \sin\theta_R - P(i,2) \cos\theta_L \cos\theta_R],
\end{equation}
\begin{equation}
C_{2} = {\lambda \over \sqrt{2}}P(i,3)\sin\theta_L \cos\theta_R + {g \over \sqrt{2}}[P(i,1)\cos\theta_L \cos\theta_R -P(i,2) \sin\theta_L \sin\theta_R].
\end{equation}
The formulae for $\Gamma(H^{\pm} \rightarrow W h_{i})$ are just as above for $\Gamma(h_i \rightarrow W H^{\pm})$ in  \eqref{hWHpmNMSSM} but with $m_{h_i} \leftrightarrow m_{H^{\pm}}$. Similarly the  formulae for $\Gamma(H^{\pm} \rightarrow W A_{i})$ are just as above for $\Gamma(A_i \rightarrow W H^{\pm})$ in \eqref{AHpmWNMSSM} but with the replacement $m_{A_i} \leftrightarrow m_{H^{\pm}}$.

\subsection{Neutralino Decays}

\begin{equation} \label{neutqLRqNMSSM}
\Gamma(\tilde{Z}_i \rightarrow \tilde{q}_{L/R} \bar{q}) = {N_{c} g^2 \over 32 \pi |m_{\tilde{Z}_i}|} \tilde{\lambda}^{1 \over 2}(m_{\tilde{Z}_i},m_{\tilde{q}_{L/R}}, m_{q}) \mathcal{C_{L/R}}^2 (m_{\tilde{Z}_i}^2 - m_{\tilde{q}_{L/R}}^2 + m_{q}^2),
\end{equation}
where
\begin{align}
C_{L} &= \begin{cases}
		-\sqrt{2}[{2 \over 3} c(1) \sin\theta_W + ({1 \over 2} - {2 \over 3}\sin^2 \theta_W) {c(2) \over \cos\theta_W}] $, for $\tilde{u}_L$,$ \\
		\sqrt{2}[{1 \over 3} c(1) \sin\theta_W + ({1 \over 2} - {1 \over 3}\sin^2 \theta_W) {c(2) \over \cos\theta_W}] $, for $\tilde{d}_L$,$ \\
		\sqrt{2}[-c(1) \sin\theta_W + ({1 \over 2} - \sin^2 \theta_W) {c(2) \over \cos\theta_W}] $, for $\tilde{l}_L$,$ \\
 		{-c(2) \over \sqrt{2}} \cos\theta_W $, for $\tilde{\nu}_L$,$ \\
 		\end{cases} \\
C_{R} &= \begin{cases}
		-\sqrt{2}{2 \over 3} \sin\theta_W \Big[c(2) \tan\theta_W - c(1)\Big] $, for $\tilde{u}_R $,$\\
		\sqrt{2}{1 \over 3} \sin\theta_W \Big[c(2) \tan\theta_W - c(1)\Big] $, for $\tilde{d}_R $,$\\
		\sqrt{2} \sin\theta_W \Big[c(2) \tan\theta_W - c(1)\Big] $, for $\tilde{l}_R $,$\\
		0 $, for $\tilde{\nu}_R.\\
 		\end{cases}
\end{align}
Here $N_{c} = 3$ for quarks and $1$ for leptons and $c(1)$ and $c(2)$ are given by:
\begin{equation} \label{C1C2NMSSM}
c(1) = N_{1i}\cos\theta_W + N_{2i}\sin\theta_W, \quad \quad 
c(2) = -N_{1i} \sin\theta_W + N_{2i}\cos\theta_W.
\end{equation}
For the third generation, the generalisation is as expected but with extra Yukawa interactions:
\begin{equation} \label{neutsttNMSSM}
\Gamma(\tilde{Z}_i \rightarrow \tilde{t}_{1/2} \bar{t}) = {3 g^2 \over 32\pi |m_{\tilde{Z}_i}|} \tilde{\lambda}^{1 \over 2}(m_{\tilde{Z}_i},m_{\tilde{t}_{1/2}}, m_{t})\left[(c_{1}^2 + c_{2}^2)(m_{\tilde{Z}_i}^2 - m_{\tilde{t}_{1/2}}^2 + m_{t}^2) + 4m_{t}m_{\tilde{Z}_i}c_{1}c_{2}\right],
\end{equation}
where:
\begin{align}
c_{1} &= \begin{cases}
		\sqrt{2}\cos\theta_t[-{2 \over 3}c(1) \sin\theta_W + (-{1 \over 2} + {2 \over 3}\sin^2 \theta_W){c(2) \over \cos\theta_W}] - \sin\theta_t {f_{t} \over g} N_{4i} $, for $\tilde{t}_1$,$ \\
		-\sqrt{2}\sin\theta_t[-{2 \over 3}c(1) \sin\theta_W + (-{1 \over 2} + {2 \over 3}\sin^2 \theta_W){c(2) \over \cos\theta_W}] - \cos\theta_t {f_{t} \over g}N_{4i} $, for $\tilde{t}_2$,$ \\
		\end{cases} \\
c_{2} &= \begin{cases}
		-\sqrt{2}\sin\theta_t{2 \over 3}\sin\theta_W [c(2)\tan\theta_W - c(1)] - \cos\theta_t {f_{t} \over g} N_{4i} $, for $\tilde{t}_1$,$ \\
		-\sqrt{2}\cos\theta_t{2 \over 3}\sin\theta_W [c(2)\tan\theta_W - c(1)] + \sin\theta_t {f_{t} \over g} N_{4i} $, for $\tilde{t}_2$,$ \\
		\end{cases}
\end{align}
\begin{equation} \label{neutsbbNMSSM}
\Gamma(\tilde{Z}_i \rightarrow \tilde{b}_{1/2} \bar{b}) = {3 g^2 \over 32\pi |m_{\tilde{Z}_i}|} \tilde{\lambda}^{1 \over 2}(m_{\tilde{Z}_i},m_{\tilde{b}_{1/2}}, m_{b})\left[(c_{1}^2 + c_{2}^2)(m_{\tilde{Z}_i}^2 - m_{\tilde{b}_{1/2}}^2 + m_{b}^2) + 4m_{b}m_{\tilde{Z}_i}c_{1}c_{2}\right],
\end{equation}
where:
\begin{align}
c_{1} &= \begin{cases}
		\sqrt{2}\cos\theta_b [{1 \over 3}c(1) \sin\theta_W + ({1 \over 2} - {1 \over 3}\sin^2 \theta_W){c(2) \over \cos\theta_W}] - \sin\theta_b {f_{b} \over g} N_{3i} $, for $\tilde{b}_1$,$ \\
		-\sqrt{2}\sin\theta_b[{1 \over 3}c(1) \sin\theta_W + ({1 \over 2} - {1 \over 3}\sin^2 \theta_W){c(2) \over \cos\theta_W}] - \cos\theta_b {f_{b} \over g}N_{3i} $, for $\tilde{b}_2$,$ \\
		\end{cases} \\
c_{2} &= \begin{cases}
		\sqrt{2}\sin\theta_b{1 \over 3}\sin\theta_W [c(2)\tan\theta_W - c(1)] - \cos\theta_b {f_{b} \over g} N_{3i} $, for $\tilde{b}_1$,$ \\
		\sqrt{2}\cos\theta_b{1 \over 3}\sin\theta_W [c(2)\tan\theta_W - c(1)] + \sin\theta_b {f_{b} \over g} N_{3i} $, for $\tilde{b}_2. \\
		\end{cases}
\end{align}
\begin{equation} \label{neutstautauNMSSM}
\Gamma(\tilde{Z}_i \rightarrow \tilde{\tau}_{1/2} \bar{\tau}) = {g^2 \over 32\pi |m_{\tilde{Z}_i}|} \tilde{\lambda}^{1 \over 2}(m_{\tilde{Z}_i},m_{\tilde{\tau}_{1/2}}, m_{\tau})\left[(c_{1}^2 + c_{2}^2)(m_{\tilde{Z}_i}^2 - m_{\tilde{\tau}_{1/2}}^2 + m_{\tau}^2) + 4m_{\tau}m_{\tilde{Z}_i}c_{1}c_{2}\right],
\end{equation}
where:
\begin{align}
c_{1} &= \begin{cases}
		\sqrt{2}\cos\theta_{\tau}[c(1) \sin\theta_W + ({1 \over 2} - \sin^2 \theta_W){c(2) \over \cos\theta_W}] - \sin\theta_{\tau} {f_{\tau} \over g} N_{3i} $, for $\tilde{\tau}_1$,$ \\
		-\sqrt{2}\sin\theta_{\tau}[c(1) \sin\theta_W + ({1 \over 2} - \sin^2 \theta_W){c(2) \over \cos\theta_W}] - \cos\theta_{\tau} {f_{\tau} \over g}N_{3i} $, for $\tilde{\tau}_2$,$ \\
		\end{cases} \\
c_{2} &= \begin{cases}
		\sqrt{2}\sin\theta_{\tau}\sin\theta_W [c(2)\tan\theta_W - c(1)] - \cos\theta_{\tau} {f_{\tau} \over g} N_{3i} $, for $\tilde{\tau}_1$,$ \\
		\sqrt{2}\cos\theta_\tau \sin\theta_W [c(2)\tan\theta_W - c(1)] + \sin\theta_{\tau} {f_{\tau} \over g} N_{3i} $, for $\tilde{\tau}_2. \\
		\end{cases}
\end{align}
Neutralino decays to a chargino and $W$ boson:
\begin{equation} \label{neutWchNMSSM}
\begin{aligned}
\Gamma(\tilde{Z}_i \rightarrow W \tilde{W}_{1}) = {g^2 \over 32\pi |m_{\tilde{Z}_i}|} \tilde{\lambda}^{1/2}(m_{\tilde{Z}_i},m_{\tilde{W}_1},m_{W})\Big[& -12m_{\tilde{Z}_i} m_{\tilde{W}_j}c_{L}c_{R} + (c_{L}^2 + c_{R}^2)\{(m_{\tilde{W}_1}^2 + m_{\tilde{Z}_i}^2 - m_{W}^2) \\ & + (m_{\tilde{Z}_i}^2 + m_{W}^2 - m_{\tilde{W}_j}^2)(m_{\tilde{Z}_i}^2 - m_{W}^2 - m_{\tilde{W}_j}^2){1 \over m_{W}^2} \} \Big],
\end{aligned}
\end{equation}
where:
\begin{align}
c_{L} &= {-1 \over \sqrt{2}} N_{4i} \cos\theta_R + N_{2i}\sin\theta_R, \\  
c_{R} &= {1 \over \sqrt{2}} N_{3i} \cos\theta_L + N_{2i}\sin\theta_L.
\end{align}

For $\tilde{W}_2$ just take $m_{\tilde{W}_1} \rightarrow m_{\tilde{W}_2}$, $\cos\theta_R \rightarrow \sin\theta_R$,  $\cos\theta_L \rightarrow \sin\theta_L$, $\sin\theta_R \rightarrow -\cos\theta_R$ and $\sin\theta_L \rightarrow -\cos\theta_L$.

Neutralino decays to a chargino and charged Higgs boson:
\begin{equation} \label{neutHpmchNMSSM}
\Gamma(\tilde{Z}_i \rightarrow H^{\pm} \tilde{W}_{1}) = {1 \over 32\pi |m_{\tilde{Z}_i}|} \tilde{\lambda}^{1/2}(m_{\tilde{Z}_i},m_{\tilde{W}_1},m_{H^{\pm}})\Big[ (\mathcal{C}_{L}^2 + \mathcal{C}_{R}^2)\{(m_{\tilde{W}_1}^2 + m_{\tilde{Z}_i}^2 - m_{H^{\pm}}^2) + 4\mathcal{C}_{L} \mathcal{C}_{R} m_{\tilde{Z}_i} m_{\tilde{W}_1}\}\Big],
\end{equation}
where:
\begin{equation}
\mathcal{C}_{L} = \lambda\cos\beta N_{5i} \cos\theta_L  - {\sin\beta \over \sqrt{2}} [g' N_{1i} + gN_{2i}]\cos\theta_L + g\sin\beta N_{3i}\sin\theta_L,
\end{equation}
\begin{equation}
\mathcal{C}_{R} = \lambda\sin\beta N_{5i} \cos\theta_R  + {\cos\beta \over \sqrt{2}} [g' N_{1i} + gN_{2i}]\cos\theta_R + g\cos\beta N_{4i}\sin\theta_R.
\end{equation}
Again for $\tilde{W}_2$ just take $m_{\tilde{W}_1} \rightarrow m_{\tilde{W}_2}$, $\cos\theta_R \rightarrow \sin\theta_R$,  $\cos\theta_L \rightarrow \sin\theta_L$, $\sin\theta_R \rightarrow -\cos\theta_R$ and $\sin\theta_L \rightarrow -\cos\theta_L$.

Neutralino decays to a lighter neutralino and $Z$ boson:
\begin{equation} \label{neutZneutNMSSM}
\begin{aligned}
\Gamma(\tilde{Z}_i \rightarrow Z \tilde{Z}_j) = {g^2 \over 32 \pi |m_{\tilde{Z}_i}|} \tilde{\lambda}^{1 \over 2}(m_{\tilde{Z}_i},m_{\tilde{Z}_j},m_{Z})\Big[ & -12m_{\tilde{Z}_i} m_{\tilde{Z}_j} c_{LZ}c_{RZ} + (c_{LZ}^2 + c_{RZ}^2)\{(m_{\tilde{Z}_i}^2 + m_{\tilde{Z}_j}^2 - m_{Z}^2) \\ & + (m_{\tilde{Z}_i}^2 - m_{\tilde{Z}_j}^2 + m_{Z}^2)(m_{\tilde{Z}_i}^2 - m_{\tilde{Z}_j}^2 - m_{Z}^2){1 \over m_{Z}^2}\}\Big],
\end{aligned}
\end{equation}
here we have:
\begin{equation}
c_{LZ} = -c_{RZ} = {1 \over 2 \cos\theta_W} [N_{3i}N_{3j} - N_{4i}N_{4j}].
\end{equation}
Neutralino decays to a lighter neutralino and CP even neutral Higgs boson:
\begin{equation} \label{neuthneutNMSSM}
\Gamma(\tilde{Z}_i \rightarrow h_{k} \tilde{Z}_j) = {1 \over 4 \pi |m_{\tilde{Z}_i}|} \tilde{\lambda}^{1 \over 2}(m_{\tilde{Z}_i},m_{h_k}, m_{\tilde{Z}_j}) \mathcal{C}_{\tilde{Z}_i \tilde{Z}_j h_{k}}^2\left[m_{\tilde{Z}_i}^2 + m_{\tilde{Z}_j}^2 - m_{h_k}^2 + 2m_{\tilde{Z}_i}m_{\tilde{Z}_j}\right],
\end{equation}
where
\begin{equation} \label{neutneuthNMSSM}
\begin{aligned}
\mathcal{C}_{\tilde{Z}_i \tilde{Z}_j h_{k}} = & {\lambda \over 2 \sqrt{2}}\Big[S(k,1)(N_{3i}N_{5j} + N_{3j}N_{5i}) + S(k,2)(N_{4i}N_{5j} + N_{4j}N_{5i}) \\ &  + S(k,3)(N_{3i}N_{4j} + N_{4i}N_{3j})\Big] - \sqrt{2} \kappa S(k,3) N_{5i}N_{5j} \\ &  + {g' \over 2}\Big[-S(k,1)(N_{1i}N_{4j} + N_{1j}N_{4i}) + S(k,2)(N_{1i}N_{3j} + N_{1j}N_{3i})\Big] \\ &  + {g \over 2}\Big[S(k,1)(N_{2i}N_{4j} + N_{2j}N_{4i}) - S(k,2)(N_{2i}N_{3j} + N_{2j}N_{3i})\Big].
\end{aligned}
\end{equation}
Neutralino decays to a lighter neutralino and CP odd neutral Higgs boson:
\begin{equation} \label{neutAneutNMSSM}
\Gamma(\tilde{Z}_i \rightarrow A_{k} \tilde{Z}_j) = {1 \over 4 \pi |m_{\tilde{Z}_i}|} \tilde{\lambda}^{1 \over 2}(m_{\tilde{Z}_i},m_{A_k}, m_{\tilde{Z}_j})[\mathcal{G}_{\tilde{Z} \tilde{Z} A_k}^2][m_{\tilde{Z}_i}^2 + m_{\tilde{Z}_j}^2 - m_{A_k}^2 - 2m_{\tilde{Z}_i}m_{\tilde{Z}_j}],
\end{equation}
where
\begin{equation}
\begin{aligned}
\mathcal{G}_{\tilde{Z}_i \tilde{Z}_j A_k} = & {-\lambda \over 2 \sqrt{2}}\Big[P(k,1)(N_{3i}N_{5j} + N_{3j}N_{5i}) + P(k,2)(N_{4i}N_{5j} + N_{4j}N_{5i}) \\ &  + P(k,3)(N_{3i}N_{4j} + N_{4i}N_{3j})\Big] - \sqrt{2} \kappa P(k,3) N_{5i}N_{5j} \\ &  - {g' \over 2}\Big[-P(k,1)(N_{1i}N_{4j} + N_{1j}N_{4i}) + P(k,2)(N_{1i}N_{3j} + N_{1j}N_{3i})\Big] \\ &  - {g \over 2}\Big[P(k,1)(N_{2i}N_{4j} + N_{2j}N_{4i}) - P(k,2)(N_{2i}N_{3j} + N_{2j}N_{3i})\Big].
\end{aligned}
\end{equation}
Note the $\mathcal{C}$ and $\mathcal{G}$ couplings here are similar to those
given for the reverse decays of Higgs bosons to neutralinos earlier, with
$i$, $j$, $k$ permuted accordingly. 

\subsection{Decays into Neutralinos}
For the first two generations of quarks and squarks:
\begin{equation} \label{qLRqneutNMSSM}
\Gamma(\tilde{q}_{L/R} \rightarrow q \tilde{Z}_i) = {g^2 \over 16\pi m_{\tilde{q}_{L/R}}} B_{\tilde{q}_{L/R}}^2 \tilde{\lambda}^{1 \over 2}(m_{\tilde{q}_{L/R}},m_{\tilde{Z}_i},m_q) [m_{\tilde{q}_{L/R}}^2 - m_{\tilde{Z}_i}^2 - m_{q}^2],
\end{equation}
where:
\begin{equation}
B_{\tilde{q}_{L/R}} = \begin{cases}
	-\sqrt{2}[{2 \over 3}c(1)\sin\theta_W + ({1 \over 2} - {2 \over 3}\sin^2 \theta_W){c(2) \over \cos\theta_W}] $, for $\tilde{u}_L$ type, $ \\
	-\sqrt{2} {2 \over 3} \sin\theta_W [c(2) \tan\theta_W - c(1)] $, for $\tilde{u}_R$ type, $ \\
	\sqrt{2}[c(1){1 \over 3}\sin\theta_W + ({1 \over 2} - {1 \over 3}\sin^2 \theta_W){c(2) \over \cos\theta_W}] $, for $\tilde{d}_L$ type, $ \\
	\sqrt{2} {1 \over 3} \sin\theta_W [c(2)\tan\theta_W - c(1)] $, for $\tilde{d}_R$ type$.\\
	\end{cases}
\end{equation}
The decay formulae for decays of the first two generations of sleptons to leptons and neutralinos are the same as for the squarks here but with the coupling change $B_{\tilde{q}_{L/R}} \rightarrow B_{\tilde{l}_{L/R}}$ and squark masses exchanged for slepton masses and quark masses for lepton masses:
\begin{equation}
B_{\tilde{l}_{L/R}} = \begin{cases}
	-{c(2) \over \sqrt{2} \cos\theta_W} $, for $\tilde{\nu}_L$ type, $ \\
	0 $, for $\tilde{\nu}_R$ type (as no RH sneutrinos), $ \\
	\sqrt{2}[c(1)\sin\theta_W + ({1 \over 2} - \sin^2 \theta_W){c(2) \over \cos\theta_W}] $, for $\tilde{l}_L$ type, $ \\
	\sqrt{2} \sin\theta_W [c(2)\tan\theta_W - c(1)] $, for $\tilde{l}_R$ type. $\\
	\end{cases}
\end{equation}
For the third generation:
\begin{equation} \label{sttneutNMSSM}
\Gamma(\tilde{t}_{1/2} \rightarrow t \tilde{Z}_i) = {g^2 \over 16 \pi m_{\tilde{t}_{1/2}}}\left[(d_{1}^2 + d_{2}^2)(m_{\tilde{t}_{1/2}}^2 - m_{t}^2 - m_{\tilde{Z}_i}^2) - 4d_{1}d_{2}m_{t}m_{\tilde{Z}_i}\right]\tilde{\lambda}^{1 \over 2}(m_{\tilde{t}_{1/2}},m_{\tilde{Z}_i},m_{t}),
\end{equation}
where
\begin{align}
d_{1} &= \begin{cases} 
		\cos\theta_t \sqrt{2} [-{2 \over 3}c(1) \sin\theta_W + (-{1 \over 2} + {2 \over 3}\sin^2 \theta_W){c(2) \over \cos\theta_W}] - \sin\theta_t {f_t \over g} N_{4i} $, for $\tilde{t}_1$,$ \\
		-\sin\theta_t \sqrt{2} [-{2 \over 3}c(1) \sin\theta_W + (-{1 \over 2} + {2 \over 3}\sin^2 \theta_W){c(2) \over \cos\theta_W}] - \cos\theta_t {f_t \over g} N_{4i} $, for $\tilde{t}_2$,$ \\
		\end{cases} \\
d_{2} &= \begin{cases}
		-{2 \over 3} \sin\theta_t \sqrt{2} \sin\theta_W [c(2)\tan\theta_W - c(1)] - \cos\theta_t {f_t \over g}N_{4i} $, for $\tilde{t}_1 $,$\\
		-{2 \over 3} \cos\theta_t \sqrt{2} \sin\theta_W [c(2)\tan\theta_W - c(1)] + \sin\theta_t {f_t \over g}N_{4i}  $, for $\tilde{t}_2.\\
		\end{cases}
\end{align}
\begin{equation} \label{sbbneutNMSSM}
\Gamma(\tilde{b}_{1/2} \rightarrow b \tilde{Z}_i) = {g^2 \over 16 \pi m_{\tilde{b}_{1/2}}}\left[(f_{1}^2 + f_{2}^2)(m_{\tilde{b}_{1/2}}^2 - m_{b}^2 - m_{\tilde{Z}_i}^2) - 4f_{1}f_{2}m_{b}m_{\tilde{Z}_i}\right]\tilde{\lambda}^{1 \over 2}(m_{\tilde{b}_{1/2}},m_{\tilde{Z}_i},m_{b}),
\end{equation}
where
\begin{align}
f_{1} &= \begin{cases}
		\cos\theta_b \sqrt{2} [{1 \over 3}c(1) \sin\theta_W + ({1 \over 2} - {1 \over 3}\sin^2 \theta_W){c(2) \over \cos\theta_W}] - \sin\theta_b {f_b \over g} N_{3i} $, for $\tilde{b}_1$,$ \\
		-\sin\theta_b \sqrt{2} [{1 \over 3}c(1) \sin\theta_W + ({1 \over 2} - {1 \over 3}\sin^2 \theta_W){c(2) \over \cos\theta_W}] - \cos\theta_b {f_b \over g} N_{3i} $, for $\tilde{b}_2$,$ \\
		\end{cases} \\
f_{2} &= \begin{cases}
		{1 \over 3} \sin\theta_b \sqrt{2} \sin\theta_W [c(2)\tan\theta_W - c(1)] - \cos\theta_b {f_b \over g}N_{3i} $, for $\tilde{b}_1$,$ \\
		{1 \over 3} \cos\theta_b \sqrt{2} \sin\theta_W [c(2)\tan\theta_W - c(1)] + \sin\theta_b {f_b \over g}N_{3i}  $, for $\tilde{b}_2. \\
		\end{cases}
\end{align}
\begin{equation} \label{stautauneutNMSSM}
\Gamma(\tilde{\tau}_{1/2} \rightarrow \tau \tilde{Z}_i) = {g^2 \over 16 \pi m_{\tilde{\tau}_{1/2}}}\left[(g_{1}^2 + g_{2}^2)(m_{\tilde{\tau}_{1/2}}^2 - m_{\tau}^2 - m_{\tilde{Z}_i}^2) - 4g_{1}g_{2}m_{\tau}m_{\tilde{Z}_i}\right]\tilde{\lambda}^{1 \over 2}(m_{\tilde{\tau}_{1/2}},m_{\tilde{Z}_i},m_{\tau}),
\end{equation}
where
\begin{align}
g_{1} &= \begin{cases}
		\cos\theta_{\tau} \sqrt{2} [c(1) \sin\theta_W + ({1 \over 2} - \sin^2 \theta_W){c(2) \over \cos\theta_W}] - \sin\theta_{\tau} {f_{\tau} \over g} N_{3i} $, for $\tilde{\tau}_1$,$ \\
		-\sin\theta_{\tau} \sqrt{2} [c(1) \sin\theta_W + ({1 \over 2} - \sin^2 \theta_W){c(2) \over \cos\theta_W}] - \cos\theta_{\tau} {f_{\tau} \over g} N_{3i} $, for $\tilde{\tau}_2$,$ \\
		\end{cases} \\
g_{2} &= \begin{cases}
		\sin\theta_{\tau} \sqrt{2} \sin\theta_W [c(2)\tan\theta_W - c(1)] - \cos\theta_{\tau} {f_{\tau} \over g}N_{3i} $, for $\tilde{\tau}_1$,$ \\
		\cos\theta_{\tau} \sqrt{2} \sin\theta_W [c(2)\tan\theta_W - c(1)] + \sin\theta_{\tau} {f_{\tau} \over g}N_{3i}  $, for $\tilde{\tau}_2. \\
		\end{cases}
\end{align}
Remember the $c(1)$ and $c(2)$ were given previously in \eqref{C1C2NMSSM}.

Sneutrino decays into neutralinos are given by:
\begin{equation} \label{snutuanutauneutNMSSM}
\Gamma(\tilde{\nu}_{\tau_{1/2}} \rightarrow \nu_{\tau} \tilde{Z}_i) = {g^2 \over 16 \pi m_{\tilde{\nu_{\tau_{1/2}}}}} \tilde{\lambda}^{1 \over 2}(m_{\tilde{\nu_{\tau_{1/2}}}},0, m_{\tilde{Z}_i})(m_{\tilde{\nu_{\tau_{1/2}}}}^2 - m_{\tilde{Z}_i}^2)\Big({c(2) \over \sqrt{2} \cos\theta_W}\Big)^2.
\end{equation}
For chargino decays into neutralinos and charged Higgs bosons the partial width is given by:
\begin{equation} \label{chHpmneutNMSSM}
\begin{aligned}
\Gamma(\tilde{W}_1 \rightarrow H^{\pm} \tilde{Z}_j) = & {g^2 \over 32\pi |m_{\tilde{W}_1}|} \tilde{\lambda}^{1 \over 2}(m_{\tilde{W}_1},m_{\tilde{Z}_j},m_{H^{\pm}})\Big[({c1}_{\tilde{W}_1 H^{\pm} \tilde{Z}_j}^2 + {c2}_{\tilde{W}_1 H^{\pm} \tilde{Z}_j}^2)(m_{\tilde{Z}_j}^2 + m_{\tilde{W}_1}^2 - m_{H^{\pm}}^2) \\ & + 4{c1}_{\tilde{W}_1 H^{\pm} \tilde{Z}_j}^2 {c2}_{\tilde{W}_1 H^{\pm} \tilde{Z}_j}^2 m_{\tilde{Z}_j} m_{\tilde{W}_1}\Big],
\end{aligned}
\end{equation}
where:
\begin{equation}
{c1}_{\tilde{W}_1 H^{\pm} \tilde{Z}_j} = {1 \over g} [\lambda \sin\beta N_{5j} \cos\theta_R + {\cos\beta \over \sqrt{2}}(g' N_{1j} + gN_{2j})\cos\theta_R + g\cos\beta N_{4j} \sin\theta_R],
\end{equation}
\begin{equation}
{c2}_{\tilde{W}_1 H^{\pm} \tilde{Z}_j} = {1 \over g}[\lambda \cos\beta N_{5j} \cos\theta_L - {\sin\beta \over \sqrt{2}}(g' N_{1j} + gN_{2j})\cos\theta_L + g\sin\beta N_{3j}\sin\theta_L].
\end{equation}
For $\tilde{W}_2$ the formulae are the same, just make the replacements $m_{\tilde{W}_1} \rightarrow m_{\tilde{W}_2}$ , $\cos\theta_{L/R} \rightarrow \sin\theta_{L/R}$ and $\sin\theta_{L/R} \rightarrow -\cos\theta_{L/R}$.
\begin{equation} \label{chWneutNMSSM}
\begin{aligned}
\Gamma(\tilde{W}_1 \rightarrow W \tilde{Z}_j) = {g^2 \over 32\pi |m_{\tilde{W}_1}|} \tilde{\lambda}^{1 \over 2}(m_{\tilde{W}_1},m_{\tilde{Z}_j},m_{W}) &\Big[-12 m_{\tilde{W}_1} m_{\tilde{Z}_j} {cL}_{\tilde{W}_1 W \tilde{Z}_j} {cR}_{\tilde{W}_1 W \tilde{Z}_j} \\ & + ({cL}_{\tilde{W}_1 W \tilde{Z}_j}^2 + {cR}_{\tilde{W}_1 W \tilde{Z}_j}^2)	 \{(m_{\tilde{W}_1}^2 + m_{\tilde{Z}_j}^2 - m_{W}^2) \\ & + (m_{\tilde{W}_1}^2 + m_{W}^2 - m_{\tilde{Z}_j}^2)(m_{\tilde{W}_1}^2 - m_{\tilde{Z}_j}^2 - m_{W}^2){1 \over m_{W}^2}\}\Big],
\end{aligned}
\end{equation}
where:
\begin{align}
{cL}_{\tilde{W}_1 W \tilde{Z}_j} &= -{1 \over \sqrt{2}} N_{4j}\cos\theta_R + N_{2j}\sin\theta_R, \\
{cR}_{\tilde{W}_1 W \tilde{Z}_j} &= {1 \over \sqrt{2}} N_{3j}\cos\theta_L + N_{2j}\sin\theta_L.
\end{align}
Again for $\tilde{W}_2$ the formulae are the same, just make the replacements $m_{\tilde{W}_1} \rightarrow m_{\tilde{W}_2}$, $\cos\theta_{L/R} \rightarrow \sin\theta_{L/R}$ and $\sin\theta_{L/R} \rightarrow -\cos\theta_{L/R}$.

\section{QCD Corrections to Decays} \label{appendix:QCDcorrdec}
Note, for the decays of neutral Higgs bosons to quarks or gluons, the possibility of including QCD corrections is included in the program, by default the QCD corrections are on. The formulae are those provided in {\tt HDECAY-3.4} in {\tt SUSYHIT} \cite{Djouadi:2006bz,Djouadi:1997yw} and {\tt NMSSMTools-4.2.1} in {\tt NMHDECAY} \cite{Ellwanger:2004xm,Ellwanger:2006ch}. With QCD corrections incorporated our formulae become as follows:
\begin{equation} \label{hqqQCDcorr}
\begin{aligned}
\Gamma(h \rightarrow qq)_{QCDcorr}= \Gamma(h-> qq)_{tree} \left(1 + {4 \alpha_s(m_h) \over 3 \pi} [{A(\tilde{\beta}) \over \tilde{\beta}} + {3 + 34 \tilde{\beta}^2 - 13 \tilde{\beta}^4 \over 16 \tilde{\beta}^3} \log{1+\tilde{\beta} \over 1 - \tilde{\beta}} + {3 \over 8 \tilde{\beta}^2} (7 \tilde{\beta}^2 - 1)]\right)
\end{aligned}.
\end{equation}
This formula applies for all the CP even neutral Higgs bosons, whether in the MSSM or NMSSM, the difference between the MSSM and NMSSM comes in the tree-level formula. Note $\alpha_s$ is evaluated at the mass of the decaying Higgs boson. Also note that $\tilde{\beta}$ and $A(\tilde{\beta})$ are given by:
\begin{equation}
\tilde{\beta} = \sqrt{1 - 4{m_{q}^2 \over m_{h}^2}},
\end{equation}
\begin{equation}
\begin{aligned}
A(\tilde{\beta}) = & (1+\tilde{\beta}^2)\Big[4Li_2 ({1-\tilde{\beta} \over 1+\tilde{\beta}}) + 2Li_2 ({\tilde{\beta}-1  \over \tilde{\beta}+1}) - 3\log({1+\tilde{\beta} \over 1-\tilde{\beta}})\log{2\over 1+\tilde{\beta}} - 2\log({1+\tilde{\beta} \over 1-\tilde{\beta}})\log\tilde{\beta}\Big] \\ & - 3\tilde{\beta} \log{4 \over 1-\tilde{\beta}^2} - 4\tilde{\beta} \log\tilde{\beta}.
\end{aligned}
\end{equation}
This is exactly as given in Eqs. (16) and (25) of \cite{Spira:2016}. $Li_2$ is
the di-logarithm function (Spence's function). 

For the CP odd Higgs bosons we have:
\begin{equation} \label{AqqQCDcorr}
\begin{aligned}
\Gamma(A \rightarrow qq)_{QCDcorr} = \Gamma(A-> qq)_{tree} \left(1 + {4 \alpha_s(m_A) \over 3 \pi} [{A(\tilde{\beta}) \over \tilde{\beta}} + {19 + 2 \tilde{\beta}^2 + 3 \tilde{\beta}^4 \over 16 \tilde{\beta}} \log{1+\tilde{\beta} \over 1 - \tilde{\beta}} + {3 \over 8} (7 - \tilde{\beta}^2 )]\right).
\end{aligned}
\end{equation}
$\tilde{\beta}$ and $A(\tilde{\beta})$ are as given above but with the change $m_h \rightarrow m_A$ as appropriate.
This formula is as given in Eqs.~(25) and (26) of \cite{Spira:2016}. It should be noted that when QCD corrections are applied one should use the pole quark masses (as we do within {\tt SOFTSUSY} here), rather than the running masses, as otherwise the formulae double count $O(\alpha_s)$ effects \cite{Drees:1990}. 

The QCD corrections for $h \rightarrow gg$ are more complicated as they involve both standard QCD corrections due to gluons being radiated, gluons in the loop, tops, bottoms and other quarks, and additional SUSY-QCD corrections due to gluinos, stops, sbottoms and other squarks. This complicates matters as whilst the usual ``fermionic" QCD (FQCD) corrections  apply to all particles in the loop, the SUSY-QCD (SQCD) corrections only apply to the scalar squark contributions, therefore rather than multiply the whole width by a correction factor (as was the case for $h \rightarrow qq$) we must now correct the SM and SUSY loop contributions separately.
The usual MSSM equation for $h \rightarrow gg$ with no corrections is:
\begin{equation} \label{phiggnorm}
\Gamma(\phi \rightarrow gg)_{1-loop} = {\alpha_{s}^2 (m_\phi) \over 128 \pi^3} {G_{F} \over \sqrt{2}} m_{\phi}^3{9 \over 8}|\Sigma I_{loop}^\phi|^2.
\end{equation}
Here the $\alpha_s$ is run to the mass of the decaying Higgs boson.
The $I_{loop}$ can be split into $I_{quark}$ and $I_{squark}$ loop contributions.
So $I_{looptot}^\phi = I_{qtot}^\phi + I_{sqtot}^\phi$, where $I_{qtot}^\phi = I_t^\phi + I_b^\phi + I_c^\phi$ and $I_{sqtot}^\phi = I_{\tilde{t}_1}^\phi + I_{\tilde{t}_2}^\phi + I_{\tilde{b}_1}^\phi + I_{\tilde{b}_2}^\phi + I_{\tilde{c}_L}^\phi + I_{\tilde{c}_R}^\phi + I_{\tilde{s}_L}^\phi + I_{\tilde{s}_R}^\phi + I_{\tilde{u}_L}^\phi + I_{\tilde{u}_R}^\phi + I_{\tilde{d}_L}^\phi + I_{\tilde{d}_R}^\phi$.

To account for the usual QCD corrections, i.e. ``FQCD" corrections, as these affect all the loop contributions, the whole partial width is multiplied by $\delta_{FQCD}$:
\begin{equation}
\delta_{FQCD}^{CP even Higgs} = 1 + {\alpha_s(m_\phi) \over \pi}({95 \over 4}-{7 \over 6}N_{f}),
\end{equation}
or
\begin{equation}
\delta_{FQCD}^{CP odd Higgs} = 1 + {\alpha_s(m_\phi) \over \pi}({97 \over 4}-{7 \over 6}N_{f}).
\end{equation}
$N_{f}$ is the number of active fermion flavours.
The SUSY QCD corrections, i.e. ``SQCD" corrections, apply only to the squark loop contributions. Therefore to incorporate these in the final partial width you must multiply both the squark loop squared contributions and the interference terms of the squark loops with the quark loops by the correction factor. Therefore the $|I_{loop}|^2$ (which comes from the matrix element squared) above with both FQCD and SQCD corrections included becomes:
\begin{equation}
|I_{looptot}^\phi|^2 = \delta_{FQCD}^\phi|I_{looptot}^\phi|^2 + Re[({I_{looptot}^\phi})^* I_{sqtot}^\phi]\delta_{SQCD}.
\end{equation}
To be clear the $({I_{looptot}^\phi})^*$ here means the complex conjugate of ${I_{looptot}^\phi}$, given this is an interference term.
The $\delta_{SQCD}$ correction factor is the same for CP even and CP odd neutral Higgs bosons and is given by \eqref{SQCD} below, note that $\alpha_s$ is run to the mass of the decaying Higgs boson:
\begin{equation} \label{SQCD}
\delta_{SQCD} = {17 \alpha_s(m_\phi) \over 6 \pi}.
\end{equation}
Consequently in the MSSM the overall formula for the QCD and SUSY-QCD corrected $h \rightarrow gg$ decay (at 2-loop) is:
\begin{equation} \label{phiggQCDcorr}
\Gamma(\phi \rightarrow gg)_{1-loop + QCDcorr} = {\alpha_{s}^2 (m_\phi) \over 128 \pi^3}{G_{F} \over \sqrt{2}} m_{\phi}^3{9 \over 8}\left[\delta_{FQCD}^\phi|I_{looptot}^\phi|^2 + Re[({I_{looptot}^\phi})^* I_{sqtot}^\phi]\delta_{SQCD}\right].
\end{equation}
$\phi$ is a CP even neutral Higgs here as CP odd Higgs bosons do not have squark loop contributions because of CP invariance of the decays.
For the CP odd Higgs $A$ in the MSSM we therefore only have the quark loops and FQCD corrections:
\begin{equation} \label{AggQCDcorr}
\Gamma(A \rightarrow gg)_{1-loop + QCDcorr} = {\alpha_{s}^2 (m_A) \over 128 \pi^3}{G_{F} \over \sqrt{2}} m_{A}^3{9 \over 8}\left[\delta_{FQCD}^A|I_{qtot}^A|^2\right].
\end{equation}
The FQCD corrections for CP even and CP odd neutral Higgs bosons in the MSSM
are as given in Ref.~\cite{Djouadi:1996} in Eqs.~(7) and (15) but with the log terms dropped as the scale of alphas is run to the masses of the decaying Higgs.
As the corrections are purely coloured and the NMSSM only alters the Higgs and neutralino sectors, the form of the QCD corrections is exactly the same in the NMSSM\@. The alterations to the Higgs sector in the NMSSM however result in the couplings of the neutral Higgs bosons to other particles, and therefore the leading order (i.e.\ 1-loop) formula for the loop contributions to $h \rightarrow gg$, being altered. The formula in the uncorrected NMSSM (i.e.\ at 1-loop) is (as detailed previously) as follows, with the $\alpha_s$ evaluated at the scale of the decaying Higgs boson:
\begin{equation} \label{hggNMSSMnorm}
\Gamma(h_{i} \rightarrow gg)_{1-loop} = {G_{F} m_{h_i}^3 \alpha_{s}^2 (m_{h_i}) \over 64 \sqrt{2} \pi^3} |M_{gg}|^2,
\end{equation}
where
\begin{equation}
\begin{aligned}
|M_{gg}^\phi|^2 = & [J_{t}^{r} + J_{b}^{r} + J_{c}^{r} + J_{\tilde{u}_L}^{r} + J_{\tilde{u}_R}^{r} + J_{\tilde{d}_L}^{r} + J_{\tilde{d}_R}^{r} + J_{\tilde{c}_L}^{r} + J_{\tilde{c}_R}^{r} + J_{\tilde{s}_L}^{r} + J_{\tilde{s}_R}^{r} + J_{\tilde{t}_1}^{r} + J_{\tilde{t}_2}^{r} + J_{\tilde{b}_1}^{r} + J_{\tilde{b}_2}^{r}]^2 \\ & + [J_{t}^{i} + J_{b}^{i} + J_{c}^{i} + J_{\tilde{u}_L}^{i} + J_{\tilde{u}_R}^{i} + J_{\tilde{d}_L}^{i} + J_{\tilde{d}_R}^{i} + J_{\tilde{c}_L}^{i} + J_{\tilde{c}_R}^{i} + J_{\tilde{s}_L}^{i} + J_{\tilde{s}_R}^{i} + J_{\tilde{t}_1}^{i} + J_{\tilde{t}_2}^{i} + J_{\tilde{b}_1}^{i} + J_{\tilde{b}_2}^{i}]^2,
\end{aligned}
\end{equation}
where the $J_{X}$ contributions are different to those in the MSSM as the
couplings are different. Here the $^{r}$ and $^{i}$ were used as shorthand for
real and imaginary parts. The $|M_{gg}|^2$ is therefore just the mod square of
the sum of the complex loop contributions. 

In order to incorporate the FQCD and SQCD corrections we again group the loop contributions into quark and squark so that $J_{qtot}^\phi = J_{t}^\phi + J_{b}^\phi + J_{c}^\phi$, $J_{sqtot}^\phi = J_{\tilde{u}_L}^\phi + J_{\tilde{u}_R}^\phi + J_{\tilde{d}_L}^\phi + J_{\tilde{d}_R}^\phi + J_{\tilde{c}_L}^\phi + J_{\tilde{c}_R}^\phi + J_{\tilde{s}_L}^\phi + J_{\tilde{s}_R}^\phi + J_{\tilde{t}_1}^\phi + J_{\tilde{t}_2}^\phi + J_{\tilde{b}_1}^\phi + J_{\tilde{b}_2}^\phi$ and $J_{looptot}^\phi = J_{qtot}^\phi + J_{sqtot}^\phi$.
Then $|M_{gg}|^2$ becomes:
\begin{equation}
|M_{gg}^\phi|^2 = \left[\delta_{FQCD}^\phi|J_{looptot}^\phi|^2 + Re[({J_{looptot}^\phi})^* J_{sqtot}^\phi]\delta_{SQCD}\right],
\end{equation}
because the FQCD and SQCD corrections apply to the loop contributions exactly as in the MSSM, however the loop contributions themselves have changed between the MSSM and NMSSM\@.
So overall in the NMSSM, the QCD corrected partial width for neutral Higgs decays to gluons is as follows, again note the $\alpha_s$ is evaluated at the scale of the mass of the decaying Higgs boson:
\begin{equation} \label{hggQCDcorrNMSSM}
\Gamma(h_{i} \rightarrow gg)_{1-loop + QCDcorr} = {G_{F} m_{h_i}^3 \alpha_{s}^2 (m_{h_i}) \over 64 \sqrt{2} \pi^3} \left[\delta_{FQCD}^\phi|J_{looptot}^\phi|^2 + Re[({J_{looptot}^\phi})^* J_{sqtot}^\phi]\delta_{SQCD}\right].
\end{equation}	
Again $\phi$ here is a CP even neutral Higgs boson as CP odd Higgs bosons do not have squark loop contributions because of CP invariance of the decays, as in the MSSM\@. Therefore CP odd Higgs bosons have only quark loop contributions and so only receive FQCD corrections, without corrections the formula was:
\begin{equation} \label{AggnormNMSSM}
\Gamma(A_i \rightarrow g g)_{1-loop} = {g^2 \alpha_s^2 (m_{A_i}) m_{A_i}^3 \over 128 \pi^3 m_{W}^2} |M_{A_i gg}|^2,
\end{equation}
remembering that the $\alpha_s (m_{A_i})$ means $\alpha_s$ evaluated at the mass of the decaying CP odd Higgs boson $A_i$. Here $|M_{A_i gg}|^2$ is:
\begin{equation}
|M_{A_i gg}|^2 = (\mathcal{R}_{t}^{r} + \mathcal{R}_{b}^{r} + \mathcal{R}_{c}^{r})^2 + (\mathcal{R}_{t}^{i} + \mathcal{R}_{b}^{i} + \mathcal{R}_{c}^{i})^2.
\end{equation}
The corrections are incorporated by multiplying by $\delta_{FQCD}^A$, so with the QCD corrections the CP odd Higgs decays in the NMSSM are given by:
\begin{equation} \label{AggQCDcorrNMSSM}
\Gamma(A_i \rightarrow g g)_{1-loop + QCDcorr} = {g^2 \alpha_s^2 (m_{A_i}) m_{A_i}^3 \over 128 \pi^3 m_{W}^2} |M_{A_i gg}|^2 \delta_{FQCD}^A.
\end{equation}
Throughout, the formulae used are those given in {\tt HDECAY-3.4} \cite{Djouadi:1997yw} (and hence {\tt SUSYHIT} \cite{Djouadi:2006bz}) and {\tt NMSSMTools-4.2.1}~\citep{Allanach:2001kg,Ellwanger:2012dd,Ellwanger:2006ch}.

\bibliography{decays.bib}
\bibliographystyle{elsarticle-num}
\end{document}